\documentclass[sn-basic]{sn-jnl}

\usepackage{graphicx}%
\usepackage{multirow}%
\usepackage{amsmath,amssymb,amsfonts}%
\usepackage{amsthm}%
\usepackage{mathrsfs}%
\usepackage[title]{appendix}%
\usepackage{xcolor}%
\usepackage{textcomp}%
\usepackage{manyfoot}%
\usepackage{booktabs}%
\usepackage{algorithm}%
\usepackage{algorithmicx}%
\usepackage{algpseudocode}%
\usepackage{listings}%

\chardef\us=`\_

\begin{document}

\title[Coronal dimmings and what they tell us about solar and stellar coronal mass ejections]{Coronal dimmings and what they tell us about solar and stellar coronal mass ejections}


\author*[1,2]{\fnm{Astrid M.} \sur{Veronig}}\email{astrid.veronig@uni-graz.at}
\equalcont{These authors contributed equally to this work.}

\author*[3]{\fnm{Karin} \sur{Dissauer}}\email{dissauer@nwra.com}
\equalcont{These authors contributed equally to this work.}

\author[4]{\fnm{Bernhard} \sur{Kliem}}

\author[5]{\fnm{Cooper} \sur{Downs}}

\author[6,7]{\fnm{Hugh S.} \sur{Hudson}}

\author[8]{\fnm{Meng} \sur{Jin}}

\author[9,10]{\fnm{Rachel} \sur{Osten}}

\author[11]{\fnm{Tatiana} \sur{Podladchikova}}

\author[12,13]{\fnm{Avijeet} \sur{Prasad}}

\author[14]{\fnm{Jiong} \sur{Qiu}}

\author[15]{\fnm{Barbara} \sur{Thompson}}

\author[16]{\fnm{Hui} \sur{Tian}}

\author[17]{\fnm{Angelos} \sur{Vourlidas}}

\affil[1]{\orgdiv{Institute of Physics}, \orgname{University of Graz}, \orgaddress{\street{Universit\"atsplatz 5}, \city{Graz}, \postcode{8010}, 
\country{Austria}}}

\affil[2]{\orgdiv{Kanzelh\"ohe Observatory of Solar and Environmental Research}, \orgname{University of Graz}, \orgaddress{\street{Kanzelh\"ohe 19}, \city{Treffen}, \postcode{9521},  
\country{Austria}}}

\affil[3]{\orgdiv{North West Research Associates}, 
\orgaddress{\street{3380 Mitchell Lane}, \city{Boulder}, \postcode{80301}, 
\country{CO, USA}}}

\affil[4]{\orgdiv{Institute of Physics and Astronomy}, \orgname{University of Potsdam}, \orgaddress{\street{Karl-Liebknecht-Str. 24-25}, \city{Potsdam}, \postcode{14476}, 
\country{Germany}}}

\affil[5]{\orgdiv{Predictive Science Inc.}, 
\orgaddress{\street{Mesa Rim Rd., Suite 170}, \city{San Diego}, \postcode{92121}, \country{CA, USA}}}

\affil[6]{\orgdiv{School of Physics and Astronomy}, \orgname{University of Glasgow}, \orgaddress{\street{Kelvin Building}, \city{Glasgow}, \postcode{G12 8QQ}, 
\country{UK}}}

\affil[7]{\orgdiv{Space Sciences Laboratory}, \orgname{U.C. Berkeley}, \orgaddress{\street{7 Gauss Way}, \city{Berkeley}, \postcode{94720}, 
\country{CA, USA}}}

\affil[8]{\orgdiv{Lockheed Martin Solar and Astrophysics Lab}, 
\orgaddress{\street{Street}, \city{Palo Alto}, \postcode{94304}, 
\country{CA, USA}}}

\affil[9]{\orgdiv{Space Telescope Science Institute}, 
\orgaddress{\street{3700 San Martin Drive}, \city{Baltimore}, \postcode{21218}, 
\country{MD, USA}}}

\affil[10]{\orgdiv{Center for Astrophysical Sciences}, \orgname{Johns Hopkins University}, 
\orgaddress{\street{Street}, \city{Baltimore}, \postcode{21218}, 
\country{MD, USA}}}

\affil[11]{\orgdiv{Skolkovo Institute of Science and Technology}, 
\orgaddress{\street{Bolshoy Boulevard 30, bld. 1}, \city{Moscow}, \postcode{121205}, 
\country{Russia}}}

\affil[12]{\orgdiv{Rosseland Centre for Solar Physics}, \orgname{University of Oslo}, 
\orgaddress{\street{Blindern Postboks 1029}, \city{Oslo}, \postcode{0315}, 
\country{Norway}}}

\affil[13]{\orgdiv{Institute of Theoretical Astrophysics}, \orgname{University of Oslo}, 
\orgaddress{\street{Blindern Postboks 1029}, \city{Oslo}, \postcode{0315}, 
\country{Norway}}}

\affil[14]{\orgdiv{Department of Physics}, \orgname{Montana State University}, 
\orgaddress{\street{P.O. Box 173840}, \city{Bozeman}, \postcode{59717}, 
\country{MT, USA}}}

\affil[15]{\orgdiv{NASA Goddard Space Flight Center}, 
\orgaddress{\street{8800 Greenbelt Rd.}, \city{Greenbelt}, \postcode{20771}, 
\country{MD, USA}}}

\affil[16]{\orgdiv{School of Earth and Space Sciences}, \orgname{Peking University}, 
\orgaddress{\street{5 Yiheyuan Road}, \city{Beijing}, \postcode{100871}, 
\country{China}}}

\affil[17]{\orgdiv{Applied Physics Laboratory}, \orgname{Johns Hopkins University}, 
\orgaddress{\street{Street}, \city{Laurel}, \postcode{20723}, 
\country{MD, USA}}}


\abstract{Coronal dimmings associated  with coronal mass ejections (CME) from the Sun have gained much attention since the late 1990s when they were first observed in high-cadence imagery of the SOHO/EIT and Yohkoh/SXT instruments. \
They appear as localized sudden decreases of the coronal emission at extreme ultraviolet (EUV) and soft X-ray (SXR) wavelengths, 
that evolve impulsively during the lift-off and early expansion phase of a CME. Coronal dimmings have been interpreted as ``footprints'' of the erupting flux rope and also as indicators of the coronal mass loss by CMEs. However, these are only some aspects of coronal dimmings and how they relate to the overall CME/flare process. The goal of this review is to summarize our current understanding and observational findings on coronal dimmings, how they relate to CME simulations, and to discuss how they can be used to provide us with a deeper insight and diagnostics of the triggering of CMEs, the magnetic connectivities and coronal reconfigurations due to the CME
as well as the replenishment of the corona after an eruption. In addition, we go beyond a pure review by introducing a new, physics-driven categorization of coronal dimmings based on the magnetic flux systems involved in the eruption process. Finally, we discuss the recent progress in studying coronal dimmings on solar-like and late-type stars, and to use them as a diagnostics for stellar coronal mass ejections and their properties.}




\maketitle

\newpage
\tableofcontents
\newpage

\section{Introduction}
     \label{Introduction} 

Coronal dimmings are an important eruptive phenomenon that is frequently observed in association with coronal mass ejections (CMEs) from the Sun. 
Dimmings appear as a
sudden, localized decrease of the coronal brightness
in extreme ultra\-violet (EUV) and soft X-ray (SXR) solar images 
that impulsively develop during the initial stages of an eruption \cite[e.g.,][]{Hudson:1996,Thompson:1998,Zarro:1999}. The dimming regions are mostly interpreted to map to the footpoints of the erupting flux-rope and to be a signature of field line opening and CME mass loss \cite[e.g.,][]{Sterling:1997,Webb:2000}. However, as we will show here, these are only some of the aspects of dimmings and do not give full recognition to this intriguing phenomenon. 
The diagnostics potential of coronal dimmings is tremendously manifold, and is yet far from being explored to its full extent. 
What makes coronal dimmings so unique and scientifically valuable, is the fact that they appear during the \emph{entire} evolution of a solar
CME, from the pre-eruption phase to the post-event recovery of the corona. Thus coronal dimmings allow us
to address many important aspects on the physics of CMEs like the CME triggering, magnetic connectivity and mass loss as well as the coronal reconfiguration and mass replenishment following the eruption. Finally, they allow us also to connect the properties of an eruption close to the Sun far out to interplanetary space, where we can compare it to   
the properties of its interplanetary counterpart, the ICME, measured in-situ at Earth orbit or other locations in the heliosphere.

Coronal dimmings caused by solar CMEs have been also observed in Sun-as-a-star EUV spectra and lightcurves by the Extreme-ultraviolet Variability Experiment (EVE) onboard the Solar Dynamics Observatory (SDO)
\citep[e.g.,][]{Mason:2014}. 
Thus, they provide us with a unique possibility to make the step to stellar coronal mass ejections, and can act as a means for the detection and quantitative characterization of CMEs occurring in the coronae of  late-type and solar-like stars
\citep{Veronig:2021,Loyd:2022}. While stellar flares are observed ubiquitously \citep[e.g.,][]{Audard2000,Maehara2012,Hawley2014},  reports on stellar CME detections are still rare 
\citep{OstenWolk:2017,Moschou:2019,Leitzinger:2022}. However, CME activity has important consequences (i) for the star's evolution as they affect the stellar mass loss and angular-momentum loss \citep{Benz:2010,Aarnio2012}, and (ii) for the habitability of planets a star is hosting, since frequent CME impacts may result in the full erosion of exo-planetary atmospheres  \citep{Lammer:2007,Khodachenko:2007,Airapetian2020I}. 

Here we present the first review dedicated to coronal dimmings. 
The review on flare magnetism with SDO by \cite{Kazachenko:2022} includes a brief overview on dimming observations with AIA. Our review is intended to give the reader an introduction to coronal dimmings, a comprehensive description of the observational findings and properties of solar dimmings, but also to discuss the physical scenarios in the framework of different categories of coronal dimmings and how they relate to CME models and simulations. Finally, we discuss the role coronal dimmings play for the detection and characterization of stellar CMEs.

The review is structured as follows. 
Sect.~\ref{History} briefly reviews the history of the research leading to the discovery of solar coronal dimmings and key results of their early investigation. 
Sect.~\ref{Observations} gives an extensive account of the observations. We start with a schematic overview on the life cycle of coronal dimmings and how it relates to the associated CME and flare evolution (Sect.\ \ref{sec:overview}). Sect.~\ref{sec:detection} presents the different approaches and algorithms for the detection of coronal  dimmings.  Sect.~\ref{sec:main} discusses the observational dimming characteristics during the main phase including their global properties 
(Sect.~\ref{sec:properties}), outflows in dimming regions (Sect.~\ref{sec:flows}) and the inference of their plasma parameters
(Sect.~\ref{sec:plasma}). This is followed by a discussion of the recovery phase of dimmings (Sect.~\ref{sec:recovery}) 
and of pre-eruption dimmings, i.e.\ dimming signatures that are sometimes observed already before the actual eruption started (Sect.~\ref{sec:pre-eruption}). In Sect.~\ref{sec:relation}, we discuss the quantitative relations of characteristic dimming parameters to decisive properties of the associated CME and flare, as well as to ICMEs measured at 1~AU. 
Sect.~\ref{Terminology} proposes a new categorization of coronal dimmings based on the magnetic flux systems and processes involved and considers their spatio-temporal evolution in relation to the erupting magnetic structure and the flare ribbons. 
Sect.~\ref{sec:simulations} summarizes our current understanding of dimmings from MHD simulations. 
In Sect.~\ref{sec:stellar}, we make the step from the Sun to the stars. Here, we discuss the various approaches considered for detecting stellar CMEs, including a discussion of the first reports of CME-induced coronal dimmings on late-type stars. We close the review by an outlook of relevant future studies and opportunities to explore the potential of coronal dimmings for the research of solar and stellar CMEs (Sect.~\ref{sec:conclusions}).

\section{History} 
      \label{History}    
The recognition of coronal holes \citep[so-called  ``koronale L{\"o}cher";][]{1956ZA.....38..219W} laid the groundwork for the dynamical processes we now observe as coronal dimmings. 
Waldmeier's synoptic charts of coronal brightness clearly revealed the presence of the polar coronal holes in particular, and showed them to have enough persistence to allow following them through solar rotation.
Subsequent X-ray imagery, originally from \textit{Skylab}, confirmed their nature and disclosed the presence of coronal holes at more equatorial latitudes as well. 
We now understand these semi-permanent features as the loci of large-scale ``open'' magnetic fields, upon which the solar-wind flow can simply \textit{exhaust} the corona.
A coronal hole thus represents a coronal domain with much-reduced plasma density. 
The dipole-like pattern of striations in the solar-minimum corona nicely explains the absence of coronal material in the context of a solar-wind outflow \citep{1958PhRv..110.1445P}.
The dipole magnetic pattern, as noted visually in total solar eclipses, was already well-known to William Gilbert in 1600 AD \citep{1893...Mottelay..book...G}.

The coronal dimmings physically represent transient losses of mass from the (lower) corona, as opposed to the quasi-steady state represented by the coronal holes.
Coronagraphic observations of Thomson scattering give us the most direct way to monitor coronal mass, via the number of free electrons on a given line of sight; this measurement provides linear data ($\propto n_e$ directly) with the straightforward bias towards the ``plane of the sky'' resulting from the angular dependence of the Thomson cross section \citep[\textit{e.g.},][]{1966gtsc.book.....B}.
The introduction of ``K-coronameter scanning photometry '' \citep{1957AJ.....62R..39W}, in which a simple non-imaging detector scans around the corona at a fixed height, led to the discovery of ``abrupt depletions'' \citep{1974PASP...86..500H}.
\cite{1957PASP...69..421E} had actually observed these depletions (he called them ``gaps'') and also what would later be called CMEs, in early Sacramento Peak coronagraphy. 
The gaps or depletions showed up as day-on-day decreases in the isophotes provided by the K-coronameter observations; these gaps often appeared at low heliographic latitudes.
Finally, the X-ray observations, initially from \textit{Skylab}, provided enough time resolution to identify these ``abrupt depletions'' as ``transient coronal holes''  \citep{1983SSRv...34...21R}.
The modern term ``coronal dimming'' suggests no particular inference about physical mechanisms, and this article discusses a rich variety of interesting variations of the basic theme.

\begin{figure}[htbp]
\centering {\includegraphics[width=0.8\textwidth]{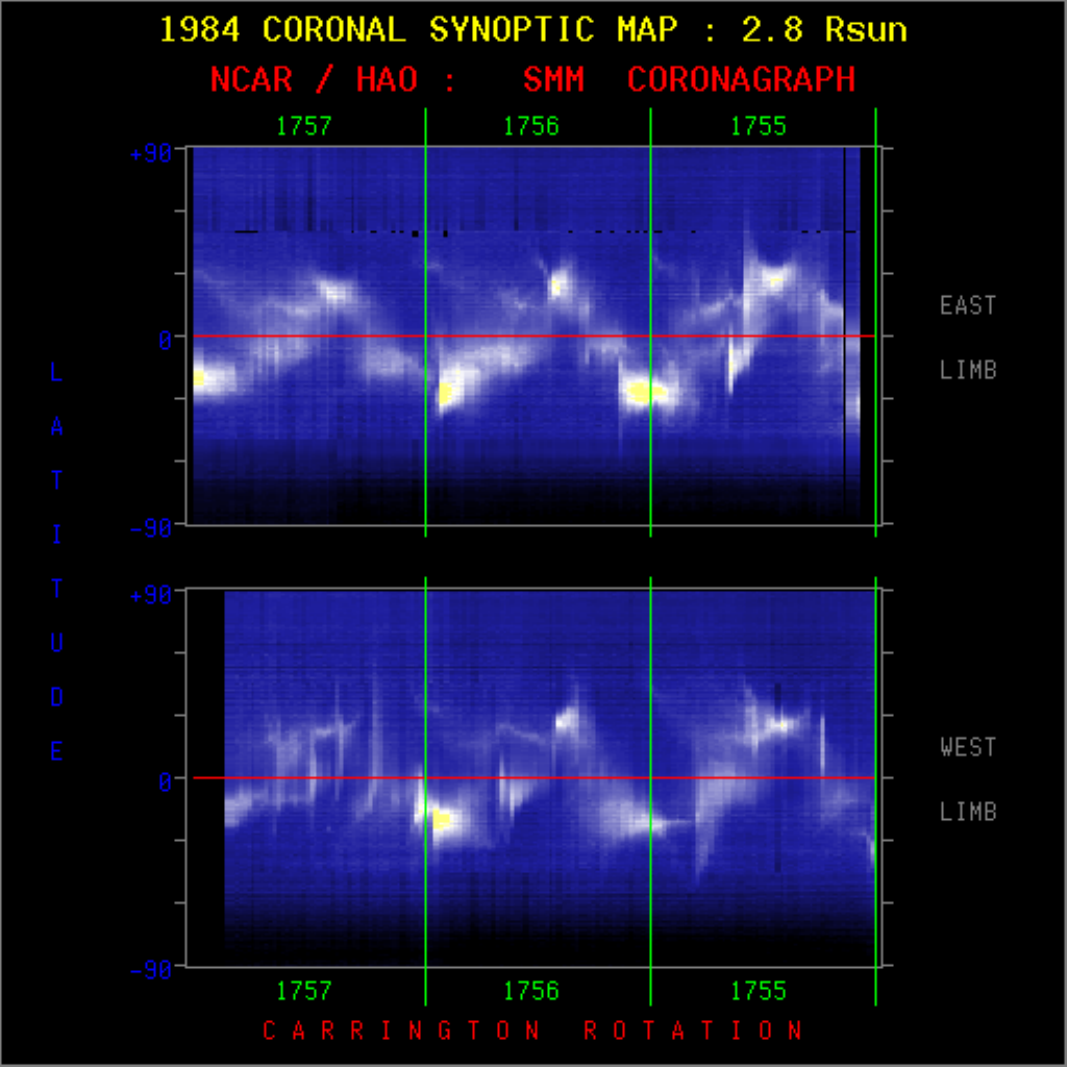}} 
\caption{Examples of streamer blowouts taken from synoptic coronagraph data. Each chart consists of the brightness in an annulus (in this case at 2.8~R$_\odot$) stacked from right to left as the Sun's rotation varies the Carrington longitude in the ``plane of the sky.'' Often a coronal streamer can be seen over many days, and a streamer blowout has a recognizeable bugle-like appearance, the horn of the bugle to the left, as the streamer swells and then suddenly disappears.
That marks the occurrence of a CME, whose eruption esentially destroys the streamer. It may reform, but on a much longer time scale. Here the upper and lower panels refer to E limb and W limb, respectively, for three Carrington rotations in 1984.} \label{fig:bugles}
\end{figure}

Routine coronagraph data, when displayed synoptically, revealed a class of events termed ``streamer blow-outs'' \citep{1982SSRv...33..219S}, as Fig.~\ref{fig:bugles} illustrates.
In these cases a coronal depletion clearly resulted from the semi-permanent destruction of a coronal streamer, which indeed would often grow and brighten prior to its eruption.
The coronagraph, of course, viewed these events from limb observations and could not readily relate them to structures visible on the disk (indeed, coronagraphic observations have essentially equal sensitivity to events originating on the back side).
We now recognize though that streamers overlie ``filament cavities,'' and that the ``disparition brusque'' \citep{1948LAstr..62..343O} represents a physical eruption of this entire cavity, complete with filament, into the heliospheric void.
The streamer blowouts represent one particular class of CME sources, and often reflect the \textit{disparitions brusques} of filaments located in the quiet Sun, even in the polar-crown filament band \citep[e.g.,][]{Schmieder:2000}.
But more often CMEs originate in active regions, where a much stronger magnetic structure may erupt with great force, clearly originating in the very low corona rather than in a large-scale streamer \citep{1997SoPh..175..601D}.

The \textit{Skylab} X-ray telescopes (two of them; SO--54 and SO--56) achieved high spatial resolution and excellent sensitivity, but had the serious limitation of film readout; this cumbersome medium distinctly limited the time resolution and coverage achieved.
Nevertheless their pioneering observations certainly defined many of our concepts; please see the books of the \textit{Skylab} workshop series for detailed information \citep{1977chhs.conf.....Z,1980sfsl.work.....S,1981sars.work....1O}.

The first comprehensive soft X-ray observations of the solar corona came with \textit{Yohkoh}/SXT, a focusing-optics telescope equipped with a CCD readout \citep{1991SoPh..136...37T,2018SoPh..293..137A}.
These observations provided systematic coverage over nearly one complete 11-year sunspot cycle, plus the crucial advantage of high time resolution.
Such an instrument integrates soft X-ray energy fluxes defined by filters at low energies (1--2 keV), with a limit at high energies essentially defined by the reflectivity of the grazing-incidence optics \citep{1991SoPh..136...37T}.
In temperature space, this instrument provided sensitivity both to the quiet corona and also to hotter sources in solar flares. 
Generally a broad-band soft X-ray instrument of this sort responds to all of the thermal emission of the corona,
with an $n^2$ density weighting and no plane-of-the-sky bias \citep{Phillips:2008}.
Note that the K-corona, as opposed to this emission corona, also reveals all of the coronal mass but without any temperature weighting.
Emission lines (optical to EUV) of course have sharply defined temperature weightings, the X-ray and coronagraph observations do not have this diagnostic power but do show the coronal mass comprehensively.
The soft X-ray data strongly favor hotter plasmas.

The SXT observations revealed sudden dimming events of various kinds \citep{1997GMS....99...27H}: large-scale loop disappearances, double dimming regions within sigmoid-to-arcade events, filament channels in the quiet Sun, and large coronal regions when seen at the limb in conjunction with arcade events.
\cite{Sterling:1997} described observations of dimmings at the origin of SOL1997-04-07, a major flare/CME event, and in general it became clear that the dimmings could constitute a major fraction of the CME mass later observed above the coronagraph's occulting disk and provide a means of determining where the ejected mass actually originated.
\cite{Kahler:2001} directly identified some of these with the transient coronal holes first identified in the \textit{Skylab} soft X-ray images. 
This association clearly pointed to an intimate relationship between flare/CME dynamics, in three dimensions, and field connectivity.
This indeed is the central focus of this entire review and became an observational reality with the discovery of complex connectivities by \cite{Mandrini:2007}.

\begin{figure}[htbp]
\centering {\includegraphics[width=0.9\textwidth]{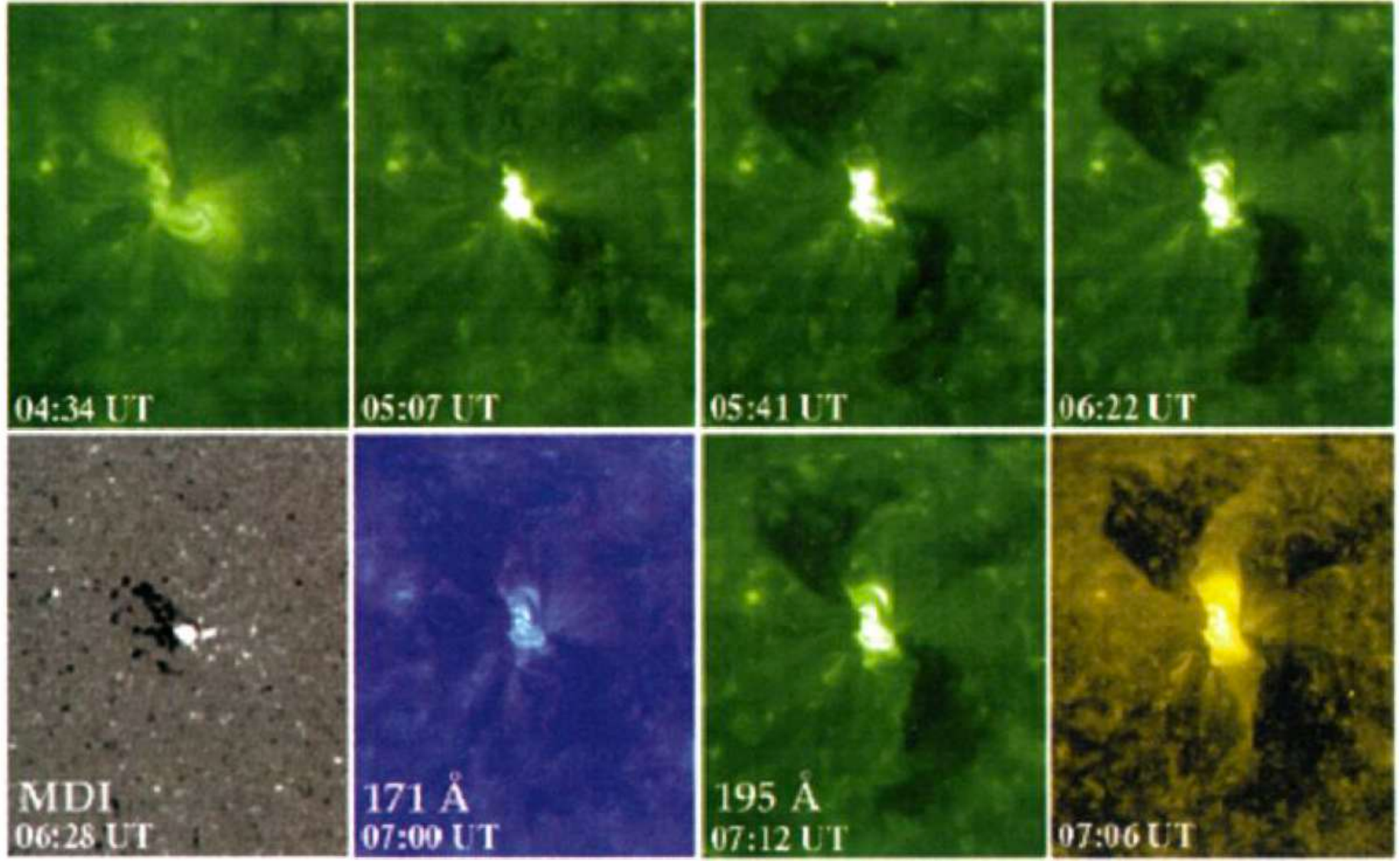}} 
\caption{Twin coronal dimming event associated with the halo CME of SOL1997-05-12. The top row shows the pre-event corona (left) and the dimming evolution in SOHO/EIT 195~{\AA} imagery. The bottom panel shows the SOHO/MDI line-of-sight magnetogram of the source AR (left) and snapshots in all three coronal EIT filters (195, 171, 284~{\AA}) about two hours after the eruption. From \cite{Thompson:1998}. The movie in the online supplement shows the evolution of the event in the EIT 195~{\AA} filter.
} 
\label{fig:thompson98}
\end{figure}

With the regular full-disk imaging of the solar corona by the Extreme-ultra\-violet Imaging Telescope \citep[EIT;][]{ Delaboudiniere:1995} onboard the Solar and Heliospheric Observatory \cite[SOHO;][]{Domingo:1995}, observations of coronal dimmings associated with CMEs became ubiquitous \citep[e.g.,][]{Thompson:1998,Zarro:1999,Thompson:2000}. Figure \ref{fig:thompson98} shows the evolution of the famous dimming associated with the halo CME of SOL1997-05-12. In this textbook example, the twin dimming is clearly seen even in the direct EUV images, located at both sides of the magnetic inversion line and the ends of the post-flare arcade. The dimming appears most pronounced in the EIT 195 and 284 {\AA} filters, but shows also an imprint on the 171 {\AA} filtergrams. This simultaneous  dimming appearance in the multi-wavelength EUV imagery already hinted to the interpretation that the decrease in emission results rather from a decrease in density than a change in temperature \citep[][]{Thompson:1998,Zarro:1999}.

\section{Observations} \label{sec:observations}
\label{Observations}     

\subsection{Overview} \label{sec:overview}

\begin{figure}
    \centering
\includegraphics[width=1.0\textwidth]{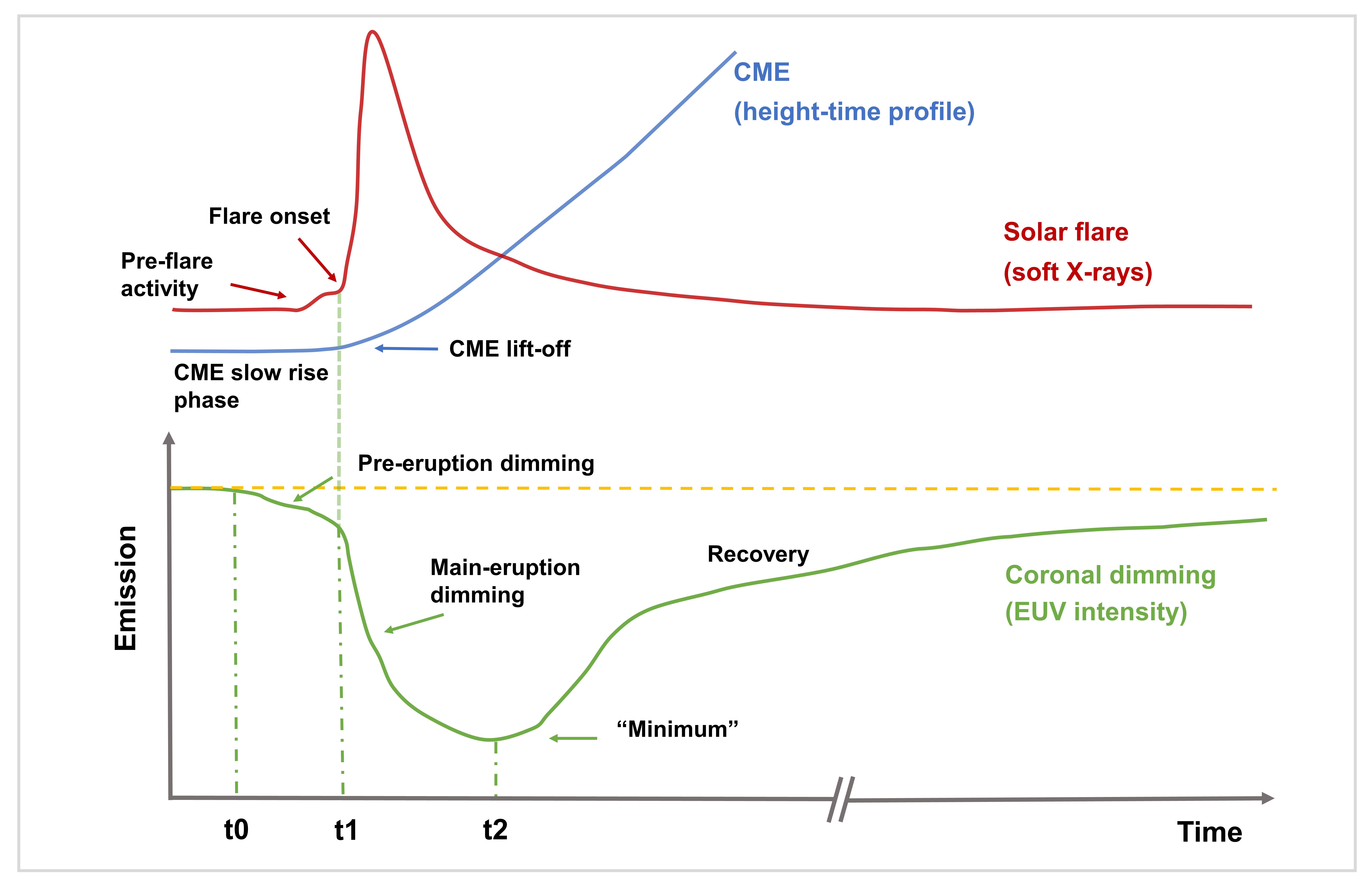}
    \caption{Simplified sketch of the time evolution of
a coronal dimming intensity profile from spatially resolved observations (green) along with the associated CME height-time profile (blue) and the flare soft X-ray light curve 
(red).  Important stages of evolution are outlined for each phenomenon. $t_0$ marks the beginning of pre-eruption dimming, $t_{1}$ the start of the dimming 
main phase which usually coincides with the start of the flare impulsive phase, and $t_{2}$ marks the minimum in the dimming intensity, respectively.
This diagram illustrates the various features that may be observed associated with a coronal dimming observed in EUV wavelengths such as 193 and 211 {\AA}.  The purpose of this diagram is to provide context for the following discussion of the associated phenomena. Some of the features are not always observed, and the time axis is shown to indicate relative 
timing of the different phenomena.
\label{fig:dimming_sketch}
}
\end{figure}

Coronal dimmings are closely linked to the CME causing it and also to the evolution of the associated flare. In Fig.~\ref{fig:dimming_sketch}, we show a simplified sketch of the time evolution of a coronal dimming in spatially resolved EUV observations along with the associated CME and flare evolution. The impulsive onset of the dimming in general coincides with the CME lift-off low in the corona and the start of the flare impulsive phase \citep[e.g.][]{Dissauer:2018b,Dissauer:2019}. During the CME's main acceleration phase, the intensity in the dimming region is typically rapidly decreasing and its area is growing \citep[e.g.,][]{Dissauer:2019}. The minimum of the dimming intensity mostly occurs after the peak of the associated GOES flare, i.e.\ during the flare's decay phase. 
This main evolutionary phase where the intensity in the dimming region is strongly decreasing, is then followed by a longer and more gradual phase of recovery of the dimming signatures, which is related to the recovery and replenishment of the corona after an eruption. In some events, the dimming was observed to gradually start already before the main acceleration phase of the CME and the main dimming evolution. These so-called pre-eruption dimmings have been related to the slow expansion of the pre-eruptive structure \citep{WangW&al2017}. We stress that the scenario illustrated in Fig.~\ref{fig:dimming_sketch} is a sketch capturing  general trends between dimming, CME and flare evolution, but the details may vary from event to event. According to this overall evolutionary scenario, we can distinguish between the \emph{dimming main phase}, its \emph{recovery phase} and \emph{pre-eruption dimmings} that are sometimes present. 

Coronal dimmings observed on the solar disk are traditionally categorized into two types: (i) \emph{core} (or \emph{twin} or \emph{double}) dimmings \citep{Webb:2000} and (ii) \emph{secondary} dimmings \citep[][]{Thompson:2000, Mandrini:2007}. The core dimmings refer to localized regions of strongly reduced coronal emission, which impulsively drops by up to 90\% with respect to the pre-event emission on time scales of some 10 minutes \citep{Vanninathan:2018}, starting during the main acceleration phase of the CME. Although there are cases of single or complex core dimmings \citep{Dissauer:2018b}, typically a pair is observed, rooted in opposite magnetic polarities on both sides of the erupting structure, often within the two hooks of a sigmoid \citep{Sterling:1997, Zarro:1999, Thompson:1999, Miklenic:2011}. These findings led to the interpretation that the core dimmings mark the footprints of the ejected flux rope \citep[][]{Sterling:1997, Webb:2000}. When the flux rope erupts, the magnetic field rooted in the footprints is expected to ``open'' to the interplanetary space, leaving the dark core dimming regions behind as low coronal traces of the escaped plasma. In addition to core dimmings, there are also more wide-spread secondary dimming regions. 
They appear to correspond to the spatial extent of the CME low in the corona \citep{Thompson:2000, Attrill:2009}, and have been interpreted as a result of the stretching and expansion of the overlying field associated with the eruption \citep[e.g.,][]{Dissauer:2018b} and/or of magnetic reconnection of the erupting structure with the ambient coronal field \citep[e.g.,][]{Mandrini:2007, Attrill:2009}. In general, they are less dark, form less impulsively, and tend to show a faster recovery than the core dimmings \citep{Vanninathan:2018}. 

In addition to the on-disk dimmings described above, coronal dimmings are also regularly observed above the limb \cite[e.g.,][]{Aschwanden:2009, Chikunova:2020}. 
Due to the superposition of off-limb structures along the line of sight in the optically thin corona, here it is usually impossible to infer the magnetic connections of the dimming structures seen.
However, since the emission in the line-of-sight does not get mixed with that from the underlying on-disk corona and transition region, off-limb dimmings may provide a better insight into the mass loss due to the eruption and also how the dimming relates to the expansion of the CME.

We emphasize that there are also other
dimmings associated with transient disturbances: so-called \emph{obscuration dimmings} due to absorption by erupted and subsequently slowly draining filament material \citep{Mason:2014,Xu:2024}, 
short-lived \emph{wave dimmings} in the rarefaction region trailing a compressional coronal EUV wave \citep{cohen09, Muhr:2011, Podladchikova:2019}, and \emph{thermal dimmings} due to the change of ionization states by heating and the associated response in EUV passbands with different peak formation temperatures \citep{Robbrecht:2010,Mason:2014}.
These types of dimmings will not be addressed in this review; instead, we concentrate on the coronal dimmings that result from the expansion and mass loss of CMEs.

In the following description of the observational dimming characteristics (Sects.\ \ref{sec:main} to \ref{sec:pre-eruption}), we group the findings along the different phases, i.e.\ the \emph{dimming main phase}, the \emph{recovery phase} and the sometimes observed \emph{pre-eruption dimmings}. Before that, we give an overview on the methods for detecting and segmenting coronal dimming regions in Sect. \ref{sec:detection}.

\subsection{Detection methods}
\label{sec:detection}

\begin{figure}[htbp]
\centering
\includegraphics[width=0.96\textwidth]{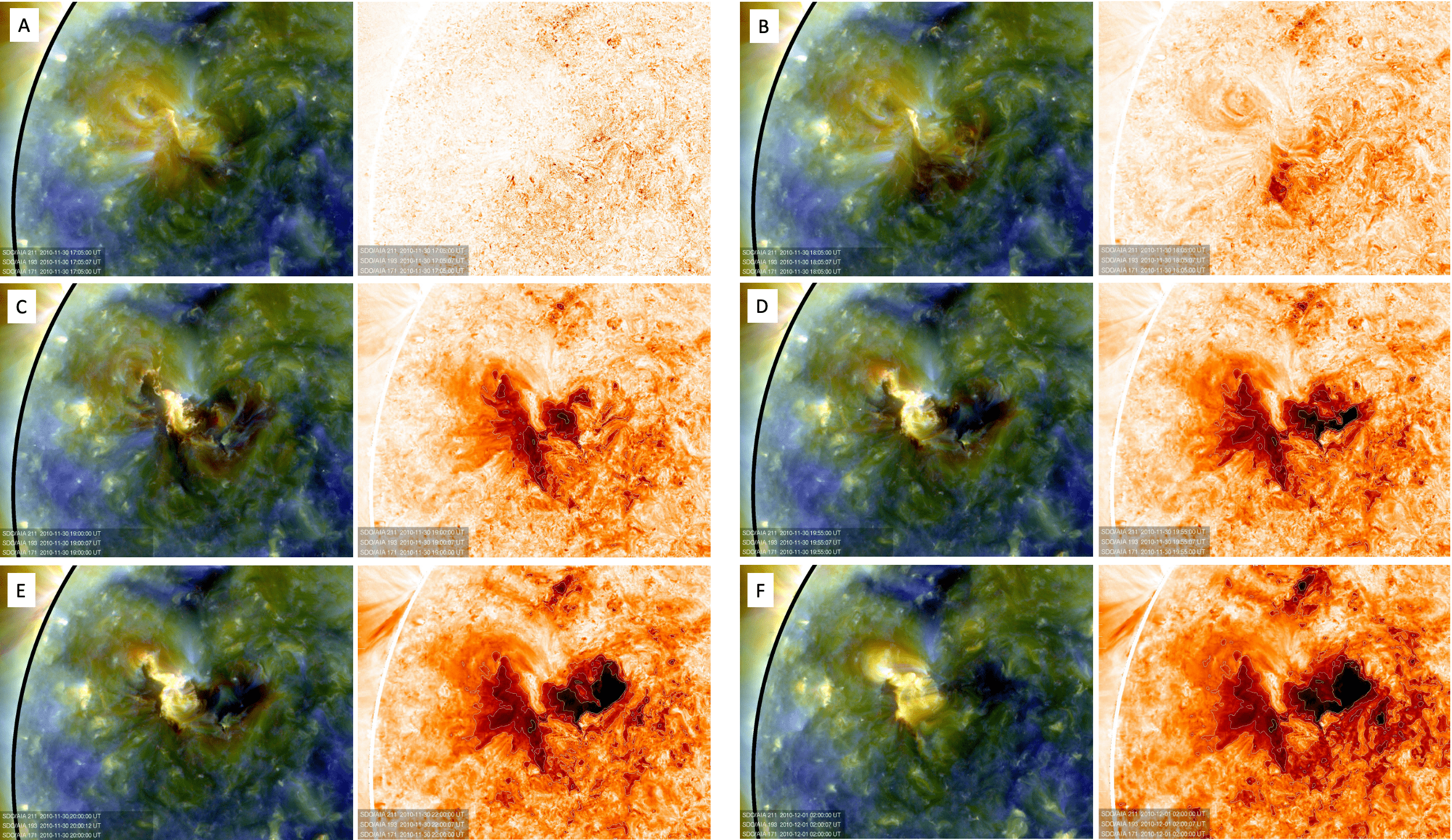}
\caption{Persistence maps derived from SDO/AIA images
over the course of the dimming event SOL2010-11-30. 
Panel~A is a pre-eruption frame at 17:00 UT, and subsequent frames are 18:05 UT, 19:00 UT, 19:55 UT, 20:00 UT, and 02:00 UT from the following day. The left sides of the panels are three-color images with AIA 211~{\AA} in the red layer, 193~{\AA} in the green layer, and 171~{\AA} in the blue layer. The right sides of the panels are persistence maps, consisting of the minimum value of each pixel evaluated from 17:00 UT to the time of each frame. The persistence technique focused only on the decreasing regions, showing the full extent of the dimmings as they develop, and excluding the expanding flare loops.  Adapted from \cite{Thompson:2016}.
The movie in the online supplement shows the evolution of the event in the AIA 211-193-171 three-color persistence maps.
}
\label{fig:persistence_map}
\end{figure}

There exist several approaches to identify and to segment coronal dimming regions in solar EUV and SXR imagery. 
In general, the detection of coronal dimmings is not trivial, since they show a nonuniform development and level of intensity decrease over time. 
In most cases, base-difference images are used to detect coronal dimming regions \citep[e.g.][]{Sterling:1997,Reinard:2008,Attrill:2010a}. In this approach, a pre-event reference image is subtracted from each subsequent frame, in order to emphasize the regions of transient changes. 
In contrast to these generally used approaches, \cite{Krista:2013} extract coronal dimmings using a local intensity thresholding applied to direct SDO/AIA 193 \AA~images that were transformed to Lambert cylindrical equal-area projection maps.
\cite{Thompson:2016} use persistence maps for the dimming detection (see Fig.~\ref{fig:persistence_map}), in which each pixel is represented by its minimum intensity that was reached over the course of the event.

The fully automatic detection of coronal dimmings is of special interest, for scientific studies of their characteristic properties but also for the detection and forecast of CMEs \citep{Robbrecht:2005}. \cite{Podladchikova:2005} and \cite{Podladchikova:2012} developed the Novel EIT wave Machine Observing (NEMO) algorithm which is based on statistical properties of the distribution of pixels. 
Two masks are extracted, the first one consists of seed pixels, representing the darkest pixels of the image, while the second mask contains pixels that decreased in intensity and are detected using a lower threshold. The final dimming region is formed by using the seed pixels for a region growing method and keeping the condition of a simply connected region \citep{Podladchikova:2005}.
Similar automatic algorithms, based on the statistical properties of the images, were developed by \cite{Attrill:2010a} and \cite{Martens:2012}.

\begin{figure}[tbp]
    \centering
    \includegraphics[width=1.0\textwidth]{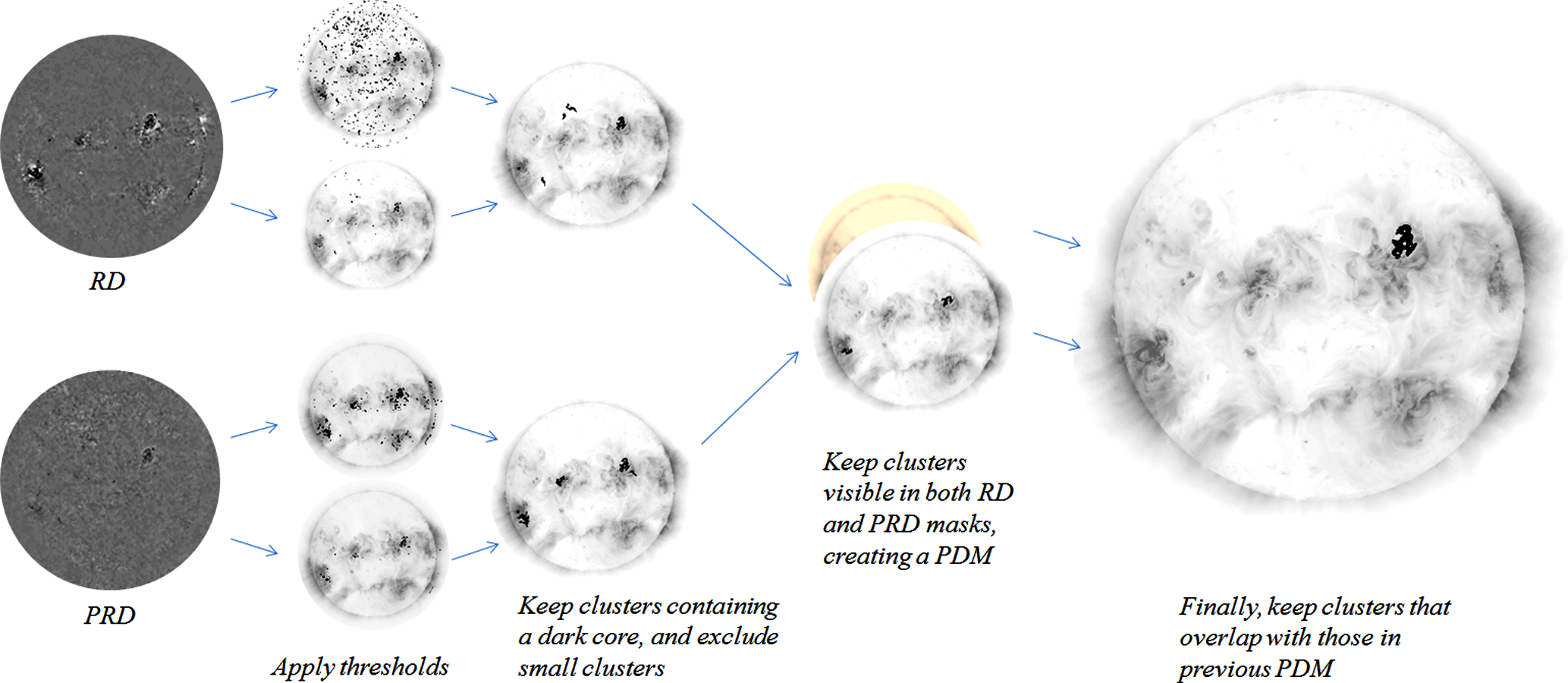}
    \caption{Illustration of the procedure used by \textit{SolarDemon} to detect coronal dimmings based on running-difference (RD) and percentage running-difference (PRD) images.
    From \cite{Kraaikamp:2015}.}
    \label{fig:solardemon}
\end{figure}

Figure~\ref{fig:solardemon} illustrates the \textit{SolarDemon} \citep{Kraaikamp:2015} 
dimming detection method which uses a combination of high and low thresholding on running-difference and percentage running difference images, by only keeping clusters identified by both masks, to  
identify coronal dimming regions. \textit{SolarDemon} is currently the only dimming detection algorithm that is used in an operational mode to detect solar dimmings in near real-time.\footnote{\url{https://www.sidc.be/solardemon/dimmings.php}}
The thresholding 
algorithm by \cite{Dissauer:2018a} detects coronal dimmings on logarithmic base-ratio images, which allows the detection of both types of dimmings: the localized core dimmings which may map to the 
footpoints of the erupting magnetic structure, as well as the more shallow secondary dimmings, mapping the overlying magnetic field that is expanding and erupting. A variant of this algorithm, which considers both base-ratio and base-difference images, was developed in \cite{Chikunova:2020} to provide also robust detections of coronal dimmings above the solar limb.

\begin{figure}[htbp]
    \centering
\includegraphics[width=0.99\textwidth]{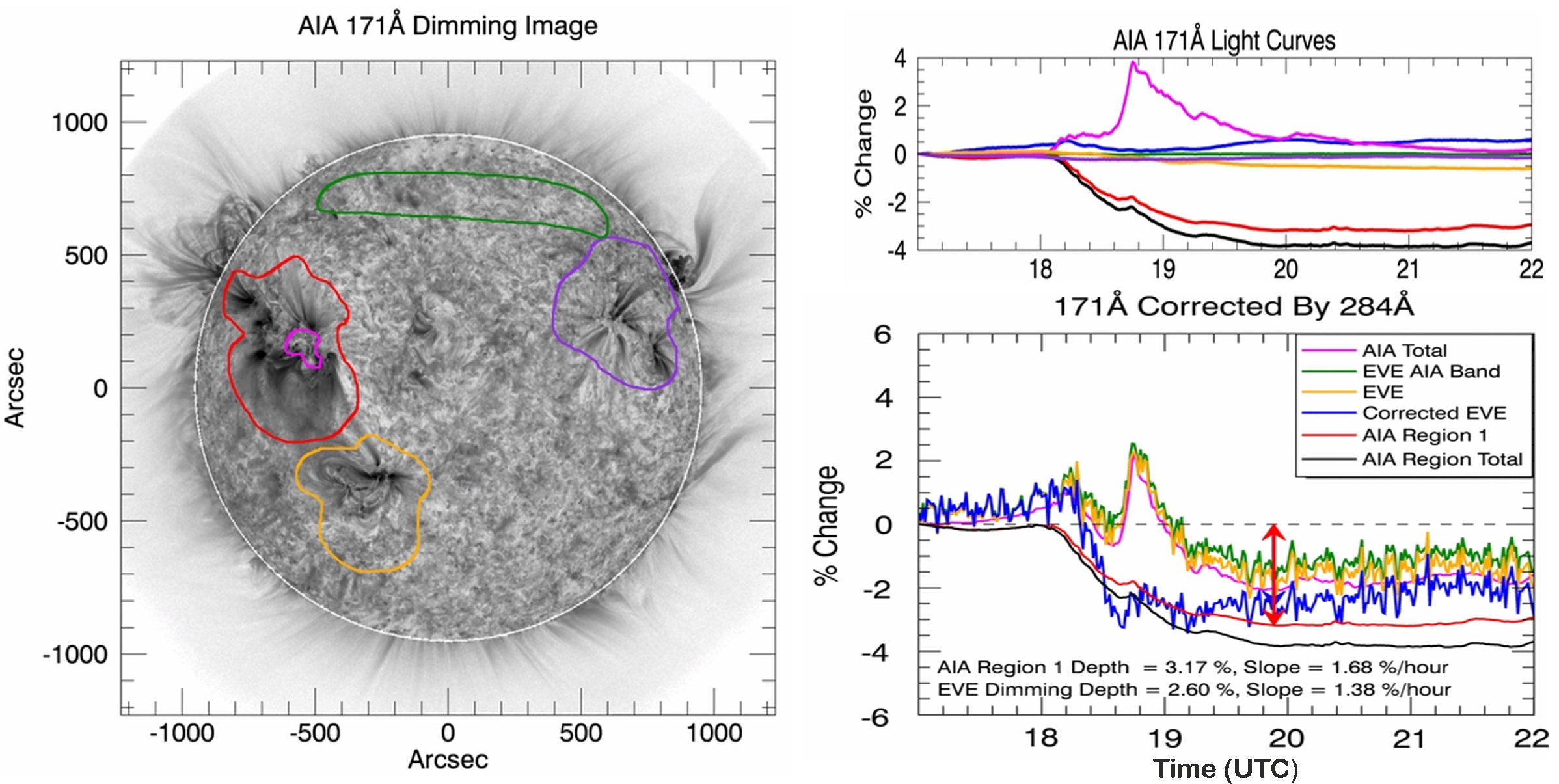}
    \caption{From  
    imaging to ``Sun-as-a-star'' observations of SOL2010-08-07: Coronal dimmings identified in regions-of-interests (colored contours and corresponding time profiles) in full-disk integrated SDO/AIA EUV
    images (left and top right panels) 
and in SDO/EVE spatially integrated irradiance measurements (bottom right panel). Adapted from \cite{Mason:2014}.}
\label{fig:eve_detection}
\end{figure}

Coronal dimmings can also be identified from spatially unresolved SDO/EVE irradiance light curves \citep[][]{Woods:2011,Sun:2013,Mason:2014,Mason:2016,Harra:2016,Veronig:2021}. An example is illustrated in Fig.~\ref{fig:eve_detection}.
These observations of dimmings in such ``Sun-as-a-Star" EUV light curves are very important in the attempt 
to  relate the solar observations to the stellar case, and to explore the possibility of coronal dimmings as a proxy of stellar CMEs (see Sect.\ \ref{Sec:SunAsStar}). It is important to note that irradiance light curves contain both the dimming and the flare contributions. Separating the two components is therefore important but at the same time challenging due to limited spectral information.

\subsection{Main dimming phase}\label{sec:main}

As outlined in Sect.~\ref{sec:overview} and schematically represented in Fig.~\ref{fig:dimming_sketch}, we distinguish in the overall  evolution between the dimming main phase, its recovery phase and sometimes observed pre-eruption dimmings. The main phase starts with the impulsive onset of the dimming during the CME-lift off, and occurs often roughly simultaneous with the start of the flare's impulsive phase. Its end is defined by the time when the dimming reveals its deepest decrease, thereafter it enters into the  recovery phase (see Fig.~\ref{fig:dimming_sketch}). In this Section, we give detailed account of the dimming observations during its main phase, which is actually the phase on which most of the dimming studies focus on.

\subsubsection{Global dimming properties}\label{sec:properties}

An immediate question that came up when coronal dimmings were first observed in image sequences by the narrow-band EUV filters by SOHO/EIT and the broad-band SXR imagery by Yohkoh/SXT, was whether the decrease in emission is due to density depletion (indicating mass loss and/or volume expansion) or due to a variation in plasma temperature. Already these early observations showed that dimmings appeared in all three of the coronal EIT channels (171, 195, 284~\AA) sampling plasma at temperatures between about 1--3~MK \citep[e.g.,][see also Fig.\ \ref{fig:thompson98}]{Thompson:1998,Chertok:2003,Zhukov:2004}, and were also simultaneously observed in SXRs by Yohkoh/SXT sensitive to temperatures $T \gtrsim 3$~MK \citep[e.g.,][]{Zarro:1999}. These findings 
suggested that the observed dimming of the coronal emission is dominantly due to density depletion caused by the expanding CME, and only to a smaller degree due to changes in temperature, which would lead to emission decreases in some filters but increases in others. In addition, as noted already in \cite{Hudson:1996}, the radiative and conductive cooling time scales of the corona are too long to explain the sudden emission decrease in dimmings. These findings have soon after also been supported by spectroscopic observations of the plasma properties  \citep{Harrison:2000} and outflows from dimming regions \citep{2007PASJ...59S.801H}. Meanwhile detailed plasma diagnostics are available for a number of  dimming events, which are discussed in Sect.\ \ref{sec:plasma}.

\begin{figure}
\centering {\includegraphics[width=0.78\textwidth]{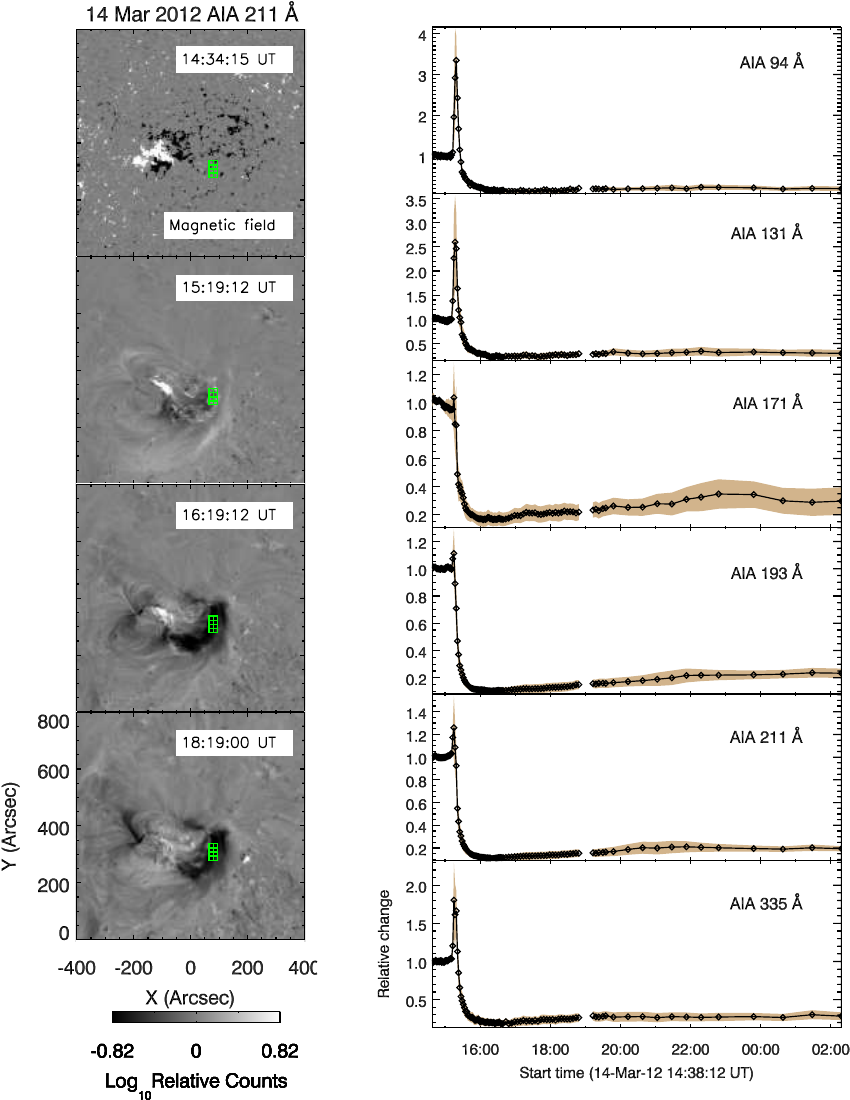}} 
\caption{Left: SDO/HMI LOS magnetogram (top) and snapshots of logarithmic base-ratio images visualizing the relative changes in the emission in the SDO/AIA 211 {\AA} filter during the flare/CME SOL2012-03-14 (M2.8). 
Right: Normalized light curves obtained from the six coronal SDO/AIA channels in a small area inside the Western dimming region indicated by the green box in the left panels. Adapted from \cite{Vanninathan:2018}.
} \label{fig:vanninathan_2018_lc}
\end{figure}

Figure \ref{fig:vanninathan_2018_lc} shows as an example the time evolution of the emission extracted from a localized area inside a dimming region for the GOES M2.8 flare/CME event SOL2012-03-14 studied in \cite{Vanninathan:2018}. 
The light curves (right column) show the median intensity derived within the box inside the negative polarity dimming region (indicated in the left panels) and normalized to the pre-event intensity for the six coronal EUV filters of the Atmospheric Imaging Assembly \cite[AIA;][]{Lemen:2012} onboard the Solar Dynamics Observatory (SDO). There is a number of important dimming characteristics that can be observed  in these curves: 

\begin{itemize}
\item[(i)] The dimming shows up clearly in each of the coronal SDO/AIA filters, which sample coronal plasma over a broad temperature range from about 0.7 to 10 MK \citep{Lemen:2012},  demonstrating the multi-thermal nature of the dimming.
\item[(ii)] The emission drop of the dimming occurs very impulsively, over a time scale of the order of 10 min.
\item[(iii)] Each light curves reveals a substantial drop of the emission during the dimming, by 80--90\% of the preflare emission. 
\item[(iv)] The emission stays at this strongly reduced level for the full time range under study, i.e., $\gtrsim$10 hrs, showing no indication of recovery yet. 
\item[(v)] Immediately before the dimming is observed, the region shows an impulsive emission spike due to coverage by flare-ribbons.  
\end{itemize}

\begin{figure}[tbp] 
\centering {\includegraphics[width=0.92\textwidth]{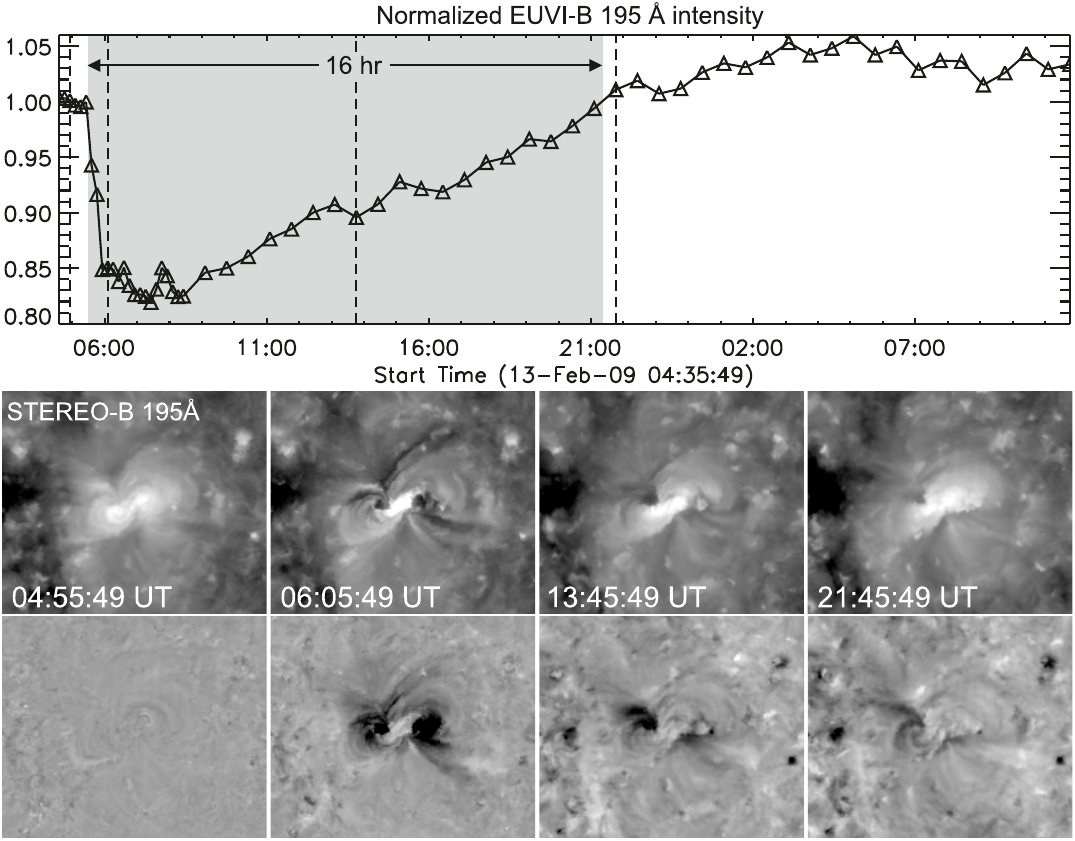}} 
\caption{Evolution of the coronal dimming associated with the SOL2009-02-13 CME. Top: Normalized light curve derived from the STEREO-B EUVI 195 \AA\ filter for the FOV shown in the bottom panels.  
The shaded area marks the duration of the dimming, i.e.\ from its start until the EUV emission recovered back to the pre-event level. 
Middle: Selected EUVI 195 \AA\
images covering a $600''\times 500''$ FOV centered on the erupting AR. The recording times of the images shown are marked by vertical lines in the top panel. 
Bottom: Corresponding base difference
images. The movie in the online supplement shows the first three hours of the event evolution in STEREO-B EUVI 195~{\AA} images. From \cite{Miklenic:2011}.
} \label{fig:miklenic_2011_F2}
\end{figure}

Figure \ref{fig:miklenic_2011_F2} shows STEREO-B EUVI 195 {\AA} observations of a dimming event on 13 February 2009 which was caused by the eruption of a coronal sigmoid that evolved into a slow CME. Here, the light curve is extracted for a field-of-view (FOV) that covers the full Active Region (AR), in contrast to Fig. \ref{fig:vanninathan_2018_lc} where only a small area inside a dimming region was considered. Therefore, the overall emission drop due to the dimming is smaller but still reaches up to about 17\%. In this full AR light curve one can see that the recovery of the overall AR corona starts about 3--4 hours after the start of the dimming, and after about 16 hours the  emission integrated over the FOV reaches again the pre-event level, indicating replenishment of the AR corona over this time scale. However, we also note that the Eastern core dimming region is not yet back to the pre-event state after these 16 hrs, as can be seen in the direct and base difference images in Fig.~\ref{fig:miklenic_2011_F2}.

Analysing observations of dimmings with SOHO/EIT, 
\cite{Chertok:2003} pointed out that the dimmings are best  observed in the 171 and 195~{\AA} filters (with peak formation temperatures at 0.9 and 1.6 MK, respectively), whereas the 284~{\AA} filter sensitive to hotter plasmas ($T \sim 2$ MK) shows mostly the deepest dimming regions only. Similar findings were also reported in  \cite{Robbrecht:2010} who studied dimmings in the multi-band STEREO/EUVI observations. Systematic studies and statistics of the visibility of the overall dimming regions observed with the seven EUV filters of SDO/AIA, were done in \cite{Dissauer:2018b} and \cite{Vanninathan:2018}. These studies show that the dimmings are best visible in the SDO/AIA~193 and 211~{\AA} passbands, which sample coronal plasma between about 1--2 MK. These findings suggest that the matter ejected by the CME comes mostly from
the ambient corona rather than from the active region itself \cite[see also][]{Harra:2016}. However, in areas inside core dimming regions, also the AIA 94~{\AA} filter (peak formation temperature $T \sim 7$~MK) may reveal distinct emission decreases (for an example, see
Fig.~\ref{fig:vanninathan_2018_lc}) indicating that in this event also hot AR loops are ejected as part of the CME \citep{Vanninathan:2018}.

\begin{figure}[htbp] 
\centering {\includegraphics[width=0.5\textwidth]{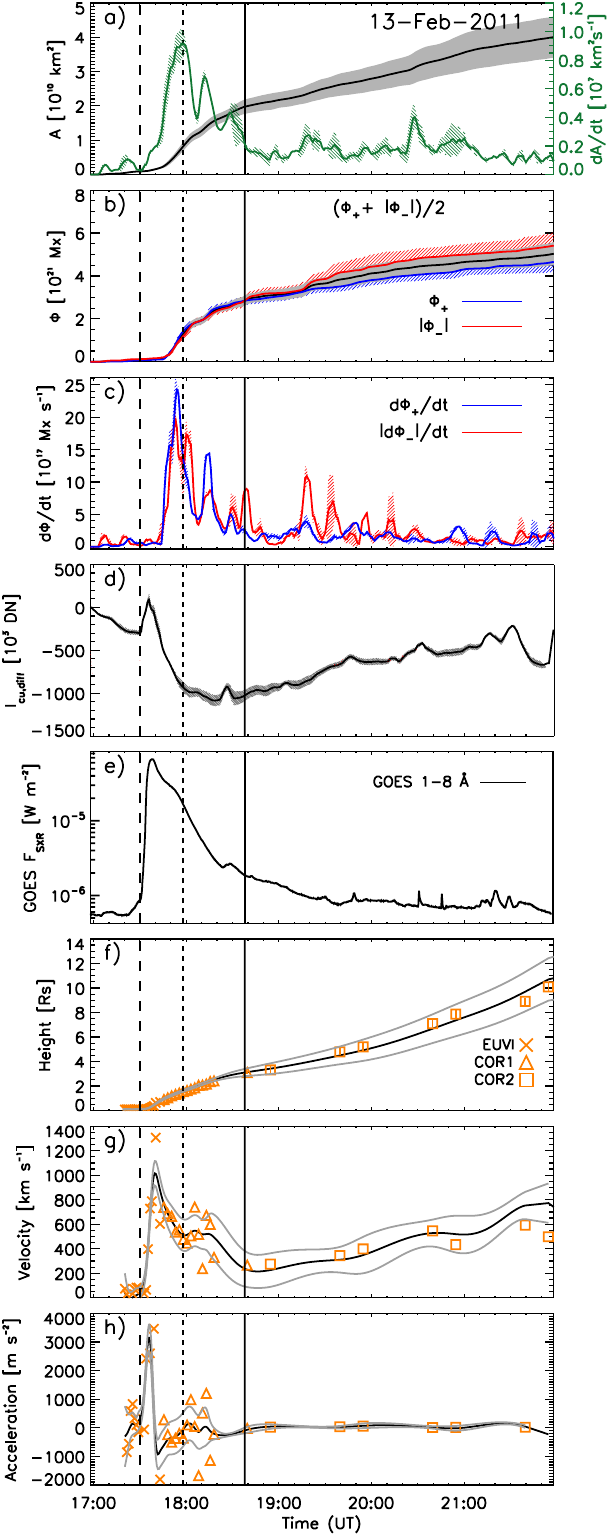}} \caption{Evolution of the coronal dimming parameters associated with the flare/CME SOL2011-02-13 (M6.7) together with the flare evolution and CME kinematics: (a) cumulative dimming area (black) and its  growth rate (green), (b) positive (blue), negative (red), and total (black) unsigned magnetic flux underlying the dimming region, (c) corresponding 
magnetic flux rates,
(d)  total dimming brightness. 
(e) GOES 1--8~{\AA} SXR flux of the associated flare. (f)--(h) CME height–, velocity- and acceleration-time profile
derived from smoothed curves (black lines) and direct measurement points (orange symbols). 
The vertical 
lines mark the start, peak, and end time of the impulsive dimming phase.
Adapted from \cite{Dissauer:2019}.
} \label{fig:dissauer_2019_fig4b}
\end{figure}

\cite{Dissauer:2018b} studied the dimmings associated with 62 CMEs in the multi-band SDO/AIA imagery. All events showed a dimming signature in the 193 and 211~{\AA} filters. In the 171 and 335~{\AA} filters, a dimming was detected in 92\% and 94\% of the cases, respectively. In filters sensitive to high temperature
plasma (i.e., 94 and 131~{\AA}), coronal dimmings could be identified in 63\% and 47\% of the events under study. Though we note that the emission decrease in the 131~{\AA} filter may be also related to the ``cool" transition region contribution to the AIA 131~{\AA} passband.
Notably, 15\% of the events revealed coronal dimmings also in chromospheric/transition region 
plasma probed by the He II 304~{\AA} line. This is in accordance with cases where a ``coronal" dimming associated with a CME was also detected in ground-based H$\alpha$ filtergrams \citep{Jiang:2003,Jiang:2006,Jiang:2007a,Jiang:2007b} to occur co-spatial and co-temporal with the dimmings observed in EUV. These findings indicate that also chromospheric and lower transition region plasma may be involved and contribute to the CME mass loss.

Figure~\ref{fig:dissauer_2019_fig4b} shows as an example the evolution of characteristic dimming parameters together with the flare and CME evolution for the SOL2011-02-13 (M6.7) event. The dimming was studied on-disk by SDO/AIA, whereas STEREO was in quasi-quadrature and observed the CME source region close to the limb \citep[][]{Dissauer:2019}. This combination of the multi-point observations minimizes the projection effects in the CME measurements as it is observed close to the plane-of-sky. From panel (a), one can see that the dimming area evolution $A(t)$ is first characterized by an impulsive growth rate profile $dA/dt$, 
followed by a gradual further growth. The plotted area  is calculated as the cumulative area of all the newly detected dimming pixels over the time series. The photospheric magnetic fluxes $\Phi$ in the dimming areas are roughly balanced for the two polarities (panel b), and show a growth rate profile $d\Phi/dt$ (panel c) that is closely related to the dimming area growth rate $dA/dt$. The flux growth rate $d\Phi/dt$ quantifies the increase of magnetic flux covered by the dimming area during its expansion.   
The vertical lines in each panel indicate the start, peak and end times of the impulsive phase of the dimming, which is defined by the dimming area growth rate  plotted in panel a (for exact definitions of various dimming quantities, we refer to \cite{Dissauer:2018a}). The dimming brightness (panel d) is calculated from the total dimming mask derived over its impulsive phase. It shows first an increase due to flaring pixels inside this mask, and then drops to a global minimum that is reached within the impulsive dimming phase. As one can see from panels e--g, the GOES SXR flare evolution and the CME velocity profile (impulsive acceleration phase) also fall into this interval of the impulsive dimming phase, giving evidence of a close connection between all three processes, i.e. flare, CME eruption and associated coronal dimming.

\begin{figure}[tbp] 
\centering {\includegraphics[width=0.9\textwidth]{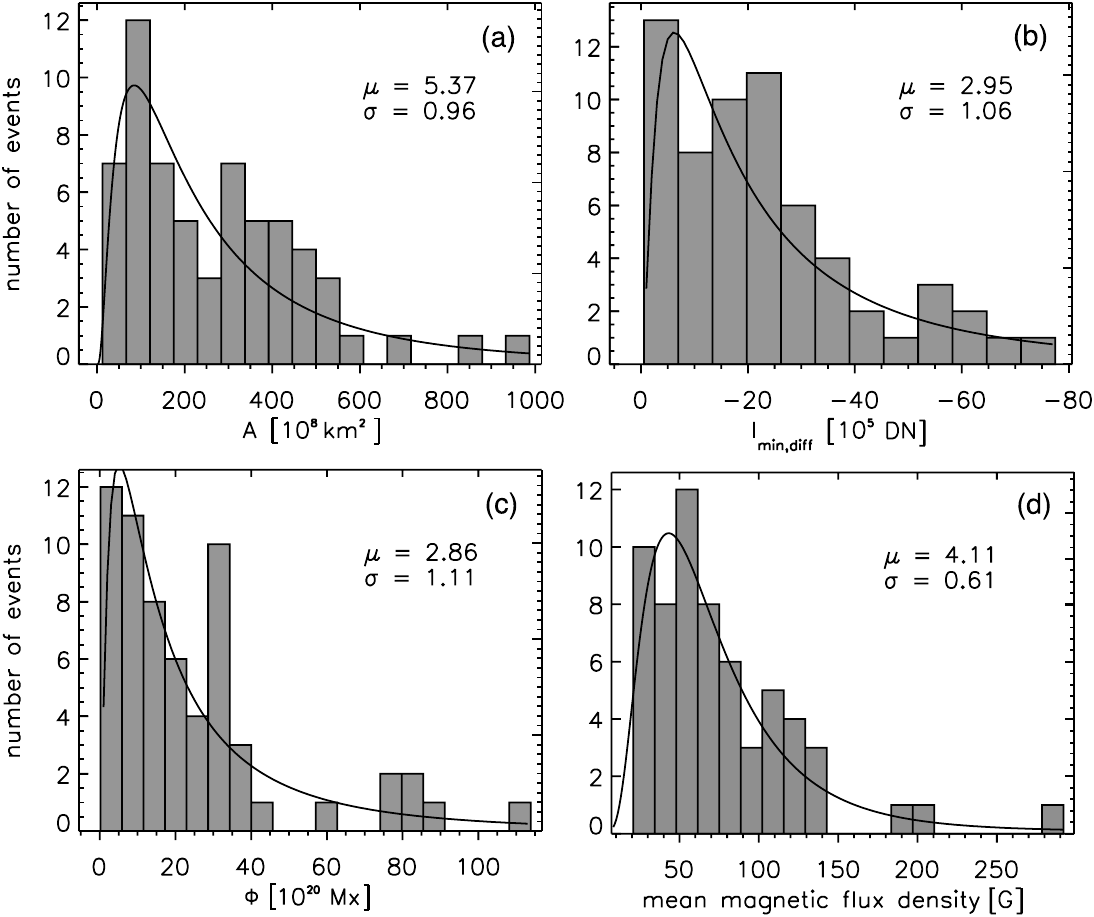}} \caption{Characteristic dimming properties derived from SDO/AIA 211~{\AA} filtergrams and co-registered SDO/HMI LOS magnetograms. Distributions of a) dimming area,  b)  minimum brightness, c) unsigned magnetic flux, d) mean unsigned magnetic flux density. The black line shows the lognormal probability density
function for each distribution, and the insets give the corresponding median and 68\% confidence interval. 
Adapted from \cite{Dissauer:2018b}.
} \label{fig:dissauer_2018b_fig1}
\end{figure}

Figure~\ref{fig:dissauer_2018b_fig1} shows histograms of different characteristic dimming properties derived from the set of 62 events studied in \cite{Dissauer:2018b,Dissauer:2019}: the total dimming area, its minimum brightness, the unsigned magnetic flux covered by the dimming and the corresponding mean flux density. All distributions are asymmetric with a
tail toward the high end, and can be fitted by a lognormal probability density
function. The dimming areas show a large spread over about two orders of magnitude from about $10^9$ to $10^{11}$ km$^2$, covering photospheric magnetic fluxes of $10^{19}$ to $10^{22}$~Mx; the unsigned flux densities vary from about 20 to 300 G \citep{Dissauer:2018b}.

So far, only a few statistical studies have studied the duration of coronal dimmings, as this is a somewhat tricky parameter to extract, due to their potentially long lifetimes of $>$10 hours (see, e.g., Figs.~\ref{fig:vanninathan_2018_lc} and \ref{fig:miklenic_2011_F2}) and faint appearance, which has to be extracted on top of the general variations in the corona on such time scales. \cite{Dissauer:2018b} concentrated on the impulsive phase of the dimming, which is closely related to the CME lift-off phase and its initial dynamics. 
The duration of the impulsive dimming phase was defined by the start/end of the dimming area growth rate $dA/dt$, and found to lie mostly within 10--120 min. The rise tends to be impulsive, and for the majority of events the rise time is $<$40 min \citep{Dissauer:2018b}. These time scales are consistent with the time scales of the impulsive acceleration phase of CMEs \cite[]{Zhang:2001,Vrsnak:2007,Bein:2011}. 

\begin{figure}[tbp] 
\centering {\includegraphics[width=0.98\textwidth]{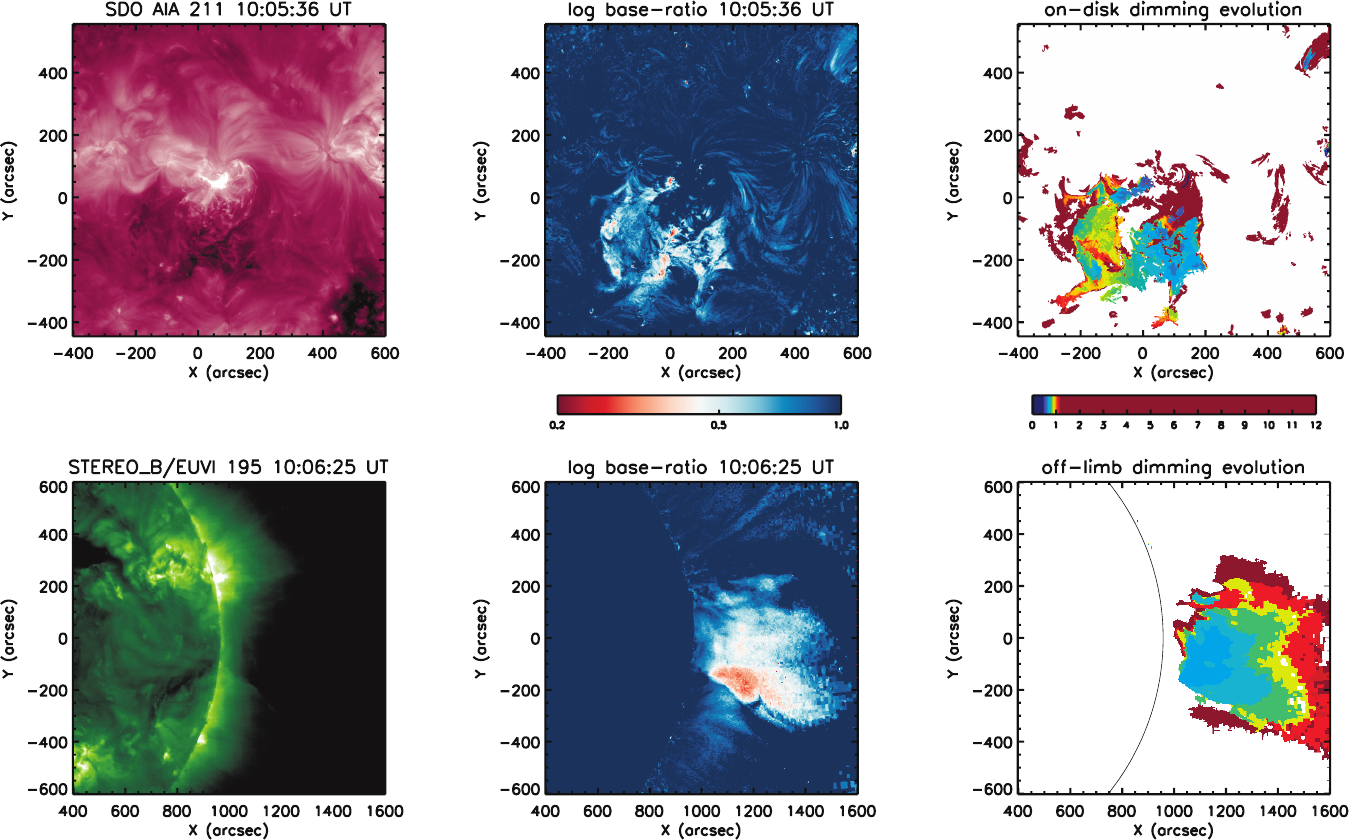}} 
\caption{Simultaneous observations of the coronal dimming SOL2011-10-01 observed from two different viewpoints. Top: dimming observed against the solar disk by the SDO/AIA 211 \AA\ filter. Left: direct image, middle: logarithmic base ratio image, right: timing map which indicates when each dimming pixel was detected for the first time
(in hours after the flare onset). Bottom: Same but for the dimming as observed above the limb by the 195 \AA\ filter of EUVI onboard STEREO-B, on that day located at a longitudinal separation of 97.6$^\circ$ from the Sun-Earth line.
From \cite{Chikunova:2020}.
} \label{fig:chikunova_2020_F1}
\end{figure}

\cite{Reinard:2008} studied the total dimming duration (life time) derived from SOHO/EIT images, defined as the time span between the start and end of the dimming area evolution $A(t)$ or the intensity evolution $I(t)$ (e.g., top panel of Fig.~\ref{fig:miklenic_2011_F2}). Their distribution is described by a mean duration of $8\pm 0.4$ hours, with the majority of the dimming life times between 3 and 12 hours. These numbers are similar to the dimming duration derived from a statistical set of coronal dimmings observed in full-Sun integrated SDO/EVE data \citep{Veronig:2021}, with a mean of $7.1\pm 3.1$~hours. It should be noted that for both studies the derived dimming life times are lower estimates. Despite the scarcity of studies of total dimming duration and the difficulty of their determination, the dimming life time is an important property: the core dimming life times may carry information on how long the CME/flux rope footpoints are connected to the Sun, and the life times of secondary dimmings may be related to the time scales on which the corona reforms and is replenished with plasma after a CME was ejected. These aspects are discussed in more detail in Sect.~\ref{sec:recovery} dedicated to the dimming recovery.

\begin{figure}[tbp] 
\centering {\includegraphics[width=0.99\textwidth]{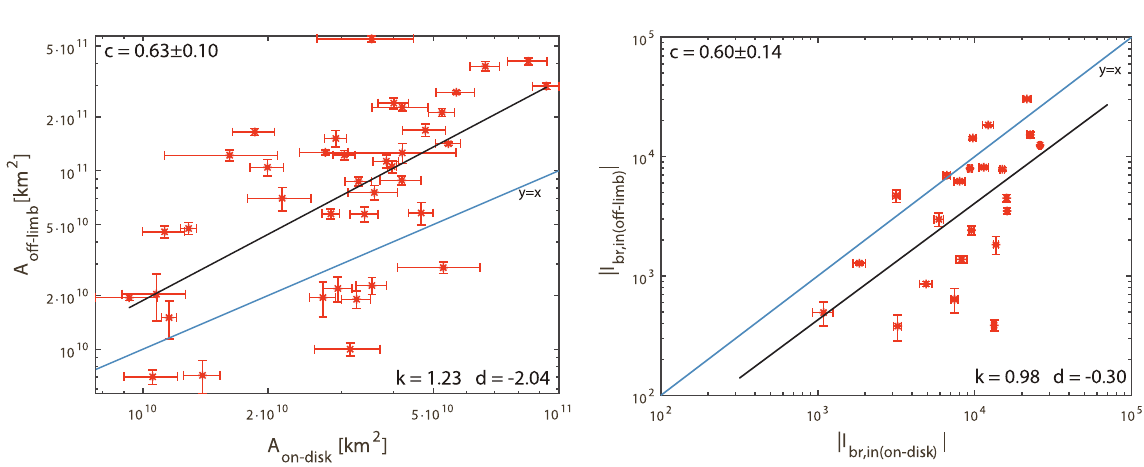}} 
\caption{Comparison of dimming properties derived from the on-disk (SDO/AIA) and off-limb view (STEREO/EUVI). Left: dimming area, right: dimming brightness from base difference images. The black line shows the least-squares linear fit to the data, the blue line the 1:1 correspondence line. Adapted 
from \cite{Chikunova:2020}.
} \label{fig:chikunova_2020_F2}
\end{figure}

Most statistical studies on dimming properties (as discussed above) are based on coronal dimmings observed against the disk. \cite{Chikunova:2020} did an off-limb study of the same set of events studied in \cite{Dissauer:2018b} that were simultaneously observed against the disk by SDO/AIA and above the limb by STEREO/EUVI. Figure~\ref{fig:chikunova_2020_F1} shows one example, at top the SDO view (AIA 211~\AA\ filter) and at bottom the STEREO-B view (EUVI 195~\AA\ filter). 
One can see the 
different on-disk vs.\ off-limb appearance of the dimming shown at a time close to its maximum extent. 
Figure~\ref{fig:chikunova_2020_F2} plots the statistical relations between the dimming area and its brightness derived from the SDO/AIA on-disk and the STEREO/EUVI off-limb view. The areas derived from the different views are related but show a large scatter around the trend line. The brightness is also correlated but shows systematically smaller values when the dimmings are observed off-limb. This systematic difference can be explained by the emission from the underlying transition region plasma that is contributing to the on-disk dimming observations.

\subsubsection{Outflows in dimming regions}
\label{sec:flows}

EUV spectroscopy is a powerful tool for the detection of plasma motions in the line of sight (LOS) and for the plasma diagnostics of solar eruptions. However, unlike many imagers, spectrographs usually can only scan a small region at a low time cadence, making it difficult to catch the transient and localized coronal dimmings. Due to this observational limitation, there has been only a few investigations of coronal dimmings through EUV spectroscopy.

One important finding in dimming regions is the outflows of coronal and transition region (TR) plasma. These plasma outflows were first detected through observations from the Coronal Diagnostic Spectrometer \citep[CDS,][]{1995SoPh..162..233H} on board the SOHO satellite. \cite{2001ApJ...561L.215H} reported CDS observations of two dimming regions. One dimming region was observed at the solar limb, showing a blue shift of $\sim$30~km~s$^{-1}$ in coronal emission lines such as Fe~{\sc{xvi}} 360.8~{\AA} and Mg~{\sc{ix}} 368.1~{\AA}. 
In the other dimming region observed on the solar disk, the authors detected a large blue shift of $\sim$100 km~s$^{-1}$ in the O~{\sc{v}} 629.7 {\AA}~line, suggesting the removal of TR material from the dimming region. Unfortunately, the intensities of available coronal emission lines in this on-disk dimming region were too low to allow an accurate measurement of the Doppler shifts. So it is unclear whether the coronal lines also show such high velocities.  

The EUV Imaging Spectrometer \citep[EIS,][]{2007SoPh..243...19C} on board Hinode has a much higher spatial resolution ($\sim$3$^{\prime\prime}$) compared to SOHO/CDS. From 2006 December 14 to 15, EIS performed repetitive raster scans of NOAA AR 10930.  These raster scans covered almost the entire duration of the CME-associated flare as well as the pre-eruption phase. Using this unique dataset, \citet{2007PASJ...59S.801H} found an obvious blue shift in not only the strong TR line He~{\sc{ii}} 256.32 {\AA} but also several strong coronal lines from the ions of Fe~{\sc{x}} through Fe~{\sc{xv}} in the CME-induced dimming region. These coronal lines are formed in the temperature range of 1--2 MK (assuming ionization equilibrium). The blue shifts are generally of the order of 20 km~s$^{-1}$, and appear to be long-lasting and highly nonuniform in the dimming region. These blue shifts were interpreted as steady plasma outflows that resulted from the CME eruption. 

This dimming event, together with another dimming event captured by EIS in the same AR on 2016 December 13, was also analyzed by \citet{2009ApJ...702...27J} and \citet{2010SoPh..264..119A}. \citet{2009ApJ...702...27J} found that significant outflows are concentrated at locations of strong photospheric magnetic field, suggesting that the strongest outflows are located at the footpoints of the erupting flux rope or coronal loops that stretch or open up as part of the eruption. 
They also found a positive correlation between the blue shift and the magnitude of the dimming, i.e., higher outflow velocities at locations of more mass depletion. \citet{2010SoPh..264..119A} also found that the concentrated outflows are located at the footpoints of coronal loops, which are present before, and re-established after the eruption. The outflow velocities decrease as the coronal intensities in the dimming regions gradually recover to the pre-eruption levels. 

\citet{2009ApJ...693.1306M} examined the same dimming event observed on 2006 December 14 to 15, and found that the nonthermal line width of the Fe~{\sc{xii}} 195.12 ~{\AA} line is obviously enhanced in the dimming region. This excess line broadening was interpreted as being caused by the growth of the amplitude of Alfv\'en waves propagating along the temporarily opened magnetic field lines. As the dimming gradually recovers, the nonthermal width gradually decreases to the pre-eruption level. In a subsequent study of the same event, \citet{2010SoPh..265....5M} examined the asymmetry of the Fe~{\sc{xiii}} 202.04~{\AA} line profiles, and found weak blueward asymmetry at a few localized sites within the dimming region. The blueward asymmetry was interpreted as the on-disk signature of a weak and fast outflow underneath the CME ejecta.

\begin{figure}[t] 
\centering 
\includegraphics[width=0.99\textwidth]{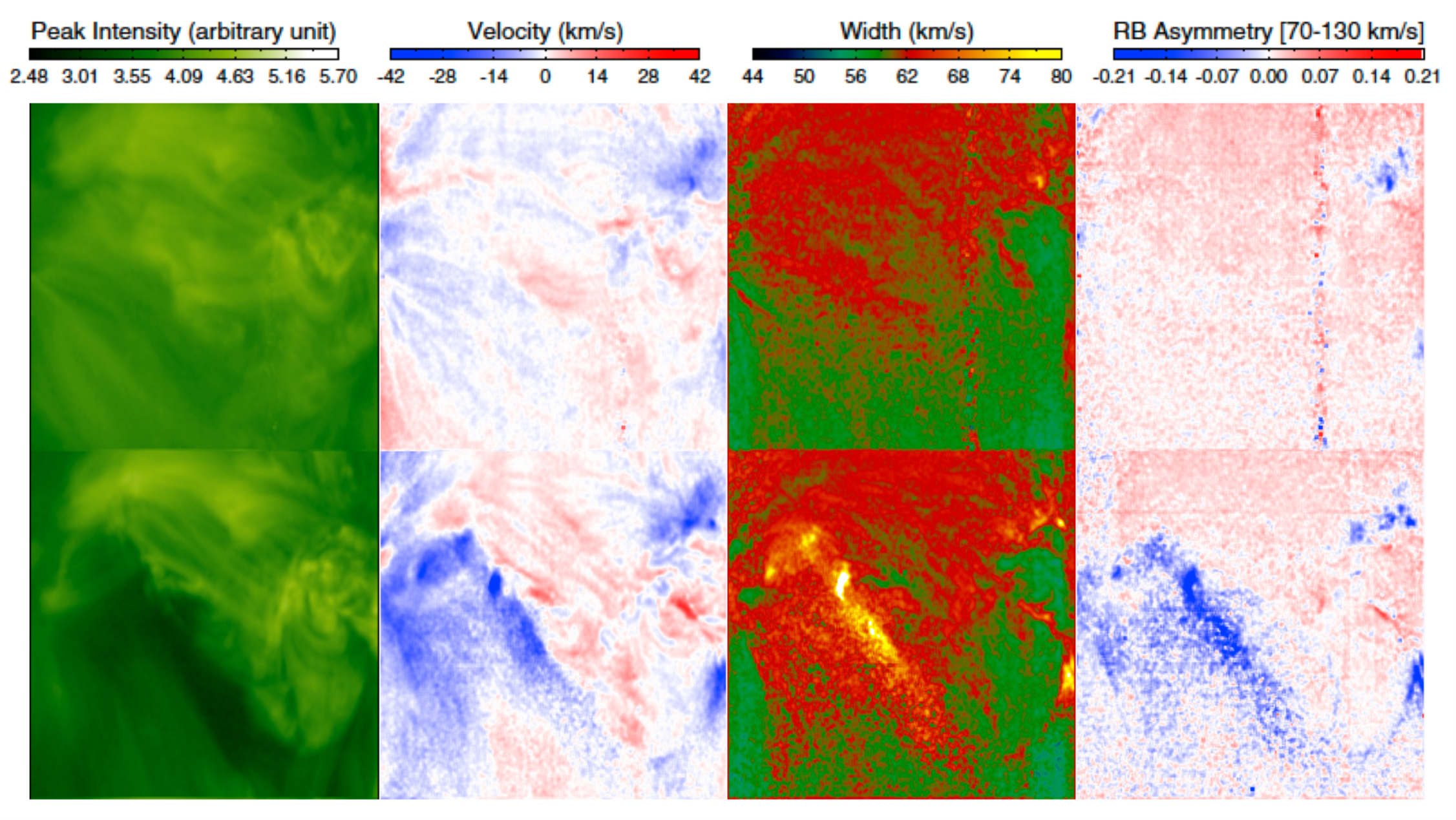}
\caption{Spatial distributions of the peak intensity, Doppler shift and line width derived from the single Gaussian fit and the average RB asymmetry in the velocity interval of 70--130~km s$^{-1}$ for the Fe~{\sc{xiii}} 202.04 {\AA}~line in the 2006 December 14--15 observations. The upper and lower panels show the EIS line parameters before and after the eruption, respectively. Adapted from \cite{2012ApJ...748..106T}.} \label{f1}
\end{figure}

\citet{2012ApJ...748..106T} performed a comprehensive study of four dimming events observed with EIS, including the above-mentioned two events observed during 2006 December 13--15. They first applied a single Gaussian fit (SGF) to each line profile, and obtained the line intensity, Doppler shift and line width. After that, they performed the red-blue (RB) asymmetry analysis for each line profile. An RB asymmetry profile is constructed by taking the difference between the emission of the two line wings as a function of spectral distance (expressed in the velocity unit) from the line center \citep{2009ApJ...701L...1D,2011ApJ...738...18T}. As an example, Fig.~\ref{f1} presents images of the peak intensity, Doppler velocity and line width derived from the SGF and the average RB asymmetry in the velocity interval of 70--130~km s$^{-1}$ for the Fe~{\sc{xiii}} 202.04 {\AA}~line, before and during the eruption on 2006 December 14--15. The blueward asymmetry is much more prominent and prevalent within the dimming region than that reported by \citet{2010SoPh..265....5M}, which is mainly due to the more appropriate determination of the line center by \citet{2012ApJ...748..106T}. A significant correlation is found between each pair of the SGF blue shift, line width and blueward asymmetry. 

\begin{figure}[tbp]
\centering {\includegraphics[width=0.99\textwidth]{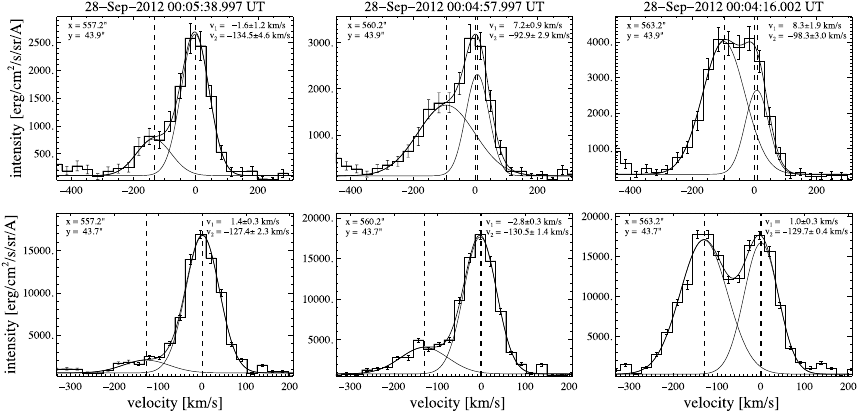}}
\caption{Selected Hinode/EIS spectra of Fe~{\sc{xiii}} 202.04~{\AA} (top row) and Fe~{\sc{xv}} 284.16~{\AA} (bottom row) in a dimming region observed for SOL2012-09-28. A double-Gaussian fit is shown for each observed spectrum, and the
velocities of the two components are displayed in the upper-right corners. Adapted from \cite{2019ApJ...879...85V}.
The movie in the online supplement shows the event evolution in SDO/AIA 94, 304 and 211~{\AA} direct and base-ratio images (from \cite{2019ApJ...879...85V}). The Hinode/EIS spectra shown were obtained in the triangle-shaped western dimming region.} \label{f2} 
\end{figure}

Using the properties of each RB asymmetry profile as a guide \citep{2011ApJ...738...18T}, they also applied a double Gaussian fit to each corresponding line profile. Their detailed analysis suggests that there are at least two emission components in the coronal dimming region: a nearly stationary background component and a secondary high-speed outflow component (also see Fig.~\ref{f2}). Taking the Fe~{\sc{xiii}} 202.04 {\AA}~line for example, the intensity ratio of the two components is usually around 10\% and can reach $\sim$40\% at some locations. The velocity of the outflow component is usually in the range of 50--150 km s$^{-1}$. These results suggest that only part of the coronal plasma sampled along the line-of-sight column in the dimming region flows outward at a velocity of $\sim$100 km s$^{-1}$, indicating that it might be inappropriate to derive the velocity and line width through a SGF as the SGF assumes everything moving at a uniform speed. Consequently, the outflow speed in dimming regions is likely not around 20~km~s$^{-1}$ but can easily reach $\sim$100~km~s$^{-1}$ in the lower corona. The enhanced line width is thus largely caused by the superposition of different emission components \citep[also see][]{2011ApJ...730..113D}. 

An expanding dimming following an EUV wave was caught by EIS on 2007 May 19. Typical characteristics including SGF blue shift and enhanced line width are again found in the dimming region. In addition, there is a ridge of enhanced line width that corresponds to the outer edge of the dimming region \citep{2010ApJ...720.1254C}. An RB asymmetry analysis and RB-guided double Gaussian fit also suggest a blueward asymmetry and thus the presence of high-speed upflow along the whole ridge \citep{2012ApJ...748..106T}. 

\citet{2019ApJ...879...85V} report the presence of high-speed outflows in the dimming region associated with a fast halo CME that occurred around 23:57 UT on 2012 September 27, supporting the findings of \citet{2012ApJ...748..106T}. In this observation, coronal emission lines with formation temperatures of 
$\log(T/{\rm K}) = 5.8-6.3$ reveal distinct double-component spectra at the growing dimming border, indicating the superposition of a stationary and a fast upflowing plasma component there with speeds up to 130~km s$^{-1}$. Interestingly, the upflowing plasma component is even comparable to or dominant over the stationary one at some pixels that were mapped by the Hinode/EIS rastering slit exactly at the time at which the dimming was formed in this location (see Fig.~\ref{f2}). Using the Fe~{\sc{xiii}} lines at 196.55/202.04~{\AA}, \citet{2019ApJ...879...85V} estimated the electron density of the outflow component, which turns out to be of the order of $\log (N_e/{\rm cm}^{-3}) = 9$. This density is two orders of magnitude larger than the lower limit of the outflow density estimated by \citet{2012ApJ...748..106T}.

As discussed in \cite{2012ApJ...748..106T}, the properties of these high-speed outflows appear to be highly similar to those observed at AR boundaries \citep[e.g.,][]{2011ApJ...738...18T}. There are suggestions that the coronal outflows at AR boundaries likely contribute to the slow solar wind \citep[e.g.,][]{2007Sci...318.1585S}. Could the outflows in dimming regions contribute to the solar wind as well? Since CME eruptions are associated with opening of the coronal magnetic field lines, CME-induced dimming regions are essentially transient coronal holes. Thus, very likely there are solar wind streams from dimming regions \citep{2007PASJ...59S.801H,2009ApJ...693.1306M,2012ApJ...748..106T}. \cite{Lorincik:2021} report a CME/dimming event, which showed plasma motions identified in SDO/AIA time-distance plots along multiple funnels located in the dimming region. They lasted for $>$5 hrs, did not change their characteristics along the funnel, and were therefore interpreted as a signature of outflows related to CME-induced solar wind flows along open fields that are rooted in dimming regions. 
Considering the durations of dimmings, these solar wind streams may generally last for a few hours. It is still unknown whether these solar wind streams are fast or slow. Since these solar wind streams just follow the CME ejecta, their kinematics may affect the propagation of CMEs. As mentioned by \cite{2010SoPh..265....5M}, the momentum flux resulting from the high-speed outflows might be a secondary momentum source to the CMEs. 

With the moderate spectral resolution of Hinode/EIS, the high-speed outflow detected in dimming regions shows no obvious temperature dependence. However, in a few dimming events temperature-dependent outflows were detected. For instance, \citet{2007PASJ...59S.793I} reported an outflow that shows a speed increasing from $\sim$10~km s$^{-1}$ at $\log(T/{\rm K}) = 4.7$ to  $\sim$150~km s$^{-1}$ at $\log(T/{\rm K}) = 6.3$. \citet{2012ApJ...748..106T} examined this type of flows, and found that they generally appear immediately outside the (deepest) dimming regions. These temperature-dependent outflows might be evaporation flows related to flare reconnection, or interaction between the opened field lines in the dimming region and the closed loops in the surrounding plage region.

\begin{figure}[tbp] 
\centering {\includegraphics[width=0.61\textwidth]{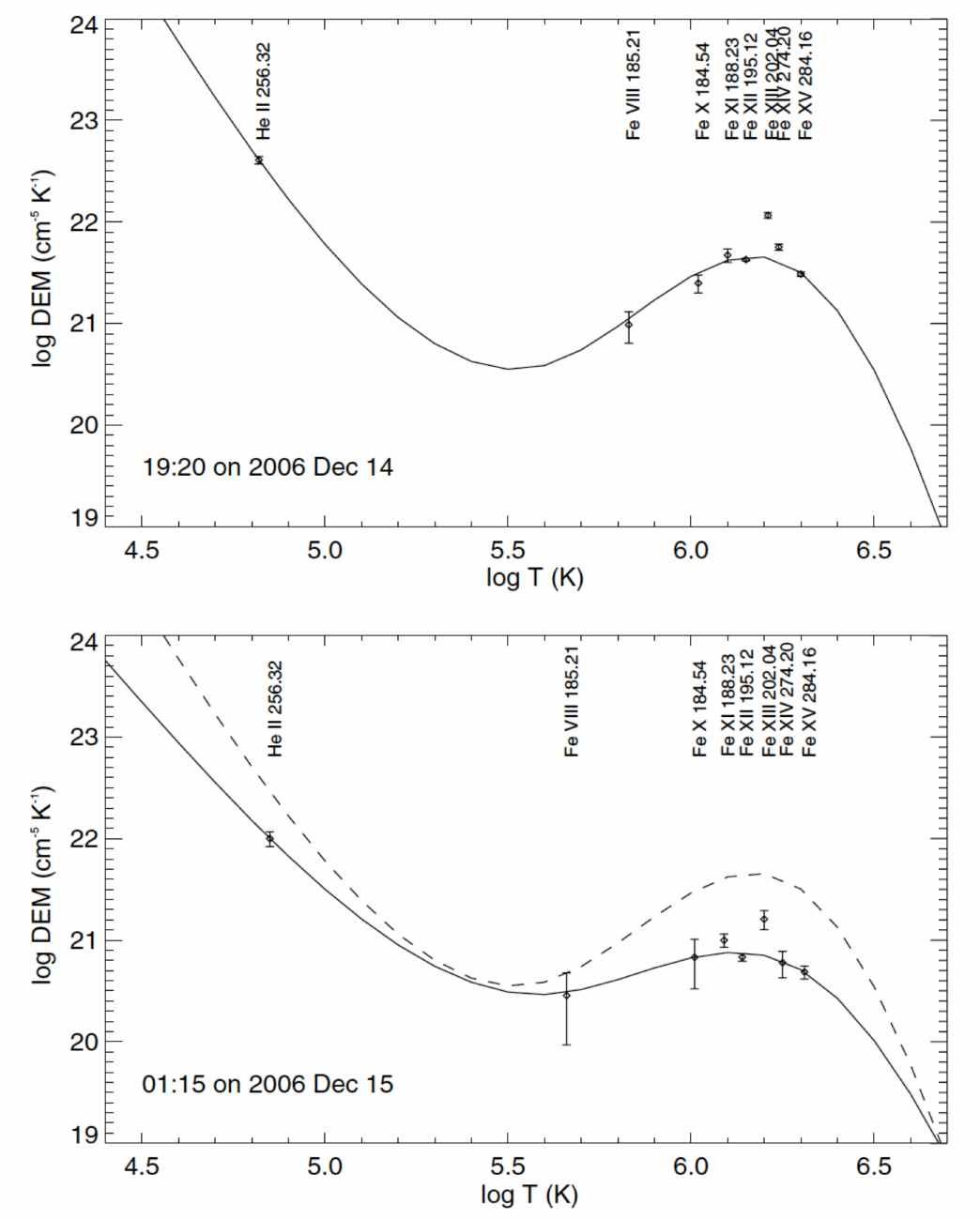}} 
\caption{Differential Emission Measure (DEM) curves before the eruption (upper panel) and in the dimming region (lower panel) in the 2006 December 14-15 observations. In the lower panel the DEM curve before the eruption (dashed line) is also overplotted for a direct comparison. Adapted from \cite{2012ApJ...748..106T}.} \label{f3}
\end{figure}

\subsubsection{Plasma parameters} 
\label{sec:plasma}

Coronal dimmings can be observed in a certain range of temperatures. For instance, \citet{Harrison:2000} analyzed EUV spectra taken by SOHO/CDS low in the off-limb corona below an ascending CME, and found a gradual darkening of the coronal Mg~{\sc{ix}} 368.1 {\AA}~line. The dimming is not evident in lower-temperature lines. This result suggests that the dimming is mainly due to depletion of the million-Kelvin plasma. However, a subsequent study of off-limb dimming events by \citet{2003A&A...400.1071H} found that the TR line O~{\sc{v}} 629 {\AA}~also reveals significant dimming. With a better performance over CDS, EIS observations allow a better temperature diagnostics for dimming regions. From EIS observations, signatures of dimming are usually found in emission lines with formation temperatures of 
$\log(T/{\rm K}) = 5.8-6.3$
\citep{2007PASJ...59S.801H,2012ApJ...748..106T,2019ApJ...879...85V,2023SoPh..298..130T}. In addition, dimming signatures can also be found in the strong TR line He~{\sc{ii}} 256.32 {\AA}. 

From EIS observations of multiple emission lines formed over a wide range of temperatures, \citet{2012ApJ...748..106T} obtained Differential Emission Measure (DEM) curves at the pre-eruption phase and of the dimming region. From Fig.~\ref{f3} we can see that the main difference between the two curves is the reduced emission measure at typical temperatures of the AR corona ($\log(T/{\rm K}) = 6.0-6.3$) in the dimming region, suggesting that the dimming is mainly due to the escape of material with a temperature of $\log(T/{\rm K}) = 6.0-6.3$ when the magnetic field lines open up. From the DEM curves, we see no obvious change of the peak temperature ($\log(T/{\rm K}) \approx 6.15$). However, it is worth noting that the lower-temperature part of the DEM curves is less constrained due to the lack of strong TR lines in the EIS observations. With a better temperature coverage, the Spectral Imaging of the Coronal Environment \cite[SPICE;][]{Andersen2020} instrument on board Solar Orbiter may provide a more accurate understanding of the temperature structures of dimmings.

Diagnostics of the electron density can be achieved through EUV spectroscopy. Using the Si~{\sc{x}} 356/347 {\AA}~line pair observed with CDS, \citet{Harrison:2000}
found a decrease of the electron density by 37\%. A similar density drop was also found by \citet{2003A&A...400.1071H} using the same line pair. \citet{2012ApJ...748..106T} chose both the line pairs Fe~{\sc{xii}} 186.88/195.12 {\AA}~and Fe~{\sc{xiii}} 203.82/202.04 {\AA}~for density diagnostics for the 2006 December 14--15 observations. They found a decrease of the average electron density from $\log (N_e/{\rm cm}^{-3}) =$ 8.89 to 8.67 using the Fe~{\sc{xii}}~diagnostics, and from $\log (N_e/{\rm cm}^{-3}) =$ 8.70 to 8.58 using the Fe~{\sc{xiii}}~diagnostics. The density decrease, together with the fact that the peak temperature of the DEM curve does not change too much when the dimming occurs, strongly suggests that the dimming is mainly due to mass loss rather than temperature change.

\begin{figure}[tp] 
\centering {\includegraphics[width=0.99\textwidth]{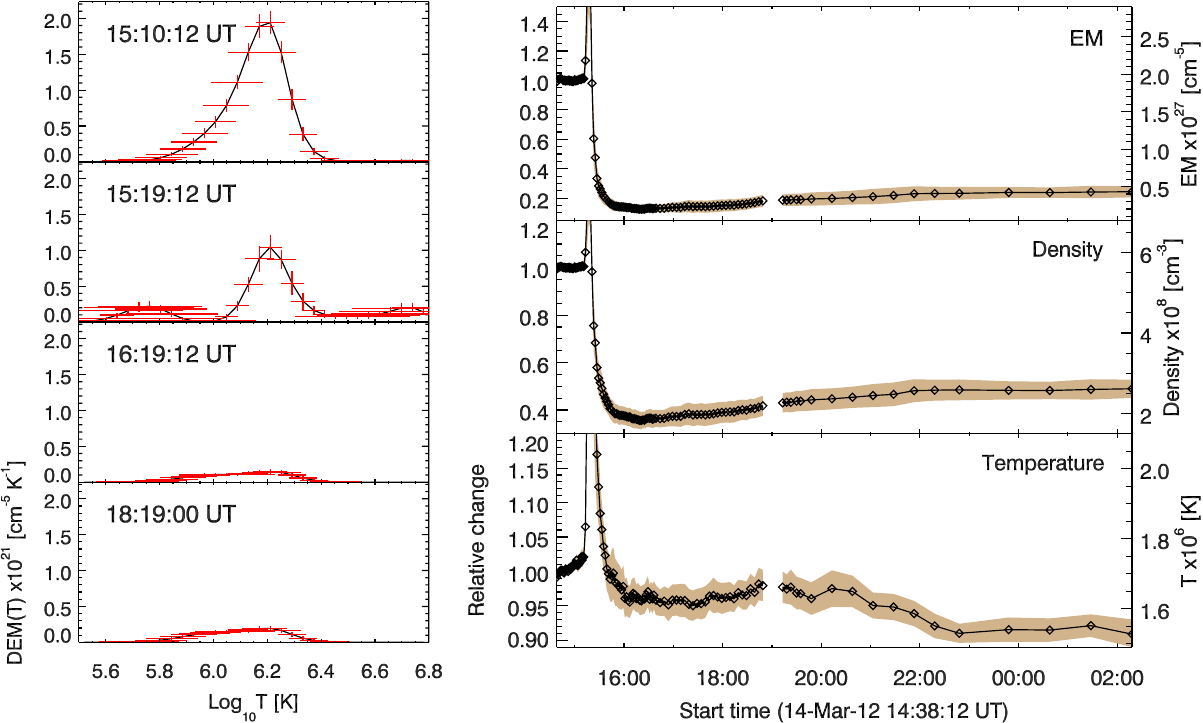}} 
\caption{Left: Reconstructed DEM profiles derived from a small area inside the Western core dimming region for SOL2012-03-14 (M2.8); see Fig.\ \ref{fig:vanninathan_2018_lc}. 
Right: Time evolution of total emission measure, density and mean plasma temperature derived from the DEM profiles over 12 hrs during the event. 
Adapted from \cite{Vanninathan:2018}.
} \label{fig:vanninathan_2018_dem}
\end{figure}

\cite{Vanninathan:2018} chose a different approach to study the plasma evolution in the dimming regions. 
They used the high-cadence multi-band EUV imagery in the six coronal AIA filters to reconstruct DEM maps and to analyze their evolution over time. Selecting small subfields within core and secondary dimming regions, they analysed the density and temperature changes in the different types of dimmings. 
Figure \ref{fig:vanninathan_2018_dem} shows reconstructed DEM profiles in the Western core dimming region during the M8.2 two-ribbon flare  SOL2012-03-14 that was accompanied by a relatively slow CME ($v = 410$~km~s$^{-1}$). Snapshots of the event evolution and the analysed region (green box) are shown in Fig. \ref{fig:vanninathan_2018_lc}. The first DEM profile plotted in Fig.~\ref{fig:vanninathan_2018_dem} (left column) shows the state immediately before the eruption. One can see a broad DEM profile with the bulk of the emission in the temperature range $\log(T/{\rm K}) \approx 6.0-6.4$ ($T \approx 1.0-2.5$ MK), indicative of quiet-Sun and AR region plasma. The following three time steps represent reconstructed DEMs during the impulsive phase of the dimming and 3 hrs later, showing that in the core dimming the overall DEM curve strongly diminishes.  

The right column in Fig.~\ref{fig:vanninathan_2018_dem} shows the corresponding time evolution of the resulting total EM, plasma density and DEM-weighted temperature over a period of 12 hours. When the CME lifts off, the total EM inside the core dimming area decreases suddenly by about 90\%, the density by more than 60\%, and the temperature drops by less than 10\%. These sudden changes occur over time scales of the order of some 10 minutes.
This behavior is representative for the evolution in the core dimming region in all six events studied in \cite{Vanninathan:2018}. All the events show an impulsive and strong decrease of density by 50\%--70\%, and then staying at these reduced levels for the overall time of analysis.  
In five of the events also an associated decrease in temperature is observed, by 5\%--25\%. The secondary dimming regions also show a
distinct decrease in density, but less strong (10\%--45\%), and no systematic changes in temperature. In the core as well as in the secondary dimming
the density changes are much larger than the temperature changes, which demonstrates that the dimming regions are mainly caused by plasma evacuation.

Already very early on, coronal dimmings have been explored for estimating the mass loss due to the CME (see, e.g., the pioneering studies by \cite{Sterling:1997,Zhukov:2004}). In Sect. \ref{sec:CMErelation}, we give a comprehensive overview of the different studies deriving CME mass from dimmings and how they compare to the CME mass calculated from coronagraphic white-light observations. Here we give brief account on efforts that have been made to estimate the mass loss due to the CME eruption from spectroscopic observations of dimmings.
The simplest method is to multiply the density change and the emission volume, and the total mass loss \textit{M} can be expressed as follows \citep{2003A&A...400.1071H,2012ApJ...748..106T}:
\begin{equation} 
	M = \delta N S L m_p
\end{equation}
where $\delta$\textit{N}, \textit{S}, \textit{L} and $m_p$ represent the change of electron number density, dimming area, depth of the dimming region and proton mass, respectively. The density change could be inferred using the line pair of Fe~{\sc{xii}} 186.88/195.12 {\AA}~\citep{2012ApJ...748..106T} or Si~{\sc{x}} 356/347 {\AA}~\citep{2003A&A...400.1071H}, whose formation temperatures are close to the DEM peak temperature in dimming regions. The area of the dimming region is obtained from simultaneous full-disk coronal images. Assuming that the emission volume is as deep as it is wide, the depth of the dimming region can be calculated as the square root of the dimming area.

Another method that was used by \citet{2009ApJ...702...27J} and \citet{2012ApJ...748..106T} takes the emissions at different temperatures into account. It first takes the formation heights of TR lines and coronal lines from the VAL3C model \citep{1981ApJS...45..635V} and the coronal model of \citet{1978SoPh...60...67M}, respectively, and then obtains the densities at these heights from an empirical density model \citep{2000asqu.book.....C}. After that, any density-sensitive line pairs available in spectroscopic observations are used to derive electron densities, which are then used to scale the model densities. From the scaled model densities and the intensity changes of many EIS lines, changes of density at different heights can be derived. The total mass loss can be expressed as 
\begin{equation} 
	M = \sum \delta N(h_i) S(h_i) \delta h_i m_p
\end{equation}
where $\delta N(h_i)$ and $S(h_i)$ are the density change and dimming area, respectively, at the formation height $h_i$ of each line, and $\delta h_i$ represents the vertical extension of the line formation volume. 

The estimated mass loss from an on-disk dimming region is generally about 20--60\% of the corresponding CME mass estimated from coronagraph observations, suggesting that a significant part of the CME material is coming from the region where dimming occurs subsequently \citep{2009ApJ...702...27J,2012ApJ...748..106T}. In other words, the dimming regions are indeed source regions, or at least part of the source regions, of the subsequent CMEs. Obviously, these estimations of mass loss are subject to significant uncertainties. However, considering the fact that the estimation of CME mass from LASCO data is also subject to uncertainties caused by several assumptions, the derived mass loss should be considered to be a reasonable number for comparison with the corresponding CME mass (see also Sect 5.1).

\subsection{Recovery phase}
\label{sec:recovery}

The dimming recovery contains various important pieces of information, namely on the post-eruption state of the solar corona, on the interplanetary connection to the ICME, as well as on the formation of the corona itself. 
Despite these manifold aspects, there exist only a limited number of studies addressing this topic.

\begin{figure}[tbph] 
\centering {\includegraphics[width=0.86\textwidth]{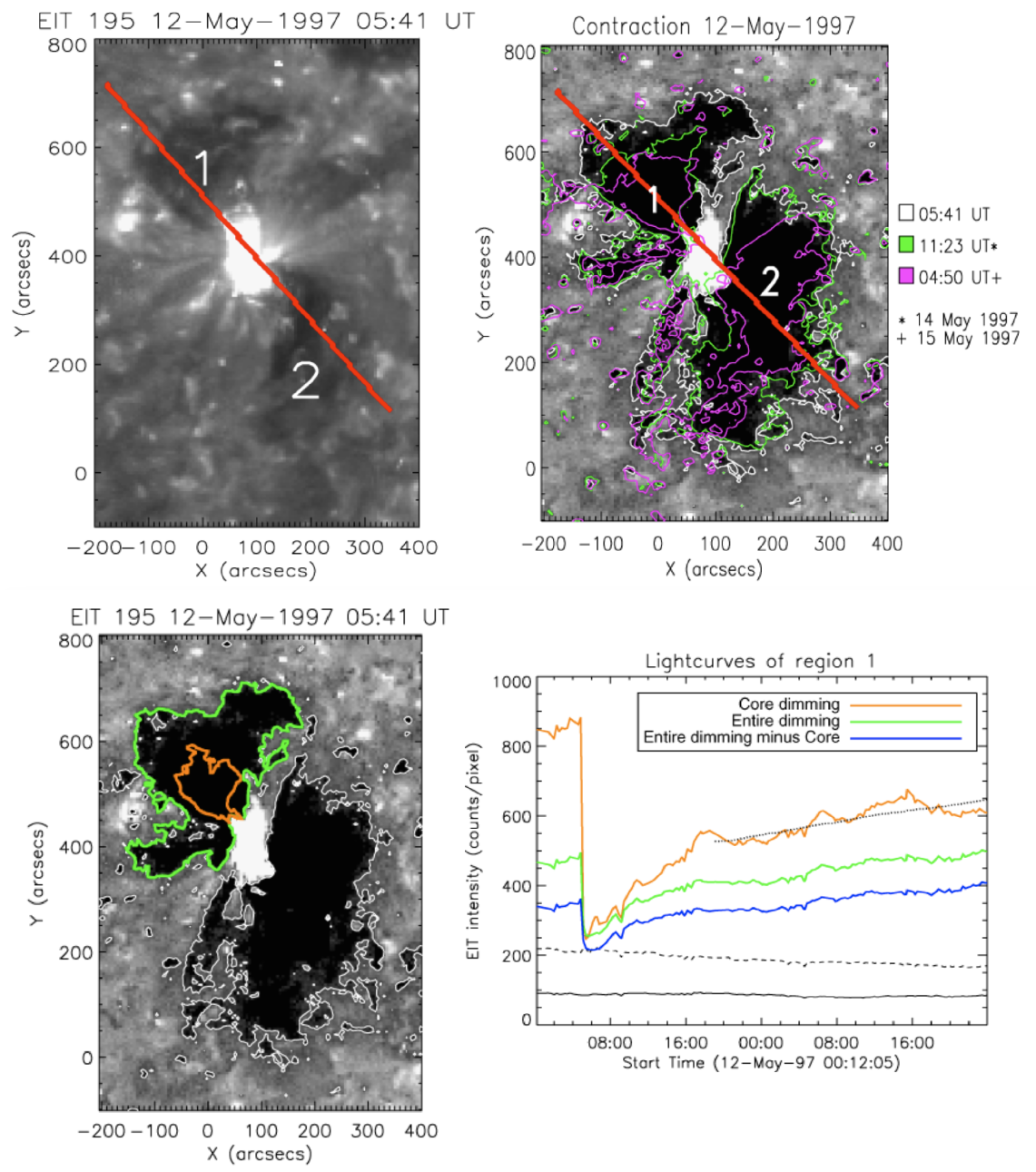}} 
\caption{Illustration of dimming recovery. Top left: SOHO/EIT 195~{\AA} image during the impulsive phase of the SOL1997-05-12 dimming. Top right: corresponding base difference image, with contours overlaid at three times as shown in the legend. Bottom left: EIT 195~{\AA} base difference image with contours showing the identified Eastern core (orange) and entire (green) dimming region. Bottom right: corresponding light curves in these regions.
Adapted from \cite{Attrill:2008}.}
\label{fig:Attril_2008}
\end{figure}

Basically there are two ways how dimmings can recover: either they shrink in area, or the emission inside the dimming volume increases again. 
The first systematic study that addressed that issue was 
\cite{Kahler:2001} who found that in all of the 19 dimmings they studied in Yohkoh/SXT images, the dimmings recovered by area contraction and not by an increase of the brightness inside the dimming region. 
\cite{Attrill:2008} studied the recovery of three cases in SOHO/EIT data, and found that they recover by both shrinkage in area and a progressive increase of the emission in the dimming region. The decrease of the spatial extent of the dimmings occurs predominantly by contraction from outward to inward, 
but in a fragmentary and inhomogeneous manner. Figure~\ref{fig:Attril_2008}
shows an example from their study. 
The contours of the dimming boundary plotted for three different time steps demonstrate the shrinkage of the dimming area (top panels). The bottom panels show the intensity evolution inside the dimming region, which quickly reaches a minimum followed by a gradual increase over more than 40 hours. 
The gradual increase of the emission during the dimming recovery is a general behavior found from EUV filtergrams \cite[e.g.,][]{Reinard:2008,Miklenic:2011,Dissauer:2018b,Chikunova:2020}, for another example see Fig.\ \ref{fig:miklenic_2011_F2}.

The apparent tension between the \cite{Kahler:2001} and \cite{Attrill:2008} viewpoints may result from the differences of X-ray and EUV observational domains these studies used.
For one of their events under study, \cite{Attrill:2008} independently confirmed  the findings of \cite{Kahler:2001}, i.e.\ that in Yohkoh/\-SXT there is no emission increase observed during the recovery of the dimming, whereas in the SOHO/EIT emission there is. They interpreted this behavior by the different temperature sensitivity of SOHO/EIT 195~{\AA} ($T \approx 1.5$ MK) and Yohkoh/SXT, which mainly samples plasmas with temperatures $T > 3$ MK. 
This suggests that the dimming recovery process does not 
heat the coronal plasma to a temperature that is high enough to be detected in Yohkoh/SXT data.

\cite{Reinard:2008} made a statistical study of 96 dimming events observed by SOHO/EIT during 1998--2000. They report that the time profiles usually show a sharp drop followed by a gradual recovery, with a mean recovery time of $(4.8 \pm 0.3)$ hrs as derived from the area evolution, and $(4.3 \pm 0.3)$ hrs derived from the brightness evolution of the dimmings. 
The median dimming duration of the sample was 7.4 hrs.
In the majority of their cases ($\approx$75\%), the recovery profiles could be fitted by a single linear slope, whereas the remaining events reveal a two-part recovery profile in which the initial slope is steeper than the later one (for an example of a two-step recovery profile, see Fig.\ \ref{fig:Attril_2008}).

If one distinguishes between the different (morphological) types of the dimmings and concentrates on small subregions, different behaviors and time scales are found. Figure \ref{fig:vanninathan_2018_dem} shows the evolution of the temperature, emission measure and density in a small subregion inside a ``core" (flux rope) dimming from the \cite{Vanninathan:2018} study. The corresponding evolution in the different SDO/AIA channels is shown in Fig.\ \ref{fig:vanninathan_2018_lc}. It is interesting to note that in the core dimming region the brightness and plasma density do not recover for at least 10 hours, whereas the secondary dimmings in the same event start to recover already after 1--2 hours. A similar behavior was observed in all of the six events studied in \cite{Vanninathan:2018}. They interpreted the unaltered brightness inside the core dimming region during the dimming recovery phase to indicate that the erupted flux rope is still connected to the Sun.
A recent dimming lifetime study by \cite{Ronca:2024} finds agreement with the results of \cite{Reinard:2008} and \cite{Vanninathan:2018} in terms of the two-step recovery process and the different recovery times for different dimming regions. Some dimming regions do not recover within the three days of the analysis interval. The authors identify the expansion of coronal loops into  the dimming region as a primary mechanism for dimming recovery.

As for the physics that lies behind the recovery of the dimmings, there have been different processes discussed that may play a role. \cite{2009ApJ...702...27J} suggested that there may be a sufficient supply of mass from the lower transition region. This interpretation is related to the strong outflows they observed in their events under study (cf.\ Sect. \ref{sec:flows}),  which may be a response of the transition region to the pressure gradient that develops when the CME
eruption evacuates the plasma in the lower corona. \cite{Attrill:2008} point out that any dimming recovery mechanism must account for the following observational facts: decrease in dimming area, increase in dimming brightness, and connection of the CME to the Sun, while the dimming region is already retracting. This last item stems from observations of unidirectional electron beams inside the associated ICME at 1 AU  more than 70 hrs after the eruption in the famous 12 May 1997 event (shown in Figs.~\ref{fig:thompson98} and \ref{fig:Attril_2008}) while the dimming has already disappeared \citep{Attrill:2008}. 
The observation of bi- or unidirectional electron beams inside ICMEs provide evidence that the erupting flux rope is still  connected to the Sun, either both legs or only one \cite[e.g.,][]{Gosling:1987}. 
\cite{Attrill:2006,Attrill:2008} propose that interchange reconnection between the ``open'' flux of the dimming region, small coronal loops and emerging flux play a key role to facilitate the recovery of dimmings. 
They argue that this process disperses the  ``open'' magnetic field that forms the dimming region out into the surrounding quiet Sun. 
This causes the decreased emission of the dimming to recover, while the CME is still magnetically connected to the Sun.
They also point out that in quiet Sun regions the time-scale for replenishment of magnetic flux is of the order of 1.5--3 days \citep{Schrijver:1997}, which may be sufficient to explain the recovery of the peripheral regions of the dimming.

Another plausible mechanism for the apparent recovery of dimming regions is the direct interaction of the erupting flux with itself, or with preexisting regions of open flux that are nearby. The former slowly shrinks the footprint of closed flux attached to the surface (see Sect.~\ref{sss:shriFRdimmings}).
The latter provides a means for open flux to become rooted near to the primary flare current sheet, creating favorable conditions for subsequent reconnection, which will disconnect the flux entirely from the Sun (see Sects.\ \ref{ss:open_flux} \ref{ss:sim_example1}). In contrast to the \citet{Attrill:2008} scenario, these mechanisms involve the erupting flux and large-scale flux systems as a means for eventual relaxation.

We note that studies on the physics of the recovery of dimmings are still scarce. Also, the observations that the dimmings already disappeared while the CME is still connected to the Sun (as evidenced by bi- and uni-directional electron beams) is only based on very few events and there exists yet no statistics on it. Other mechanisms, such as global shock waves, can also affect the beam properties and confuse the interpretation in terms of connectivity. So, it is unclear how/whether this property has to be built into the interpretations. Lastly, because the hydrodynamic timescales for the plasma to adjust to changing magnetic conditions should be relatively short (with asymptotic values set by the coronal heating rate and field line geometry), while the magnetic timescales can vary drastically as a function of height, it is likely that the recovery is key diagnostic for how the large-scale magnetic field of the corona relaxes and reforms after an event.

\subsection{Pre-eruption dimmings} 
 \label{sec:pre-eruption}
 
 Although the phenomenon of coronal dimmings has been commonly observed during or after eruptions, dimming signatures before the actual eruption have also been reported in some case studies, so-called pre-eruption dimmings. \citet{Gopalswamy1999, Gopalswamy2006} conducted a multi-wavelength study of the X-class flare SOL1998-04-27, accompanied by the CME eruption that took place before the flare start. SOHO/EIT observations revealed a two-stage coronal dimming, with a weak coronal dimming starting already two hours before the rise of the flare soft X-ray emission, followed by a strong rapid dimming 75 minutes later. The weak and gradual dimming before the eruption likely indicates quasi-steady expansion of the coronal structure before its catastrophic change, and detection and tracking of these dimming signatures can, therefore, provide important diagnostics of the CME evolution prior to its eruption. However, since such an expansion is quite slow, the associated dimming signature is rather weak and hard to detect.

\begin{figure}[tbp] 
\centering {\includegraphics[width=0.99\textwidth]{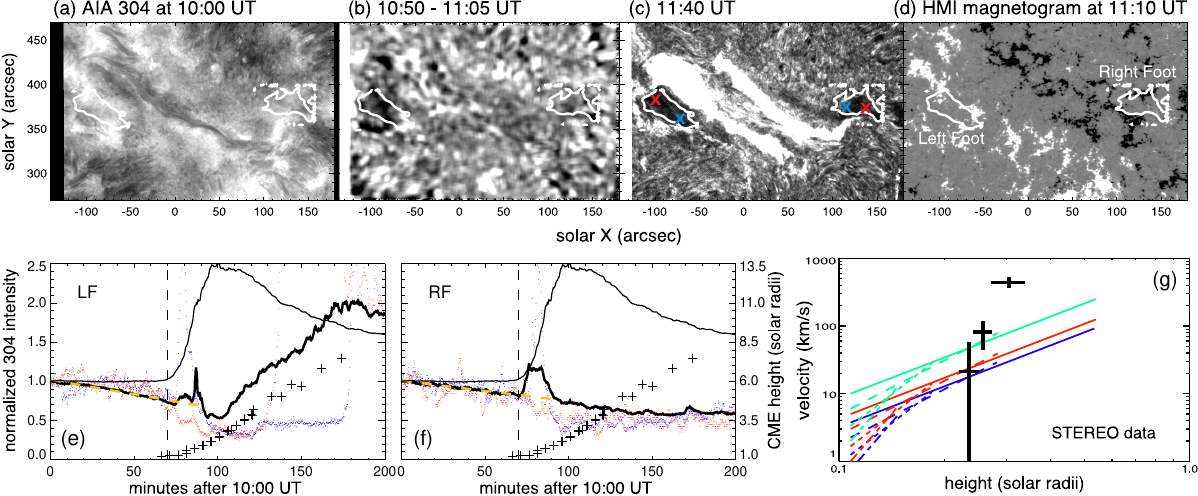}} 
\caption{Example of pre-eruption dimming. (a)--(d) Twin dimmings observed in the 304~\AA\ passband by SDO/AIA, which are located at the two ends of flare ribbons in magnetic fields of opposite polarities, prior to the CME eruption of a C5.7 two-ribbon flare on 2011 December 26. The contours indicate the two dimming regions, the left foot (LF) region and the right foot (RF) region, respectively. (e)--(f) The total light curves in the 304~\AA\ passband of the dimming region (thick black), and light curves of sample pixels (red and blue) in the left and right foot regions, in comparison with the total light curve of the active region (thin black), and the measured height in the plane of sky of the STEREO observation (symbols). The positions of the sample pixels in the LF and RF regions are denoted by crosses (red and blue) in panel (c). In panels (e) and (f), the orange dashed straight lines show the least-squared fit of the dimming depth (thick black) to a linear function of the time-lapse.  (g) Height-velocity graph estimated using the pre-eruption dimming evolution with different expansion models (indicated by different line styles) and varying dimming rate, 0.003 (blue), 0.004 (red), and 0.008 (green) per minute, respectively, together with the observed height-velocity measured from the STEREO observations (symbols).  Adapted from \cite{Qiu2017}.} \label{fig:qiu2017}
\end{figure}

 \begin{figure}[tbp] 
\centering {\includegraphics[width=0.99\textwidth]{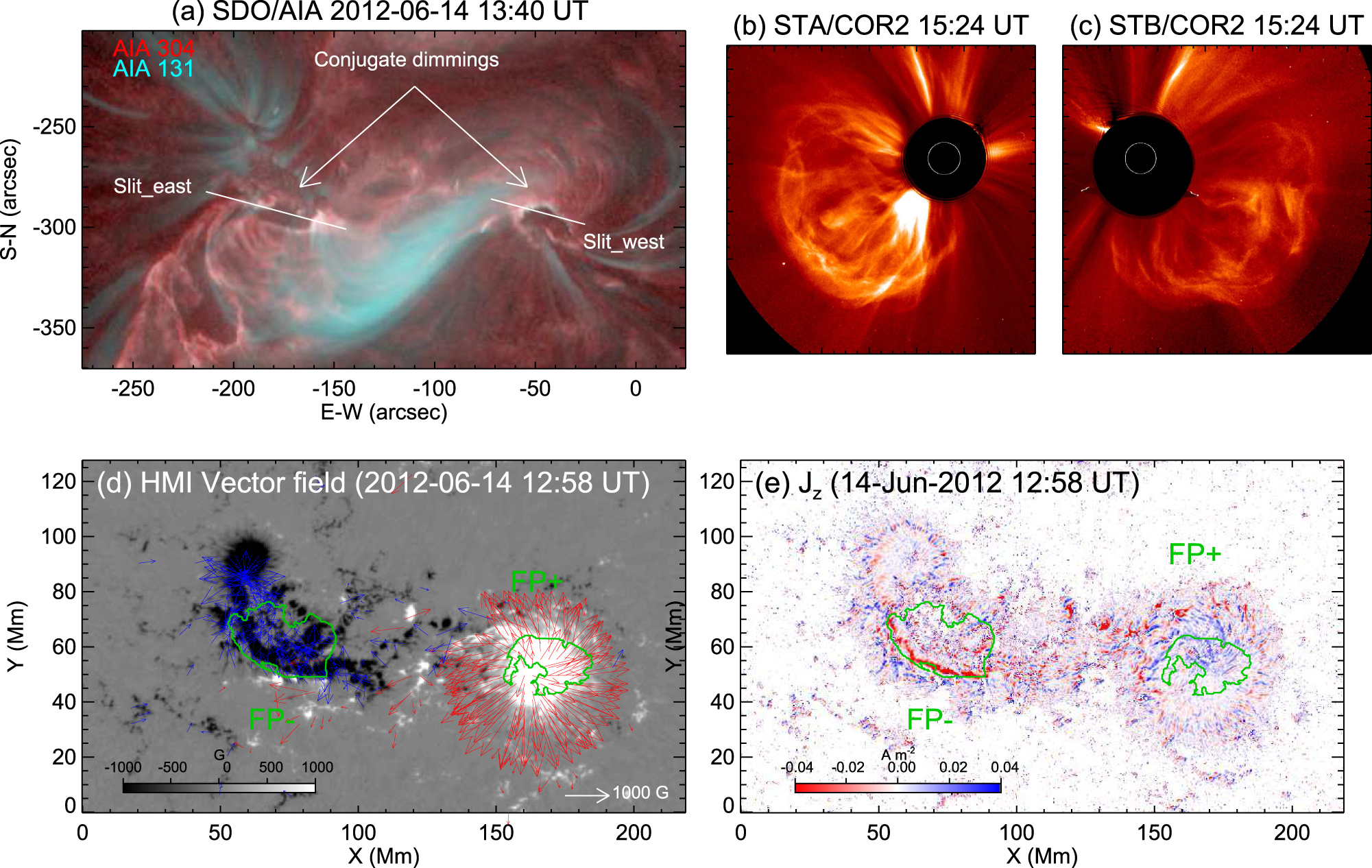}} 
\caption{(a) An M1.9 flare (SOL2012-06-14) associated with a sigmoid, and coronal dimming observed in 131~\AA\ and 304~\AA\ by SDO/AIA. (b) Associated halo coronal mass ejection observed by COR2 on STEREO A and B. (c) Vector magnetic field from SDO/HMI and (d) vertical electric current density in the host active region, superimposed with the contours that outline the coronal dimmings at the footprint of the erupting sigmoid. Adapted from \cite{Wang2019}.
The movie in the online supplement shows the event evolution in AIA 1600 and 94~{\AA} multi-color images (from \cite{Wang2019}).
} \label{fig:wang2019a} 
\end{figure}

Observations by SDO/AIA with improved tempo-spatial resolutions and sensitivity allow to detect and locate weak dimmings before the CME eruption. \citet{Qiu2017} and \citet{Zhang2017} have studied two eruptive flares respectively, and both reported pre-eruption coronal dimmings observed in multiple EUV lines at temperatures 1--3 MK, which proceeded for nearly one hour before the onset of flare reconnection (signified by brightened two flare ribbons), CME eruption (visible in STEREO/EUVI observations), and rapid strong dimming at the same locations of the pre-eruption dimming. In both events, the two-stage dimming occurred in a pair of conjugate dimming regions, adjacent to the two flare ribbons and residing in magnetic fields of opposite polarities (Figure~\ref{fig:qiu2017}). Within an hour, the brightness of the twin-dimming regions measured in the EUV 304~\AA\ passband decreased by 20\%.  
If the weak and gradual dimming is caused by the slow expansion of the overlying coronal structure, the measured dimming rate provides an estimate of the speed of the expansion of the order of a few kilometers per second (Figure~\ref{fig:qiu2017}). The eruptive event occurred in a decaying bipolar region, and the configuration of the twin-dimming and two-ribbons of the subsequent flare manifest the standard model of two-ribbon flares driven by an overlying magnetic flux rope (MFR) that evolves towards eruption \citep{Forbes2000, Moore2001, Aulanier:2012}. 

\begin{figure}[htbp] 
\centering {\includegraphics[width=0.99\textwidth]{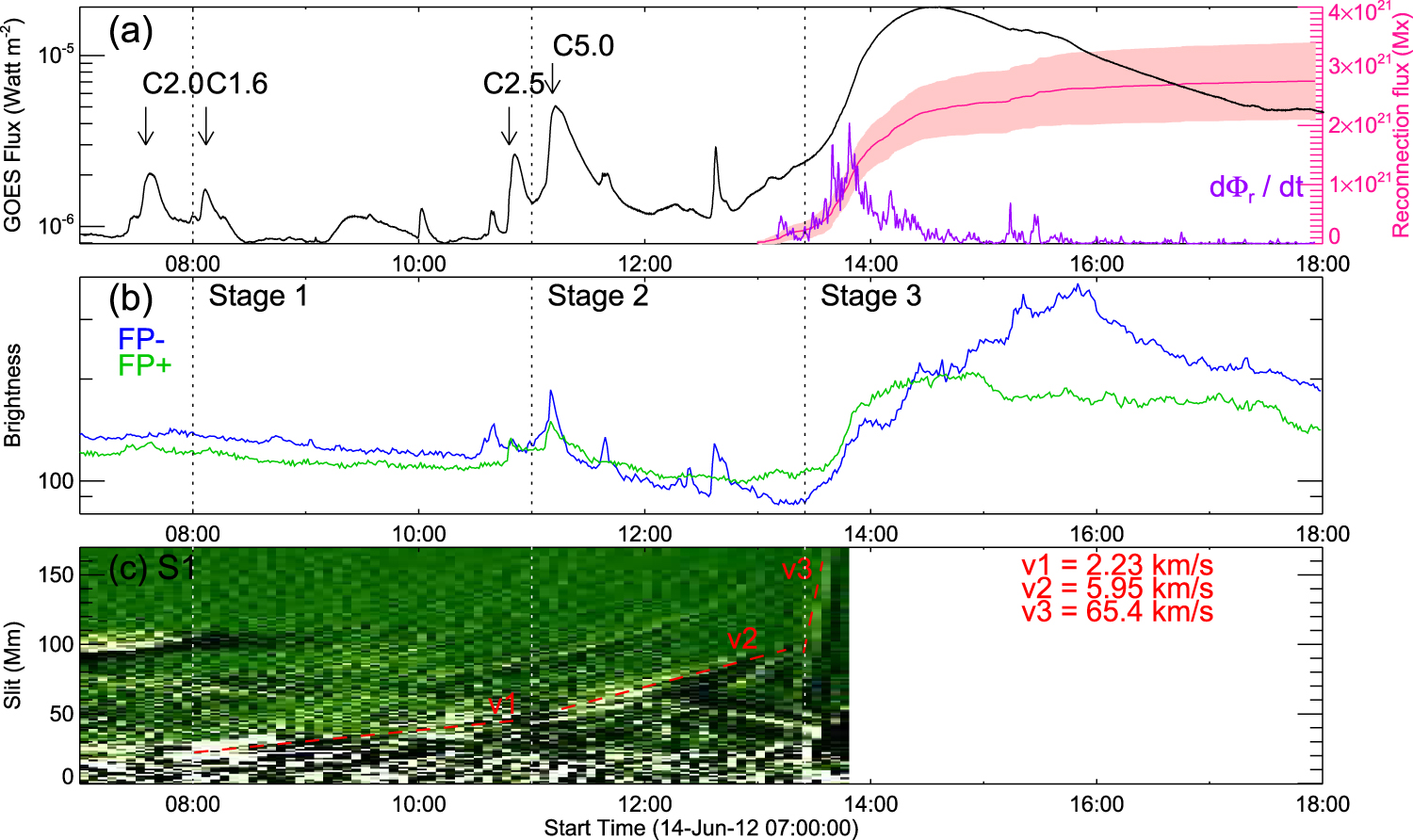}} 
\caption{(a) GOES 1--8~\AA\ soft X-ray light curve before and during the M1.9 flare. (b) Light curves of SDO/AIA 304~\AA\ emission in the twin core dimming regions showing persistent dimming prior to the flare. (c) Slow rise of a coronal structure tracked in STEREO/EUVI 195~\AA\ images. Adapted from \cite{Wang2019}.} \label{fig:wang2019b}
\end{figure}

Using SDO/AIA observations in multiple passbands, \citet{Wang2019} have also identified twin-dimming signatures starting nearly five hours prior to an M-class two-ribbon flare associated with a fast CME (Figure~\ref{fig:wang2019a}). Similar to \citet{Qiu2017}, the twin-dimming was adjacent to the two flare ribbons, and at the onset of the flare and fast CME, the weak gradual dimming rapidly turned into a strong dimming.  \citet{Wang2019} have also been able to detect and track a coronal structure in STEREO-EUVI observations, finding that, coincident with the gradual dimming, the coronal structure was slowly rising at a speed of a few kilometers per second (Figure~\ref{fig:wang2019b}).  

Importantly, dimmings were also detected in the EUV 304~\AA\ passband, which usually encompasses much smaller areas than detected in the other EUV lines, for example, the 193~\AA\ line. Since emission in the EUV 304~\AA\ passband mostly originates from the upper chromosphere/transition region, dimming signatures observed in this passband likely map the feet of the expanding or erupting coronal structure.
In fact, \citet{Wang2019} have measured strong non-neutralized vertical electric currents of opposite dominant signs in the twin-dimming regions determined from the signatures in the 304~\AA\ passband (Figure~\ref{fig:wang2019a}d, e). These observations support the scenario that the pre-eruption dimming is caused by an expanding coronal structure, such as a pre-existing CME flux rope. \citet{Wang2019} have estimated the mean twist $\tau$, or the number of turns of the field lines, in the magnetic flux rope believed to be anchored at the twin-dimming regions of strong electric currents, and found $\langle \tau \rangle \sim 2$ prior to the eruption of the flux rope, which is consistent with the estimate by \citet{James2018} who studied the same active region and CME with nonlinear force-free field extrapolation.

\citet{Wang2022} recently conducted a survey searching the database of flares, CMEs, and vertical electric currents observed by SDO for dimming signatures related to CMEs. They found 28 CMEs in active regions exhibiting significant vertical current $I_z$, which are associated with a twin-dimming signature either before or after the eruption. The 9 pre-eruption dimming events are mostly
accompanied by coronal signatures like sigmoids, expanding loops, or hot channels, and in 4 of them, strong and non-neutralized vertical electric current is present within the dimming regions, although there is usually an asymmetry between the two conjugate feet, namely, the net current in one region tends to be stronger than in the other.  

These examples of pre-eruption dimming signatures, though rare, demonstrate the potential to diagnose dynamic and magnetic properties of CME magnetic flux ropes prior to their eruption, and can help identify physical mechanisms leading to the MFR eruption. In particular, regular measurements of magnetic fields in the corona are not available by the present technical capabilities; instead, magnetic properties of MFRs can be studied if the feet of MFRs on the solar disk can be identified using dimming signatures aided with coronagraph observations from a different vantage point. 

In spite of its promising diagnostic potential, pre-eruption dimming has been rarely observed, because the dimming signatures produced by ``expansion" are relatively weak compared with those caused by ``eruption". We may estimate the sensitivity
required to observe pre-eruption dimmings. As a coronal structure expands -- eruption being an extreme case of expansion -- 
plasma density decreases. A coronal dimming observed in optically-thin lines in X-ray or EUV wavelengths
is usually caused by reduced emission measure along the observer's line of sight $\xi = \int n^2 dl$,
convolved with the contribution function for the line(s) and observing instrument's response function $I_{\lambda} = \int n^2(T) R_{\lambda}(T) dl$. The expansion
is a dynamic process, reflected in the temporal evolution of the coronal dimming. Therefore, analysis
of the time variation of the dimming depth provides approximate diagnostics of the dynamics of the
expansion. Such analyses have been conducted with observations of {\em post-eruption} coronal dimmings in a number of studies, where the expansion is usually approximated as an adiabatic process \citep{Aschwanden:2009, Schrijver:2011, Mason:2016, Cheng:2016}. In particular, \cite{Aschwanden:2016} found the time variation of the emission measure $\xi$ of expanding plasmas from the coronal dimming evolution observed in six EUV passbands by SDO/AIA, and applied forward-modeling analysis of the emission measure variation,
assumed to be caused by the CME expansion with a prescribed kinematic profile. The analysis was applied to 400 eruptive flares observed on the disk, and the derived kinetic energies of CMEs are statistically comparable with those measured using data by LASCO that observed the same events. 

We denote the volume, mean density, and mean temperature of the plasma by $V$, $n$, and $T$, respectively, ignoring the temperature and density variation inside the plasma.  
During the expansion, the total mass $nV = n_0V_0$ is conserved and 
determined by the pre-eruption initial conditions $n_0$ and $V_0$. We further relate the line-of-sight length-scale $L$ and the gas volume $V$ by $V \propto L^{\alpha}$, where $1 \le \alpha \le 3$; for an isotropic self-similar expansion, 
$\alpha = 3$, and for a linear expansion along the line-of-sight, $\alpha = 1$.
Treating the plasma as a monatomic ideal gas, we model the thermodynamic evolution of the expanding gas 
with a polytrope $Tn^{-\eta} = T_0n_0^{-\eta}$, $\eta$ being a real constant. Note that $\eta = \frac{2}{3}$ describes an adiabatic process, and $\eta = 0$ refers to an iso-thermal process. During the expansion, the mean temperature of the plasma therefore evolves as 
\begin{equation}
\frac{T}{T_0} = \left(\frac{L_0}{L}\right)^{\alpha \eta},
\end{equation}
and the observed emission (in units of data counts) at a certain wavelength relative to the pre-expansion emission is given by
\begin{equation}
\frac{I_{\lambda}}{I_{\lambda0}} \approx \left(\frac{L}{L_0}\right)^{1-2\alpha} \frac{R_{\lambda}(T)}{R_{\lambda}(T_0)}.
\end{equation}

\begin{figure}[htbp] 
\centering {\includegraphics[width=1\textwidth]{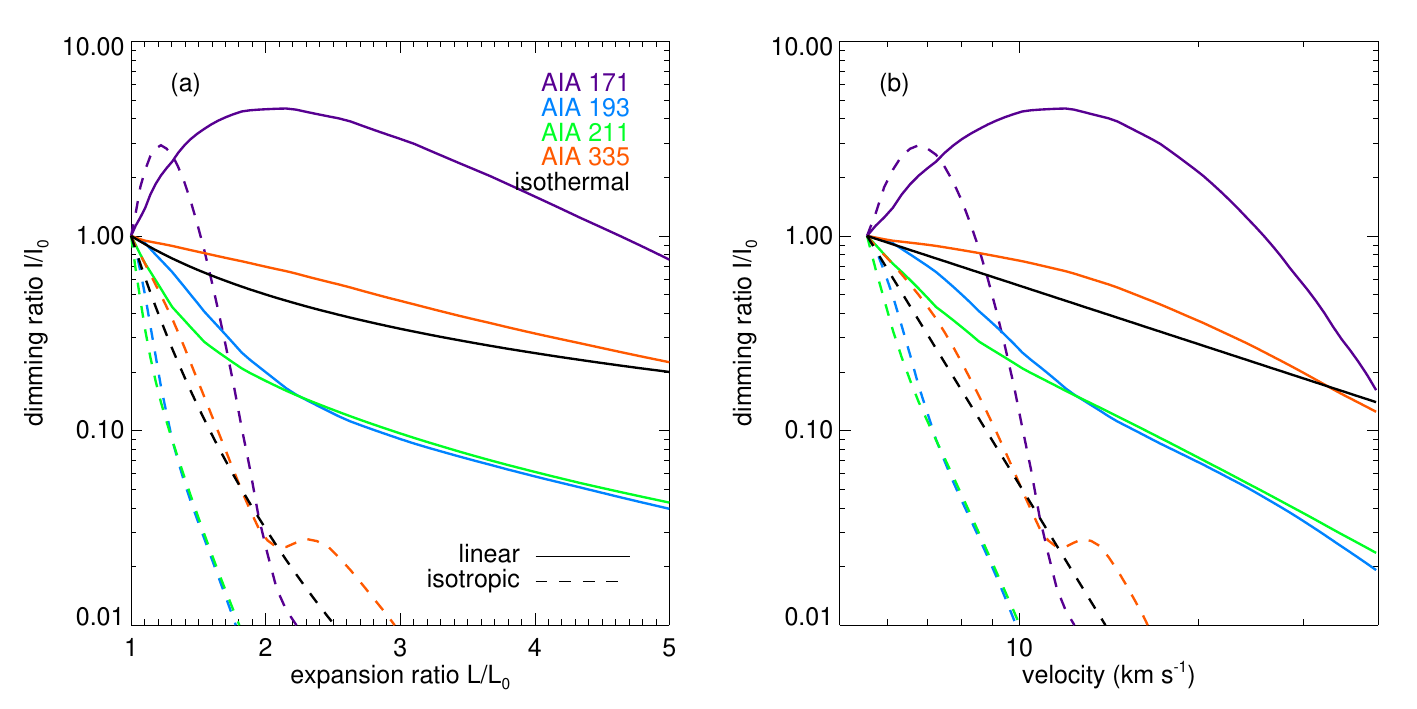}} 
\caption{(a) The expected dimming depth in terms of the base ratio observed by SDO/AIA in the 171, 193, 211, and 335~\AA\ passbands, respectively, with respect to the expansion height for isothermal (black) or adiabatic (color) expansion, assuming that the initial temperature of the corona $T_0 = 1.5$~MK. (b) The expected dimming depth with respect to the expansion velocity (see text). \label{fig:expansion}}
\end{figure}

We use these expressions to model the dimming evolution observed in SDO/AIA 171, 193, 211, and 335~\AA\ passbands, shown in Figure~\ref{fig:expansion}a. Note that these observations are made with a broad passband, so that the Doppler shift effect is not significant and therefore neglected in this simple estimate. These models assume that the initial mean temperature of the active corona is $T_0 = 1.5$~MK. Obviously, for isothermal expansion, the dimming evolution is the same in all these bands, and the variation of the dimming depth is also less significant. Depending on the expansion model, at $L  = 1.5L_0$, the dimming depth reaches $I/I_0 \approx 0.7, 0.2$ for the linear expansion and isotropic expansion, respectively. In comparison, the adiabatic expansion leads to a more pronounced dimming. With an initial temperature $T_0 = 1.5$~MK, this simplified experiment suggests that dimming variation is most significant in the 211~\AA\ passband. In particular, since the AIA 171~\AA\ response function peaks at a temperature below 1~MK, expansion initially causes increased $I$ in this passband. 
Furthermore, we may relate the dimming depth with the expansion speed denoted by $v_e \sim \dot{L}$, 
which are displayed in Fig.~\ref{fig:expansion}b, with an assumed expansion profile $L = L_0e^{t/\tau}$, with an initial value $L_0 = 10$~Mm, and time constant $\tau = 30$~min. Fig.~\ref{fig:expansion}b suggests that persistent expansion even at the speed of a few kilometers per second may produce observable dimming in several AIA passbands. With the sub-sonic expansion speed $\dot{L} < c_s$, an isothermal model is likely more suitable than an adiabatic expansion, producing less significant dimming, yet still observable with the present capabilities, if the expansion is persistent. 

These over-simplified estimates ignore the temperature stratification (i.e., differential emission measure) within the expanding plasma. For example, to maintain the constant temperature $T = T_0$, heating is required, and a more realistic scaling law can be invoked to compute the evolution of the differential emission measure, and therefore the evolution of dimming signatures. Furthermore, coronal dimmings  
can be observed in spectral lines or passbands, like the He I 1083 nm line \citep{Harvey:2002} or the AIA 304~\AA\ filter \citep{Qiu2017, Wang2019}, which are sensitive to temperatures lower than the typical coronal temperature of $\sim$1~MK. Dimming observed in these lines likely form at the base (feet) of the coronal structure. The formation mechanism of these lines is more complex and is therefore not modeled here.
Since the exact geometry and heating process are not known, the information in Fig.~\ref{fig:expansion} from the simplified model merely provides some bounds of observable dimming signatures. On the other hand, the multi-wavelength observations of the dimming evolution will help inform about the property of the gas expansion (or compression!) \citep{Schrijver:2011}.

\section{Relation to CMEs, solar flares and ICMEs} 
\label{sec:relation}

\subsection{Relation to CMEs and solar flares}
 \label{sec:CMErelation}

\begin{figure}[tbp] 
\centering {\includegraphics[width=0.7\textwidth]{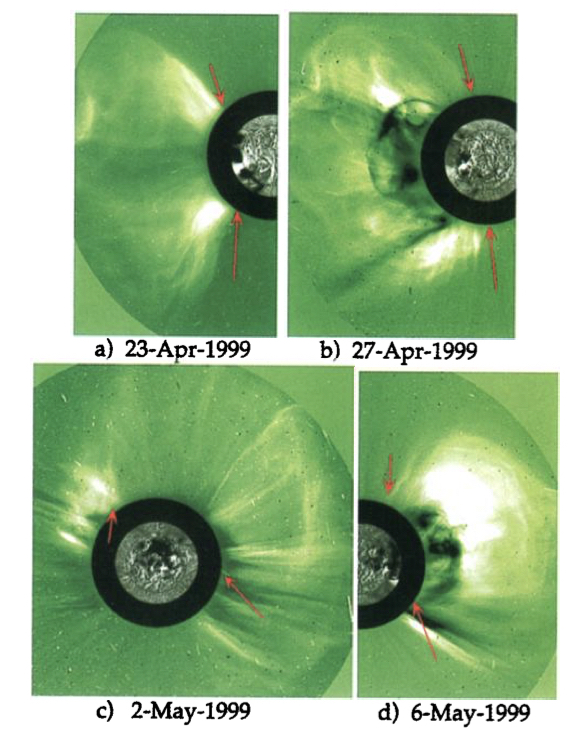}} 
\caption{SOHO/EIT percentage difference images combined with SOHO/LASCO C2 difference images illustrating the connection between coronal dimmings in the low corona and the CME. From \cite{Thompson:2000}.} \label{fig:cme_dimmings}
\end{figure}

The close association of coronal dimmings with CMEs was recognized as soon as regular EUV full-disk imaging and coronagraphic white-light observations became available from the SOHO mission in the late 1990s
\citep{1997GMS....99...27H, 
Thompson:1999, Thompson:2000, Webb:2012}. Fig.~\ref{fig:cme_dimmings} shows the first combined observations of coronal dimmings low in the corona in the EUV using SOHO/EIT and the corresponding CME observed by the SOHO/LASCO C2 coronagraph. 
These composite images allowed to identify dimmings as the CME ``footprints" in the low corona.  Since then, the relation between coronal dimmings, CMEs and flares has been addressed in several statistical studies using different methods and data. 
Those efforts are mainly related to the following questions: \\

1) Are CMEs always associated with coronal dimmings? And vice versa, are coronal dimmings always associated with CMEs (and flares)? 

2) Are the characteristic properties of CMEs (and flares) different for events that are associated with coronal dimmings compared to events which do not show dimming signatures?

3) When coronal dimmings occur, do their characteristic parameters reflect the properties of the associated CME (and flare)? \\

\noindent 
Related to the first question, the first systematic study was done by \cite{Bewsher:2008} using spectroscopic observations of SOHO/CDS in two emission lines (Mg\,{\sc ix} and Fe\,{\sc xvi}) sampling coronal plasma at temperatures of 1 MK and 2 MK, respectively. They found that 84\% (73\%) of CMEs were associated with coronal dimmings, and 48\% (55\%) of 96 dimmings (observed in both Fe and Mg) were associated with CMEs, depending on the CME catalog  used, CDAW or CaCTus.
Comparing with a probability model of the random case, i.e.\ there is no direct link between coronal dimmings and CMEs, they found that these association rates are significantly higher than the random case (by at least 2$\sigma$).
\cite{Reinard:2008} investigated all halo CMEs that occurred during the years 1998--2000 and reported a 53\% association rate of coronal dimmings with front-sided halos (57 cases), as a lower limit. \cite{Dissauer:2018b} selected all halo CMEs (from CDAW) and large-scale EUV waves (from \cite {Nitta:2013}) that occurred during 5/2010--12/2012, and found that 87\% of the 71 CMEs were associated with a visually identified dimming in the SDO/AIA 211~{\AA} images.
Recently, \citet{Veronig:2021} studied the association between coronal dimmings and CMEs related to strong flares ($\ge$ GOES class M5) during 2010--2014 (38 eruptive, 6 confined flares). They found that  84\% of the CMEs revealed a significant dimming in broadband 150--250 {\AA} SDO/EVE full-Sun light curves, and 83\% of the dimmings identified were associated with a CME. The results were similar for a larger set of 52~flares (39 eruptive, 13 confined) where the dimmings were identified in light curves from spatially-integrated  SDO/AIA 193 {\AA} images (83\% and 88\%, respectively).
The results of these studies demonstrate that coronal dimmings are a robust diagnostics of CME occurrence. The variance between the earlier and later studies may be related to the different selection criteria and also due to the better time cadence and sensitivity of SDO/AIA, SDO/EVE vs.\ SOHO/EIT, SOHO/CDS.

We note that to date there exist no systematic studies on coronal dimmings detected in confined flares, i.e. whether these are false positives or whether flares without CMEs or failed eruptions may also produce coronal dimmings under certain conditions. In a case study, \cite{Zhang2020} report a remote dimming associated with a confined circular-ribbon flare, which seems to lie at the footprint of the outer spine.
Statistically, for the 13\% (12\%) false alarm rates for dimmings found from the full-Sun integrated light curves with EVE (AIA) in the \cite{Veronig:2021} study, it was noted that the dimmings that were detected in confined large flares, were all weak. We will come back to this issue in Chapter \ref{sec:stellar}, when we discuss coronal dimmings in the context of stellar CME detections. 

We also note that events that might have been missed in the statistical studies discussed above are CMEs that occur without or only very weak signatures in the lower corona and chromosphere, in terms of flare emission, post-eruption arcades, filament eruptions, coronal waves or coronal dimmings \citep{Robbrecht:2009, Ma:2010,D'Huys:2014}. These so-called \textit{stealth CMEs} are in general faint, slow and narrow, indicating that they contain less energy than regular CMEs \citep[][]{Nitta:2021}.
Stealth CMEs are assumed to originate from larger heights in the solar corona, where the electron density and magnetic field strength is substantially decreased and thus less mass and energy available \cite[][]{Robbrecht:2009}. This  may explain why they produce no or only faint dimming and flare signatures. As reported in \cite{Nitta:2017}, stealth CMEs generally start slowly, which will result also in slowly evolving dimming signatures (if any).

\begin{figure}[tbp]
\centering {\includegraphics[width=0.64\textwidth]{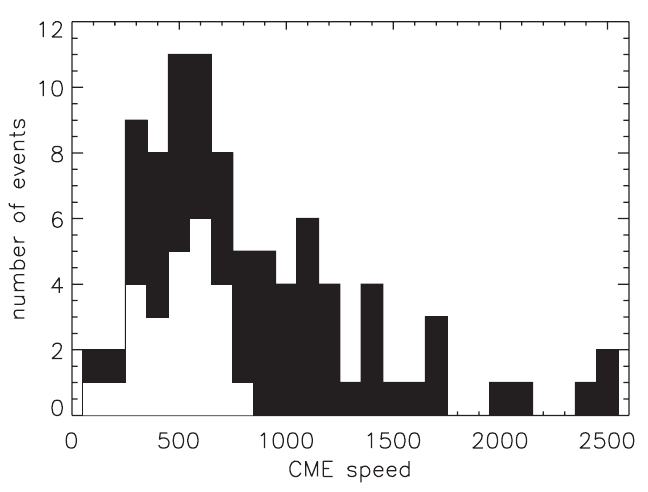}} 
\caption{ 
Histograms of CME speed for non-dimming-associated CMEs (white) and the total sample (black). From \cite{Reinard:2009}. 
} \label{fig:cme_stats}
\end{figure}

Addressing the second question, \cite{Reinard:2009} analyzed 90 CME-ICME pairs over the years 1997--2005 and checked for associated dimming signatures in SOHO/EIT images. 65 CMEs (72\%) had an associated dimming, while 25 (28\%) showed no dimming signature. On average, CMEs accompanied by dimmings were faster (964 km s$^{-1}$ versus 498 km s$^{-1}$), more likely to be accompanied by flares (72\% versus 48\%), 
and the associated flares were more energetic (mean GOES flare class M9.3 vs. M1.9).
The CME speeds for dimming versus non-dimming events represent two distinct populations (see Fig.~\ref{fig:cme_stats}). Applying the Kolmogorov-Smirnov test, the probability that they come from the same population was as small as $\sim$$10^{-5}$.  
Interestingly, the lack of dimming signatures seems to provide an upper limit on the CME linear 
speed (as reported in the CDAW catalog).
All 25 non-dimming CMEs had speeds $<$800~km~s$^{-1}$, whereas for the CMEs associated with dimmings only about half of the events had speeds $<$800~km~s$^{-1}$. This CME sample was otherwise indistinguishable from the non-dimming CMEs in terms of association to flares, flare energies and other source region properties. 

Over the past $\sim$20 years, various single-event and statistical studies 
have been performed related to the third question and these investigate the relationship between coronal dimmings and characteristic parameters of the associated CMEs and flares, such as CME speed and mass, flare peak flux as well as the relative timing between the onsets of the different phenomena. 
One has to keep in mind that those studies often use different definitions of what coronal dimmings are (core vs. secondary dimming), different methods, instruments and data types (e.g. imaging data, spectroscopic observations, spatially-unresolved irradiance lightcurves) to detect dimmings, and also different approaches how to derive characteristic parameters and their comparison (e.g. basic modeling, DEM analysis, correlation analysis).

\subsubsection{Dimming -- CME mass relationship}

Since coronal dimmings are mainly formed by the evacuation of plasma due to the CME 
(see Sect.~\ref{sec:plasma}), several case studies made the attempt
to estimate the mass loss from the observed dimming regions 
and to compare it to the mass of the associated CME  as derived  from the Thomson scattering observed by white-light coronagraphs 
\citep[e.g.][]{Harrison:2000, Wang:2002, 2009ApJ...702...27J,2003A&A...400.1071H,2012ApJ...748..106T}. 
\cite{Sterling:1997} provide the first mass estimate from coronal dimmings associated with a halo CME using Yohkoh/SXT data. Depending on the composition of plasma within the dimming region, they find 
that a mass of a few times $10^{14}$~g was ejected. \cite{Harrison:2000} analyzed the first spectroscopic observations of coronal dimmings.  
For the event under study, they derived a mass of
$\sim3\times10^{13}$~g from the coronal dimming observations, which accounts for 70\% of the CME mass. Using spectroscopic data of Hinode/EIS, \cite{2012ApJ...748..106T} and \cite{2009ApJ...702...27J} derived a mass loss from the dimmings 
that accounts for 20--60\% of the CME mass. 
\cite{Zhukov:2004} investigated both the mass loss within the deep core dimming as well as the whole dimming region, which includes also the more widespread secondary dimmings. They report that about 50\% of the estimated total dimming mass of $1.4\times10^{15}$~g results from core dimming regions, whereas the average CME mass of $4.2\times10^{15}$~g is about three times larger. 

\begin{figure}[tbp]
    \centering
  \includegraphics[width=0.98\textwidth]{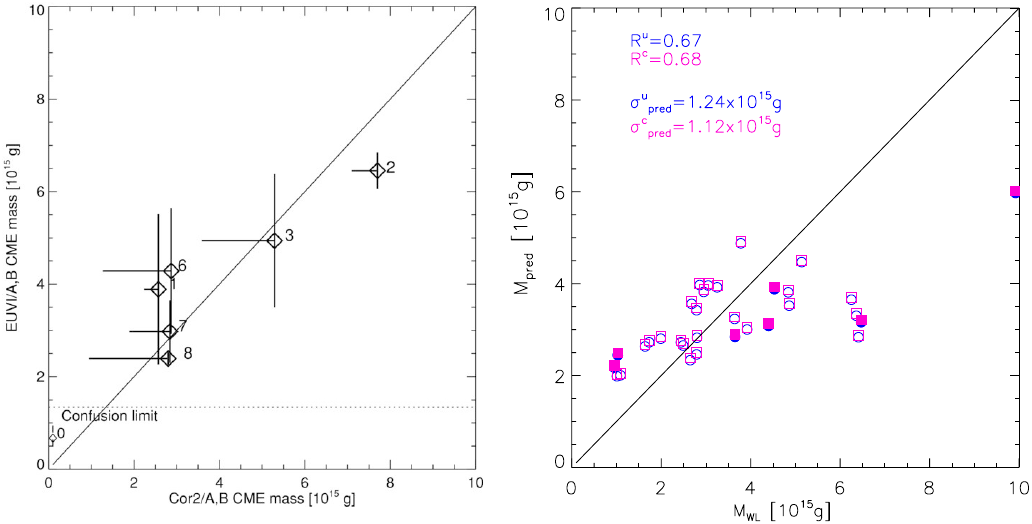}  
    \caption{Comparisons of CME masses inferred from EUV coronal dimmings and white-light coronagraph data.
    The left panel shows masses predicted from the dimming that are derived from a three-dimensional volume and density modeling in the dimming region and background corona \citep{Aschwanden:2009}; the right panel shows predicted masses from the dimming using a DEM method \citep{Lopez:2019}.  The solid line marks the 1:1 correspondence level. 
}    \label{fig:cme_mass_comp}
\end{figure}

There are several recent statistical studies that investigate the relationship between coronal dimmings and CME mass \citep{Aschwanden:2016,Krista:2017,Aschwanden:2017,Mason:2016,Dissauer:2019, Lopez:2019,Chikunova:2020}. 
A first small statistics was already done by \cite{Aschwanden:2009} who calculated the CME mass based on the stereoscopic data provided by the STEREO twin satellites. They use the simultaneous observations of coronal dimmings in STEREO/EUVI A and B, and apply three-dimensional volume and density modeling of
the dimming region and the background corona. Their results of the EUV dimming mass are in basic agreement with the CME masses obtained from the STEREO/COR2 white-light measurements (see Fig.~\ref{fig:cme_mass_comp}, left panel). 

\citet{Krista:2017} investigated the evolutionary properties of 115 EUV dimmings during the year 2013,  
using an automated tracking algorithm that operates on SDO/AIA 193 \AA\ direct images. They found moderate correlations between the CME mass and their derived dimming intensity parameters, with the Pearson correlation coefficient in the range $|r| = 0.4-0.5$.
The studies by \cite{Aschwanden:2016, Aschwanden:2017} use an analytical EUV dimming model based on radial adiabatic expansion and DEM distribution within a volume defined by the characteristic length scale of the coronal dimming. 
The mass estimates from the SDO/AIA EUV dimming observations 
are compared with the CME mass measured from white-light coronagraphic data of SOHO/LASCO (CDAW catalog). \cite{Aschwanden:2017} includes the largest sample of all the statistical studies undertaken, with a sample size of 399 events that occurred between 2010 and 2016, covering a substantial part of solar cycle 24. 
However, only a weak correlation ($r=0.3$) between the 
EUV dimming and white-light CME mass estimates was obtained.

\cite{Lopez:2019} applied the DEM method introduced in \cite{Lopez:2017} to derive the CME mass from the dimming regions observed in the six SDO/AIA coronal filters for a sample of 32 events, and compared it with the CME mass from the coronagraphic white-light images. They find a correlation of $r \approx 0.7$  and a relative error 
between predicted and measured CME masses  of $\sim$30\% (see Fig.~\ref{fig:cme_mass_comp}, right panel).

\begin{figure}[tbp]
    \centering
    \includegraphics[width=0.99\textwidth]{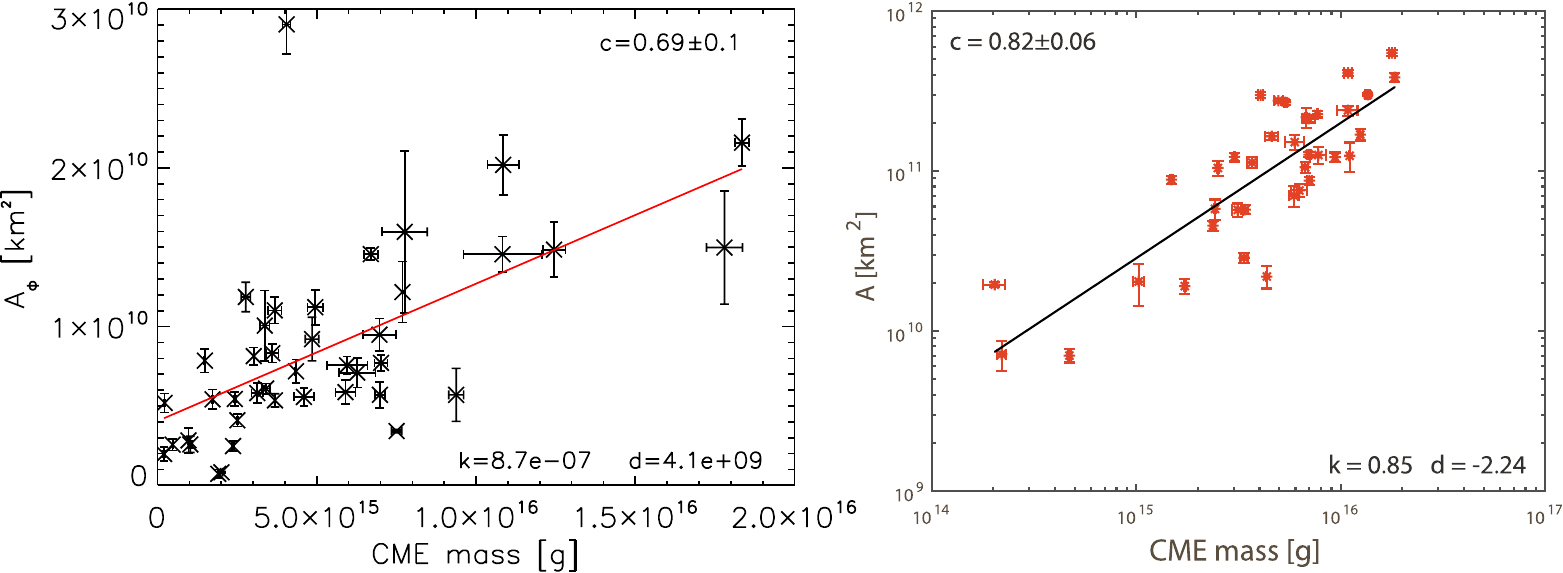}    
    \caption{Relation between coronal dimming area and CME mass for dimmings observed on-disk  by SDO/AIA (left panel; from \cite{Dissauer:2019}) and off-limb by STEREO/EUVI (right panel; from \cite{ Chikunova:2020}). The red (black) line presents the linear fit to the corresponding data points.}
    \label{fig:dimming_mass}
\end{figure}

\cite{Dissauer:2018b,Dissauer:2019} use SDO/AIA logarithmic base-ratio images to 
study coronal dimmings for 62 events associated with both CMEs and flares that occurred during the period 2010--2012, when SDO and STEREO were in quasi-quadrature formation. The  events were observed on-disk from Earth view (SDO) and close to the limb for at least one of the two STEREO satellites, thus minimizing projection effects for the CME measurements. In their study, the coronal dimming is treated as a time-integrated phenomenon, by summing newly detected dimming pixels over time. This is in contrast to other studies that derive at each time step the instantaneous dimming area, which is then the basis also for all further parameters calculated.
The CME mass shows the strongest correlations with so-called first-order coronal dimming parameters, which reflect properties of the total dimming region at its final extent, like the size of the dimming region, its total brightness and the total unsigned magnetic flux covered by the dimming area ($r\approx 0.6$--$0.7$).
\cite{Chikunova:2020} extended these studies by further exploiting 
the multi-viewpoint aspect. They studied the same set of events as  \cite{Dissauer:2018b,Dissauer:2019} but investigated the properties of the coronal dimmings observed above the limb by STEREO/EUVI. 
In the case of the off-limb dimming observations, the correlations between the dimming parameters and CME mass are  somewhat higher ($r\approx 0.7$--$0.8$) 
than for the same dimmings observed on-disk by AIA (see Fig.~\ref{fig:dimming_mass}).

\cite{Mason:2016} used a different approach. They did not extract the dimmings from EUV imagery but identified coronal dimmings as temporary minimum in spatially-unresolved Sun-as-a-star SDO/EVE irradiance profiles as response to the ejected CME. Their study covers two separate two-week periods in 2011, including a total of 37 events. They combined the information from different EVE spectral lines to remove the flare contribution in the cooler lines by using the information from the hotter lines. Applying this technique, they found a correlation of $r \approx 0.7$ between the maximum depth of the irradiance profiles and the mass of the corresponding white-light CME.

In summary, it seems that the different approaches to estimate masses from the EUV dimming observations give results in the right order of magnitude (though with a large scatter) but in general show a tendency for underestimating the white-light CME mass. These findings may indicate that the CME mass does not only consist of plasma evacuated from the low corona but also contains mass of the cooler prominence as well as material that is piled up during the CME expansion and propagation through the corona \citep{Bein:2013,Feng:2015}.

\subsubsection{Dimming -- CME kinematics relationship}

\begin{figure}[tbp]
    \centering
    \includegraphics[width=1\textwidth]{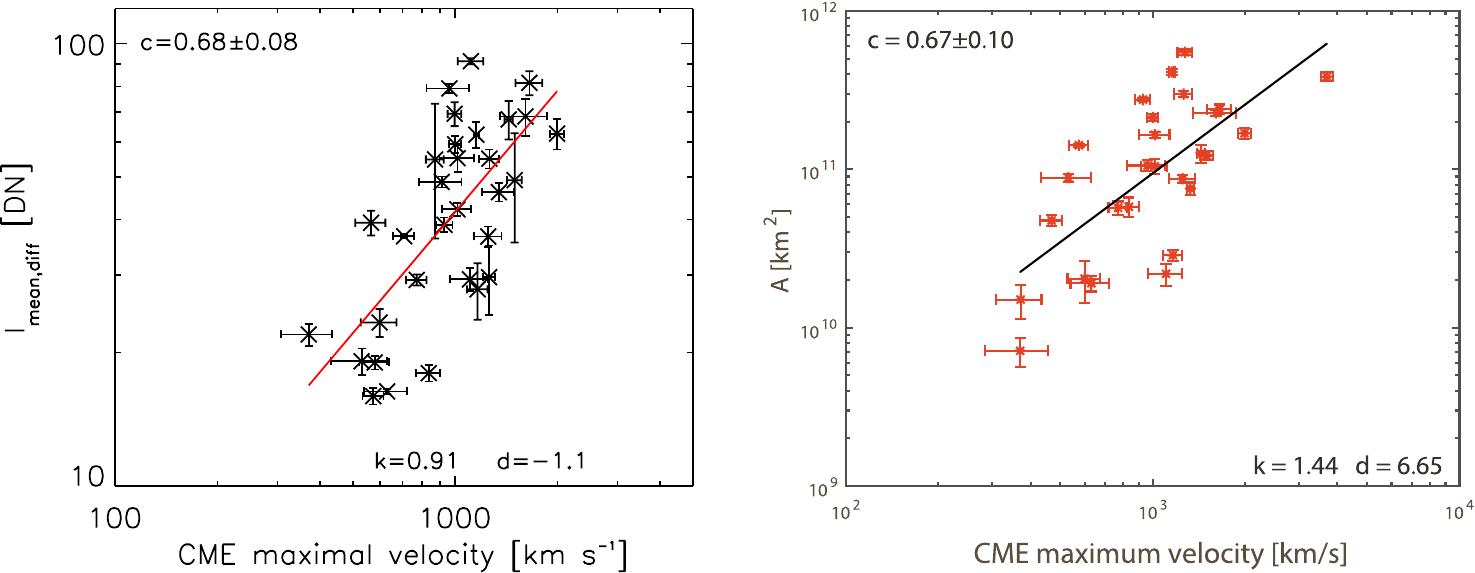}
    \caption{Examples for the strongest correlations of dimming parameters with the speed of the associated CME. 
    Left: average dimming brightness from base-difference images for dimmings observed on-disk by SDO/AIA \citep[from][]{Dissauer:2019}, right: dimming area for the same dimming events but observed off-limb by STEREO/EUVI \citep[from][]{Chikunova:2020}. The red (black) line represents the linear fit to the corresponding data points. The CME speed is derived as the maximum speed that was reached within a distance of 20 solar radii.}
    \label{fig:dimming_cme_speed}
\end{figure}

If coronal dimmings reflect the early evolution of CMEs in the low corona, we would expect that their characteristic properties are also related to the CME speed and acceleration. 
However, no significant correlations were found in the statistical analysis performed by \cite{Krista:2017}. \cite{Aschwanden:2017} 
reported a weak correlation between the speed estimated from coronal dimmings and the associated CME speed ($r=0.24$). This is in contrast to other statistical studies which do find higher correlations of the order of $r \sim 0.6-0.7$ \citep{Dissauer:2019,Chikunova:2020}, see Fig.~\ref{fig:dimming_cme_speed}. \cite{Mason:2016} report a strong correlation between the CME speed and the intensity drop rate (slope) of SDO/EVE irradiance curves 
($r=0.78$). 
This high correlation was determined from a subset of 14 events (out of a total of 37 events studied) 
and has not been yet confirmed with a larger sample analysis. 
It is worth noting that this study covered a limited time range (two periods, each covering two weeks). Therefore, homologous events originating from the same AR might contribute to the correlations. 

But why are some statistical studies successful in relating dimming parameters to the CME mass and speed, whereas others do not find any significant correlations?
We suspect that two reasons may be responsible for this mixed outcome. First, the uncertainties in the measurements of CME parameters from different vantage points and second, the definition of the dimming itself that affects how coronal dimming regions are detected. 
Measurements of characteristic properties of Earth-directed CMEs are subject to large uncertainties for satellites along the Sun-Earth line due to projection effects. One assumption for the calculation of the CME mass is that the full CME body is projected against the plane-of-sky. This is more appropriate, and hence may result in lower uncertainties for CMEs observed at the limb.
The studies by \cite{Dissauer:2018b,Dissauer:2019} and \cite{Chikunova:2020} use CME mass estimates of Earth-directed CMEs observed on the limb from STEREO, while the studies by \cite{Reinard:2009,Krista:2017,Aschwanden:2016,Aschwanden:2017}
used SOHO/LASCO CME mass estimates measured along the Sun-Earth line. 

In contrast to methods that use difference images, the direct image approach used in \cite{Krista:2017} captures mostly the evolution of the (darker) core dimmings, that usually form close to the feet of the erupting flux rope at the core of the eruption, and therefore may miss a significant part of the overall dimming volume 
associated with the expanding CME \citep{patsourakos:2012}. 
This may account for the low correlations between the dimming and the CME mass obtained
in the \cite{Krista:2017} study.

\subsubsection{Dimming – Flare relationship}
Only a few statistical studies investigate the relationship between solar flares and coronal dimmings \citep{Reinard:2009, Aschwanden:2016, Krista:2017, Dissauer:2018b, Krista:2022}. One of the first ones, by \cite{Reinard:2009} found that dimming-associated CMEs tend to have a stronger correlation with the flare size (i.e. peak of the GOES 1--8~{\AA} SXR flux) and that faster/more energetic CMEs tend to be associated with larger flares and larger dimming regions. These findings may be another form of the “big-flare syndrome” introduced by \cite{Kahler:1982}. More energy available results in more intense flares, more energetic CMEs, and larger associated dimming regions.
\cite{Aschwanden:2016} and \cite{Krista:2017} investigate the potential relationship of coronal dimmings with basic flare properties such as the flare duration or the GOES peak flux, however no significant correlations were found, concluding that the energetics of the flare shows no connection to the physical properties and
evolution of the dimmings.

\begin{figure}[tbp]
    \centering
    \includegraphics[width=0.495\textwidth]{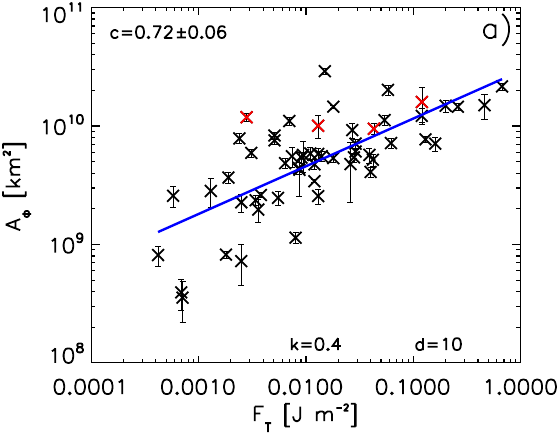}
\hfill    \includegraphics[width=0.478\textwidth]{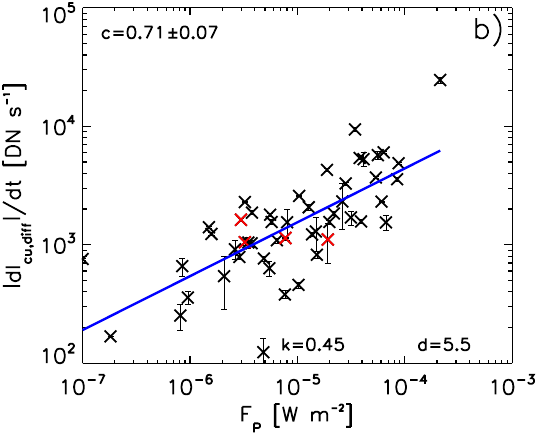}
    \caption{Exemplary relations between dimming and flare parameters. 
    Static dimming parameters (e.g., dimming area $A_{\phi}$, left panel), reflecting the total dimming extent, show the strongest correlations with the flare fluence ($F_{T}$). Dynamic parameters (e.g., maximal brightness change rate $|\dot{I}_{\text{cu,diff}}|$, right), quantifying how the dimming is changing over time, show the strongest correlations with the soft X-ray peak flux ($F_{P}$) of the corresponding flare. The blue line presents the linear fit to the corresponding data points. From  \cite{Dissauer:2018b}.}
    \label{fig:enter-label}
\end{figure}

However, these conclusions are contrasted by the findings by
\cite{Dissauer:2018b} who present one of the more detailed studies on the  dimming-flare relationship. They found that “static” parameters reflecting the total extent of the dimming, such as its area, total unsigned magnetic flux, and absolute minimum brightness show the highest correlation with the SXR fluence of the associated flares ($r \sim 0.7$; for an example see Fig.~\ref{fig:enter-label}a). The SXR fluence is a measure of the 
flare radiation loss in the 1--8 {\AA} SXR band and has been shown to be strongly correlated with the total energy released during the flare \citep[][]{Emslie:2005}.
On the other hand, 
parameters extracted from the time derivative of the  dimming evolution, thus describing the dynamics of coronal dimmings,
such as the maximal area growth rate, the maximal magnetic flux rate, and the maximal brightness change rate, show the strongest correlations with the peak of the GOES SXR flux, reflecting the flare strength ($ r\sim 0.6-0.7$; e.g. Fig.~\ref{fig:enter-label}b). The stronger the associated flare, the faster the dimming is growing and darkening, and the more magnetic flux is ejected by the CME. 
The authors conclude that dimmings correlate with both CME and flare quantities, providing further evidence for the CME-flare feedback relationship \citep[][]{Zhang:2001,Temmer:2008,Vrsnak:2008}.
The study of coronal dimmings indirectly shows that magnetic reconnection affects the dynamics of CMEs by the reduction of the tension of the overlying magnetic fields, the increase of magnetic pressure below the flux rope, as well as the supply of additional poloidal flux to the flux rope, prolonging the driving Lorentz forces \citep[][]{Vrsnak:2016}.

Important evidence for this last point is also the strong correlation found between the magnetic fluxes inferred from secondary coronal dimmings and the magnetic reconnection fluxes determined from the flare ribbons \citep[$r \sim 0.6$;][]{Qiu:2007,Dissauer:2018b}. Especially for stronger flares 
($>$M1.0), it was found that, within the uncertainties, both flux estimates give similar results  \citep[][]{Dissauer:2018b}. This empirical finding is in line with the hypothesis in \cite{Lin:2004}, 
that the same amount of magnetic flux is leaving both ends of the current sheet during a CME-related magnetic reconnection event. The upward directed component is estimated from coronal dimming observations, representing the overlying field that gets stretched during an eruption (secondary dimmings) and eventually added as poloidal flux to the erupting flux rope, whereas the downward directed component is estimated from flare ribbons that outline the footpoints of newly formed loops as a result of reconnection. These results not only emphasize the importance of studying dimming and flare signatures together, but they also point us towards key locations for particle acceleration.
\vspace{0.2cm}

\begin{figure}[tbp]
\includegraphics[width=1\textwidth]{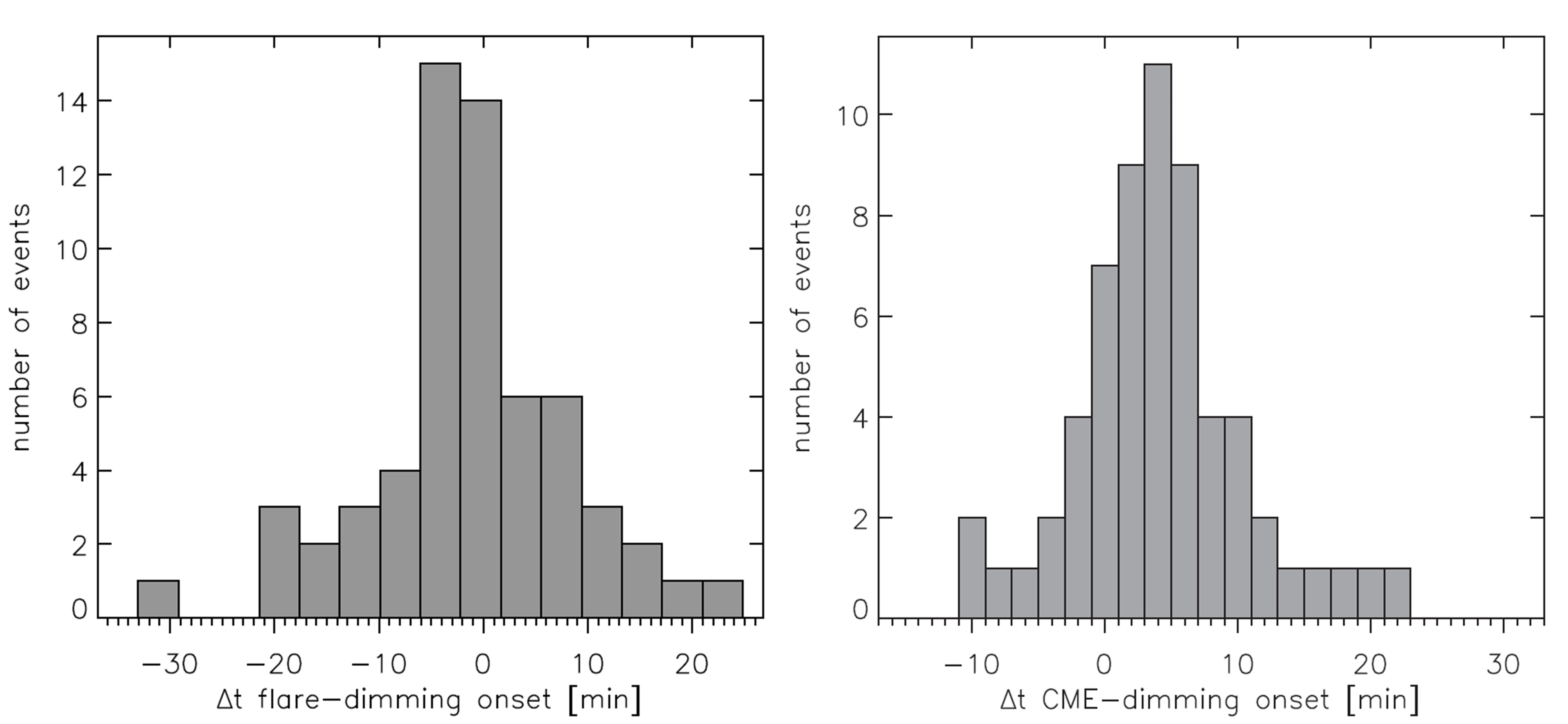}
\caption{Distributions of the time difference between the onset of the impulsive phase of the dimming and the flare start time (left) and the CME onset time (right). From \cite{Dissauer:2018b, Dissauer:2019}.}
\label{fig:timing}
\end{figure}

Coronal dimmings, CMEs and solar flares are also closely related in terms of their timing. Fig.~\ref{fig:timing} shows the distributions of the time differences between the start of the impulsive phase of the dimming and the onset of the flare (left panel) and CME (right panel). 
It was found that for more than 50\% of the events the time difference between the flare and dimming onset is $|\Delta t|<$ 5 minutes, with a mean of $-1.5\pm9.9$ 
minutes. This means that on average the flare onset occurs roughly simultaneously with the impulsive dimming onset \citep[][]{Dissauer:2018b}.
For CMEs similar results are obtained: $|\Delta t|<$ 5 minutes for 55\% and $|\Delta t|<$ 10 minutes for 85\% of the
events, with a mean of $4.2\pm6.4$ minutes, meaning that the CME onset occurs slightly later than the dimming onset \citep[][]{Dissauer:2019}. Within the given observation cadences (5 minutes for CMEs and 1 minute for flares), the distributions indicate a close synchronization of coronal dimmings with CMEs and flares, providing further evidence for their tight relationship.

\subsubsection{Dimming -- energetic particle relationship}

We know of several distinct populations of non-thermal particles in and around flare/CME activity.
These range from the impulsive-phase fast electrons, inferred to be present from their hard X-ray brems\-strahlung, to the high-energy ions observed as solar energetic particles (SEPs) at and beyond 1~AU. Particles trace the field connectivity that we see projected onto the lower atmosphere, including the dimmings that correspond to open fields.
The highest-energy particles are relativistic and therefore can reflect plasma dynamics as quickly as photons can for an observer at Earth, not counting the extra path length involved with the general Parker-spiral pattern of the large-scale heliospheric field.
Thus, as the coronal magnetic field evolves in the manner described in this review, the charged particles must generally follow this evolution immediately.

The problem in using charged particles to map out the coronal field, of course, is that atomic and nuclear processes require some dense ``target'' plasma, normally found only in the lowest solar atmosphere. 
The corona is collisionally thin to all charged particles above a few keV.
Although we can detect SEPs \textit{in situ} in the heliosphere, we presently have little direct understanding of how these particles become accelerated and either ``escape,'' or find a downward path into a collisionally thick layer.
All is not lost, however; radio emission mostly does not require dense target plasma and can be amazingly sensitive. In combination with coronal dimmings that have the potential to point us at locations of newly opened field during eruptions, they may enhance our knowledge on this important topic.

\begin{figure}[tbp]
\centering
\includegraphics[width=0.61\textwidth]{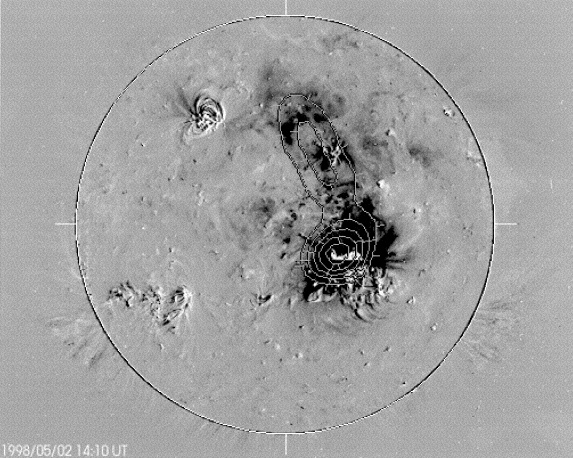}
\caption{SOHO/EIT 195~{\AA} difference image at 14:10 UT during SOL1998-05-02 showing the coronal dimming. The contours are the Nan{\c c}ay 236~MHz radio emission at 13:48~UT, which show a radio source along a transequatorial interconnecting loop. At the time of the EIT image showing the coronal dimming, this radio source has already disappeared.
Adapted from \cite{Pohjolainen:2001}.
}
\label{fig:radio}
\end{figure}

\cite{Pohjolainen:2001} combined imaging in the metric radio domain from the Nan{\c c}ay Radio Heliograph (NRH) with imagery at EUV and SXR wavelengths by SOHO/EIT and Yohkoh/SXT, respectively, to study the flare/CME of 1998 May 2. They note that this event was associated with the disappearance of a large trans\-equatorial interconnecting loop, which subsequently showed up as a coronal dimming. Interestingly, at the beginning of the eruption, radio emission along the transequatorial loop was observed by NRH (see Fig.~\ref{fig:radio}) but disappeared thereafter. 
The authors conjecture that the  observed radio continuum emission arises from electrons accelerated in the expanding large-scale loops associated with the CME, and that the large-scale radio sources therefore may provide insight into the on-disk locations of the CME source region along with information on energetic particles inside these structures. \cite{Wen:2007} studied the position and evolution of radio noise storms and their changes during four CMEs in NRH images. Radio noise storm continua are in general thought to be generated by nonthermal electrons accelerated in closed loops above ARs \citep[e.g.,][]{Mercier:2015}. \cite{Wen:2007} found that coronal dimmings observed by SOHO/EIT were located on the periphery of the noise storms, where the field lines were previously closed. The noise storms disappeared when the eruption/dimming started, probably as a consequence of the closed field lines being opened by the CME. 
These studies demonstrate the diagnostic potential of combining radio observations with coronal dimmings to better understand solar eruptions. 

We note that existing models do not satisfactorily describe the connectivity between the deep atmosphere and the heliosphere during flares.
Certainly high-energy particles do exist within flare loop environments \citep{2006ApJ...644L..93H}, and may ``escape'' to the heliosphere even though the CME-driven shock picture dominates lower-energy SEPs.
Given that ``flare reconnection'' occurs in locally closed fields, an additional mechanism must operate to allow for the escape.
\cite{2019ApJ...884..143M} suggest that this implies an independent reconnection process that could link the flare volume to open field lines.
Observations of high-energy SEPs that could illuminate this problem are generally not yet available.

\subsection{Connection with in-situ measurements} 
\label{ss:relations_in-situ}

Coronal dimmings, particularly the twin-dimming residing in magnetic fields of opposite polarities, have been considered to map the feet of magnetic flux ropes (MFRs). Therefore, the magnetic flux encompassed in the twin-dimming regions provides a measurement of the total magnetic flux in the MFR. \citet{Wang2019} demonstrate the feasibility of measuring additional magnetic properties of the MFR. At present, such measurements cannot be done any other way on the Sun. However, a few days after the eruption, when Earth-directed CMEs arrive at Earth, their magnetic and plasma properties can be measured in-situ by satellites like ACE and Wind \citep{Lepping1990}. Since the 1980s, in-situ measurements of interplanetary CMEs (ICMEs) have revealed the helical magnetic field structure of many events, called Magnetic Clouds  \citep[MC;][]{Burlaga1981}, which define the in-situ MFR. So far, nearly all of our knowledge of the magnetic structure of CME MFRs comes from in-situ measurements. 

Hundreds of MCs have been observed in the past decades. In-situ instruments measure vector magnetic fields, and plasma properties including electron and ion temperatures, densities, and velocities, but only along a single path across the MC. Assuming the MC to be a cylindrical structure, several methods have been developed \citep{Lepping1990, Riley2004, Hu2017} to reconstruct the MC, yielding the total or toroidal magnetic flux $\Phi_t$, axial current $I$, and the poloidal flux $\Phi_p$, or magnetic twist, per unit length of the MFR \citep{Lepping1990, Lynch2005, Qiu:2007, Hu2014}. It is the prevailing belief that the MFR structure originates from the Sun. Being able to identify them on the Sun and measure comparable properties, however, is non-trivial. 

\begin{figure}[tbp]
\centering {\includegraphics[width=0.98\textwidth]{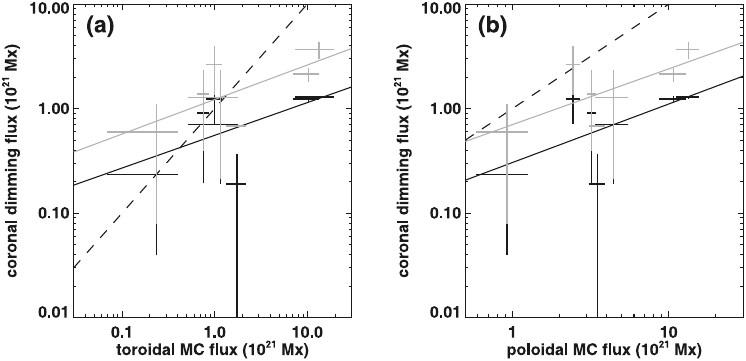}} 
\caption{Relation between coronal dimming flux and the reconstructed toroidal (a) and poloidal (b) magnetic flux of the associated ICME measured in-situ at 1 AU for 9 events. Dark and gray symbols indicate measured dimming fluxes determined by two different intensity thresholds used for the detection of the dimming regions. The solid lines show the linear least-squares fit applied to the data pairs in logarithmic scales. The dashed lines show the identity line. Adapted from \cite{Qiu:2007}.} 
\label{fig:Qiu_2007_F1}
\end{figure}

\citet{Webb:2000} made the first attempt to compare MFR properties measured on the Sun and at 1~AU. They measured the magnetic flux encompassed by the twin-dimming regions in an eruptive event from a bipolar region on 1997 May 12 (cf., Fig.~\ref{fig:thompson98}), finding it consistent with in-situ measurements of the MC observed 3 days after the coronal event.  Studying 9 CME-MC events observed on the Sun and then tracked to 1~AU, \citet{Qiu:2007} measured the magnetic flux $\Phi_d$ in dimming regions, which are less well defined as twin-dimmings, on the Sun, as well as the toroidal flux $\Phi_t$ in MCs (cf.\ Fig.~\ref{fig:Qiu_2007_F1}a), and reached a similar conclusion. 
In a case study of the 2011 October 1 event, \cite{Temmer:2017} found the dimming flux $\Phi_d$, $\Phi_t$ in the MFR reconstructed using non-linear force free modeling and $\Phi_t$ in the associated MC to coincide within a factor of 2. As the total flux of the MFR is largely conserved from the Sun to 1~AU, the results of these studies support the simple scenario that coronal dimmings map the feet of the erupting MFR.

Nevertheless, coronal dimmings may also reflect the expansion of the overlying plasma when reconnection restructures the magnetic field so that the plasma suddenly finds itself along longer, or even open, field lines. In such a scenario, coronal dimmings map magnetic reconnection, which may be responsible to form or add to the twist, i.e., the poloidal flux $\Phi_p$, of CME flux ropes \citep{Forbes2000, Demoulin1996}. The case studies by \citet{Mandrini2005} and \citet{Attrill:2006} show that the flux measured in coronal dimming regions was comparable with $\Phi_p$ measured in the MC,  lending support to the hypothesis. \cite{Qiu:2007} found  that the dimming flux $\Phi_d$ and the in-situ measured $\Phi_p$ are correlated, but that in all nine events under study $\Phi_d \lesssim \Phi_p$ (Fig.~\ref{fig:Qiu_2007_F1}b), showing that $\Phi_d$ is a lower limit to the MC $\Phi_p$.

Given the difficulty in measuring magnetic fields in the corona, now and for the foreseeable future, the ability to quantify the \emph{magnetic} configuration of the erupting MFR remains a challenge. Inferring magnetic properties by exploring coronal dimming signatures, in conjunction with in-situ measurements of MFR properties, may provide important insights into mechanisms forming and erupting MFRs.
  
On the other hand, it should be recognized that measurements of magnetic properties, such as the magnetic flux and total vertical currents, inside the dimming regions are subject to a range of uncertainties. First, it remains a challenge to define the boundary of coronal dimmings. In most observational studies, it is required that the intensity of a dimming region should drop continuously to be below a certain fraction of its pre-dimming intensity (cf., Sect.~\ref{sec:detection}). This certain fraction has been mostly determined empirically, and varies for measurements using EUV or X-ray observations in different passbands. Next, to find magnetic flux and vertical current from the dimming region \citep{Webb:2000, Wang2019}, the underlining assumption is that the dimming maps the feet of a coronal magnetic structure, and the apparent relative dimming due to projection effects, such as the removal or re-orientation of overlying coronal structures \citep[e.g.][]{2007PASJ...59S.801H} should be excluded. To address this artifact, \cite{Qiu:2007} considered that, if dimming regions map the feet of opening field lines (by erupting flux ropes), the absolute intensity of the dimming should fall below that of the quiescent Sun, and used the threshold $I_{d} \le N I_{q}$ to further constrain the dimming boundary, with $N$ varying between 1--2. These criteria may result in variations in the measured magnetic flux $\Phi_d$ by as much as a factor of 3 (see Fig.~\ref{fig:Qiu_2007_F1}). Furthermore, there are uncertainties in the magnetic field measurements as well. The majority of coronal dimmings tend to occur in regions of weak magnetic fields \citep[][also see Fig.~\ref{fig:dissauer_2018b_fig1}d]{Dissauer:2018b}, where the uncertainty in the line-of-sight magnetic field measurements (for example by MDI and HMI) is about 10--20~G \citep{Liu:2012}; this value, integrated over a large region, introduces a large uncertainty in $\Phi_{d}$ measurements. In regions much away from the disk center, the flux $\Phi_d$  may be computed by integrating the radial magnetic field, which is calculated from the line-of-sight and transverse magnetic components, hence the uncertainty is even larger. All these considered, at the present, measurements of the magnetic flux, and those of the vertical current \citep[e.g.][]{Wang2022}, should serve as semi-quantitative estimates with uncertainties up to a factor of a few.

\section{A new categorization}
  \label{Terminology}


The traditional categories of on-disk solar dimmings produced by CMEs are {\em core} and {\em secondary} dimmings. These reflect
the observed properties---location, morphology, and depth of the dimming. 
However, there are no commonly accepted definitions of these terms. Moreover, observations with improved resolution and cadence, especially from SDO, show overlap in location and size; namely, core dimmings often expand from their initial position to the outer parts of the eruption source region, reaching, over time, a size typical of secondary dimmings. Additionally, secondary dimmings forming far away from the source region may differ in nature from those forming within. Most importantly, this categorization lacks 
a clear link to the magnetic structures and physical processes involved in dimming. For example, core dimmings are usually considered to outline the footprints of the erupting flux, but it is  unclear whether and 
when secondary dimmings outline such footprints as well. 

Therefore, we propose a categorization based on the magnetic flux systems and processes involved, where magnetic reconnection plays a key role. 
Off-limb dimmings are not included because they do not allow resolving the role of all relevant flux systems. 
The proposed categorization aims 
to aid the understanding of the often complex spatial and temporal evolution of dimmings and 
to help exploit their diagnostic potential for the eruption process. We also discuss the relationship with other elements of the eruptions, in particular flare ribbons, which can help in distinguishing different categories of dimming in the observations. 

The processes immediately underlying the formation of dimmings by density depletion are 
\begin{itemize} 
 \item expansion of erupting flux, 
 \item lifting of overlying flux, 
 \item plasma outflow from the low corona into newly connected open or large-scale flux systems. 
\end{itemize} 
The first of these likely always involves reconnection (in particular, the work by \cite{Forbes2000} strongly suggests that it is required for the evolution into a CME), the second may or may not involve reconnection, and the third one 
requires it. 
In agreement with all current eruption models, we consider the erupting flux to take the structure of a flux rope, regardless of whether the rope exists already before the onset of eruption 
or forms during the eruption \citep[see, e.g.,][]{Chen:2011, Patsourakos&al2020}. 

The flux systems involved are the following: 
\begin{itemize} 
 \item the erupting flux rope, 
 \item overlying ambient flux (the ``strapping'' flux), 
 \item closed flux exterior to the immediate source region of the eruption, 
 \item open flux.
\end{itemize} 
The first two of these form the immediate source region of the eruption, assumed to exist in a force-free magnetic equilibrium prior to the eruption. Therefore, the overlying ambient flux is usually referred to as the strapping flux (or ``strapping field''), which holds down the current-carrying ``core flux''. The strapping flux usually includes a ``guide field'' (or ``shear field'') component in the main direction of current flow, which is along the polarity inversion line (PIL) of the immediate source region. In technical terms, the guide field component points in the toroidal direction (along the often partly toroidal core flux), and the strapping field points in the poloidal direction. Due to the high shear at its base, the immediate source region is always observed to be a filament channel \citep[e.g.,][]{Patsourakos&al2020}. ``Exterior flux'' is not a part of the initial equilibrium but becomes involved in the eruption. This may be flux in a complex AR that passes 
over a remote section of the PIL separate from the erupting section or over a PIL other than the erupting one, 
a neighboring AR, or generally other flux exterior to the source region of the eruption. 
In the given context of eruptions, we do not distinguish between flux that has stretched upward beyond the Alfv\'en surface, i.e., very roughly beyond $\sim\!(15-20)~R_\odot$ \cite[][]{Kasper2021}, and flux that extends to the outer heliosphere. Any plasma this flux is frozen in and any waves propagating through it will not return to the Sun, so, we will refer to it as ``open flux''. 
Although open flux is always exterior to the force-free equilibrium of a filament channel, we distinguish it here from closed exterior flux in view of the different outcomes of reconnection with erupting flux and the resulting different effects on the dimmings. 

\begin{figure} 
\centering
\includegraphics[width=0.8\textwidth]{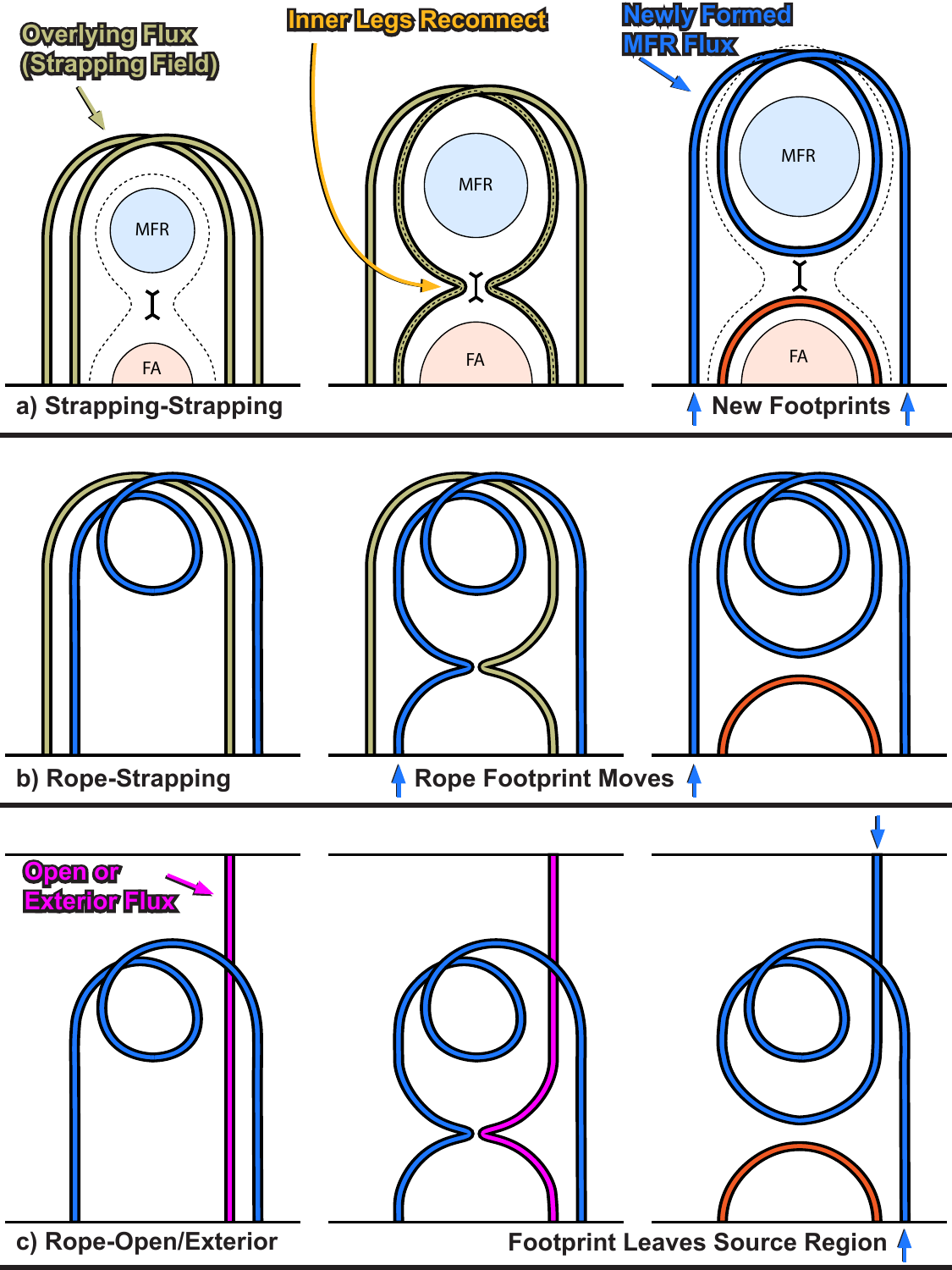}
\caption{Schematic for the basic eruption scenario, the relevant flux-systems, and their reconnection products. Representative field lines are drawn before, during, and after reconnection from left to right. The basic building blocks, including the erupting magnetic flux rope (MFR), the flare current sheet (thick black line), the flare arcade (FA), and the 2D projection of flux that is being drawn into the flare current sheet (thin dashed line) are indicated in the top row only for clarity. Blue arrows indicate the connectivity changes of the MFR. a) \emph{Strapping-strapping} reconnection for the standard flare model in 3D (Sect.~\ref{ss:strapping}). Here the legs of two overlying loops convert strapping flux (tan) into MFR flux (blue) and a post-flare loop (red). b) \emph{Rope-strapping} reconnection, where the leg of a rising MFR flux bundle reconnects with strapping flux, adding new poloidal flux to the MFR, and swaps the footprint 
(Sect.~\ref{ss:movingFR}).  c) Analogous reconnection with exterior (Sect.~\ref{ss:exterior}) or open flux (Sect.~\ref{ss:open_flux}) that originally had one leg rooted near the source region and one far from it (magenta), effectively shifting the MFR footprint to an external location which lies at infinity in case of rope-open reconnection. 
The other footprint can similarly be shifted out of the source region (not shown). The potentially involved leg-leg reconnection of the MFR is topologically similar to strapping-strapping reconnection, with both participating field lines running in the body of the flux rope. Variants of this process are discussed in Sects.~\ref{ss:FRdimmings}, \ref{ss:movingFR}, and \ref{ss:sim_example1}. 
}
\label{f:rxn_schematics}
\end{figure}

We illustrate the essentials of the basic eruption scenario and the relevant flux systems for coronal dimmings schematically in Fig.~\ref{f:rxn_schematics}. Here the colored field lines represent a given flux-system and their footprints prior to, during, and after reconnection takes place (from left-to-right). In the top row we also draw the main elements of the standard model of eruptive flares, including the erupting magnetic flux rope (MFR), the flare current sheet (thick black line), the flare arcade (FA), along with the 2D projection of flux that is being drawn into the flare current sheet (thin dashed line), which are left off the remaining panels for clarity. Because the presence of a non-zero guide field breaks the 2D symmetry of the system, magnetic field lines or flux bundles will not in general reconnect with themselves on either side of the flare current sheet in a 3D configuration. Instead we must consider two field lines and four footprints or legs in total. This includes the two interior legs that reconnect and their two counterparts on opposite sides of the PIL. 
The identity of each field line and the direction of the guide field is indicated by how they overlap one another. 

The changing identity and connectivity of a given footprint through reconnection has implications for the area, motion, and appearance of dimming regions. \emph{Strapping-strapping} reconnection  (Sect.~\ref{ss:strapping}) adds new flux to the rope footprint (Fig.~\ref{f:rxn_schematics}a), \emph{Rope-strapping} reconnection  (Sect.~\ref{ss:movingFR}) moves the footprint (Fig.~\ref{f:rxn_schematics}b), and reconnection of the rope or strapping flux with exterior flux (Sect.~\ref{ss:exterior}) or open flux (Sect.~\ref{ss:open_flux}) can shift one or both footprints of the rope out of the source region entirely (Fig.~\ref{f:rxn_schematics}c).

The expansion and lifting of flux, typically working in conjunction with reconnection in one or several of these forms, leads to dimming, which can be categorized as (several forms of) \emph{flux-rope dimming} in the footprints of the erupting flux rope, \emph{strapping-flux dimming}, \emph{exterior dimming}, and \emph{open-flux dimming}. These are typically observed as core dimming, core or secondary dimming, secondary dimming, and core dimming that can be superseded by exterior dimming, respectively. The following Sections~\ref{ss:FRdimmings}--\ref{ss:open_flux} will consider these relationships in detail. Section~\ref{ss:complex_dimmings} illustrates how the proposed categories can be a useful guide in the analysis of complex dimming events. 
A summary of the new dimming categories, including a compilation in tabular form (Table~\ref{t:categories}), and of their diagnostic potential is given in Section~\ref{ss:dimming_summary}.

\subsection{Flux-rope dimmings} 
\label{ss:FRdimmings} 

\subsubsection{Stationary flux-rope dimmings} 
\label{sss:statFRdimmings} 

\begin{figure} 
\centering
\includegraphics[height=8.3cm]{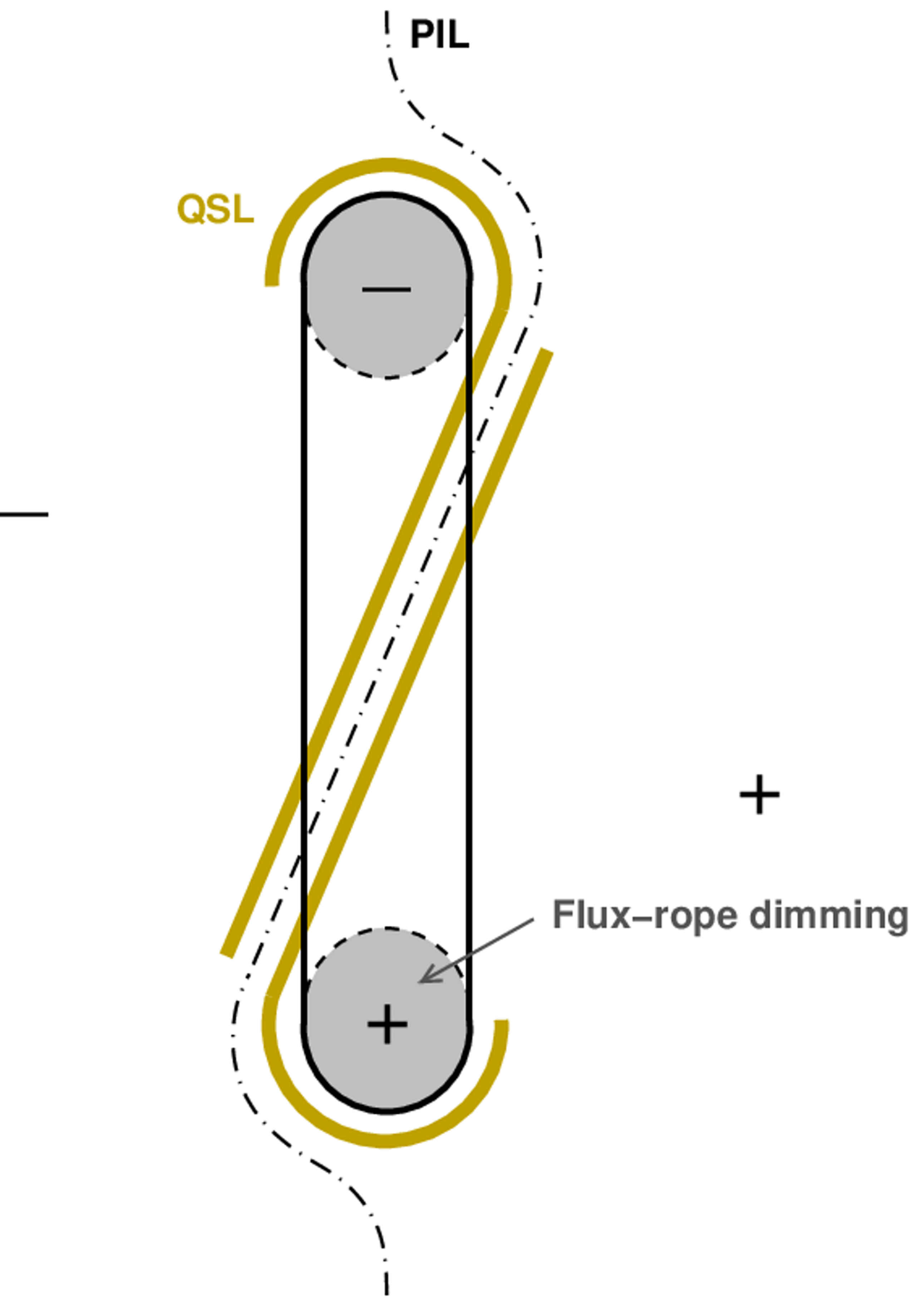}  
\includegraphics[height=8.3cm]{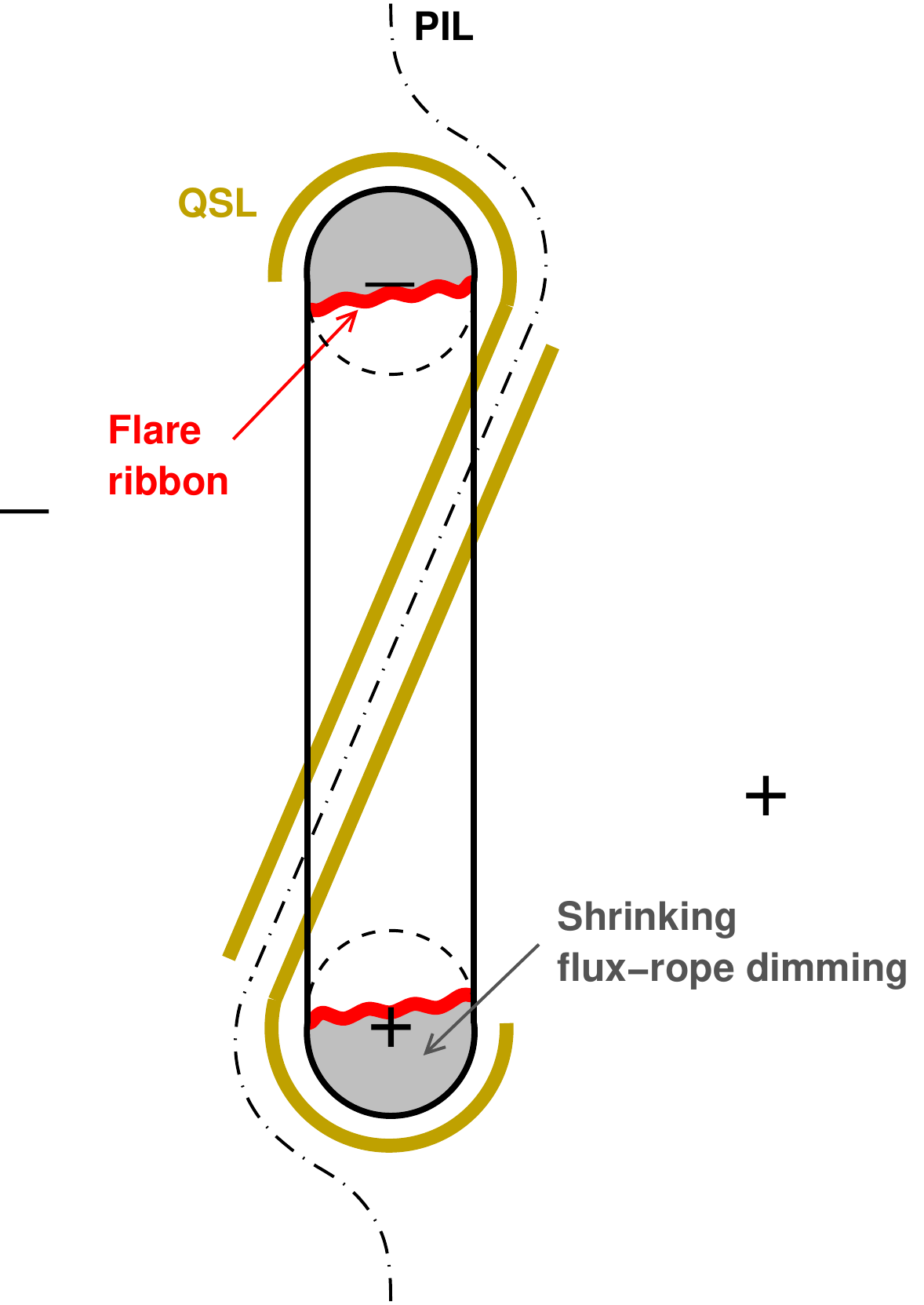} 
\caption{Left: Schematic of flux rope, double-J shaped QSLs, and \emph{stationary flux-rope dimmings}, due to the expansion of the erupting flux rope. Right: \emph{Shrinking flux-rope dimmings}, due to the expansion and primary leg-leg reconnection of the erupting flux rope. 
} 
\label{f:FRdimmings}
\end{figure}

The erupting flux rope in the classical 2D model of eruptive flares stretches the (essentially vertical) flare current sheet underneath, where ``flare'' reconnection of strapping flux with strapping flux and the acceleration of the rope establish a positive feedback, but the rope itself does not reconnect (Fig.~\ref{f:rxn_schematics}a). 
Therefore, the erupting flux rope expands into the interplanetary space, drawing plasma out, with its legs staying intact. This leaves a pair of core dimmings at its footprints (Fig.~\ref{f:FRdimmings}, left). In our classification, we refer to these dimmings as \emph{stationary flux-rope dimmings}. In many eruptions from ARs, the footprints of the erupting flux rope lie in the elbows of a soft X-ray or EUV sigmoid,  and there are indeed numerous examples of twin dimmings forming in these places \cite[e.g.,][]{Sterling:1997}. 

\begin{figure} 
\centering
\includegraphics[width=0.98\textwidth]{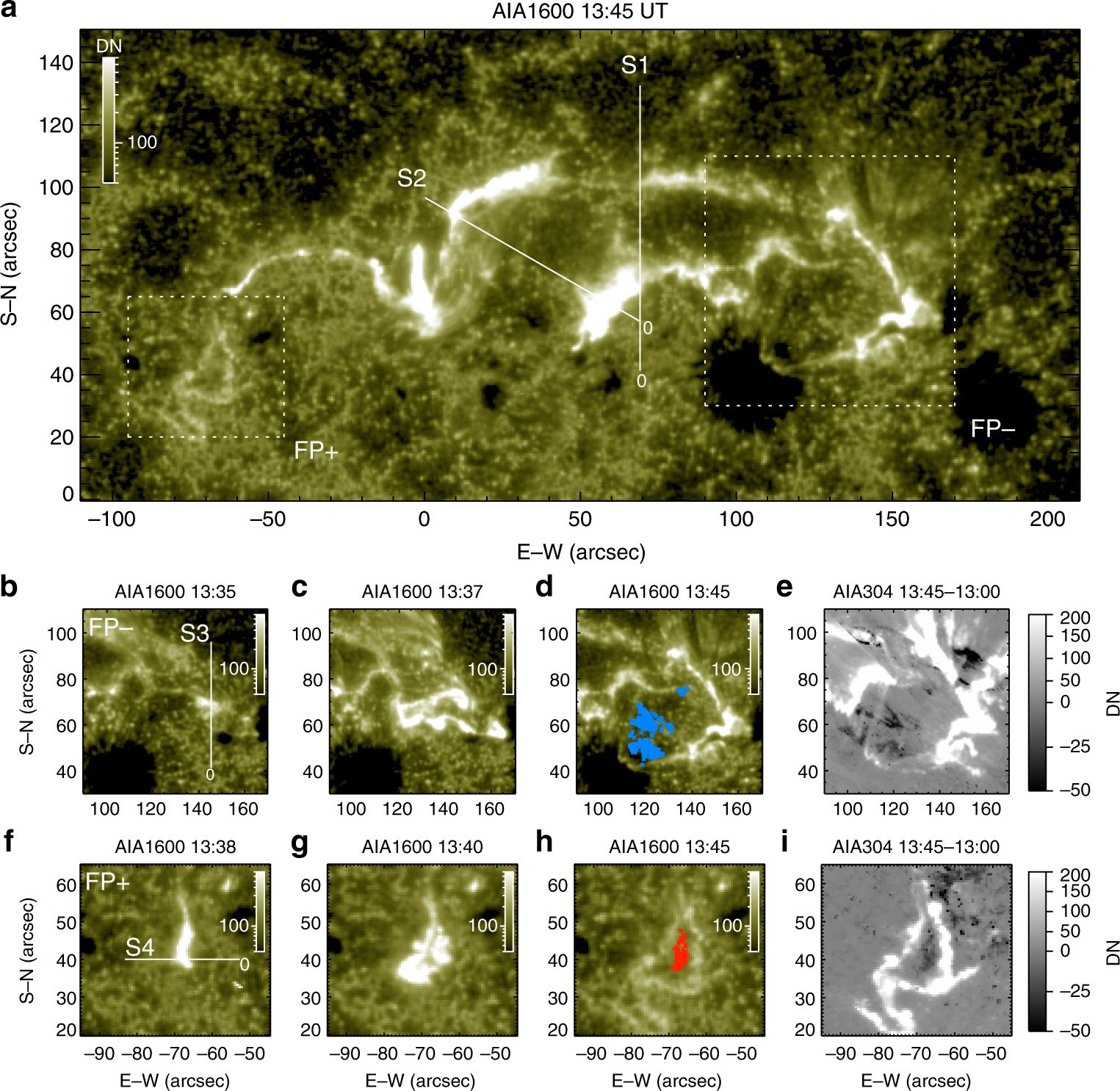}
\caption{Flare ribbons closing completely around \emph{stationary flux-rope} (twin/core) \emph{dimmings} in an eruption from NOAA AR~12443 on 2015 Nov~4. The dimmings in the SDO/AIA 304~{\AA} base-difference images in panels (e) and (i) are replotted in panels (d) and (h) in blue and red colors on top of the AIA 1600~{\AA} images, respectively. The closed ribbons also show the growth of the flux rope's footprints due to the accretion of flux by the flare reconnection of strapping flux. From \cite{WangW&al2017}. The movie in the online supplement shows the event evolution in nine SDO/AIA (E)UV channels along with GOES SXR and Fermi hard X-ray light curves \citep[from][]{WangW&al2017}.} 
\label{f:closed_ribbons}
\end{figure}

The flare ribbons in compact, sigmoidal source regions often also take a double-J shape that follows the evolving quasi-separatrix layers (QSLs), with the curved ends 
of the J's wrapping around the footprints of the erupting flux rope; see Fig.~\ref{f:FRdimmings} 
and also \citet{Demoulin1996}, \citet{Titov&Demoulin1999}, \citet{Janvier&al2014}, \citet{Savcheva&al2016}, and \citet{Aulanier&Dudik2019}. 
The ribbons may even close completely around one or both footprint areas in some cases. The footprint area with its enclosing ribbons can grow in size, indicating the flux accretion to the erupting flux rope by the strapping-strapping
reconnection 
(Fig.~\ref{f:rxn_schematics}a); this is illustrated by the example in Fig.~\ref{f:closed_ribbons}. However, 
the flare ribbons at the edges of expanding, non-reconnecting flux ropes never sweep across the stationary flux-rope dimmings. 

Dimmings of this type following eruptions from the quiet Sun can be quite extended, showing the true, often large area of photospheric flux that joins the horizontal flux in the erupting filament channel. Ribbons or ribbon-like features that partly or fully enclose these dimmings have
been referred to as endpoint brightenings \citep{WangY-M&al2009}. 
The frequently observed slow rise of the current-carrying core flux prior to the main eruption can cause weaker pre-eruption flux-rope dimmings (see Figs.~\ref{fig:qiu2017}, \ref{fig:wang2019b} and references in Sect.~\ref{sec:pre-eruption}). 

These dimmings provide a direct measure of the flux opened into the interplanetary space by the CME \cite[e.g.,][]{Attrill:2006, Qiu:2007}. The source-region (``dimming'') flux and interplanetary flux content of CMEs are found to agree in order of magnitude, in some cases up to a factor of two, but this also indicates that significant uncertainties exist in estimating their values (see Sect.~\ref{ss:relations_in-situ}). 

\begin{figure}[htbp]
\centering
\includegraphics[width=0.8\textwidth]{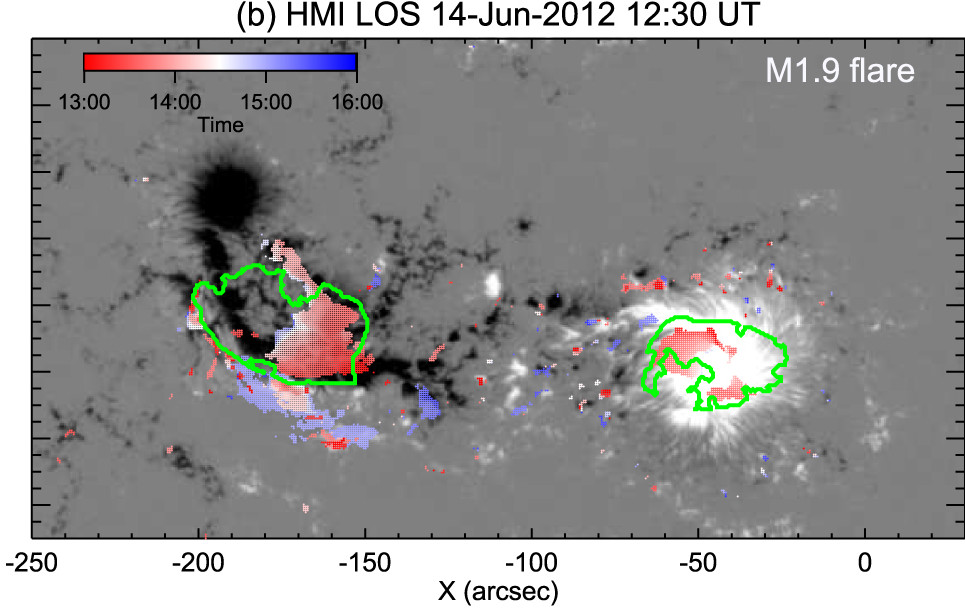}
\caption{\emph{Shrinking flux-rope dimmings} with the initial dimmings shown as green contours and the position of the flare ribbons, which partially re-close them, color coded as given in the time bar. Background image is the line-of-sight magnetogram of AR~11504 on 2012 June~14. From \cite{Wang2019}. The evolution of this event in SDO/AIA along with a movie is shown in Fig.~\ref{fig:wang2019a}. } 
\label{f:shrinkingFR}
\end{figure}

\subsubsection{Shrinking flux-rope dimmings} 
\label{sss:shriFRdimmings} 

One way for flux-rope dimmings to re-close, relevant for the stationary category, is by reconnection of the erupting flux rope's legs with each other. Here we consider the case of ``primary leg-leg'' reconnection, occurring before any other reconnection of the rope (the complementary case of ``secondary leg-leg'' reconnection will be addressed in Sects.~\ref{ss:movingFR} and \ref{ss:sim_overview}--\ref{ss:sim_example1}). Such reconnection 
has been seen in numerical simulations of highly twisted erupting flux ropes in modeling a strongly writhing eruption at the limb, which did not allow any dimming to be observed. The very strong writhing for a twist of at least three turns forced the flux-rope legs to approach each other strongly, squeeze and destroy the flare current sheet, and reconnect \citep{Kliem&al2010}. 
Such leg-leg reconnection 
leads to the reformation of a flux rope of typically weak twist low in the corona and to a fully or partly detached multi-turn toroidal flux bundle that may escape as a CME (e.g., \citealt{Gibson&Fan2008, Kliem&al2010}; Sect.~\ref{ss:sim_example1}). The initial rise of the flux rope is expected to produce stationary flux-rope dimmings, which may grow due to the flare reconnection of strapping flux. The leg-leg reconnection forms flare ribbons that sweep across these dimmings, typically from the side facing the other footprint, with the exact geometry being determined by the twist and height of the interaction. 
\emph{Shrinking flux-rope dimmings}, which are stationary, will be observed (Fig.~\ref{f:FRdimmings}). Although distinct observationally by separate sweeping flare ribbons and shrinking, these dimmings do not represent a new category of dimming formation in physical terms, but rather the re-closing of stationary flux-rope dimmings.  For an example, see Fig.~\ref{f:shrinkingFR} \cite[from][its pre-eruption dimming is discussed in Sect.~\ref{sec:pre-eruption}]{Wang2019}. 

It appears that flux ropes with a twist of at least three turns, primary leg-leg reconnection and shrinking flux-rope dimmings are only relatively rarely observed. \citet{Dudik&al2022} report primary leg-leg reconnection in a large prominence eruption on 2012 August~31, which appeared to be related to the writhing of the prominence. However, the legs interacted very close to the top of the prominence without any noticeable effect on the flux-rope dimmings, which rather appeared to belong to the moving flux-rope dimming category that is discussed in Sect.~\ref{ss:movingFR}. A pair of nearly stationary flux-rope dimmings shrinking from the inner side was observed in an eruption from AR~11577 on 2012 September~27 \citep{2019ApJ...879...85V, PanH&al2021}. However, different from the schematic in Fig.~\ref{f:FRdimmings}, the shrinking was due to the straight legs of the J-shaped main ribbon pair sweeping into the flux-rope dimmings, which suggests rope-strapping reconnection as the underlying process (also discussed in Sect.~\ref{ss:movingFR}). 

A process intermediate between primary and secondary leg-leg reconnection has been pointed out in \citet{Aulanier&Dudik2019}. Field lines in the legs of an erupting flux rope, which had shortly before been added to the rope by strapping-strapping reconnection, reconnected with each other. 
If such reconnection, once established, would continue until it also involves the original flux in the rope, then shrinking flux-rope dimmings are expected to form. However, most of the flux in the rope reconnected with strapping flux in this simulation, leading to moving flux-rope dimmings (Sect.~\ref{ss:movingFR}). 

If the flux rope legs do not reconnect completely, the remaining part of the shrinking flux-rope dimmings, not swept by flare ribbons, is a footprint area of flux that opens into the interplanetary space. Additional flux may open if the toroidal flux bundle (the upper product of leg-leg reconnection) reconnects with closed exterior flux before escaping from the Sun. A pair of exterior dimmings (Sect.~\ref{ss:exterior}) will then form in the footprints, which can serve to estimate the additionally opening flux. An upper bound to this potential contribution to the opening flux is given by the flux swept by the flare ribbons in the shrinking flux-rope dimming, which equals the flux content of the 
toroidal flux bundle.

For stationary flux rope dimmings and the remaining part of their shrinking variant, the only way to recover consists in re-establishing a coronal stratification as a consequence of the coronal heating. Because this operates in the opened flux, a coronal hole-like stratification results, which is usually very different from the original source-region stratification. However, a dimming is only a \emph{transient} coronal hole. This implies that these dimmings must either shrink completely or cannot remain stationary during the entire recovery, which are both at variance with the standard 2D flare model. The latter case indicates some form of reconnection with ambient flux and migration of the open flux's footprint as part of the large-scale coronal recovery after the eruption (Sects.~\ref{sec:recovery} and \ref{sec:simulations}). Several options exist for such reconnection; these lead to the further dimming categories characterized below. The examples above indeed show these effects. The dimmings in the event on 2012 June~14 (Fig.~\ref{f:shrinkingFR}) and on 2012 September~27 eventually shrink completely. The dimmings in the event on 2015 November~4 begin to move after the stationary phase shown in Fig.~\ref{f:closed_ribbons}, a process considered in Sect.~\ref{ss:movingFR}. If the recovery begins already during the stationary phase, then the temporal and DEM characteristics of this part of the recovery bear diagnostic potential for the coronal heating process.

\subsection{Strapping-flux dimmings} 
\label{ss:strapping}

\begin{figure}[t]
\centering
\includegraphics[height=10.6cm]{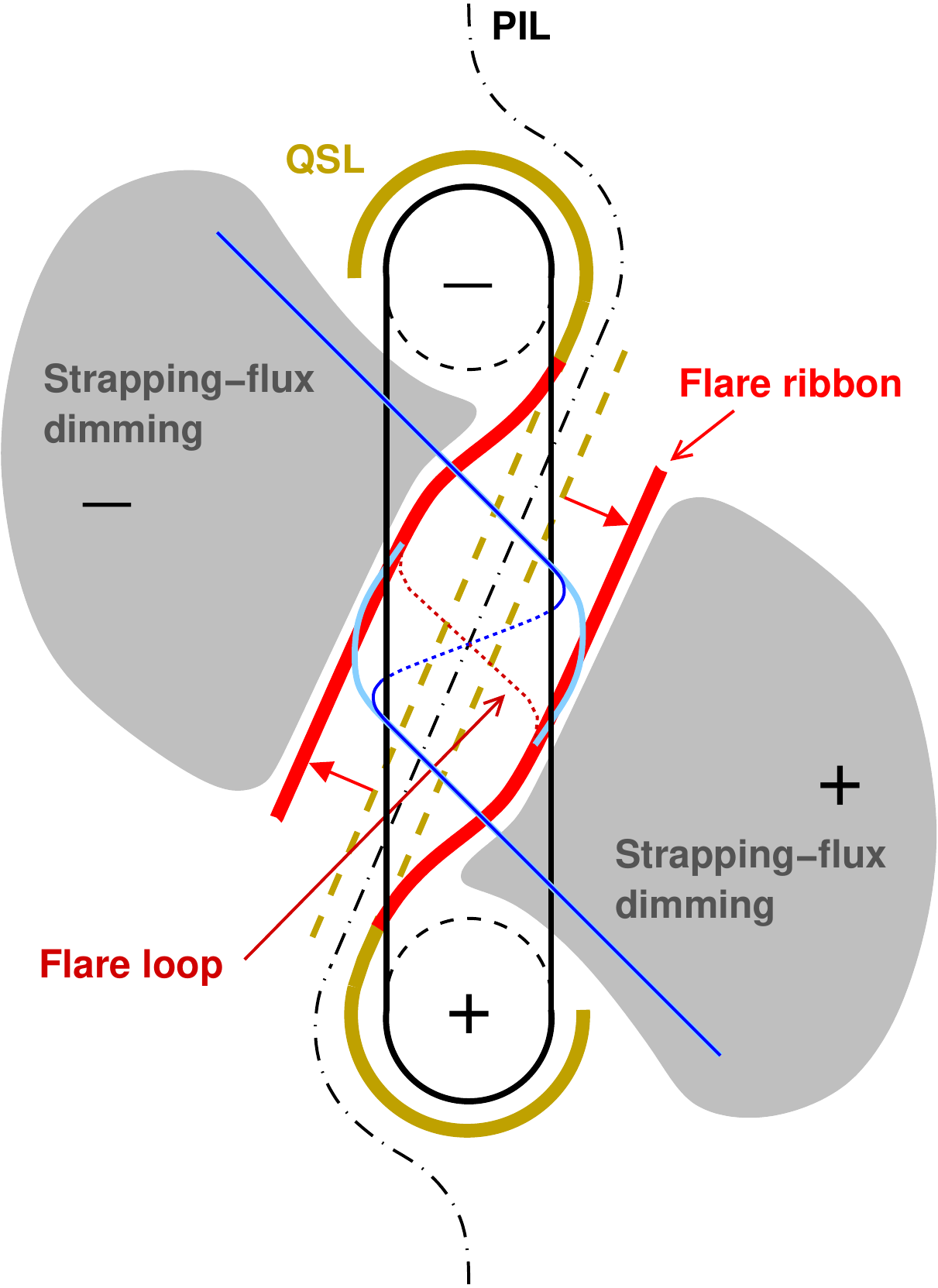} 
\caption{Schematic showing the formation of \emph{strapping-flux dimmings} in the roots of the lifted strapping flux in the presence of a guide field component and the sweeping of the flare ribbons across these dimmings (red filled arrows). The position of the straight QSL sections at the onset of flare reconnection is shown dashed. 
The underlying strapping-strapping reconnection (Fig.~\ref{f:rxn_schematics}a) is here visualized by the change of the light-blue field lines to a flux-rope field line (blue) and a flare-loop field line (red). 
The erupting flux rope is expected to form flux-rope dimmings in its footprints as well; these are not included here, to emphasize the strapping-flux dimmings.
} 
\label{f:strapping}
\end{figure}

\emph{Strapping-flux dimmings} develop in the footprint regions of the strapping flux when it is lifted by the erupting flux rope, especially in the part that is added to the erupting flux by strapping-strapping 
reconnection (Figs.~\ref{f:rxn_schematics}a and \ref{f:strapping}). 
These dimmings can appear as extended secondary but also as core dimmings. 
A pairwise occurrence in opposite polarities is typical, but the dimming may not form if the strapping flux has a very compact root, e.g., in a sunspot (see the event on 2011 October~1 described in Sect.~\ref{sss:complex_dimmings1}). The dimmings expand outward as the lifting progresses to higher-lying flux which is rooted further away from the PIL. The lifting of the strapping flux begins already in the slow-rise phase, potentially forming pre-eruption dimmings, which, however, are usually too faint to be detected (an example is briefly characterized below). 

As a result of 
strapping-strapping reconnection, which typically 
begins simultaneously with the upward acceleration of the erupting flux rope, flare ribbons sweep across the strapping-flux dimmings from the side facing the PIL (Fig.~\ref{f:strapping}), and the area behind the ribbons turns bright due to the formation of flare loops. The reconnected strapping flux becomes part of the flux rope (Fig.~\ref{f:rxn_schematics}a) and its footprint areas can dim as deeply as the dimmings in the initial footprints of the flux rope. Strapping flux passing over the end sections of the erupting flux is lifted but does not reconnect. Rather, this flux leans to the side early in the process to let the erupting flux pass through, and only a moderate stretching and minor or no contribution to the dimming result (see Sect.~\ref{ss:sim_example1}). 
Due to 
the progressing flare reconnection, the strapping-flux dimmings exhibit an apparent motion away from the PIL. In this process, the outward expansion of the area is initially often stronger than the shrinking at the inner side, because the roots of the strapping flux often include extended areas of low photospheric flux density in the outer parts of the source region. 

These dimmings can form outside the whole straight sections of the J-shaped QSLs/ribbons, 
but also adjacent to their elbows (Fig.~\ref{f:strapping}). The latter, i.e., a location near the footprints of the erupting flux rope, is to be expected if the strapping field includes a significant guide-field component, which is often indicated by a strong shear of the first flare loops. The flux-rope and strapping-flux dimmings in each polarity can then initially appear as one dimming area. The two dimming components may be distinguished by the motion of the flare ribbons if the erupting flux rope does not itself join the flare reconnection (cf.\ Sect.~\ref{ss:movingFR}); ribbons then sweep only across the strapping-flux dimmings. A discrimination is also possible if the hook-shaped end sections of the ribbons close largely or fully around the flux-rope dimmings. Because eruption source regions typically possess considerable shear, it is not easy to isolate a pure strapping-flux dimming in the observations. 

\begin{figure} 
\centering
\includegraphics[width=\textwidth]{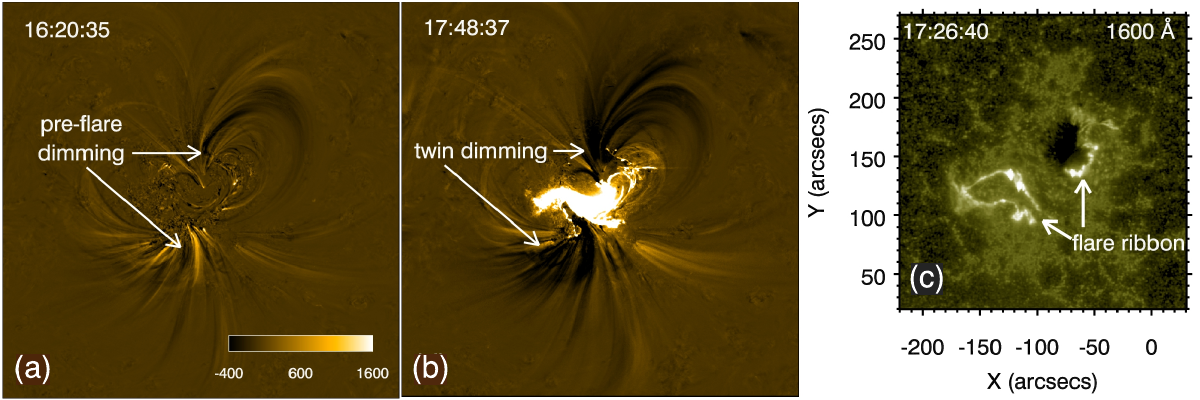} 
\caption{Pre-flare dimming (a)
and \emph{strapping-flux dimming} (here called ``twin dimming''; b) in SDO/AIA 171~{\AA} base-difference images of the eruption of 
SOL2014-09-10, and the flare ribbons at the onset of the impulsive flare phase in an SDO/AIA 1600~{\AA} image (c). Adapted from \cite{Zhang2017}.
The movie in the online supplement shows the event evolution in SDO/AIA 171~{\AA} base-difference images \citep[from][]{Zhang2017}.
}
\label{f:20140910}
\end{figure}

The dimmings in the eruption from AR~12158 on 2014 September~10, analyzed in \citet{Zhang2017}, provide an example. This event is interesting in particular because it also produced a pre-eruption dimming in the strapping flux. The pre-eruption dimming consisted of two narrow bundles of darkening overlying loops rooted near, but not in, the hooks of the double-J flare ribbons developing from the onset of the eruption (Fig.~\ref{f:20140910}a,c). 
The dimming broadened and deepened strongly in the broad temperature range covered by the 171, 193, 211, and 335~{\AA} channels from the onset of the eruption, conforming well to the expectation that a large part of the flux passing over the center of the erupting flux is lifted by the erupting flux and forced to reconnect (Fig.~\ref{f:20140910}b). 

Two further examples can be found in the sympathetic eruptions of two quiescent filaments discussed in Sect.~\ref{sss:complex_dimmings3}. Both showed strapping-flux dimmings along an extended arcade of flare loops; in one eruption, these were also clearly separate from flux-rope dimmings in the relatively weakly sheared environment of the large-scale filaments (see Fig.~\ref{f:sympathetic} below). 

The eventual recovery 
of the strapping-flux dimmings is due to two effects.  Primarily, flux that has joined the erupting flux can find new footprints
over time by further reconnection. The 
reconnection between the legs of the erupted flux and strapping flux, discussed in Sect.~\ref{ss:movingFR}, and reconnection with closed or open exterior flux (Sects.~\ref{ss:exterior}--\ref{ss:open_flux}) are candidates for such reconnection. 
The latter is favored by the dominance of area reduction from the outer side of the dimming in the recovery process for many events (\citealt{Kahler:2001, Attrill:2008}; Sect.~\ref{sec:recovery}). Secondarily, flux that leaned to the side returns to near the initial position after the event. 

Only a part of the strapping-flux dimming---the footprint of the flux that becomes part of the erupting flux rope---marks 
flux that opens into the interplanetary space. Its amount equals the flux swept by the flare ribbons, as long as the flux rope does not itself join the flare reconnection. However, observational estimates of the flux in extended dimming areas (which include strapping-flux and exterior dimmings), although agreeing with the estimated flux swept by the flare ribbons in order of magnitude, are typically smaller than this ``ribbon flux'' by a factor $\sim\!2\mbox{--}3$ 
\citep{Qiu:2007, Temmer:2017}. Reasons for this difference are reviewed in Sect.~\ref{ss:relations_in-situ}; a primary one being the underestimation of the source-region flux from the dimming area, especially for extended dimmings, whose boundaries are typically less well defined than the boundaries of flux-rope dimmings.

\begin{figure}[tb] 
\centering
\includegraphics[width=0.82\textwidth]{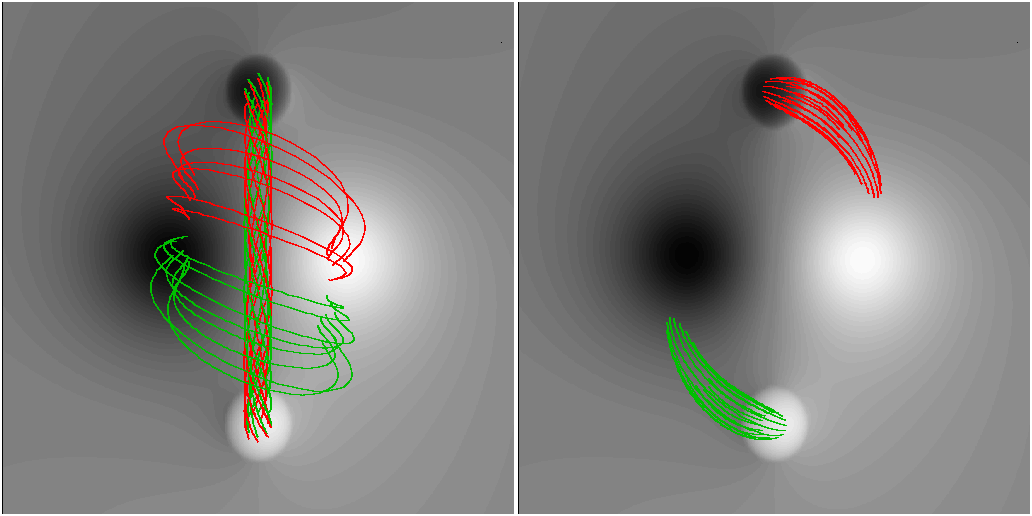}
\caption{
   Left: Field lines showing the core of the initial flux rope equilibrium of the CME simulation in Fig.~\ref{f:FRflare_reconn} (same data as in panel (a)). Green (red) field lines in the center of the rope are started in the positive (negative) footprint of the rope. The bundles of strapping flux that reconnect with the flux rope in the course of the simulation are visualized by field lines of the same color as the field lines traced from the footprint of the rope leg that reconnects with this strapping flux. The normal component of the magnetogram is displayed in grayscale. 
   Right: Field lines with identical start points as the rope field lines of the left panel in the corresponding potential field 
   (for clarity, only every second of the corresponding field lines is shown in the left panel). These field lines connect to near the footpoints of the strapping field lines in the left panel, indicating that the post-eruption core field of the source region (Fig.~\ref{f:FRflare_reconn}c,d) approaches the potential field, but does not reach it fully. 
} 
\label{f:Bpot}
\end{figure}

\subsection{Moving flux-rope dimmings} 
\label{ss:movingFR} 

In three dimensions, the classical 2D picture of the erupting flux rope not participating in the flare reconnection is at variance with the expectation that the eruption leads to an approach of the configuration to the potential-field state. The change to that state requires reconnection of the flux rope with the strapping flux, as illustrated in Fig.~\ref{f:Bpot}. Therefore, one may expect that the flux-rope (core) dimmings are often modified by ``rope-strapping'' reconnection. 
Such reconnection has been demonstrated and studied in detail in \citet{Gibson&Fan2006, Gibson&Fan2008} and \citeauthor{Aulanier&Dudik2019} (\citeyear{Aulanier&Dudik2019}; the ``ar--rf'' reconnection in their nomenclature), see Sect.~\ref{ss:sim_overview} for a summary. It has been seen in a number of other simulations as well 
\citep[e.g.,][]{Kliem&al2012, vanDriel&al2014, Downs&al2015, JiangC&al2021} 
and has also been inferred from imaging observations \citep{Zemanova&al2019, Lorincik&al2019, Dudik&al2019, ChenHC&al2019, GouT&al2023}. 

\begin{figure} 
\centering
\includegraphics[width=0.92\textwidth]{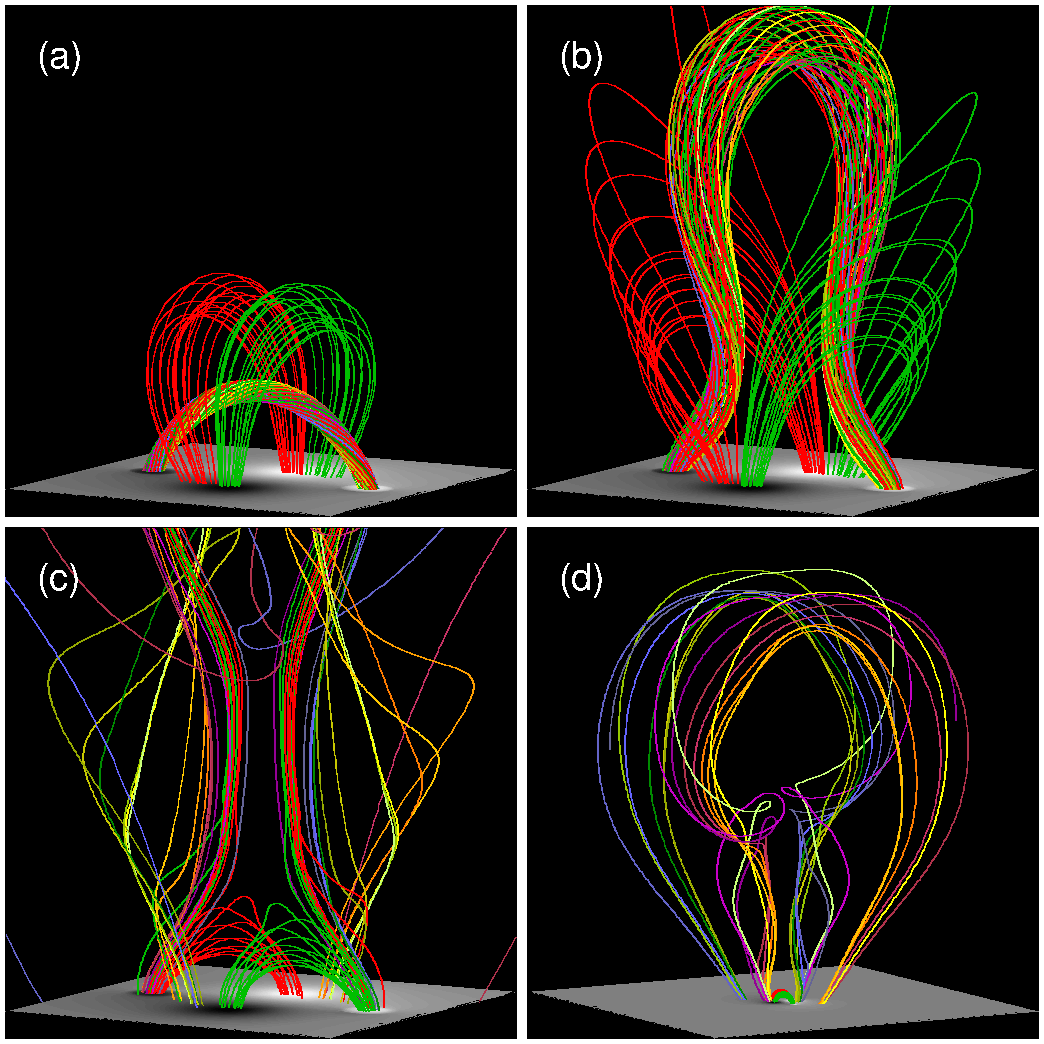}
\caption{Field lines of an erupting flux rope undergoing rope-strapping reconnection. Green (red) rope field lines are traced from points on a circle at the positive (negative) rope footprint, rainbow-colored ones initially from points on a circle at the rope apex; all of $\approx\!1/3$ the minor flux-rope radius. The latter points move with their fluid elements. Field lines of the strapping flux that reconnects with the rope around the time of panel (c) are shown in (a) and (b) in the same color as the field lines traced from the footprint of the reconnecting rope leg. The bottom plane shows the magnetogram in grayscale. 
(a) Initial, unstable force-free equilibrium \citep[from][]{Torok&Kliem2005}. 
(b) Eruption of the kink- and torus-unstable flux rope. 
(c) Reconnection with strapping flux allows approaching the potential field (compare the green and red field lines with Fig.~\ref{f:Bpot}). 
(d) Eventually, the erupting rope is completely rooted in the original roots of the strapping flux, analogous to the finding in \citet{Gibson&Fan2008}. 
} 
\label{f:FRflare_reconn}
\end{figure}

When an erupting flux rope rises, the associated flare reconnection involves inflows from both sides into the flare current sheet, initially of purely ambient (strapping) flux. However, when the legs of the flux rope have become essentially vertical, their internally driven dynamics decreases substantially, and they become more susceptible to being dragged into the flare current sheet by the ongoing inflow. The precise conditions for this to occur remain to be clarified, but observations of dimmings \citep{Kahler:2001} indicate that it happens 
frequently (see detail below). 

The downward reconnection outflow of a flux rope leg interacting with overlying (strapping) flux in the flare current sheet yields flare loops that connect the nearby footprint of the rope with the photospheric roots of the strapping field of conjugate polarity, similar to the potential field of a bipolar source region (Figs.~\ref{f:FRflare_reconn}c and \ref{f:Bpot}). These flare loops extend (lengthen) the classical arcade of flare loops from strapping-strapping reconnection. They tend to arch somewhat higher than the classical flare loops, producing a saddle shape of the upper arcade envelope, as has been pointed out by \citet{Lorincik&al2021} and can also be envisioned from the high-arching strapping field lines in Fig.~\ref{f:FRflare_reconn}b in combination with the fact that the classical strapping-strapping flare reconnection tends to proceed rather low in the atmosphere \citep{Forbes&al2018}. 

\begin{figure}[htbp]
\centering
\includegraphics[height=11.4cm]{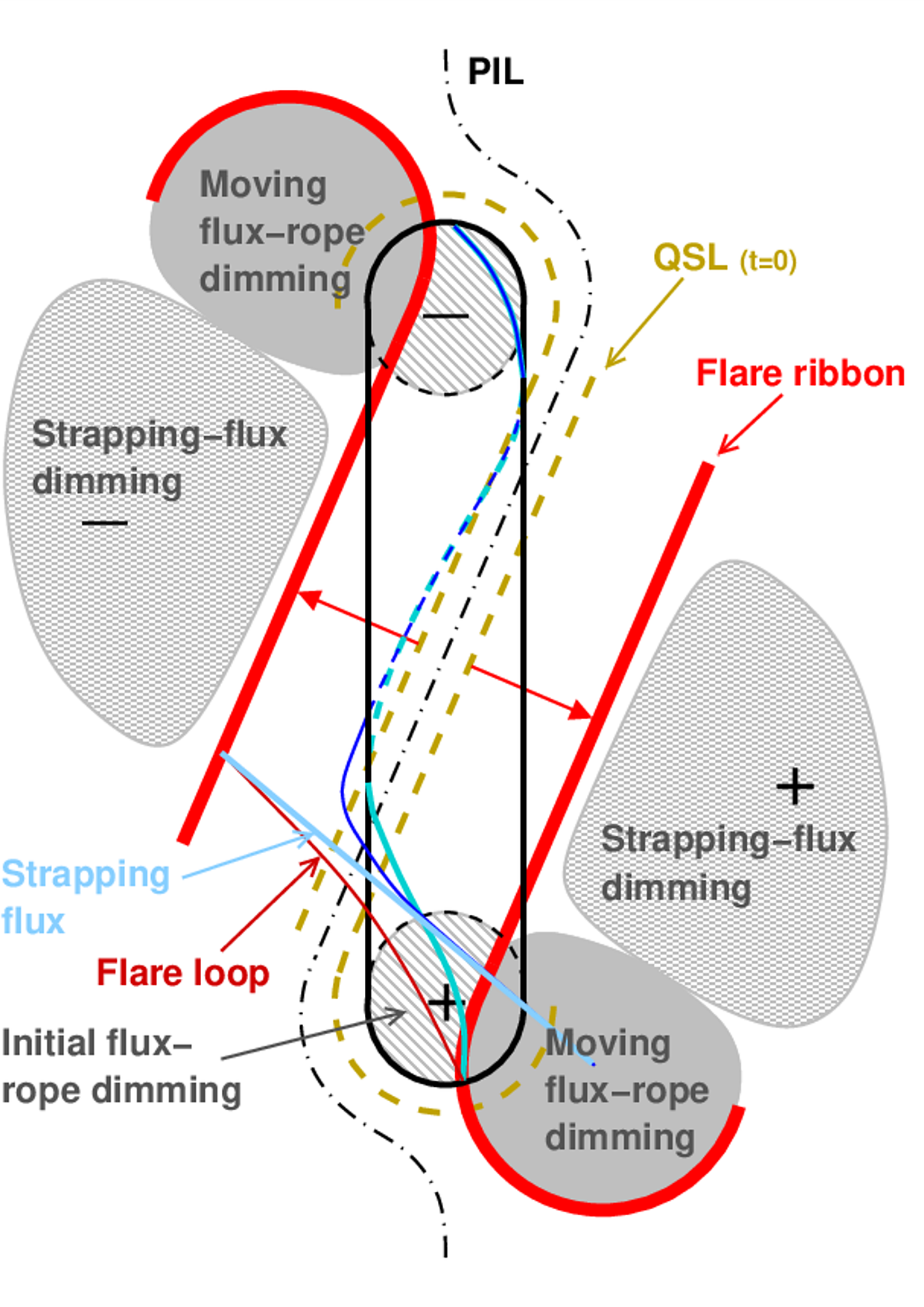} 
\caption{Schematic showing the formation of \emph{moving flux-rope dimmings}. The underlying rope-strapping reconnection (Fig.~\ref{f:rxn_schematics}b) is here visualized by the change of the cyan flux-rope field line and the light-blue strapping field line to the blue flux-rope field line and the red field line in the outward-extending flare loop arcade. For clarity, this reconnection is visualized only for the bottom flux-rope leg; it happens similarly in the other leg. The outward moving and expanding strapping-flux dimming is also shown. Filled red arrows show the motion of the flare ribbons from the initial position of the QSLs. 
} 
\label{f:mfrd}
\end{figure}

As with the downward outflow of strapping-strapping reconnection, flare ribbons propagate into the footprint area of the reconnecting flux along the footprints of the flare loops. 
The straight legs of the J-shaped ribbons sweep into the strapping flux area, and the inner edges of the round ends of the J's (``hooks'') sweep across the initial footprints of the erupting flux rope; see these associations in \citet{Lorincik&al2021} and the schematic in Fig.~\ref{f:mfrd}. 

The upper reconnection product connects the top part of the flux rope to the long section of the cut 
strapping flux (Fig.~\ref{f:rxn_schematics}b), which had arched over the reconnecting flux rope leg (Fig.~\ref{f:FRflare_reconn}b). 
The rope becomes increasingly rooted in the sources of the strapping flux (Figs.~\ref{f:FRflare_reconn}c,d 
and \ref{f:mfrd}). 
Deep dimmings form at the new footprints of the flux rope, 
because it continues to erupt and expand, opposite to the short and shrinking flare loops. 
The dimmings begin to develop from the outer edge of the original flux rope footprints/inner edge of the adjacent strapping flux, which are the outer parts and end regions of the hooks of the J-shaped QSLs (see the QSLs at $t=0$ in Fig.~\ref{f:mfrd}). 
The motion of these hook-shaped or possibly even ring-shaped end sections of the QSLs into the strapping-flux area has been demonstrated by \citet{Aulanier&Dudik2019} and \citet{JiangC&al2021}. It is not likely that flare ribbons form in this intersection of expanding flux, implying that the dimming in the original flux rope footprint and the dimming in the footprint of the reconnecting, adjacent strapping flux form a contiguous area. The dimming area moves away from the PIL into the original footprint area of strapping flux, which has turned into the new flux-rope footprint. We refer to this form of dimming as \emph{moving flux-rope dimming}. 
It develops from an initially stationary flux-rope dimming with the onset of rope-strapping reconnection. 
While this process also completes the migration of the flux rope footprint into the strapping-flux area in the simulation shown in Fig.~\ref{f:FRflare_reconn}, (secondary) leg-leg reconnection contributes to the migration in \citet{Gibson&Fan2008} (see detail in Sect.~\ref{ss:sim_overview}).

Because rope-strapping reconnection typically commences already during the impulsive flare phase, while the flux-rope dimming is still darkening (Fig.~\ref{fig:dimming_sketch}), the initial stationary phase may not stand out clearly in the observations. The two phases can best be distinguished in slow eruptions from the quiet Sun, as in the eruption of a polar-crown filament shown in Sect.~\ref{sss:complex_dimmings3} (see Fig.~\ref{f:sympathetic} below). 
Different from the combination of stationary flux-rope dimmings and moving strapping-flux dimmings, which can form two separate (although often nearby) areas, the moving flux-rope dimming appears to always form a single, contiguous area. The observations reported by \citet{Kahler:2001} and numerical simulations reported in Sect.~\ref{sec:simulations} suggest that the majority of the observed core dimmings belong to this category. They also support 
the above expectation that flare ribbons are not likely to form at the boundary 
between the two original flux components of this dimming. If the motion extends into the outer, weak-field part of the original strapping-field area, then the dimming grows in size (similar to a moving strapping-flux dimming), because the flux in the rope is conserved. 

The complete picture includes a moving strapping-field dimming in each polarity, because the eruption will generally also involve strap\-ping-strapping reconnection; actually, the flare reconnection starts in this way. As pointed out above, the moving strapping-flux dimming can be expected to form near the flux-rope footprint in the presence of a significant shear-field component, which is very often indicated by a significant shear of the first flare loops. The two moving dimmings may then be indistinguishable in the observations, except in fortunate cases that exhibit flare ribbons closing largely or completely around the flux-rope dimming or show a long delay until rope-strapping reconnection and the resulting displacement of the flux-rope dimming set in. 

\begin{figure}[t]  
\centering
\includegraphics[width=0.95\textwidth]{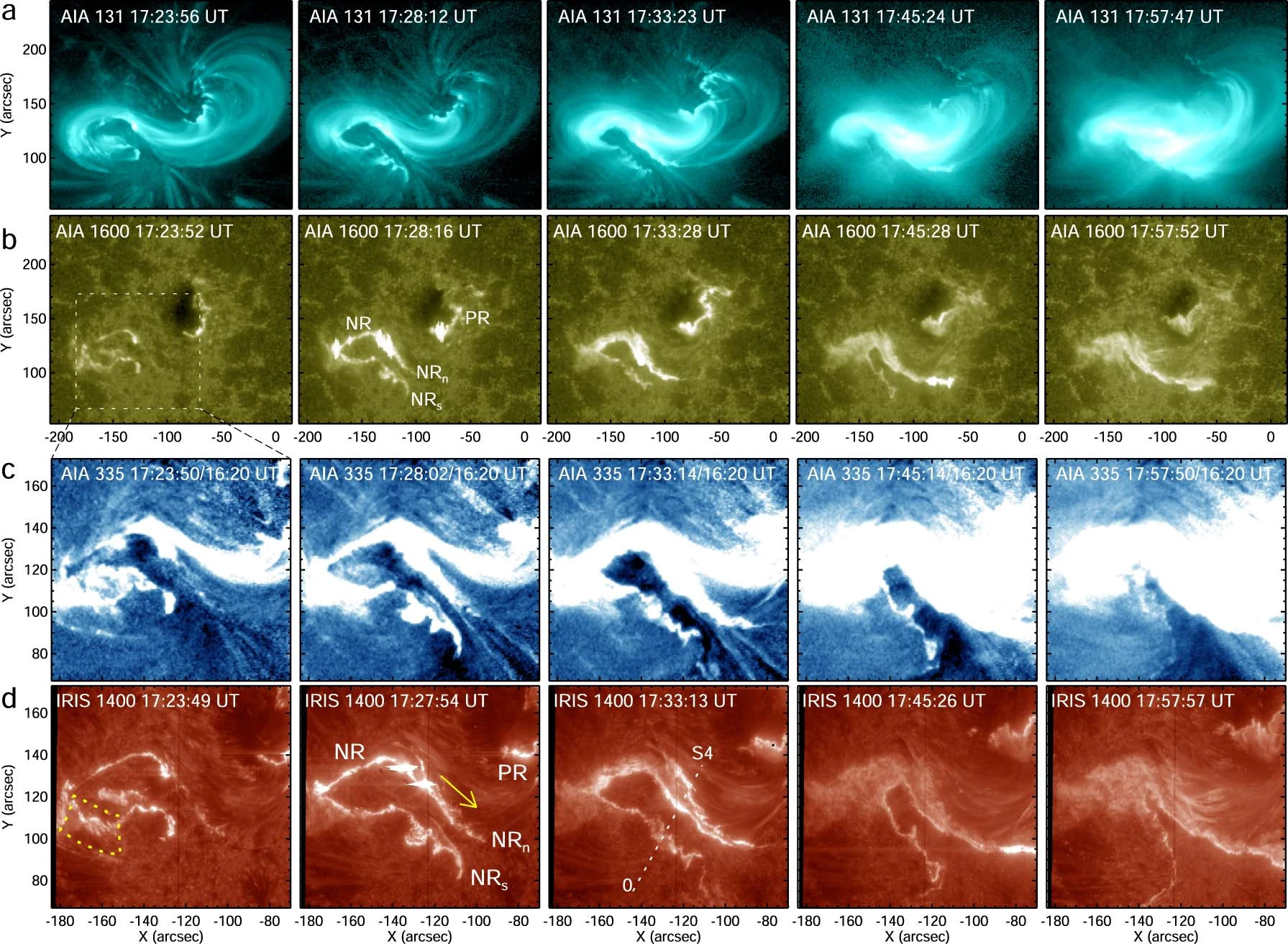}
\caption{\emph{Moving flux-rope dimming} and associated flare ribbon evolution in an eruption from AR~12158 (SOL2014-09-10). SDO/AIA~131~\AA, 1600~\AA, 335~\AA\ base-ratio and IRIS 1400~\AA\ slit-jaw images are displayed (from top to bottom). The characteristic migration of the flux rope's footprints and associated curved ends of the flare ribbons into the strapping-flux area is most clear in the negative polarity at the east side (ribbon NR), but also visible in the positive polarity at the west side in the 131~\AA\ and 1600~\AA\ images. The movie in the online supplement shows the event evolution in AIA 131~{\AA}, 1600~{\AA} and IRIS 1400~{\AA} observations. Figure and movie from \cite{GouT&al2023}. 
} 
\label{f:movingFR}
\end{figure}

Overall, twin core dimmings, simultaneously moving and expanding into the strapping-field area will be observed when the erupting flux rope joins the flare reconnection. Such a development of the core dimmings appears to be the typical case. \citet{Kahler:2001} observed it in at least 11 and up to 17 of their 19 events, and this is now also very often seen in high resolution EUV observations with the SDO/AIA instrument. 
An early demonstration of the footprint migration was given by \citet{Miklenic:2011} in the analysis of an eruption from AR~11012 on 2009 February~13 (part of which is also shown in Fig.~\ref{fig:miklenic_2011_F2}). A very clear motion of flux-rope dimmings and associated flare ribbons into the strapping-flux area has been seen in the eruption from AR~12158 on 2014 September~10, demonstrated in \citet{Dudik&al2016} and \citet{GouT&al2023}. This is shown in Fig.~\ref{f:movingFR}. 

Moving flux-rope dimmings likely often close by reconnection with closed or open exterior flux, similar to the strapping-flux dimmings. This is indicated by the dominant role of area shrinking, primarily from the outer side, in the recovery of many dimming events (Sect.~\ref{sec:recovery} and references therein). Additionally, secondary leg-leg reconnection (of the secondary flux-rope legs formed by rope-strapping reconnection) is a plausible process 
from the geometry of flare reconnection. 
The secondary flux-rope legs are formed in the strapping flux. The relaxation of the source region after the eruption attempts to relax all flux in the region. This includes the reconnection (re-closure) of all strapping flux that has been cut and stretched upward by the eruption, including the secondary flux-rope legs. Secondary leg-leg reconnection is far easier to realize than primary leg-leg reconnection, because the secondary flux-rope legs are positioned in the inflow volume of a well-established flare current sheet. 
Therefore, and because rope-strapping reconnection occurs in many eruptions, secondary leg-leg reconnection is more likely to occur than primary leg-leg reconnection. The process and its relevance for the closure of moving flux-rope dimmings have first been demonstrated by \citet{Gibson&Fan2008}, see Fig.~\ref{f:GibsonFan2008} in Sect.~\ref{sec:simulations}. Their results suggest that the dimming area would retract from the inner side, which differs from closure by reconnection with exterior flux and allows discriminating these different shrinking mechanisms in the observations. 

The moving flux-rope dimmings outline flux that opens into the interplanetary space. A part of the accompanying (and also outward moving) strapping-flux dimmings outlines opening flux as well. This part can be estimated, in principle, from the flux swept by the corresponding flare ribbons (Sect.~\ref{ss:strapping}). However, it is difficult to distinguish the sections of the flare ribbons due to strapping-strapping reconnection from the sections due to rope-strapping reconnection; these are expected to form a pair of continuous ribbons 
along the straight sections of the J-shaped QSLs \citep{Aulanier&Dudik2019}.

\subsection{Exterior dimmings} 
\label{ss:exterior}

The strong expansion of erupting flux in the corona, especially its lateral over-expansion in impulsive events \citep[e.g.,][]{Patsourakos_Vourlidas_Kliem_2010,Veronig&al2018}, often leads to interaction with neighboring flux systems that are exterior to the unstable equilibrium. These can be other parts of a multipolar source region, a neighboring AR (especially during times of enhanced solar activity), or a coronal hole, which spreads out strongly with height. 
Dimmings in the footprints of exterior flux result primarily from reconnection with closed exterior flux but are also possible when exterior flux is merely strongly lifted (an example of the latter is shown in Sect.~\ref{sss:complex_dimmings2}). When a leg of the erupting flux rope reconnects with closed exterior flux, the rope exchanges its footprint in the like polarity and becomes fully or partly rooted at one end in the exterior flux, while its top part continues to rise. 
At the new footprint, 
a secondary dimming develops due to the resulting outflow of plasma into the expanding magnetic flux. We refer to such dimmings as \emph{exterior dimming} and to the underlying interchange reconnection generally as ``rope-exterior'' reconnection (Fig.~\ref{f:rxn_schematics}(c)). Here we do not distinguish the case that the exchange of flux may start with the reconnection of strapping flux wrapped around (not yet fully reconnected into) the erupting flux rope, because the forming exterior dimming is not influenced by this difference. 

\begin{figure}[t]
\centering
\includegraphics[width=\textwidth]{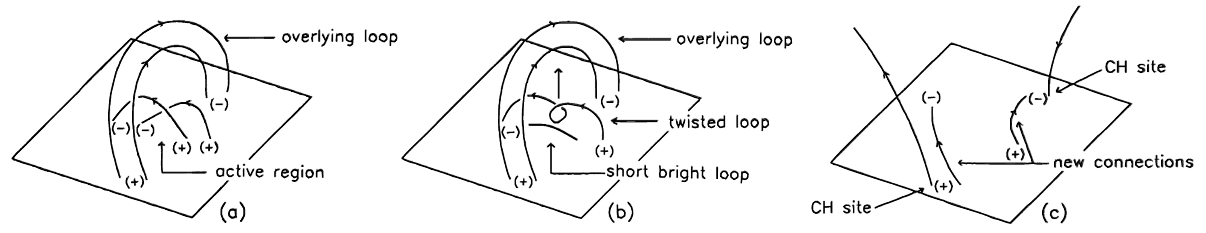}
\caption{Schematic of twin \emph{exterior dimming} formation by interchange reconnection between the legs of the erupting flux (``twisted loop'') and exterior flux (``overlying loop''). Note that each leg of the erupting flux reconnects with a different part
 of the exterior flux (indicated by two field lines). From \cite{Manoharan&al1996}.} 
\label{f:Manoharan+1996}
\end{figure}

A computation of the magnetic connectivities in the potential field allows distinguishing exterior dimmings from strapping-flux dimmings, especially for multipolar source regions, where exterior dimmings can form in relative proximity. One must be aware, however, that a different methodology of the computation or a different resolution of the magnetogram can yield a different solution (see Sect.~\ref{sss:complex_dimmings1} for an example and also Sect.~\ref{ss:sim_example2}). 
Generally, strapping-flux dimmings form immediately outside the primary flare ribbon pair, while exterior dimmings 
form more remotely. In some events, the strapping-flux dimming is completely masked by the bright flare-loop arcade by the time the exterior dimmings have developed fully (Fig.~\ref{f:Mandrini+2007} below shows an example). 

Twin exterior dimmings, suggesting the reconnection of both flux rope legs with exterior flux, were reported by \citet{Manoharan&al1996}. 
They proposed a schematic (Fig.~\ref{f:Manoharan+1996}) that illustrates some of the general features of the interchange reconnection leading to such dimmings. For completeness, the following notes should be kept in mind. First, the schematic does not explicitly distinguish the strapping flux from the higher overlying flux rooted in the remote dimmings. The indicated reconnection is topologically equivalent to rope-strapping reconnection in this particular case, because the exterior flux passes over the erupting flux (also compare Fig.~\ref{f:rxn_schematics}). However, since the strapping flux must have been rooted within the isolated AR that erupted, the ``overlying loop'' refers to exterior flux. Second, the formation of exterior dimming(s) requires rope-exterior reconnection in the \emph{leg(s)} of the erupting flux, not in its top part, which must continue to rise; otherwise, a confined eruption would result. Third, the legs reconnect with different parts of the exterior flux (therefore, two exterior field lines are drawn). 

\begin{figure}[tbph]
\centering
\includegraphics[width=1\textwidth]{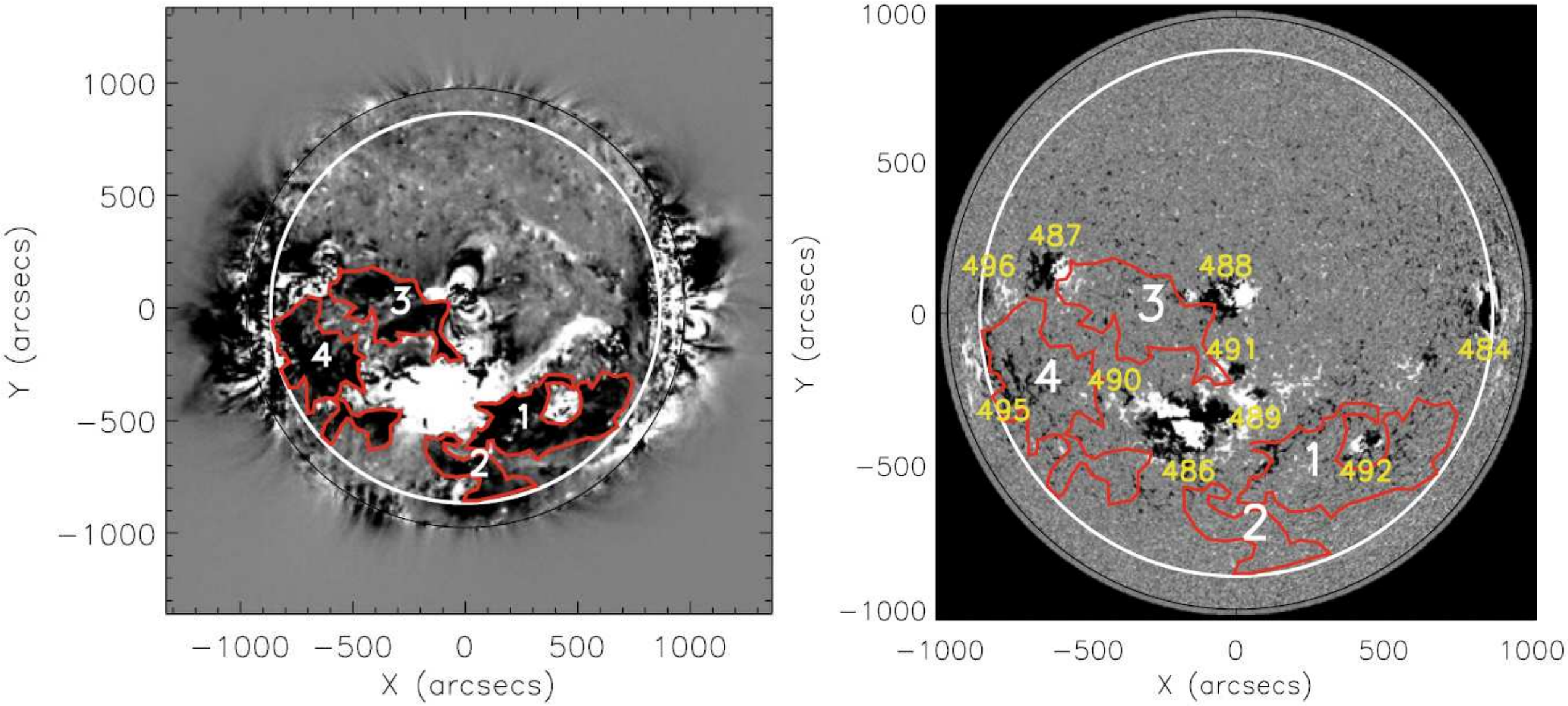}
\caption{Multiple \emph{exterior dimmings} in the major eruption SOL2003-10-28 from AR~10486 (areas 1--4, enclosed by red contours). The background images show a SOHO/EIT 195~{\AA} base difference image 50~min after the peak of the X17 flare (left) and a SOHO/MDI magnetogram converted to the radial field component within the white circle (right). From \citet{Mandrini:2007}.} 
\label{f:Mandrini+2007}
\end{figure}

\begin{figure}[t]
\centering
\includegraphics[width=0.95\textwidth]{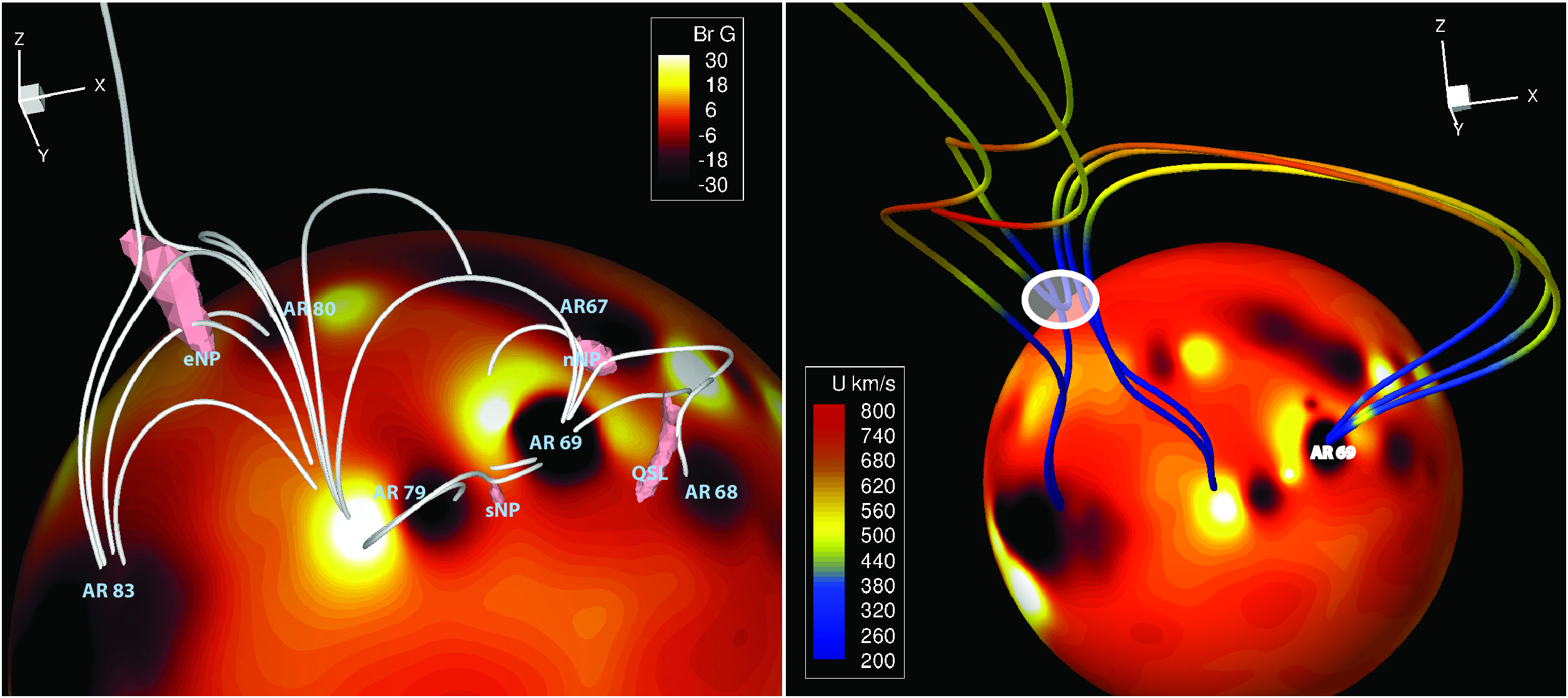}
\caption{Numerical simulation of reconnection between erupting flux (rooted in AR~10069) and closed exterior flux (rooted in AR~10079) at a coronal null point (sNP); this reconnection formed an \emph{exterior dimming} in AR~10079. Additional reconnection with open flux rooted in AR~10083 at another coronal null point (eNP) turns part of the erupting flux into open flux. From \citet{Lugaz&al2010}.} 
\label{f:Lugaz2010}
\end{figure}

Multiple exterior dimmings are preferentially formed in major eruptive events and approximately correspond to the underlying CME in angular extent \citep{Thompson:2000}. They can spread over a large part of the solar disk and encompass several magnetic flux systems \citep{Mandrini:2007, Zhukov:2007, Zhang:2007}, see Fig.~\ref{f:Mandrini+2007}. \citet{Mandrini:2007} 
introduced the term \emph{secondary dimmings} for such events. 
They suggested, following \citet{Attrill:2007} and further elaborated by \citet{vanDriel:2008}, that these dimmings form through a sequence of reconnection events as the erupting flux spreads laterally to increasingly remote regions of exterior flux and reconnects there low in the corona. 
\citet{Zhang:2007} analyzed the event shown in Fig.~\ref{f:Mandrini+2007} and three similarly complex events, and pointed out that multiple magnetic loop systems and a coronal magnetic null point were involved in each of them. This was substantiated by large-scale MHD simulations of several major CMEs that originated in ARs with complex environments comprising several further ARs \citep{Roussev:2007, Lugaz&al2009, Lugaz&al2010}. For all CMEs modeled, reconnection of the originally erupting flux with exterior flux at coronal null points or at a coronal quasi-separator (intersection of two QSLs), located between the flux systems, was indeed found to represent the interchange reconnection that leads to exterior dimmings, as well as to open-flux dimmings considered in the subsequent section. Additionally, such reconnection was found to be a key process in the initiation of the eruptions by moving to the side flux that passes over the erupting core flux. Figure~\ref{f:Lugaz2010} shows one of their results obtained in modeling the fast CME on 2002 August~22. Flux at the eastern side of AR~10069 expands due to imposed shearing; this triggers reconnection at a null point (sNP) between ARs~10069 and 10079, which, in turn, launches the eruption of the sheared flux and, subsequently, reconnection with open flux rooted in AR~10083 at a null point (eNP) between ARs~10079 and 10083. As a result, part of the flux erupting from the vicinity of AR~10069 finds a new footprint in the positive polarity of AR~10079, where a deep exterior dimming developed in EIT 195~{\AA} images over several hours, and the other part opens up into the interplanetary space. 

At least one exterior dimming always forms when a CME originates at a PIL around a parasitic polarity embedded in closed surrounding flux, i.e., under the fan-dome separatrix of the associated coronal null point, because reconnection between the erupting flux and the closed exterior flux above the fan dome is then always triggered. 
We will consider such eruptions in Sect.~\ref{sss:complex_dimmings2} and their MHD modeling in Sect.~\ref{ss:sim_example2}. 
The lifting and eventual opening of interconnecting loops between two ARs, due to an eruption in one of the ARs, has been suggested to be the cause of a deep and extended exterior dimming that formed adjacent to the other AR \citep{Delannee:1999}. Such dimmings may be termed \emph{passive exterior dimmings}. 

Exterior dimmings are not related to the primary ribbon pair of the eruption. Since their underlying reconnection typically proceeds high in the corona after considerable expansion of the erupting flux, they are often not associated with flare ribbons at all. However, when the exterior flux is part of the same AR as the erupting flux, an association with a remote ribbon is more likely. This is true also for dimmings at the footprint of the outer spine in eruptions 
from a 
null-point configuration considered in Sects.~\ref{sss:complex_dimmings2} and \ref{ss:sim_example2}. 

Exterior dimmings represent an intermediate evolutionary step in the large-scale evolution of the corona in association with CMEs: between the primary energy release leading primarily to flux-rope and strapping-flux dimmings and the final distribution of relaxed re-closed and open flux. They are expected to eventually close by reconnection with open flux (Sects.~\ref{ss:open_flux} and \ref{ss:sim_example1}). 

If a single exterior dimming forms due to reconnection, its peak magnetic flux content should be equal to that in a flux-rope dimming in the conjugate polarity (if observed) and provide a good measure of the opening flux. Similar for twin exterior dimmings in conjugate polarities. However, multiple exterior dimmings form often, exchanging flux over extended periods of time at the large scales involved. The deepening of later forming exterior dimmings and the related closure of earlier dimmings may overlap in time, with their instantaneous size and implied flux content potentially depending also on the time scale of the thermodynamic evolution. These effects introduce substantial uncertainties in flux measurements. Some associated dimmings may form at or behind the limb, preventing flux measurements at all. As a result, the inferred fluxes can be strongly imbalanced in polarity and can deviate considerably 
from estimates of the flux in the associated interplanetary CME \cite[e.g.,][]{Mandrini:2007, Temmer:2017}.

\subsection{Open-flux dimmings} 
\label{ss:open_flux} 

\begin{figure} 
\centering
\includegraphics[width=0.54\textwidth]{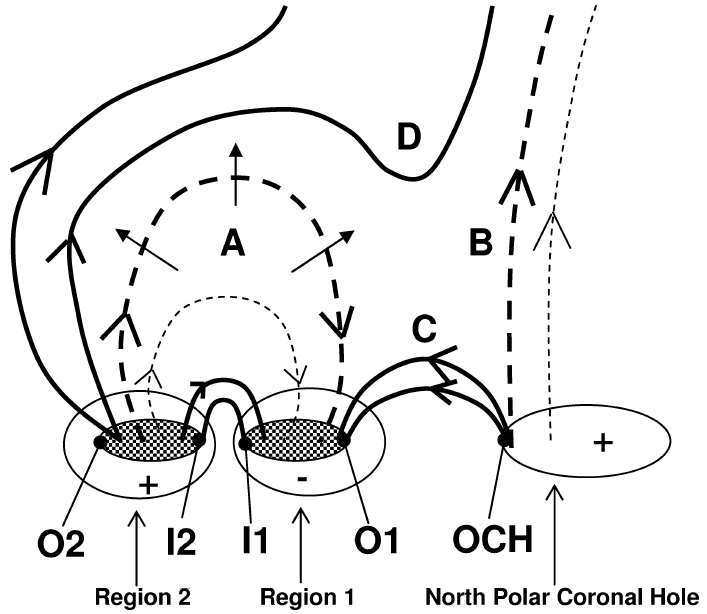}
\caption{Change from flux-rope dimming to \emph{open-flux dimming} in Region~2 by reconnection of the leg of the erupting flux-rope (\textbf{A}) rooted in Regions~1 and 2 with the flux (\textbf{B}) of a coronal hole. From \cite{Attrill:2006}.} 
\label{f:open_flux}
\end{figure}

When erupting flux reconnects with open flux, then the reconnecting leg loses its connection to the Sun and opens up into the interplanetary space. The other footprint becomes the base of open flux, see Figs.~\ref{f:rxn_schematics}c and \ref{f:open_flux}. The already existing dimming at this site is strengthened by the opening of its flux: the dimming may deepen further and its area may stay near the maximum extent for a longer time. Any of the above dimming categories, and consequently, core as well as secondary dimmings, can turn into an \emph{open-flux dimming}, being strengthened in this way. An example is given by the eruption on 1997 May~12 shown in Fig.~\ref{fig:thompson98}, whose asymmetry of the twin dimmings was suggested to result from partial reconnection of the erupting flux with the flux of the nearby polar coronal hole \cite[][Fig.~\ref{f:open_flux}]{Attrill:2006}. An example of greater complexity is the eruption on 2005 August~22 from AR~10798, an anemone-shaped region within 
a coronal hole, modeled by \citeauthor{lugaz11} (\citeyear{lugaz11}; see discussion in Sects.~\ref{sss:complex_dimmings2} and \ref{sec:simulations}). We refer to the underlying reconnection as ``rope-open'' reconnection, although strapping flux wrapped around the erupting flux rope may initially be involved, as for the rope-exterior reconnection discussed in Sect.~\ref{ss:exterior}.

Open-flux dimmings may recover after their flux has re-closed by reconnection with other exterior flux, driven by the further expansion of the erupted flux. A new footprint of open flux is formed if the exterior flux was closed; otherwise fully detached U-shaped flux escaping from the Sun is formed (see Fig.~\ref{downs_feb2009_summary}f). One can expect that such large-scale interchange reconnection repeats until reconnection with open flux of the opposite polarity ultimately completes the sequence (Sect.~\ref{sec:simulations}). Alternatively, an overall diffusive process of small-scale interchange reconnection events between open and closed flux across the periphery of the open-flux dimming has been suggested to re-close the dimming (see \citealt{Attrill:2008} and Sect.~\ref{sec:recovery}). 

Open-flux dimmings in relatively close proximity to the eruption site can be a diagnostic tool to discover open flux that allows energetic electrons to escape into interplanetary space during flare-CME events. These electrons emit groups of fast-drift radio bursts of the category ``Complex Type~III Bursts'', which often occur associated with the slow-drift (``Type~II'') radio signature of the CME-driven shock in the dynamic radio spectrum and extend to Dekameter and longer wavelengths, and, at other occasions, ``Isolated Type~III Bursts'' extending to this frequency range. In spite of the frequent association of Complex Type~III Bursts with CME-driven shocks, some evidence suggests that 
the primary acceleration of these electrons occurs in a reconnection region in the aftermath of the CME, i.e., in a region of flare energy release, not in the CME-driven shock \citep{Cane&al2002}. This is also a likely site of origin for the Isolated Type~III Bursts. However, the evidence is not conclusive \citep{Gopalswamy2004, Gopalswamy2011}.  If true, this would present 
a challenge to traditional flare models, which do not include open flux. Open-flux dimmings may help resolve this issue.

\subsection{Complex dimming events} 
\label{ss:complex_dimmings} 

Exterior dimmings of major CMEs during times of high solar activity map out some of the complexity of the large-scale coronal field \cite[e.g., Fig.~\ref{f:Mandrini+2007} and][]{Zhang:2007}. Additionally, the dimming categories proposed above often occur in combination and in a characteristic temporal sequence. This holds for strapping-flux and moving flux-rope dimmings, as the classical initial strapping-strapping reconnection often extends to rope-strapping reconnection (Sect.~\ref{ss:movingFR}). 
Exterior dimmings are next in the chain, followed by open-flux dimmings. The underlying reconnection processes often overlap in time. Consequently, secondary dimmings often represent a combination of the new dimming categories in a complex geometry.

\subsubsection{A case of combined strapping-flux and exterior dimmings} 
\label{sss:complex_dimmings1} 

\begin{figure} 
\centering
\includegraphics[width=0.92\textwidth]{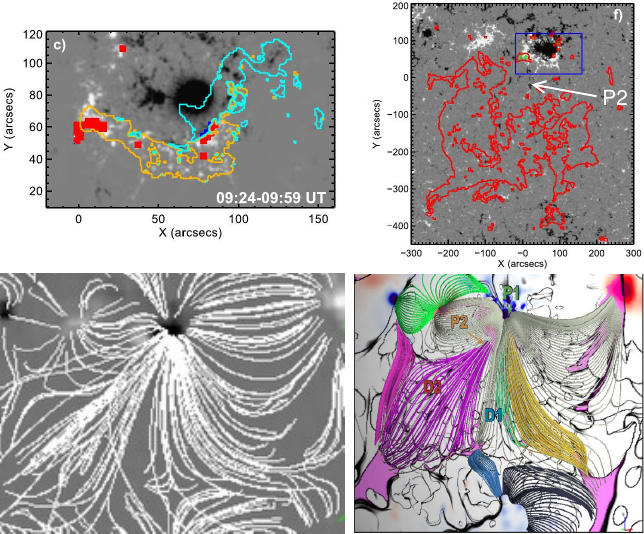}
\caption{A conjectured \emph{strapping-flux dimming} in SOL2011-10-01 observed as a large secondary dimming (red contours in top-right panel, from \citealt{Temmer:2017}) under high-lying flux rooted in a dominantly positive, extended weak-field area to the south of AR~11305 and in the main negative sunspot of the AR, according to the potential-field source-surface (PFSS) computation in the bottom-left panel \cite[from][]{Krista:2013}. However, a more detailed, high-resolution PFSS computation (bottom-right panel; C.~Downs, private commun.) suggests that part of the flux (pink field lines) is instead rooted in more distributed negative flux labeled P2 to the south of the erupting filament channel. In this case, the secondary dimming is composed of a strapping-flux dimming (D1) and an \emph{exterior dimming} (D2) that forms in closed exterior flux. \emph{Moving flux-rope dimmings} (red and blue filled contours) are shown along with the position of the flare ribbons at the peak time of the associated flare (yellow and cyan contours) in the top-left panel \cite[from][]{Temmer:2017}.} 
\label{f:2011-10-01}
\end{figure}

A large-scale 
eruption from AR~11305 on 2011 October~1, analyzed in \citet{Krista:2013} and \citet{Temmer:2017}, 
illustrates that secondary dimmings may 
be composed of different categories with different underlying physics. 
Figure~\ref{f:2011-10-01} shows the extended dimming south of the AR in weak mixed flux of dominantly positive polarity (top right panel). The eruption and this dimming are very asymmetric because the negative strapping-flux roots are highly concentrated in a 
sunspot. The event also shows 
moving flux-rope dimmings (Sect.~\ref{ss:movingFR}) and flare ribbons close to the PIL (top left panel). Field lines of a global potential field model passing over the erupting section of the PIL are rooted in the sunspot and the dominantly positive flux under the secondary dimming (bottom left panel). According to the model, this flux is (an upper) part of the strapping flux. Hence, it is lifted by the southward-rising erupting flux (traced by an erupting filament) and becomes fully or partly entrained with the CME by standard flare (strapping-strapping) reconnection, producing a strapping-flux dimming (Sect.~\ref{ss:strapping}). However, a more detailed, high-resolution
computation of the global potential field (bottom right panel) suggests that the flux above the eastern part of the secondary dimming (pink field lines) is not rooted in the sunspot, but rather in a minor negative polarity (``P2'') at the southern edge of the AR. 
Based on this field model, the eastern part of the dimming is an exterior dimming (Sect.~\ref{ss:exterior}) and only the western part is a strapping-flux dimming. The southward-erupting flux propagates above the exterior flux in this case. It is conceivable that this leads to a passive lifting of the exterior flux. However, a more likely scenario consists in reconnection of the erupting flux with the exterior flux at the large-scale QSL between them, where they have a 
nearly antiparallel direction and the formation of a current sheet is likely. 
This would connect the exterior flux with the negative sunspot, making it part of the erupting flux, and is supported by the facts that the dimming was deepest in the eastern part and that a secondary loop arcade formed between P2 and the adjacent positive flux at the main PIL (Fig.~\ref{f:2011-10-01} top left) during the late phase of the flare (these features were not shown in \citealt{Krista:2013} and \citealt{Temmer:2017}, but are obvious from the SDO/AIA data of the event).

\begin{figure} 
\centering
\includegraphics[width=0.99\textwidth]{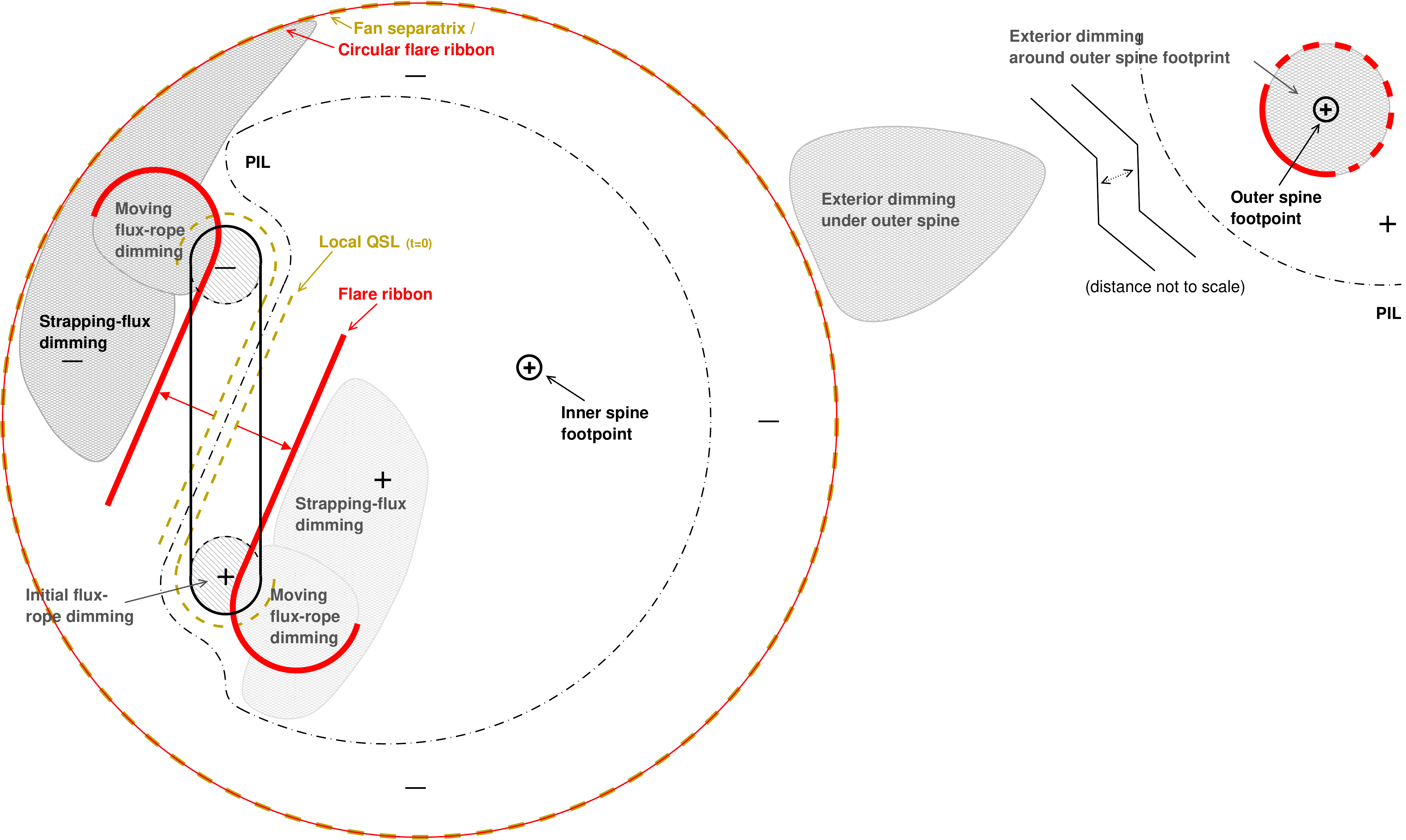}
\caption{Schematic showing the dimmings that typically form in CMEs from source regions with a null point above a parasitic polarity embedded in closed exterior flux. The \emph{flux-rope and strapping-flux dimmings} develop asymmetrically; the outer strapping-flux dimming additionally extends into an arc shape if the strapping flux is very strongly sheared. An \emph{exterior dimming} and associated additional (remote) flare ribbon form at the footprint of the outer-spine flux; the ribbon fully encloses the dimming in some events. A further exterior dimming may form in flux rooted under or near the the path of the erupting outer spine. The characteristic circular ribbon, forming at the base of the fan separatrix in such events, is also included. The corresponding magnetic connections, i.e., the spines and fan dome, 
are not included, to avoid overloading the figure. They are visualized by the yellow field lines in Fig.~\ref{fig-prasad2020}d and also, e.g., in \citet{Antiochos1998, Reid&al2012}. The changing magnetic connections underlying the dimmings are visualized by the red field lines in Fig.~\ref{fig-prasad2020}b--f. 
} 
\label{f:ed_np}
\end{figure}

\subsubsection{Dimmings in circular-ribbon events and related breakout-category configurations} 
\label{sss:complex_dimmings2} 

CMEs that originate from a filament channel along part of a closed 
PIL around a parasitic polarity systematically form complex dimmings if a magnetic null point with its fan-spine topology \cite[e.g.,][]{Antiochos1998, Reid&al2012} exists in the corona above the parasitic polarity. This is the case if the parasitic polarity contains less flux than the surrounding polarity. At the filament channel, the classical ribbon pair forms, and flux-rope and strapping-flux dimmings often form in the standard way outlined in Sects.~\ref{ss:FRdimmings}--\ref{ss:movingFR}, but develop a characteristic asymmetry. 
The eruption disrupts the fan dome, spine, and part of the overlying flux, typically resulting in additional flare ribbons in the footprint of the fan separatrix around the whole closed PIL and in the region of the outer-spine footpoint (Fig.~\ref{f:ed_np}), sometimes also at the footpoint of the inner spine. The former is commonly referred to as a ``circular ribbon''; it tends to remain less bright than the classical flare ribbon pair forming at the erupting filament channel. A detailed description of circular-ribbon flares is given by \citet{ZhangQM2024}. 

\begin{figure}[ht]
\centering
\includegraphics[width=0.65\textwidth]{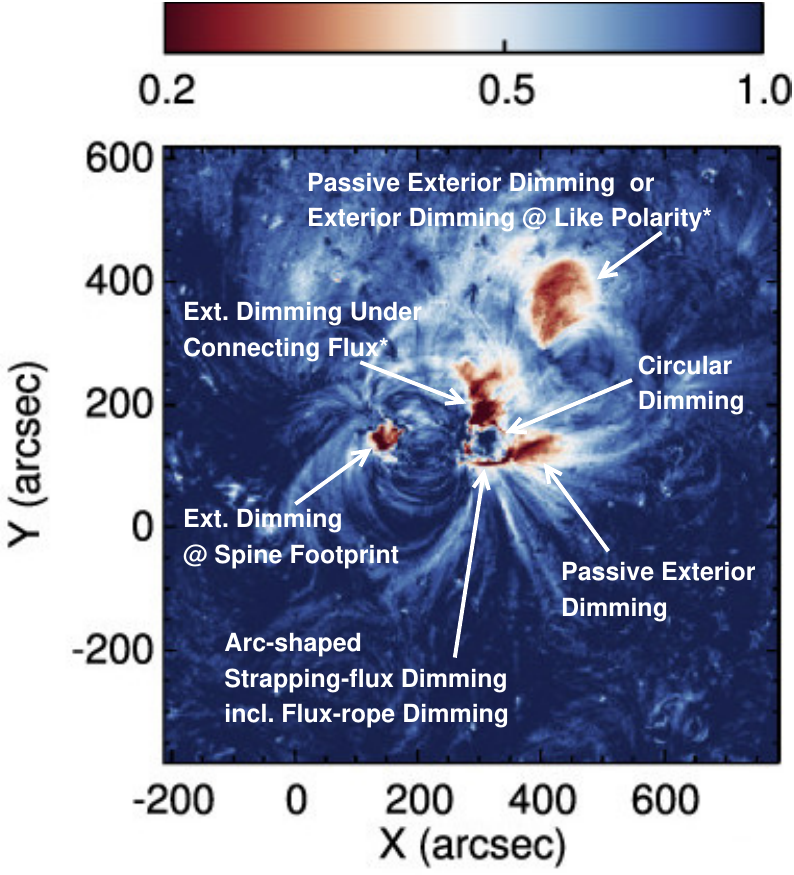}
\caption{Various dimmings in the eruption under a coronal null point in AR~11283 in SOL2011-09-06 shown in an SDO/AIA 211~{\AA} logarithmic base-ratio image (see color bar for the range of values in the image). These are interpreted as various \emph{exterior dimmings}, including \emph{passive exterior dimmings} and as an arc-shaped \emph{strapping-flux dimming}, which includes the \emph{flux-rope dimming} of the event in the polarity surrounding the parasitic polarity (compare with Fig.~\ref{f:ed_np}). Tentative interpretations are marked with an asterisk. The circular 
dimming, which formed in place of the circular ribbon, is not yet understood (a tentative interpretation is given in the text). 
The movie in the online supplement shows the event evolution in SDO/AIA 211~{\AA} direct and logarithmic base-ratio images (adapted from \cite{Dissauer:2018a}).} 
\label{f:2011-09-06T2300MaxDimm}
\end{figure}

The expanding flux of the eruption first perturbs the null point, which then stretches into a current sheet along which the inner and outer spine align in approximately opposite directions. The ensuing reconnection \cite[e.g.,][]{Pontin&Priest2022} feeds the additional ribbons. The erupting flux rope joins and amplifies this reconnection when it propagates toward this region \cite[e.g.,][]{lugaz11}, but reconnects with overlying flux at a side of the fan dome if the geometry favors this case 
(see Sect.~\ref{ss:sim_example2} for an example). 
This reconnection typically involves the outer spine and its surrounding flux \citep{lugaz11, JiangC&al2018, Prasad2020}. Just outside the fan separatrix, this flux is directed roughly oppositely to the leg of the erupting flux rope rooted in the parasitic polarity. As a result of its reconnection, the footprint of this leg then jumps to the area where the outer spine is rooted, or the erupting flux becomes open flux if the outer spine extends into the interplanetary space. In the former case, an exterior dimming forms at or near the footpoint of the outer spine and a flare ribbon often also forms (Fig.~\ref{f:ed_np}). In the latter case, the dimming at the footprint of the erupted flux in the surrounding polarity changes to an open-flux dimming. In either case, this flux-rope dimming develops strongly, because it lies at a footprint of erupting (further expanding) flux. This part of the reconnected flux is then completely rooted outside the parasitic polarity. A new null point with its fan-spine separatrices can form under this flux, allowing for homologous circular-ribbon eruptions \cite[e.g.,][]{Mitra&Joshi2021}. 
The other part of the reconnected flux simply arches over the circular PIL without giving rise to a further dimming. 
Rather, its flux-rope dimming weakens because the reconnected flux relaxes and shrinks similar to standard flare loops. Therefore, the dimming at the footprint of the erupting flux in the surrounding polarity is often the stronger one (see, e.g., 
the event shown in Fig.~\ref{f:2011-09-06T2300MaxDimm} and discussed in Sect.~\ref{ss:sim_example2} and an event from AR~12261 on 2015 January 12 analyzed in \citealt{Kumar&al2021}). 

Circular-ribbon eruptions may form three further variants of exterior dimming that are characteristic for such events. 
First, an arc-shaped 
dimming has been observed to run along 
the circular ribbon in some events and has been analyzed in the event shown in Fig.~\ref{f:2011-09-06T2300MaxDimm} \citep{Dissauer:2018a, Prasad2020}. Another example is provided by the eruption 
on 2015 January 12 \citep{Kumar&al2021}. 
The dimming forms between the outer of the double ribbons and the initial position of the circular ribbon along (part of) the double ribbon and beyond it to the side where the sheared strapping flux of the erupting filament channel is rooted. Its extension along the circular ribbon suggests a strong shear/guide-field component in the strapping flux, which appears to be the condition for the dimming to extend into an arc. This dimming can be interpreted as a strapping-flux dimming (compare Fig.~\ref{f:ed_np}). All of the strapping flux rooted between the erupting filament channel and the adjacent footprint of the fan separatrix must reconnect at the fan separatrix (preferably at the current sheet formed in place of the null point) with flux above the fan separatrix, to allow the erupting flux rope to get through. Both the original strapping flux and the rope then connect to the outer-spine flux and continue to erupt, which produces the dimming. The other half of the strapping flux that is cut at the fan separatrix, rooted on the other side of the erupting filament channel, re-closes between the inner spine and the opposite side of the fan separatrix. This explains why the strapping-flux dimming on the inner side of the erupting filament channel is much weaker and why the circular ribbon propagates outward preferably on the side opposite to the erupting filament channel. The asymmetry of the strapping-flux dimming is completely analogous to the asymmetry of the flux-rope dimming. The outer flux-rope dimming is a part of the arc-shaped dimming (Fig.~\ref{f:ed_np}). 

In the event shown in Fig.~\ref{f:2011-09-06T2300MaxDimm}, the arc-shaped dimming extended into a nearly complete circular 
dimming. This dimming formed in place of the circular ribbon after the ribbon faded. The dimming was as narrow as the circular ribbon, except for its arc-shaped section, 
whose outer edge coincided with the early formed and faded section of the circular ribbon. 
It is not clear how magnetic flux could open in such a narrow strip along a large part of the fan separatrix. 
For this reason and because of the spatial and temporal relationship to the circular ribbon, it will be worth investigating whether this dimming might have a thermal origin.

Second, a dimming can form in the surrounding polarity between the main eruption site and the remote dimming and ribbon in the area of the outer spine's footpoint. This is under the outer spine, which connects these areas and becomes a leg of the erupting flux (Fig.~\ref{f:ed_np}). The dimming may thus simply be due to the lifting of flux by the rising outer spine \citep{JiangC&al2018}. A more efficient dimming mechanism is the reconnection of flux under the rising outer spine, which entrains the reconnected flux with the erupting flux. The reconnecting flux can be located slightly to the side of the outer spine. Such reconnection has been found in the numerical simulations by \citet{lugaz11} and \citet{JiangC&al2018}. 

Third, a second remote dimming similar to the one in the area of the outer spine's footpoint, i.e., in flux of the same polarity as the parasitic flux, has been observed in the event illustrated in Fig.~\ref{f:2011-09-06T2300MaxDimm} and is discussed in Sect.~\ref{ss:sim_example2}. Slipping reconnection between flux rooted in the two areas has been suggested to cause the second remote dimming \citep{JiangC&al2018}, but it also appears possible that its flux is directly involved in the reconnection of the erupting flux rope with ambient flux above the fan separatrix. 

The event likely also included 
\emph{passive exterior dimmings} (Fig.~\ref{f:2011-09-06T2300MaxDimm}). This is suggested by the close synchronization of their onset with the passage of an erupting filament that propagated from the parasitic polarity across these dimmings along a very inclined path (Sect.~\ref{ss:sim_example2} provides more detail). 
Overall, the event illustrates that exterior dimmings are 
an integral part of CMEs originating under a coronal null point and can take a complex structure even in a relatively simple source-region configuration.  

In contrast, 
the side lobes in a quadrupolar ``breakout'' configuration \citep{Karpen&al2012}, although classical examples of exterior flux, are not likely to host dimmings. The side-lobe flux is not lifted. It can enter the flare reconnection after all flux in the erupting center lobe has been reconnected, however, the resulting upper reconnection product remains beneath the erupting flux rope. Only a modest upward stretching of this flux by the upward reconnection outflow is expected \citep{DeVore&Antiochos2008, ChenJ&al2023}, i.e., no or at most weak exterior dimmings. Flux-rope and strapping-flux dimmings will of course form in the center lobe, as in any bipolar source region. 
The term ``breakout reconnection'' has in recent years been used in a wider sense to include reconnection at the null line in pseudostreamer configurations. These are topologically very similar 
to the null point configuration discussed above, but the surrounding polarity always hosts open flux, so that the outer spine and its surrounding flux open 
into the interplanetary space as a pseudostreamer. An eruption in this configuration leads to a similar lifting of flux and similar topology changes as discussed above for the closed null-point configuration.

\begin{figure}
\centering
\includegraphics[width=0.46\textwidth]{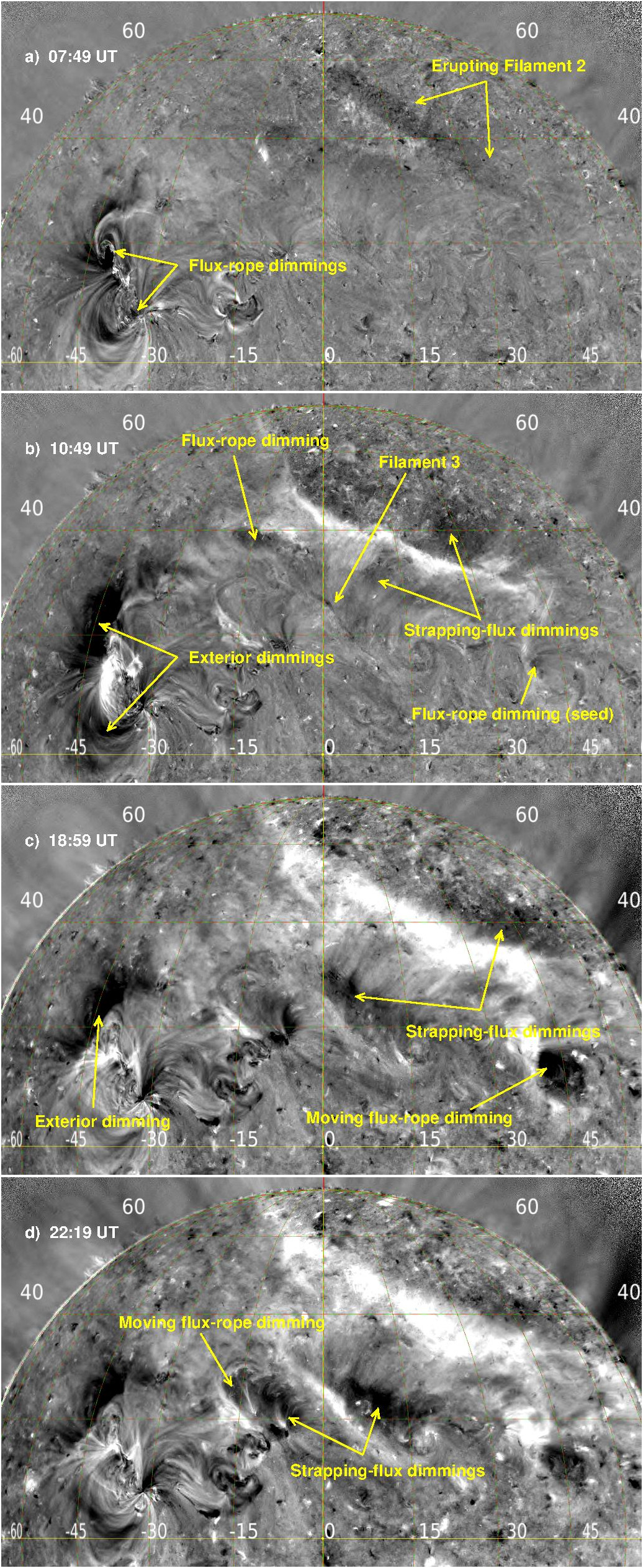}
\caption{Various dimmings in the final three major filament eruptions in the sympathetic eruptions in SOL2010-08-01. SDO/AIA 211~\AA\ base-difference images, with the base image at 06~UT, are shown. The movie in the online supplement shows the evolution in  AIA 211~{\AA} base difference images (left) along with AIA 304~{\AA} direct images (right).} 
\label{f:sympathetic}
\end{figure}

\subsubsection{Dimmings in sympathetic events} 
\label{sss:complex_dimmings3} 

Pseudostreamer are also a favorable location for the occurrence of sympathetic eruptions, which naturally produce complex dimming configurations. \citet{Schrijver&Title2011} describe the main elements of such an event on 2010 August 1--2, consisting of at least four filament eruptions and associated flare and dimming signatures spread over a full hemisphere of the Sun. Most of the dimmings on the solar disk were flux-rope and strapping-flux dimmings of the individual eruptions, but at least one exterior dimming is also produced (Fig.~\ref{f:sympathetic}). The modeling of the final three large filament eruptions in the event \citep{Torok:2011}, with the second and third one originating from the lobes of a pseudostreamer configuration, revealed that the first erupting filament triggered reconnection at the null line of the pseudostreamer, in basic agreement with the findings in \citet{Zhang:2007} and \citet{Roussev:2007}. This initiated the second eruption (of a polar-crown filament) by removing overlying flux, similar to the breakout model \citep{Antiochos&al1999}. Different from other models for the triggering of eruptions, however, the third eruption was triggered by the flare reconnection under the second eruption, which weakened the flux passing over the third filament located between the other two. Noteworthy dimmings in the event are very deep exterior dimmings in the first of these filament eruptions, extended, clear strapping-flux dimmings and a stationary turning into a moving flux-rope dimming in the second, and clear strapping-flux dimmings in the final one. These are illustrated in Fig.~\ref{f:sympathetic}.

\subsection{Summary of dimming categories and diagnostic potential} 
\label{ss:dimming_summary}

The basic properties of the magnetic-flux-based dimming categories introduced in this section are summarized in Table~\ref{t:categories}. All three categories of flux-rope dimming are observed as core dimming. Strapping-flux and open-flux dimmings can form as core or as secondary dimming. Exterior dimmings 
belong to the secondary dimming category. 

The dimmings form in a characteristic temporal sequence, beginning with stationary flux-rope and strapping-flux dimmings, of which the flux-rope dimming often turns into a moving flux-rope dimming relatively early in the evolution. These are often followed by exterior dimmings. Open-flux dimmings are a feature of the late stage of CME evolution that mark a transition to a full opening or even a detachment of the erupting flux. 

\begin{table}[tbp]
\caption{Basic dimming categories (FR = flux rope)}
\label{t:categories}
\centering
\begin{tabular}{p{1.9cm}p{2.3cm}       p{1.2cm} p{1.3cm}p{1.5cm}         p{1.55cm}}
\hline
Category                & Process      & Pre- \- eruption dimming 
                                                & Ribbons sweeping across
                                                       & Dimming motions & Footprint of 
                                                                           interplane\-tary~FR    \\ 
\hline
Stationary FR~dimming   & FR expansion & Y      & N    & N               & Y                      \\[15pt]

Shrinking FR~dimming    & FR expansion \& leg-leg reconn.
                                       & N/A    & Y    & N               & Y, if not fully closed \\[15pt]

Strapping-flux dimming  & Lifting \& strap\-ping-strapping reconnection 
                                       & Y (rarely) 
                                                & Y    &Expanding, moving& Y, part of             \\[15pt]

Moving FR~dimming       & FR expansion \& rope-strapping reconnection 
                                       & N/A    & Y    &Moving, expanding& Y                      \\[15pt]

Exterior dimming        & Reconn. w. closed exterior flux
                                       & N      & Y in some cases  
                                                       & Can jump 
                                                                         & Y                      \\[15pt]

Open-flux dimming       & Reconnection w. open flux
                                       & N/A    & N    & N               & Y                      \\  

\hline
\end{tabular}
\end{table}

The first four categories in Table~\ref{t:categories} derive from considering the two magnetic flux systems that make up a simple bipolar force-free equilibrium, their expansion, and possible forms of reconnection. Even in this simple configuration it is difficult to distinguish moving flux-rope dimmings from strapping-flux dimmings in the observations if the shear-field component is comparable in strength to the strapping-field component, as is often the case on the Sun. Both dimmings then form in a similar location, move away from their initial position jointly with the flare ribbons, and expand into the footprint area of the strapping flux (Sects.~\ref{ss:strapping}--\ref{ss:movingFR}). 
Additionally, the source region of the eruption can be complex, further affecting the appearance of dimmings. Three typical cases are the following. The current-carrying core flux may contain two (a ``double-decker'') or even several flux ropes \citep{LiuR&al2012, Awasthi&al2018}. The eruption may proceed in steps, sequentially involving several sections of an unstable filament channel \cite[e.g.,][]{Patsourakos&al2016}; a flux-rope footprint can then be replaced by a new one further away. 
Two nearby located filament channels, e.g., in a pseudostreamer configuration, may be sequentially destabilized, leading to sympathetic eruptions in quick succession \citep{Torok&al2011b, Torok&al2018}. Complex dimming geometries and dynamical sequences can result. 
Finally, exterior dimmings can take many different geometries and dynamical sequences as well, due to the many options for interaction with a complex magnetic environment, as often exists on the Sun (Sect.~\ref{ss:exterior}). Consequently, it is not straightforward to interpret complex dimming events in terms of the categories in Table~\ref{t:categories}, although current experience suggests that core dimmings very often represent moving flux-rope dimmings, often combined with a strapping-flux dimming component, 
and secondary dimmings are very often exterior dimmings. The computation of the potential field is a first, useful, and sometimes necessary step to infer the exterior flux system(s) that may be involved in the formation of secondary dimmings and to distinguish exterior from strapping-flux dimmings. Numerical modeling is an even more powerful tool to infer the dimming categories (Sect.~\ref{sec:simulations}). 

The interpretation of observed dimmings within the proposed framework of Table~\ref{t:categories} 
carries diagnostic potential 
for inferring the participating flux system(s) and their interactions. 
This includes: 
\begin{enumerate} 
\item an observational means to infer the footprints of the originally erupting flux independent of magnetic extrapolation; 
\item inferring connectivity changes in the coronal magnetic field due to eruptions, including the conditions for their occurrence; in particular: 
      \begin{enumerate} 
      \item rope-strapping reconnection in the generalized 3D flare model; 
      \item rope-exterior reconnection; 
      \item rope-open reconnection; 
      \item formation of the core structure of a subsequent eruption (e.g., SOL2012-03-07 discussed in Sect.~\ref{sec:stellar_overview} and Sect.~\ref{Sec:SunAsStar}); 
      \end{enumerate} 
\item inferring the final footprints of the CME; 
\item an estimate of the amount of flux opened into the interplanetary space; 
\item inferring potential pathways for flare-accelerated energetic particles into the interplanetary space; 
\item addressing the relaxation and recovery of the large-scale coronal magnetic field from the migration of flux rope footprints and from the recovery of dimmings by shrinking; 
\item addressing the time scale/efficiency of coronal heating from the temperature-dependent recovery of brightness (e.g., EUV vs.\ SXR) inside dimmings that is not due to area shrinking. 
\end{enumerate} 
Additionally, 
\begin{enumerate} 
\setcounter{enumi}{7} 
\item pre-eruption dimmings provide a means to infer the onset of a slow-rise phase, which often marks the onset of the final evolution toward eruption. 
\end{enumerate} 


\section{Simulations} 
\label{sec:simulations}

As highlighted in Sect.~\ref{Terminology}, coronal dimmings offer a unique probe of the underlying mechanisms of solar eruptions. On the surface their relationship is simple; an erupting flux-system stretches, reconnects with, and temporarily opens the overlying field. Dimmings inform us about the how, the when, and the where. On the other hand, as evidenced by the litany of ways dimmings are manifested observationally (Sect.~\ref{Observations}) and the growing complexity of Figs~\ref{f:FRdimmings}, \ref{f:strapping}, \ref{f:mfrd}, \& \ref{f:ed_np} as we attempt to categorize them, it is clear that it can often be challenging to connect our observations to theory directly.

Numerical simulations and modeling offer one way to bridge this gap, and in the following section we overview several examples of how numerical simulations of CMEs have been used to better understand coronal dimmings. Increasing realism and complexity has also been incorporated over time into CME models, and we highlight two case-studies that illustrate how the richness of such simulations can be used to study dimmings observed on the Sun.

\subsection{Overview}
\label{ss:sim_overview}

With the advent of modern supercomputing resources and computational techniques, the literature of numerical simulations of flare and CMEs has become relatively vast. At the same time, only a relatively small fraction of these works are either designed to study dimmings directly or discuss their implications for dimmings. On the other hand, simulations of solar eruptions can be incredibly useful to study: a) How the simulated magnetic field of an erupting MFR relates to our interpretations of dimmings, and b) How the simulated plasma evolution actually creates dimming. The former is studied by examining how the field is perturbed or topologically changed via reconnection, and the latter is tackled by examining the hydrodynamic evolution (heating, mass-loss, etc) and/or by forward modeling synthetic observables (e.g. EUV emission). 

\begin{figure} 
\centering
\includegraphics[height=0.65\textheight]{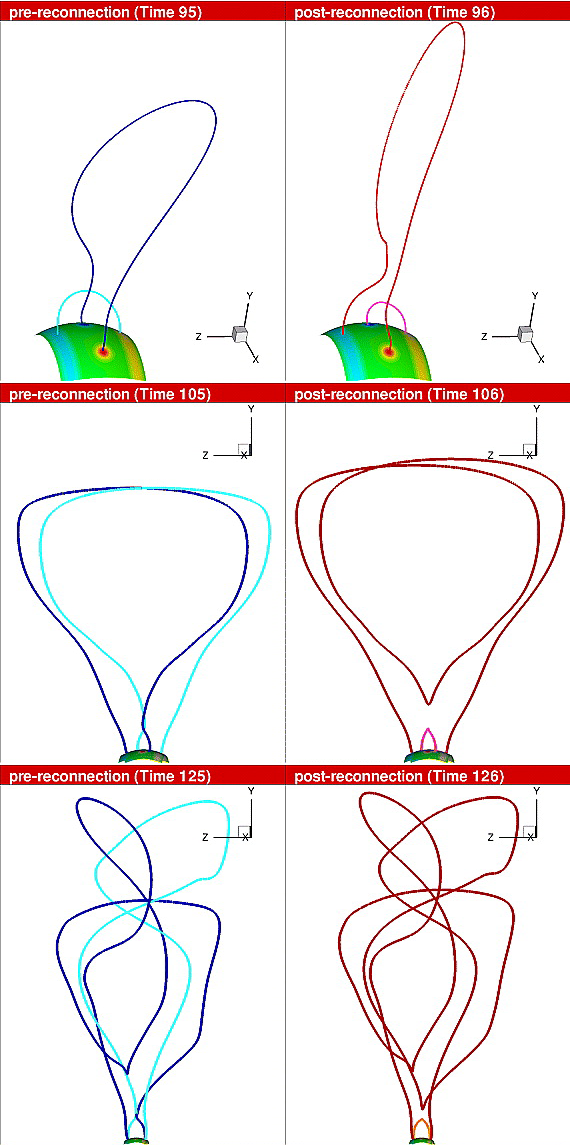}~~~
\includegraphics[height=0.65\textheight]{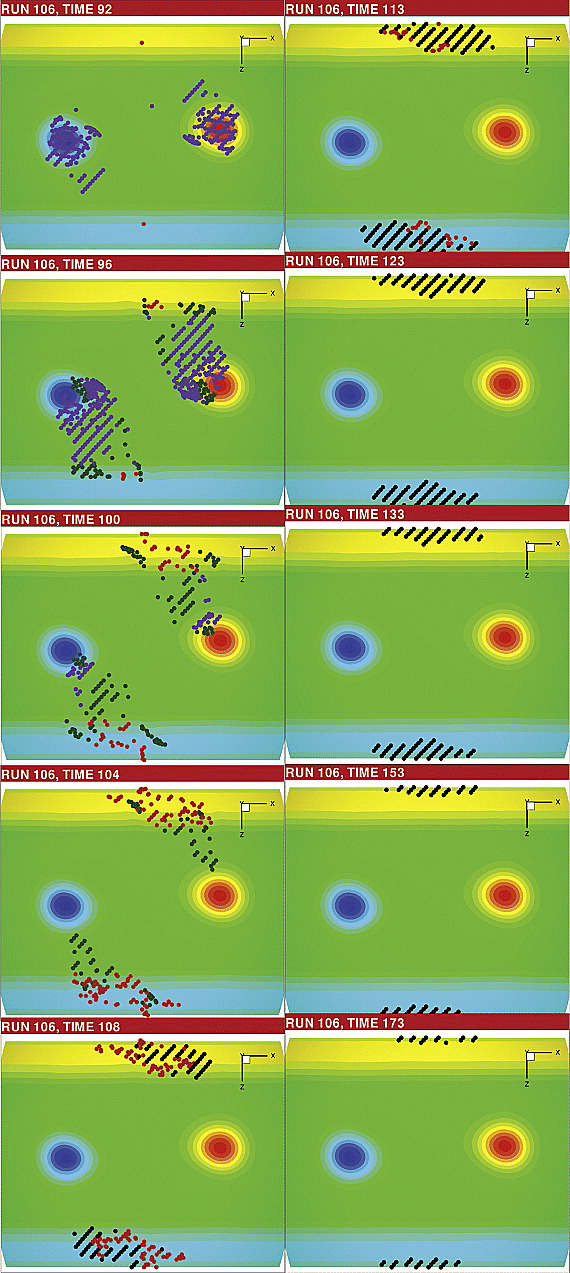}
\caption{\emph{(left panels)} Illustration of rope-strapping reconnection (top row) and two forms of secondary leg-leg reconnection (middle and bottom row) in the simulation of \citet{Gibson&Fan2008}. 
\emph{(right panels)} Footpoints of erupting flux in various snapshots during the simulation representing a proxy of dimmings. The footpoints show the migration into the strapping-flux area (yellow and light blue stripes), due to rope-strapping and the first form of leg-leg reconnection ($t\lesssim110$), and the detachment of the erupting flux (closure of dimmings), due to the second form of leg-leg reconnection ($t\gtrsim110$). 
} 
\label{f:GibsonFan2008}
\end{figure}

Because the nature of dimming (and thus its categorization) depends strongly on the properties of the eruptive flux system, the strapping field, and the external flux systems, simulations involving idealized configurations with simple and/or symmetric flux-distributions can be incredibly useful for isolating individual mechanisms. 
A pioneering example is \citet{Gibson&Fan2008}, where a zero-beta MHD simulation was used to study how flux-ropes evolve as they emerge into the corona and erupt and what this implies for dimming (Fig.~\ref{f:GibsonFan2008}). In this specific study, they focused on the implications of changing field line connectivity for a CME detected in interplanetary space, describing various ways in which rope-strapping field reconnection and leg-leg reconnection will lead to a change in flux-rope connectivity at various stages of the event. Specifically they detail how rope-strapping field reconnection can effectively `move' the footprint of the MFR away from the source region (i.e. dimming motion) 
and how subsequent reconnection (in their case secondary leg-leg [to be distinguished from primary leg-leg reconnection introduced in Sect.~\ref{sss:shriFRdimmings}]) can completely erase the connectivity to the original footprint of the rope, shifting the connectivity from the core of the AR to the periphery. 
They also distinguish this type of secondary leg-leg reconnection, which occurs early on and simultaneously shifts the footprint while removing surface connected flux, from secondary leg-leg reconnection that occurs at a later stage when the external footprint of the MFR is established, which serves to further detach the MFR from the surface in time, i.e., closes the dimming. 

Another excellent example of an idealized simulation with key relevance for dimming and the related flare ribbons is \citet{Aulanier&Dudik2019}, where the combination of a 3D zero-beta MHD model of an eruptive flare and field line mappings of QSLs is used to map out a pattern 
of MFR connectivity as a function of time. 
Following
our categorization, Table~\ref{t:categories} describes the various ways in which the strapping fields and MFR fields reconnect and lead to a growth (strapping-strapping = ``aa--rf''), drift (rope-strapping = ``ar--rf''), and shrinkage (leg-leg = ``rr--rf'') of the final MFR footprint (here the relation to the nomenclature of \citeauthor{Aulanier&Dudik2019} is also given, which uses ``a'' for arcade, ``r'' for rope, and ``f'' for flare loop). These processes were analyzed in great detail. 
Their work also illustrates how such reconnection will naturally follow the QSL layers in the magnetic configuration, which delineate the essential magnetic building blocks of the configuration. The evolution of the QSLs resulting from 3D flare reconnection was found to closely resemble the ribbon evolution observed in many eruptions.

These two simulations \citep{Gibson&Fan2008, Aulanier&Dudik2019}, as well as the one shown in Fig.~\ref{f:FRflare_reconn} \citep{Torok&Kliem2005}, demonstrate that ``flare'' reconnection can naturally extend from the classical ``strapping-strapping'' reconnection to also involve ``rope-strapping'' reconnection, which leads to a migration of the MFR footprints and their associated dimmings and flare ribbons. However, the sequence of subsequent reconnection processes differs between these simulations, resulting in potential differences in the way the dimmings recover. The origin of the differences might lie in the different geometry of the photospheric flux distribution, but this conjecture requires confirmation through further study. The areas of the strapping flux are fully separate from the footprints of the flux rope and have a larger separation than the footprints in \citet{Gibson&Fan2008}. In contrast, the roots of the strapping flux and flux rope form a continuous distribution in each polarity without any obvious dividing line between them in the vertical magnetogram component in \citet{Aulanier&Dudik2019}, which is often seen in young, compact active regions. The flux distribution in \citet{Torok&Kliem2005} is intermediate: continuous with clearly separable rope and strapping-flux components (Fig.~\ref{f:FRflare_reconn}). This is more typical of older, dispersed active regions and eruption source regions in the quiet Sun \cite[e.g.,][]{Martin1998}. The latter two configurations share the property that the separation of the polarities of the strapping flux is smaller than the rope's footprint separation---opposite to the configuration in \citeauthor{Gibson&Fan2008}. This difference could be the reason for the prompt onset of secondary leg-leg reconnection while each leg of the rope still undergoes rope-strapping reconnection in \citeauthor{Gibson&Fan2008} (Fig.~\ref{f:GibsonFan2008} left panels, middle row). Secondary leg-leg reconnection is not dynamically important in the CME simulation in \citet{Torok&Kliem2005}, 
and the same appears to be true for the simulation in \citeauthor{Aulanier&Dudik2019}, which, however, did not follow the flux rope to very large heights. The latter authors find that part of the flux added to the erupting rope by strapping-strapping reconnection detaches later through leg-leg reconnection (a sequence different from the one in \citeauthor{Gibson&Fan2008}), but this appears to encompass only a small amount of flux. Secondary leg-leg reconnection continues in \citeauthor{Gibson&Fan2008} when both legs of the flux rope have migrated to the strapping-flux area, closing the dimmings by shrinking from the inner side. 

While idealized modeling is generally the go-to approach for isolating and studying specific mechanisms, the inherent complexity of observed CME events and their corresponding 3D coronal magnetic field configurations can often lead to some ambiguity in the interpretation or categorization of what actually happened on the Sun. Case-study modeling of specific events, where boundary conditions are derived from photospheric magnetograms, and thus the basic 3D magnetic configuration is captured, is one such way to overcome this gap. \citet{cohen09} describes one of the first case-study simulations of a real CME event that focused (in-part) on dimming. Here they used a polytropic MHD model and a strongly out-of-equilibrium rope to study how a wave is launched, an MFR is ejected, and density changes develop behind. They used the simulation to argue how the large, secondary dimming region(s) within the vicinity of the AR are due to true mass-loss by the CME. They also discuss how the stretching of overlying field and reconnection of rope with overlying field could lead to secondary dimming. Some of the secondary dimming in this simulation was also attributed to the overlying field being significantly perturbed and reconnecting with other external sources (i.e. another form of exterior dimming). Using a similar eruptive MFR setup for another CME event, \citet{lugaz11} also illustrated how a remote flux-rope footprint can be formed via reconnection of the rope with external coronal field topologies, again leading to a shifting of the rope footprint (i.e. the exterior dimming described in Table~\ref{t:categories}). One should keep in mind, however, that such modeling still depends on the realism of the initial configuration and the degree of approximation in the equations and numerical scheme employed. This is illustrated by three case-study simulations of the same event described in Section~\ref{ss:sim_example2}.

Another encouraging prospect for understanding coronal dimming through case-study modeling is with the use of modern data-driving techniques for evolving the boundary magnetic field in coronal MHD models to better match  observed magnetic field morphologies and evolution. A first application of this approach to coronal dimming was touched on by \citet{fan24}, who used a data-driven thermodynamic MHD calculation to model the 2011 February 15 CME at high fidelity. Here the model captured the formation of a sheared structure at the main PIL, which forms a complex set of MFRs via tether-cutting reconnection during the primary eruption, and this erupting flux undergoes connectivity changes that likely involve several of the reconnection scenarios outlined in Sect.~\ref{Terminology}. As a check, they compared the relative lengthening of field lines at the inner boundary (e.g. the erupting MFR/CME) to the dimming regions observed in SDO/AIA 211 \AA\ observations close to the primary PIL and at secondary sites, finding reasonable agreement between the two.

Idealized modeling and case-study modeling are not mutually exclusive either. For dimmings, \citet{Jin:2022} is one such modern example, where the complex 3D magnetic field topology of the corona on 2011 February 15 is used as the background for a parametric study. Focusing on the consequences of an eruption by inserting out-of-equilibrium MFRs, they demonstrate how varying MFR properties can influence the appearance of dimming. They connect the formation of strong dimming in the total irradiance of EUV lines to mass-loss caused by the eruption, which depends on MFR orientation and evolution. They also describe how remote dimmings may (or may not) be preferentially formed depending on the MFR properties. Most importantly, this type of sun-as-a-star analysis can help make the stellar connection Sect.~\ref{sec:stellar}.

\subsection{Example 1: Eruption properties and dimming observables}
\label{ss:sim_example1}

Here we describe a case that illustrates how the combination of comprehensive eruption modeling and the synthesis of forward modeled observables can be used to connect the formation and evolution of dimmings to their underlying physical mechanisms. One drawback of the aforementioned \citet{cohen09}, \citet{lugaz11}, and \citet{Jin:2022} case-study simulations was the use of a strongly out-of-equilibrium analytic MFR model, which is inserted into the source region and erupts immediately. This makes it challenging, particularly near onset, to distinguish between the magnetic and hydrodynamic signatures induced by the MFR insertion itself and the physical evolution relevant to coronal dimming. In modeling the relatively isolated CME that occurred on 2009 February 13, \citet{Downs&al2015,downs21,downs2023_dimming_inprep} addressed this limitation by inserting a stable MFR into the source region and then relaxing towards a stable equilibrium. The model was then quasi-statically driven at the boundary, transitioning the initially stable configuration towards eruption onset. Another focus of this study was to include a reasonable prescription for coronal heating, thermal conduction, and radiative losses, such that EUV observables could be forward modeled and subsequently compared to observations\footnote{See \citet{downs21} for details on the numerical setup, eruption model, and synthetic forward modeled observables.}. 

\begin{figure}[tbp]
\centerline{\includegraphics[width=1\textwidth,clip=]{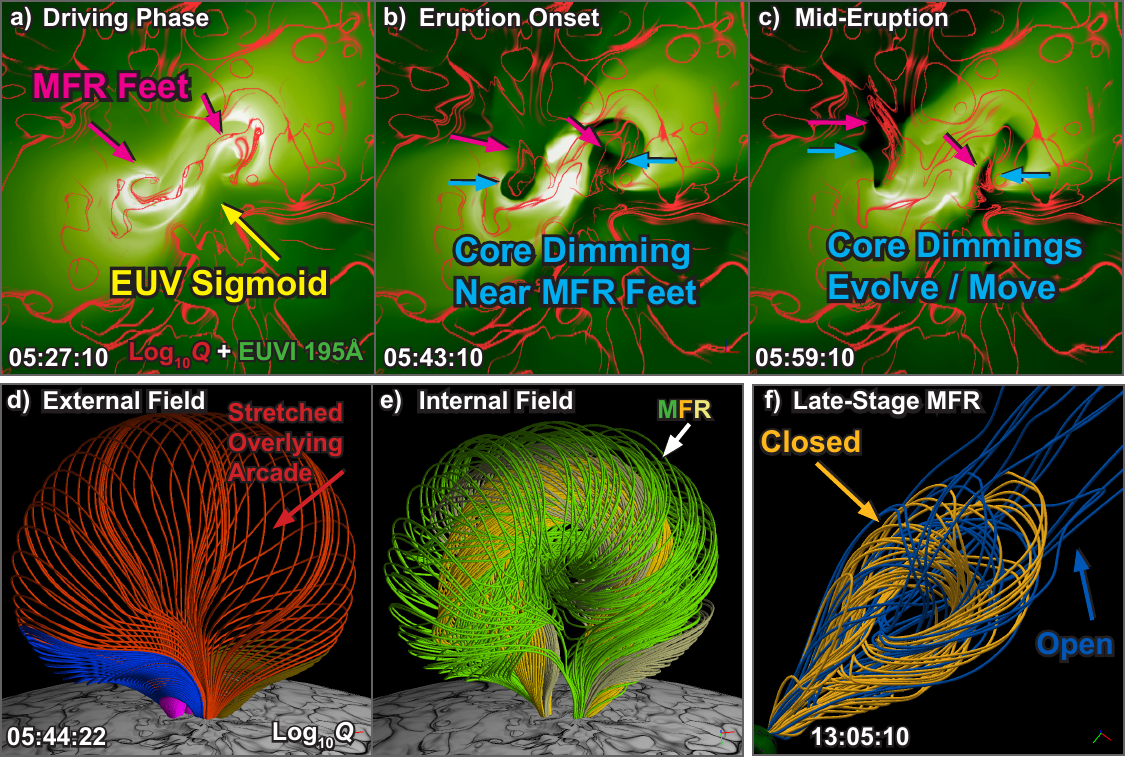}}
\caption{Representative snapshots of the SOL2009-02-13 CME simulation from \citet{downs2023_dimming_inprep}. Together these show core dimming and motion of the dimming region (a-c), stretched arcade loops (d) which undergo leg-arcade reconnection to become part of the MFR (e), and interchange reconnection between the leg and open flux (f). See the text for more details. A movie of the simulations and synthetic STEREO/EUVI 195 {\AA} data is shown in the online supplement. STEREO/EUVI 195 {\AA} observations of this event can be found in Fig.~\ref{fig:miklenic_2011_F2}.}
\label{downs_feb2009_summary}
\end{figure}

The highlights of this simulation with respect to dimming are summarized in Fig.~\ref{downs_feb2009_summary}. Panels \ref{downs_feb2009_summary}a-c show the evolution of the forward modeled STEREO/EUVI 195\AA\ emission with an overlay of the high-$Q$ lines \citep{titov07} that delineate true separatrix surfaces and QSLs. The pre-event sigmoid configuration is associated with the feet of the MFR delineated by QSL `hooks'. As the eruption proceeds, core-dimming regions form near the MFR feet, which in this case are caused by a combination of the MFR rising, lifting of the strapping field, strapping-strapping reconnection (this case has a strong guide field), and rope-strapping reconnection. Later in the eruption, the core dimmings are observed to move and shift to the periphery of the region along with the MFR footprints (red lines) as rope-strapping and leg-leg reconnection become the dominant processes. It is also evident from this case that while the observed dimming region (transient dark patches in the green 195~\AA\ image) is well correlated with the location of the MFR footprint in the photosphere, they do not have a one-to-one correspondence due to projection effects and the volumetric nature of coronal emission.

The way in which the overlying strapping field is perturbed by the eruption and its relation to the MFR field is also shown in Panels \ref{downs_feb2009_summary}d-e. In this case a fraction of the overlying field is truly stretched and reconnected (red field lines), while the rest (blue field lines) are deflected to the side and the MFR effectively `squeezes' through the middle---a process that can only be realized in 3D configurations. In this simulation, only the plasma along the reconnected field lines dimmed appreciably, while the deflected flux was perturbed but not qualitatively changed. This is consistent with the observational association of deep dimmings with the reconnected flux.

The late stage of the eruption is also examined in the model (Panel \ref{downs_feb2009_summary}f). For this case, the global magnetic configuration prior to onset included a pair of equatorial coronal holes (open flux) at the periphery of the relatively compact bipolar source region of the CME. At this later stage $\approx\! 8$ hr after onset, the MFR footprint has completely migrated outside of the overlying strapping field via rope-strapping reconnection and the MFR has begun interchange reconnecting with these neighboring open flux patches. This leads to an MFR comprised of mixed closed (gold) and open (blue) connectivity. The migration of the MFR footprint (and thus dimming region) towards the boundary of adjacent coronal holes makes logical sense as all 
CMEs must eventually overcome their overlying streamer fields and rope-strapping reconnection 
presents an obvious means to do so. The mixing of open flux with the MFR footprint via interchange reconnection also leads to favorable conditions for subsequent reconnection. This is because after the initial interchange reconnection, the new footprint of open flux is rooted in the opposite leg of the MFR, 
which can again place this flux within close proximity to the primary flare current sheet and lead to subsequent reconnection.  In the simulation, this leads to an increasing amount of disconnected flux within the MFR as function of time, illustrating a viable mechanism for the detachment of the erupting flux and eventual relaxation of the corona to pre-event conditions (as discussed in Sect.~\ref{ss:open_flux}).

Lastly, that MFR/ICME connectivity may eventually end up near or inside pre-existing coronal hole boundaries may also resolve another puzzling aspect of dimming observations: that dimmings generally begin to recover to pre-event level within several hours after a CME event \citep{Kahler:2001}, but interplanetary measurements of bi-directional electrons indicate that the CME field lines are still connected to the Sun 3--4 days later (see also Sect. \ref{sec:recovery}).

\subsection{Example 2: Dimmings in a circular-ribbon eruption}
\label{ss:sim_example2}

In this section, we summarize the key findings of a study by \citet{Prasad2020}, who carried out high-resolution data-constrained MHD modeling of an eruption under a coronal null point to investigate the relation between the coronal dimmings, flare ribbons and magnetic reconnection in such a configuration. The event originated from AR~11283 on 2011 September~6 and produced a fast halo CME accompanied by an X2.1 flare. The SDO/AIA multi-wavelength observations revealed features like a confined C1 precursor flare, whose hot sigmoidal loops suggest the formation of a flux rope, the fast rise of a set of diffuse loops at the erupting filament channel synchronously with the onset of the main flare, the rapid expansion of various faint loops across the whole eruption source region during the impulsive flare phase, the formation of double, circular, and remote flare ribbons, and the formation of multiple, complex dimmings. The dimmings were studied in \citet{Dissauer:2018a}, \citet{Vanninathan:2018}, and \citet{Prasad2020}.
They are briefly characterized in Sect.~\ref{sss:complex_dimmings2}, see Fig.~\ref{f:2011-09-06T2300MaxDimm}. 

Additionally, the event comprised two filament eruptions. The first filament erupted slightly delayed, at about the peak time of the flare, from the southern section of the closed PIL, where the sigmoid had brightened and one part of the double ribbons was running. The second filament erupted about 20~min later from the other part of the double ribbons at the eastern section of the closed PIL. Both filaments erupted considerably inclined toward the northwest. An overview of the event evolution was given by \citet{Janvier&al2016}, also see references therein. 

\begin{figure}[htbp]   
\centerline{\includegraphics[width=0.98\textwidth,clip=]{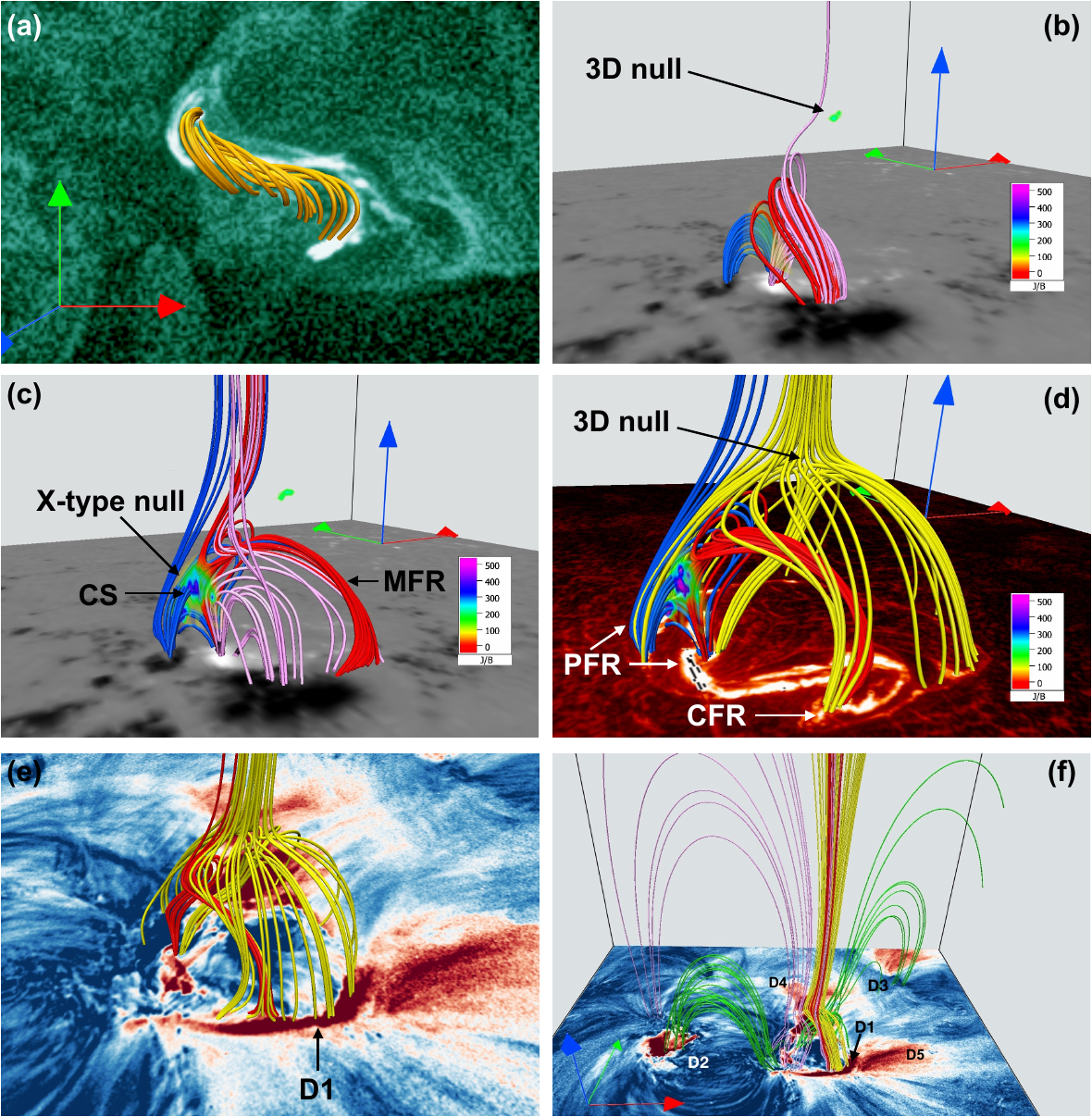}}
 \caption{
 Illustration of the magnetic reconnection and field 
topology changes associated with an erupting MFR and their observational counterparts in flare ribbons and dimming regions in the eruption from AR~11283 (SOL2011-09-06). Panel a shows a hot sigmoid in the SDO/AIA 94~\AA\ channel and the co-spatial highly sheared, weakly twisted orange field lines from the extrapolation. 
Panel b shows the position of the magnetic null point as a green dot and the initial position of the field lines that form the erupting MFR in red. 
Panel c shows the reconnection of the MFR (red field lines) and its envelope (purple field lines) with ambient field lines (blue) that were initially running just above the fan separatrix associated with the null point. The reconnection proceeds at an X-type reconnection site, where a current sheet (CS) with high values of J/B is formed. Panel d shows that the footprints of the MFR coincide with the end regions of the parallel (double) flare ribbons (PFRs). 
The circular flare ribbon (CFR) observed in the AIA 304~\AA\ channel can be seen tracing the footpoints of the field lines constituting the fan dome of the null point. Panel e overlays these magnetic field lines onto the arc-shaped 
dimming region D1 (indicated by the black arrows). This dimming region is thus seen to form along the circular flare ribbon. Panel f depicts the global connectivity in the NFFF, modified around the outer spine by reconnection, 
and its association with the remote dimming regions marked as D2 and D3. 
Panels e, f show AIA 211~\AA\ logarithmic base-ratio images in the bottom plane, where white-to-red colors denote decreases in emission, i.e.\ dimming regions. The red, green, and blue arrows represent the $x$, $y$, and $z$ directions, respectively. Adapted from \citet{Prasad2020}.  A movie of the simulations is shown in the online supplement (from \citet{Prasad2020}). } 
   \label{fig-prasad2020}
   \end{figure}

The simulation was initialized using an extrapolated non-force-free magnetic field (NFFF), derived from a photospheric vector magnetogram of the active region taken by SDO/HMI just before the rise of the precursor flare. The NFFF extrapolation method \citep{Hu2010} used in the study effectively captured the highly sheared and weakly twisted magnetic field lines above the section of the PIL,  where the eruption was observed to start (Fig.~\ref{fig-prasad2020}a), and detected topological features like a coronal magnetic 
null point close to the flaring region (Fig.~\ref{fig-prasad2020}b). These sheared and twisted field lines correlate well with the pre-flare sigmoidal brightening observed in the AIA 94~\AA\ filter (Fig.~\ref{fig-prasad2020}a). 

The Lorentz forces in the initial field, concentrated near the bottom boundary, were critical in generating self-consistent flows that initiated the dynamics and triggered magnetic reconnection. 
The evolution in the simulation starts with the rise of a flux bundle (red field lines in Fig.~\ref{fig-prasad2020}b) situated above the flux that corresponds to the observed sigmoid (orange field lines in Fig.~\ref{fig-prasad2020}a). Reconnection starts between them, leading to a decrease of shear and twist in the lower, stable flux bundle and a simultaneous increase of twist in the rising flux, which develops the structure of an MFR. The outcome of this reconnection bears similarities to the outcome of flare reconnection in the standard flare model, which produces less and less sheared flare loops below the reconnection region and adds poloidal flux (i.e., twist) to the rising MFR. 
The rising MFR in the simulation corresponds to the observed set of diffuse loops, whose onset of rapid rise was synchronized with the onset of the main flare's impulsive phase, the formation of the main (double) flare ribbons, and the formation of the flare loop arcade rooted in these ribbons. The positions of the MFR footprints are in good agreement with the footprints of the diffuse loops and with the extent of the double flare ribbons (Fig.~\ref{fig-prasad2020}d). 

During the rise of the MFR, reconnection also commences at the null point. This is consistent with the strong upward and lateral expansion of flux during the impulsive flare phase, as indicated by multiple rapidly expanding faint loops observed all over and even around the source region, which must have perturbed the null point above the source region. Null points then transform into a local current sheet, where reconnection ensues, likely to be of the spine-fan reconnection category \citep{Pontin&al2013}. This is consistent with the simultaneous formation of the circular ribbon early in the impulsive flare phase and its location along the footprint of the fan separatrix in the extrapolated field (Fig.~\ref{fig-prasad2020}d). 

Next, the MFR hits the fan separatrix at a point at the side of the fan dome. As a result, the current density steepens, and reconnection commences, also in this place. Here the MFR reconnects with ambient flux just above the fan separatrix and is cut completely in the process. As also summarized in Sect.~\ref{sss:complex_dimmings2}, the MFR leg rooted in the parasitic (positive) polarity is nearly oppositely directed to the ambient flux and reconnects, opening up the MFR flux, which is rooted in the surrounding (negative) polarity, along the outer spine (red field lines in Fig.~\ref{fig-prasad2020}c--f). It is expected that this turns a flux-rope dimming at the footprint of the MFR in the surrounding polarity into an open-flux dimming, i.e., that a deep dimming develops at this location, which lies within the deep arc-shaped dimming D1. The arc-shaped dimming can be interpreted as the outer strapping-flux dimming, with the peculiar shape representing a high shear of the strapping flux in this source region (Fig.~\ref{f:ed_np}). The reconnection also produces a flux bundle simply arching from the MFR footprint in the parasitic polarity to nearby surrounding flux outside the circular ribbon (additional closed blue field lines in Fig.~\ref{fig-prasad2020}c--d). This flux relaxes, shrinking like flare loops. Correspondingly, the flux-rope dimming observed at this footprint remains weaker than the conjugate flux-rope dimming. It is worth noting that the reconnection of the erupting MFR at the side of the fan dome here results in fully similar topological changes as the reconnection of the erupting MFR at the null point in \citet{lugaz11}. 

Together with the main eruption causing the CME and X-class flare, the two filament eruptions constitute a three-step eruption yet too complex for a data-constrained modeling to match all main elements. However, one can infer directly from the data an interpretation for the dimmings D3 and D5 in Fig.~\ref{fig-prasad2020}f. Both dimmings commence when the first erupting filament, propagating toward their location, sweeps over D5. Exterior flux connecting D3 and D5 might have been blown open by the erupting filament, similar to the process proposed in \citet{Delannee:1999}. The dimmings would in this case be categorized as \emph{passive exterior dimmings}. The second filament eruption occurred when all major dimmings had already started, but it may have contributed to the enlargement and deepening of D3 and D5 after its propagation across D3. 

\begin{figure} 
\centering
\includegraphics[width=0.7\textwidth]{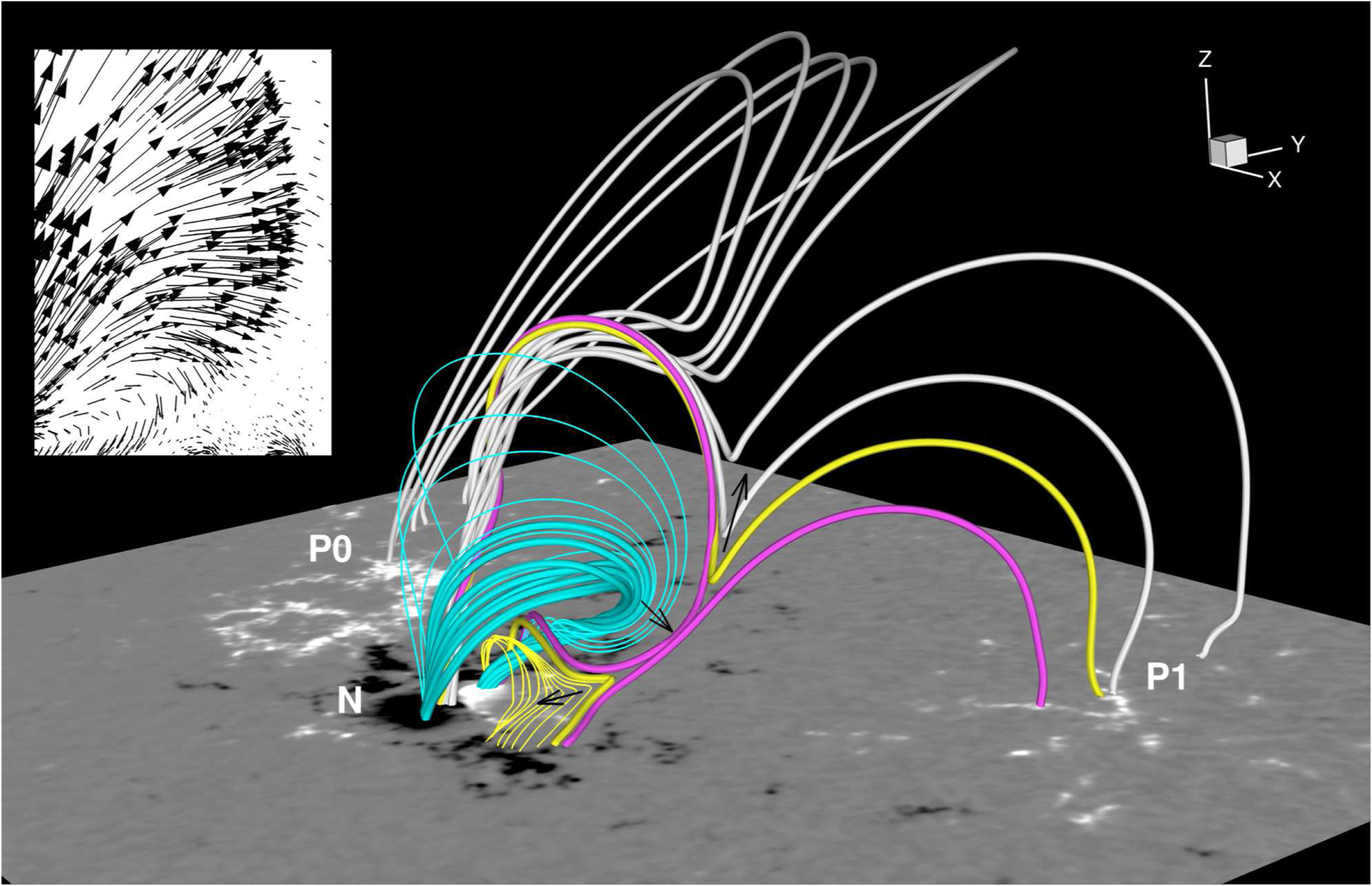}
\caption{Reconnection of an erupting MFR (cyan field lines) with flux around the outer spine (white-yellow-pink field lines rooted in positive flux P1 (corresponding to dimming D3 in Fig.~\ref{fig-prasad2020}f) at the null point of the modeled AR~11283. Black arrows show the reconnection flows. A movie of the simulation is shown in the online supplement. Figure and movie from \cite{JiangC&al2018}. 
}
\label{f:JiangC&al2018}
\end{figure}

An alternative interpretation of the remote dimming D3 was proposed by \citet{JiangC&al2013, JiangC&al2018}, who also performed data-constrained MHD simulations of the event. These authors used an extrapolated nonlinear force-free field (NLFFF) as their initial condition. This model of AR~11283 contains an MFR in the torus-unstable height range, which initiated the evolution in the simulation. While the basic fan-spine topology associated with a coronal null point is similar to the NFFF used in \citet{Prasad2020}, an important difference consists in the magnetic connection of the outer spine. The outer spine in the NLFFF connects to a region of positive polarity, P1, at the position of dimming D3, but in the NFFF, the outer spine is part of the open flux (i.e., connects to the top boundary of the volume). 
However, both models contain flux rooted in the surrounding (negative) polarity and connecting to the region D3 (as well as flux connecting to the dimming region D2), see 
Fig.~\ref{fig-prasad2020}f. The erupting MFR, propagating very obliquely in this simulation, as the erupting filament, reconnects at the null point in a manner similar to the reconnection of the MFR in the NFFF, see Fig.~\ref{f:JiangC&al2018}. The upper reconnection product, consisting of the original MFR leg rooted in the surrounding (negative) polarity and the outer spine and its surrounding flux, continues to erupt. This corresponds to the dimming D3 around the footpoint of the outer spine in the positive-polarity region P1 and a related (weak) remote flare ribbon at the inner edge of D3. This also leads to the lifting of flux from the low corona just outside the fan dome, where the newly reconnected field lines are strongly bent, driving a strong upward reconnection outflow. The latter corresponds in location 
to the strong dimming D4 just north of the circular dimming D1 in Fig.~\ref{fig-prasad2020}e--f. Additionally, reconnection under this now large-scale rising MFR adds further flux to the MFR, thereby enhancing its twist. This reconnection is basically similar to the second phase of reconnection in \citet{lugaz11} and likely to further strengthen this dimming. However, in the observational data, the dimming D4 forms simultaneously with the onset of activity (formation of a ribbon, quickly followed by dimming) at the position of dimming D2, which is earlier than the onset of dimming D3. The erupting flux enters a further phase of reconnection in the simulation, which involves flux rooted in a further region of positive polarity (P0), where the dimming D2 in Fig.~\ref{fig-prasad2020}f formed. The erupting flux rope, still rooted very near the original negative footprint of the initial MFR, then splits, with its parts connecting to the polarities P0 and P1 in Fig.~\ref{f:JiangC&al2018}. This corresponds to the formation of dimmings in both polarities P0 and P1 (dimmings D2 and D3, respectively). 

Finally, a slightly modified interpretation of the dimmings D2 and D3, with their relationship to the outer spine reversed, is suggested by a series of NLFFF models of AR~11283 constructed by flux-rope insertion in \citet{Janvier&al2016}. These models also reveal magnetic connections from the surrounding negative polarity to P0 and P1, but the outer spine in them is rooted in P0. The observations suggest that this location of the spine is closest to reality, because some of the large-scale diffuse loops rising from the early phase of the eruption, as well as many post-eruption loops, connect the dimming D2 with the region of the parasitic polarity. Moreover, a remote flare ribbon encircling D2 is far more prominent than the ribbon at the inner edge of D3, and it brightens well synchronized with the circular ribbon early in the impulsive phase of the flare, while the ribbon at D3 brightens when the first filament eruption is underway shortly after the peak of the flare. One can expect that the lifting of flux and reconnection occur under the rising outer spine in this model as well. This would bring the formation time of the dimming D4 in agreement with the observations, but it remains unclear whether its location is also consistent with this interpretation. \citet{Janvier&al2016} studied the evolution of their models (relaxation to equilibrium vs.\ instability) in the magnetofrictional simplification of MHD, which cannot realistically model reconnection in a current sheet. It is remarkable that the inserted MFR in their unstable model nevertheless showed an evolution similar to the one observed in \citet{Prasad2020} and \citet{JiangC&al2013, JiangC&al2018}: the MFR broke in the leg rooted in the parasitic polarity at the eastern side of the fan separatrix to connect its leg rooted in the surrounding polarity with the outer spine and its surrounding flux. 

The interpretations of the dimmings in the 2011 September~6 eruption, suggested by the observational data and the numerical modeling reviewed in this section, are summarized in Fig.~\ref{f:2011-09-06T2300MaxDimm}. The three technically very different models support each other in the following essential aspects. 
(1) All find an erupting MFR that is consistent with the formation and extent of the double flare ribbons. 
(2) All find that the MFR leg rooted in the parasitic polarity reconnects through the fan separatrix with the outer spine and its surrounding flux. This MFR, built from the other leg of the erupting MFR and the outer-spine flux, continues to erupt. This is consistent with the formation of a deep flux-rope dimming (or even open-flux dimming) at the MFR footprint in the surrounding polarity and with the formation of an exterior dimming and associated remote ribbon at the spine-flux footprint (either D2 or D3). 
(3) Both MHD simulations find reconnection at the null point, consistent with the formation of the circular flare ribbon and the subsequent formation of a circular thermal dimming in its place. 
However, there are also essential aspects that differ between the models and illustrate their limitations. 
(4) The eruption is initiated by a primary MFR found in both NLFFF models of the active region, but by a secondary MFR that forms from sheared flux in the NFFF. 
(5) The three extrapolations yield three different end points of the outer spine. 
Therefore, a careful comparison of such model results with the observations is essential.

\section{From the Sun to stars} 
\label{sec:stellar}

The Sun is of course a star, but we observe it very differently from how the more remote stars can be accessed with current instrumentation. 
This leads to innumerable difficulties in translating solar observational methods to stellar targets, despite the assumption that the underlying physics is the same.  
Decades of research in the solar-stellar connection shows that magnetic activity on main sequence cool stars is of fundamentally the same nature as that studied in detail on the Sun \citep{Schrijver:1991,Gudel:2002,Benz:2010,Linsky:2017}, with many magnetically active stars exhibiting levels far in excess of what the Sun demonstrates. 
Thus there is the expectation that stellar flares should have coronal mass ejections associated with them.  

Thanks to high precision visible light photometric space missions designed to find transiting exoplanets, our knowledge of stellar types observed to flare has expanded in the last decade, and flaring appears to be a common feature of stars in the cool half of the HR diagram \citep{Yang:2019}.
Flaring correlates with other signatures of magnetic activity \citep{Osten:2016}. 
In particular, high flaring rate stars/those with enhanced magnetic activity tend to be either young \citep[due to the well-known correlation between age and rotation;][]{Noyes:1984}, nearly to full convective \citep[possibly due to changes in dynamo generation of magnetic fields;][]{Kochukhov:2021}, or in close binary systems where rapid rotation is enforced by tidal synchronization. 
A common measure of magnetic activity used is the ratio of the star's X-ray luminosity $L_X$ to its bolometric luminosity $L_{\rm bol}$, as this ratio describes the efficiency of coronal heating. Stars with the highest level of magnetic activity show $L_{X}/L_{\rm bol}$ values near 10$^{-3}$ \citep{Pizzolato:2003}, which seems to be the maximum amount of quiescent coronal heating that a star can maintain; for reference, the Sun has a value of $\log (L_{X}/L_{\rm bol})$ ranging from $-5.7$ to $-6.8$ over the course of its activity cycle in the X-ray energy band commonly used by stellar astronomers \citep{Judge:2003}. In Table~\ref{tbl:star_param}, we list the coronal parameters $L_X$ and $L_{X}/L_{\rm bol}$ for individual stars that are discussed in this chapter.

\begin{table}[htb]
    \centering
     \caption{Stellar Coronal Parameters for Individual Stars Discussed }
    \label{tbl:star_param}  
    \begin{tabular}{l|l|l|l}
    Star    &  L$_{X}$ & reference & L$_{X}$/L$_{\rm bol}$\\
       & (erg s$^{-1}$) & & \\
    \hline
$\varepsilon$ Eri   &   2$\times$10$^{28}$  & \citet{Coffaro:2020}   &        2$\times$10$^{-5}$ \\
Kappa Cet  & 10$^{28.79}$   & \citet{Wood2014}        &   2$\times$10$^{-5}$ \\
EK Dra    &  7.5$\times$10$^{29}$ & \citet{Dorren1995}  & 2.5$\times$10$^{-4}$ \\
AB Dor      & 10$^{29.8}$     &  \citet{LalithaSchmitt2013}  & 2.6$\times$10$^{-4}$ \\

EV Lac     & 3$\times$10$^{28}$    &  \citet{Favata2000}   &      5.7$\times$10$^{-4}$ \\

Proxima    & 0.4$\times$10$^{27}$   & \citet{Fuhrmeister2022}  &6.9$\times$10$^{-5}$ \\

AU Mic       &10$^{29.7}$      & \citet{Pallavicini1990}  &   10$^{-3}$\\

AD Leo &3.2$\times$10$^{28}$ & \citet{Favata2000_adleo}& 3.7$\times$10$^{-4}$\\
EQ Peg &4.7$\times$10$^{28}$ &\citet{Liefke2008} & 5.9$\times$10$^{-4}$\\
Algol &1.1$\times$10$^{31}$ &\citet{Ness2002} &1.5$\times$10$^{-5}$ \\
    \hline

    \end{tabular}
 
\end{table}

Over the last decade, the topic of CMEs on magnetically active stars has gained increased interest \cite[e.g.,][]{Lynch:2023}, with an appreciation that stellar CMEs and their space weather consequences may have substantial adverse impacts on the habitability of exoplanets orbiting those stars \citep{Lammer:2007,Khodachenko:2007,Airapetian2020I,Alvorado2022}. 
This is in addition to any effects on the planet of ionizing radiation from the flare itself. However, the observational constraints on stellar CME occurrence and properties are still scarce. 

\begin{figure*} 
\centering {\includegraphics[width=0.99\textwidth]{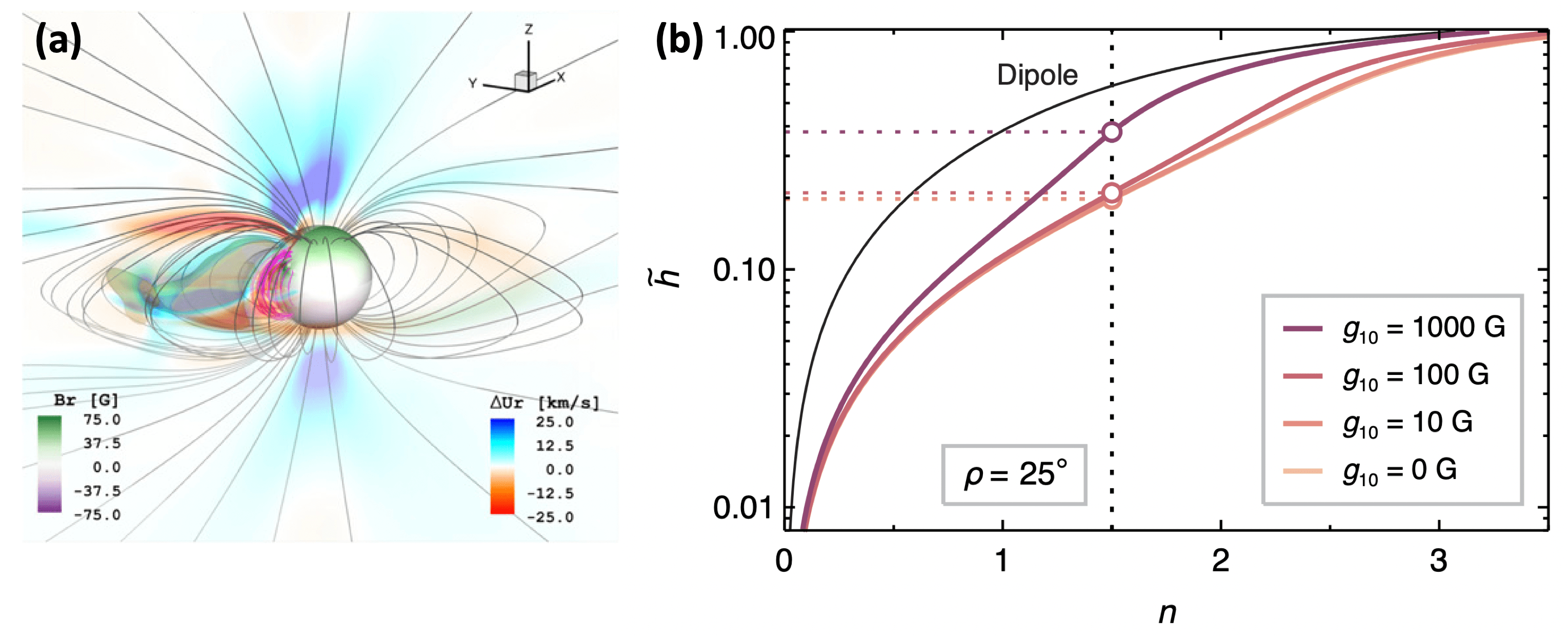}} 
\caption{(a) A confined Gibson-Low flux rope eruption after 1 hour simulation. The identified CME ejecta is shown as a translucent yellow shade. The background shows a transversal plane crossing the source AR with color contours representing the radial wind speed difference with respect to the pre-eruption condition \citep{AlvaradoGomez:2018}. (b) Decay index $n$ as a function of height (in units of the star radius) for potential field model (source surface at 2.5 star radius) with both starspot and dipole. The bipolar starspot size $\rho$ is fixed at 25$^\circ$. The thin black curve shows the dipole-only case and the colored curves represent different dipole strength g$_{10}$. The vertical dotted line shows the critical decay index $n_{c}$ = 1.5 and the horizontal dotted lines indicate the critical heights \citep{Sun:2022}. A movie of the simulations shown in panel (a) can be found in the online supplement (from  \cite{AlvaradoGomez:2018}). }
\label{fig:AlvaradoGomez_Sun} 
\end{figure*}

The elevated stellar magnetic activity arises from enhanced magnetic fields, and this  would affect the conclusion of a relationship between stellar flares and CMEs, as  stronger magnetic confinement  could lead to more confined flares or failed stellar eruptions than on the Sun. 
\citet{AlvaradoGomez:2018} show in a global MHD model that a dipole field of 75 Gauss is able to fully suppress CMEs with up to 3$\times$10$^{32}$ ergs initial energy (comparable to a large solar X-class flare/CME; Figure~\ref{fig:AlvaradoGomez_Sun}a). Using a potential field source surface model, \citet{Sun:2022} estimate the Torus-stable zone above a bipolar stellar active region embedded in a global dipole field and find that in  active cool star conditions, the Torus-stable zone can extend significantly higher compared to solar conditions (Figure~\ref{fig:AlvaradoGomez_Sun}b). This effect is believed to be one of the reasons causing the sparseness of detections of CMEs on stars \citep{OstenWolk:2015,Drake:2016,Odert:2017}. 

Due to the lack of direct observations, stellar CME occurrence rates have been derived from stellar flare occurrence rates and energies together with solar-stellar scalings of the flare-CME relationship. 
This approach results in largely overestimated CME occurrences and associated stellar mass losses that are orders of magnitude higher than the present-day solar mass loss rate \citep{OstenWolk:2015,Drake:2016,Odert:2020}. The largest uncertainty in these models comes probably from the flare-CME association rate that does not properly account for the likely larger magnetic confinement of the flaring corona in active stars \citep{Odert:2017}.

In this chapter, we first give an overview on the different methods and approaches that have been considered for detecting CMEs on other stars than the Sun. 
One of the desirements for an observational method to detect and study 
stellar CMEs are that 
the signature being sought should be attributable to the CME 
and not require special flare circumstances in order to explain;
additionally it should also allow some inference of CME properties
(some combination of mass, velocity, energy) to be determined. 
The next section describes studies of solar disk-integrated dimmings to examine applicability
of coronal dimmings to stellar cases.
Following this, we describe a few recent studies applying  CME-associated coronal dimmings as signatures of stellar CMEs both in observations and modeling.

\subsection{Overview on detection methods for stellar CMEs}\label{sec:stellar_overview}

A comparison of the observational methods for detecting CMEs is useful. 
As Table~\ref{tbl:cme_compare} illustrates,
there are a number of observational techniques that have been either used or  proposed to diagnose the presence and characteristics of coronal mass ejections on the Sun. Most of these are potentially applicable for stellar targets, and in fact there are individual detections of stellar CMEs claimed using some of these techniques. The goal is to go beyond singular detections of the phenomenon in stars, to explore the systematics of transient stellar mass loss.

 \begin{table}
    \centering
     \caption{Observational Methods to Detect Coronal Mass Ejections on the Sun and Stars}
    \label{tbl:cme_compare}  
    \begin{tabular}{l|l|l|l}
    Observational Signature     & Sun & Stars  &Reference for Stellar Work\\
    \hline
Resolved Thomson scattering\\ via coronagraph & $\checkmark$ & X \\
\hline
Type II burst &    $\checkmark$ & ?& \parbox[t]{2in}{\citet{Crosley:2016,CrosleyOsten:2018a,CrosleyOsten:2018b}}\\
\hline
Resolved nonthermal emission \\from CMEs & $\checkmark$ & ?& \\
\hline
Scintillation of point radio sources & $\checkmark$ & ? &\\
\hline

Type IV burst & $\checkmark$ & $\checkmark$ & \citet{Zic:2020} \\
\hline

Negative radio bursts & $\checkmark$ & ? &  \\
\hline

Mass-loss coronal dimming \\during a flare & $\checkmark$ & $\checkmark$ & \citet{Veronig:2021,Loyd:2022}\\
\hline

\multirow{2}{*}{\parbox[t]{2in}{High-velocity outflows in coronal\\ emission lines during a flare}} & $\checkmark$ & $\checkmark$ & \parbox[t]{2in}{ \citet{Argiroffi:2019,Chen2022}} \\
   & & & \\
   \hline
\multirow{2}{*}{\parbox[t]{2in}{High-velocity outflows in \\chromospheric emission lines \\during a flare}} & $\checkmark$ & $\checkmark$ & \parbox[t]{2in}{\citet{Odert:2020,Leitzinger:2020,Muheki:2020, Namekata:2022,Inoue2023}}\\
   & & & \\
   \hline
   
\multirow{2}{*}{\parbox[t]{2in}{Absorption dimming: increase in\\hydrogen column density during a flare}} & ? &$\checkmark$ & \citet{Moschou:2017}\\
   & & & \\
   \hline
Pre-flare dips & $\checkmark$ & ?  & \parbox[t]{2in}{\citet{Giampapa:1982,Leitzinger:2014}}\\
 \hline
Effect of CMEs on stellar \\environment & $\checkmark$ & ? &\citet{Osten:2013}\\ 
\hline
Association with flares & $\checkmark$ & $\checkmark$ &\parbox[t]{2in}{\citet{Aarnio2012,Drake:2013,OstenWolk:2015,Odert:2017}}\\ 
\hline
 
\multicolumn{4}{l}{$\checkmark$ = has been used and demonstrated as a technique
    to probe CMEs}\\
\multicolumn{4}{l}{X = not possible to use} \\
\multicolumn{4}{l}{? = Possible technique but not used to date or no positive results}\\
   \end{tabular}  
\end{table}

The observational techniques listed in Table~\ref{tbl:cme_compare} are wide-ranging, and for the most part there is a concurrence between solar and stellar studies. 
A notable departure  is the use of coronagraphic measurements of Thomson scattering of photospheric photons off coronal electrons. While this technique is the workhorse of solar CME observations, the spatial resolution and contrast requirements needed in order to apply this to the nearest stars are far out of reach: consider that the disk of a commonly occurring M-dwarf star at 5 pc spans $\sim$1 milli-arcsec. 

Apart from spatially resolving the evolution of the ejected coronal plasma (which is currently unfeasible for distant stars given current technology), type II radio bursts
are the next most conclusive signature. On the Sun these 
 are frequently observed to accompany fast CMEs \cite[with an assocation rate of $>$40\% for CMEs with speeds $\gtrsim$1000~km~s$^{-1}$ and approaching almost 100\% for CMEs with speeds $\gtrsim$2000 km s$^{-1}$;][]{Gopalswamy2008}. 
 Type II bursts are produced not by the CME itself, but by the MHD shock produced as a super-Alfv\'{e}nic ejection travels through the outer atmosphere. 
 Type II bursts have been searched for on flaring stars \citep{Leitzinger:2010,Boiko:2012,Crosley:2016,CrosleyOsten:2018a,CrosleyOsten:2018b} but no convincing signature has been found.
A possible explanation for these non-detections may be that the higher stellar magnetic fields in stars  correspond to higher coronal Alfv\'en speeds that need to be overcome for a shock to be formed. 
\citet{Alvarado-Gomez:2020} simulated CME events on a young Sun-like star and a mid-activity M~dwarf, 
studying the distribution of coronal Alfv\'en speeds to investigate shock-prone regions.
The super-Alfv\'enic transition occurs at larger stellar radii compared to the Sun. This study 
indicates that weakly and partially confined CMEs can generate shocks in the corona, but as they are
at much larger distances, the plasma frequency (and hence optimal observing frequency) is lower than the ionospheric cutoff. This precludes ground-based radio observations from being able to detect the shock and
hence infer the presence of the CME.

In principle, direct nonthermal emission from CMEs could be probed in a stellar context, analogous to the method \citet{Bastian:2001} used, but since the necessary spatial resolution is not possible, this would require modelling of stellar radio emission to include flare emission and the likely much smaller potential CME contribution. Scintillation of point radio sources due to foreground propagation of a CME within the astrosphere of another nearby system is also potentially achievable.
Models of stellar astrospheres \citep{Wood:2005} suggest that the angular extent should be large for the nearest stars.
For instance, the suggested size of the astrosphere for the nearby flare star EV~Lac is of order 1 arcminute, and recent results have expanded the number
of nearby M dwarfs which have detected astrospheres \citep{Wood:2021}. 
Monitoring of a network of flat-spectrum radio quasars, some whose lines of sight intersect the astrosphere and others far enough removed to serve as a reference, could in principle be used to relate flaring on the central star and scintillation events.
Additionally, note that recent observational evidence on the steady winds from nearby M dwarf stars \citep{Wood:2021} finds winds that are weaker or comparable in strength to the Sun, with $\dot{M}\leq \dot{M_{\odot}}$, in general agreement with the modelling results for transient mass loss. 

\begin{figure*} 
\centering 
\includegraphics[width=0.54\textwidth]{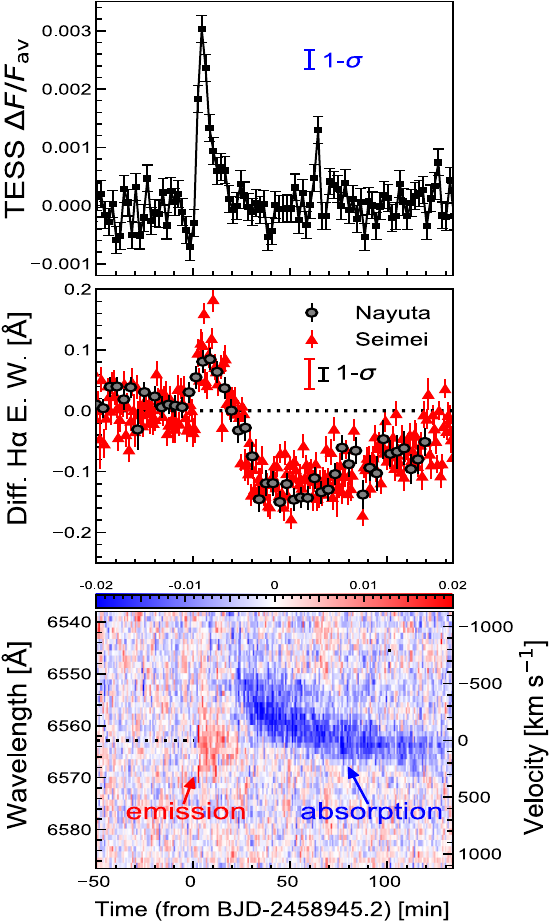}
\caption{Observations of a stellar filament eruption associated with a superflare on EK Draconis. From top to bottom: Pre-flare subtracted TESS light curve in white light, H$\alpha$  light curves integrated over $\pm$10 {\AA} around the line center from the Nayuta and Seimei telescopes,  and two-dimensional Seimei H$\alpha$ spectra.  Adapted from \cite{Namekata:2022}.} \label{fig:Namekata2022_F1}
\end{figure*}

Line-of-sight plasma motions of an outward moving CME and/or the embedded erupting prominence would result in either Doppler-shifted spectral lines, or excess emission in line wings, and is another potentially more direct signature.
There have been reports of high-velocity plasma motions in several types of flaring stars, from pre-main sequence to main sequence to evolved stars.
These events have been derived from a broad array of spectra,
ranging from optical wavelengths \citep[in particular Hydrogen Balmer lines;][]{Houdebine:1990,Gunn:1994,Guenther:1997,Fuhrmeister:2004,Vida:2016,Koller:2021,Namekata:2022,Lu:2022}; ultraviolet chromospheric and transition region lines \citep{Mullan:1989,Leitzinger:2011}; as well as coronal lines in the X-ray \citep{Argiroffi:2019,Chen2022}.
 Recently, \citet{Namekata:2022}  studied high-cadence spectroscopic observations of the young solar-type star EK Draconis (G1.5V; coronal parameters in Table~\ref{tbl:star_param}) in the H$\alpha$ line showing evidence for a stellar filament eruption with a 
velocity of 510 km s$^{-1}$ that occurred in association with a superflare
(see Figure~\ref{fig:Namekata2022_F1}).

While the occurrence of such events is  promising, it is important to keep in mind that these are a limited number of positive detections. 
This is particularly true in light of numerous systematic studies 
searching for high velocity outflows or line asymmetries in optical chromospheric lines as signatures of stellar CMEs. These studies have mainly resulted in non-detections.
An additional concern about interpretation of velocities in favor of CMEs is the potential for high velocities to be achieved in the flaring plasma, apart from any relation to ejected material. 
A more stringent test would be comparison to the escape speed of the star. 
While solar CMEs have a distribution of velocities from 
$\sim$300--3500~km~s$^{-1}$ \citep[e.g.,][]{Yashiro:2004,yurchyshyn:2005,rodriguez-gomez:2020}, only those exceeding the local escape speed are eruptive. Due to the effective-temperature scaling of stellar
mass and radius for stars on the main sequence, the escape speed at the stellar surface for stars on the main sequence is roughly similar to solar at $\approx$ 620 km s$^{-1}$. Reports of plasma upflows at X-ray wavelengths by \citet{Chen2022} at temperatures of 3 MK and 5--10 MK were accompanied
by a decreasing plasma density in the M dwarf flare star EV Lac.
This was suggested to originate from a stellar filament or prominence eruption, but with observed blueshifts less than 130~km~s$^{-1}$. 
In a similar vein \citet{Argiroffi:2019} interpreted a 90~km~s$^{-1}$ blueshift from cool (about 4~MK) coronal plasma
late in the phase of a large flare event on an evolved star as evidence of outward moving CME material.
\citet{Leitzinger:2020} found no CME signatures from 3700 hours of observations on 425
main sequence stars of spectral type F through K, while
\citet{Muheki:2020} focussed on 2000 spectra of the highly active flaring M dwarf AD~Leo,
with no asymmetries having velocities exceeding the stellar escape velocity. 
\citet{Koller:2021} investigated optical spectra provided by the Sloan Digital Sky Survey (SDSS). They selected all F, G, K, and M main-sequence type stars, resulting in a sample of more than 630,000 stars, on which they found six cases that showed excess flux in Balmer line wings, potentially indicative of CME-related mass flows.
\citet{Odert:2020} investigated the potential for such a signal from Balmer line asymmetries during flares, which requires a significant investment in spectroscopic monitoring time ($>$100 hrs) and must address activity level concerns as well as observational biases like integration time.

Recently, \citet{Zic:2020} observed a series of intense, coherent radio bursts at the same time as a bright, long-duration optical flare from the fully convective Proxima Centauri. 
The authors identified this as a stellar equivalent to a type IV burst based on the polarization and temporal structure of the radio burst, and noted that 
these events have a high degree of accompaniment to CMEs or energetic particle events in solar eruptive events. 
So-called ``negative radio bursts" \citep{1953JRASC..47..207C} are also in principle an applicable technique for the stellar case, as they can probe absorption; \citet{Grechnev2008} presents an analysis of an eruptive event on the Sun accompanied by a blast wave
and absorption phenomena seen at 195 \AA\ as well as at radio wavelengths. 
The opacity of the absorber will change with observing frequency, enabling a determination of plasma parameters for multi-frequency observations of  bremsstrahlung emission. Stellar radio flares, unfortunately, are generally produced from gyrosynchrotron emission or coherent plasma emissions, making this a less optimistic technique.

\citet{Moschou:2017} interpreted previously reported variations in Hydrogen column density N$_{H}$ seen in a large X-ray flare on the evolved semidetached binary system Algol 
as a dimming event. This general technique of ``absorption dimming" (as opposed to the mass-loss dimmings on which we focus in this review) was proposed by \citet{Mason:2014} when the CME obscures the underlying flaring emission, causing a temporary increase in the obscuration.  As the CME material expands, the amount of obscuration would then decrease back to the pre-eruption level. \citet{Moschou:2017} found that the column density temporal variations were consistent with a quadratically decreasing trend with time, which would be consistent with a self-similar expansion at constant velocity of a CME front. Obscuration dimmings by solar filament eruptions accompanying CMEs have recently been reported in Sun-as-a-star observations in the EUV \citep{Xu:2024} and metric/decimetric radio domain \citep{Hou2025}. 
 
The idea of absorption dimming at optical wavelengths due to passage of a stellar CME has been raised in previous studies, but without a confident conclusion as to its likelihood. \citet{Giampapa:1982} had noted a remarkably large preflare diminution from a U-band flare observed on the M dwarf binary EQ Peg. 
The quiescent flux dropped to 75\% of its normal value for about 2.7 minutes, just prior to a large flare event which peaked at three times the quiescent flux. 
\citet{Giampapa:1982} offered two hypotheses for the phenomenon: an increase in H$^{-}$ opacity from enhanced heating of the outer stellar atmosphere due to a flare blast wave (based on an earlier suggestion by \citet{Grinin:1976}), or a filament undergoing destabilization and dissipation. 
In this scenario, an off-limb filament deposits material into disk lines of sight, thereby increasing chromospheric line and continuum opacity which would provide the necessary attenuation. The dissipation of kinetic energy of the falling material would produce the consequent flare-like brightening. 
The paper noted that these events were rare; this conclusion was confirmed several decades later by \citet{Leitzinger:2014} in a search for optical absorption dimmings from stars in the young open cluster Blanco-1, where two ``dimmings" were seen over the course of 140 star-hours. The long-term monitoring of white-light emission from stars used for transiting exoplanet searches from Kepler, K2, and TESS have resulted in numerous papers on flares, but none apparently revealing a 25\% drop over the course of several minutes as noted in \citet{Giampapa:1982}'s U-band measurements. 
 
For recent overviews on stellar CME candidates and different detection methods, we also refer to \cite{Moschou:2019}, \cite{Leitzinger:2022}, and \citet{Osten2022IAUS}.

\subsection{Sun-as-a-star observations of CME-induced dimmings} 
\label{Sec:SunAsStar}

\begin{figure} 
\centering 
{\includegraphics[width=0.99\textwidth]{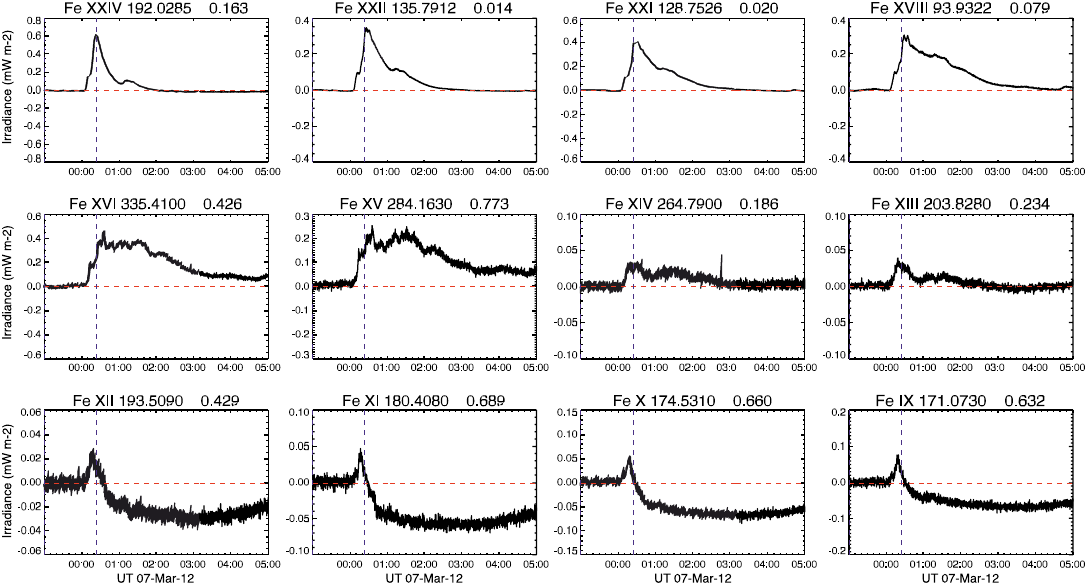}} 
\caption{SDO/EVE full-Sun light curves for the flare/CME SOL2012-03-07.
The different panels show iron lines of different ionization stages from Fe~{\sc{xxiv}}  ($\log(T/{\rm K}) = 7.3$; top left panel) to Fe~{\sc{ix}} ($\log(T/{\rm K})= 5.8$; bottom right panel). The curves show 10-s sampling and are normalized to the pre-flare level, as indicated by the red dashed line and labeled in the panel titles. The vertical line indicates the flare peak time in the GOES 1--8~{\AA} soft X-ray band. Adapted from \cite{Harra:2016}.} \label{fig:Harra2016_10}
\end{figure}

Several studies have shown that coronal mass-loss dimmings can be also observed in Sun-as-a-star EUV irradiance curves measured by the Extreme-ultraviolet Variability
Experiment \cite[EVE;][]{woods:2012} onboard SDO  \citep{Woods:2011,Mason:2014,Mason:2016,Harra:2016,Veronig:2021,Xu:2022}. Fig.~\ref{fig:Harra2016_10} shows as an example the strong coronal dimming associated with the X5.1 flare/CME of 7 March 2012, and the associated dimming from the \cite{Harra:2016} study. The different panels show SDO/EVE full-Sun light curves in iron lines at different ionization stages from 
Fe~{\sc{ix}} to Fe~{\sc{xxiv}}.
The flux curves clearly show that at high temperatures the flare is dominant, whereas at the lower ionization stages (Fe~{\sc{ix}} to {\sc{xii}}; 
corresponding to formation temperatures $\log(T/{\rm K})$ = 5.8 to 6.2), a (small) flare enhancement is followed by a distinct dimming that is ongoing for many hours. The temperature dependence in Fig.~\ref{fig:Harra2016_10} again suggests that the dimming is mostly due to the loss of material from the ambient corona.
Though we note that in some events the coronal dimming signatures could be detected at spectral lines formed at a temperature as low as $\log(T/{\rm K})= 5.4$ \citep{Xu:2022}.

\cite{Mason:2016} studied the flare/CME activity and associated dimmings of two Active Regions. In total, they identified 37 coronal dimmings, whereof 23 could be associated with a CME, and the corresponding GOES flare classes covering a broad range from B to X class. Notably, they found distinct correlations between dimming properties, like the dimming depth and the slope of the emission decrease, with the speed and mass of the CME. However, so far these correlations have not been shown for larger samples of events originating from different ARs.
\cite{Harra:2016} investigated a set of 42 X-class flares observed by SDO/EVE. 
They report that all the 33 eruptive X-class flares in their sample revealed a coronal dimming in the SDO/EVE irradiance spectra, most notably in spectral lines of ionized iron sampling plasmas at temperatures between about 1--2 MK, that starts within about one hour after the flare peak. These observations of coronal dimmings in EUV irradiance data are important, as they allow us to connect the solar observations to the stellar case.

\begin{figure}
\centering {\includegraphics[width=0.99\textwidth]{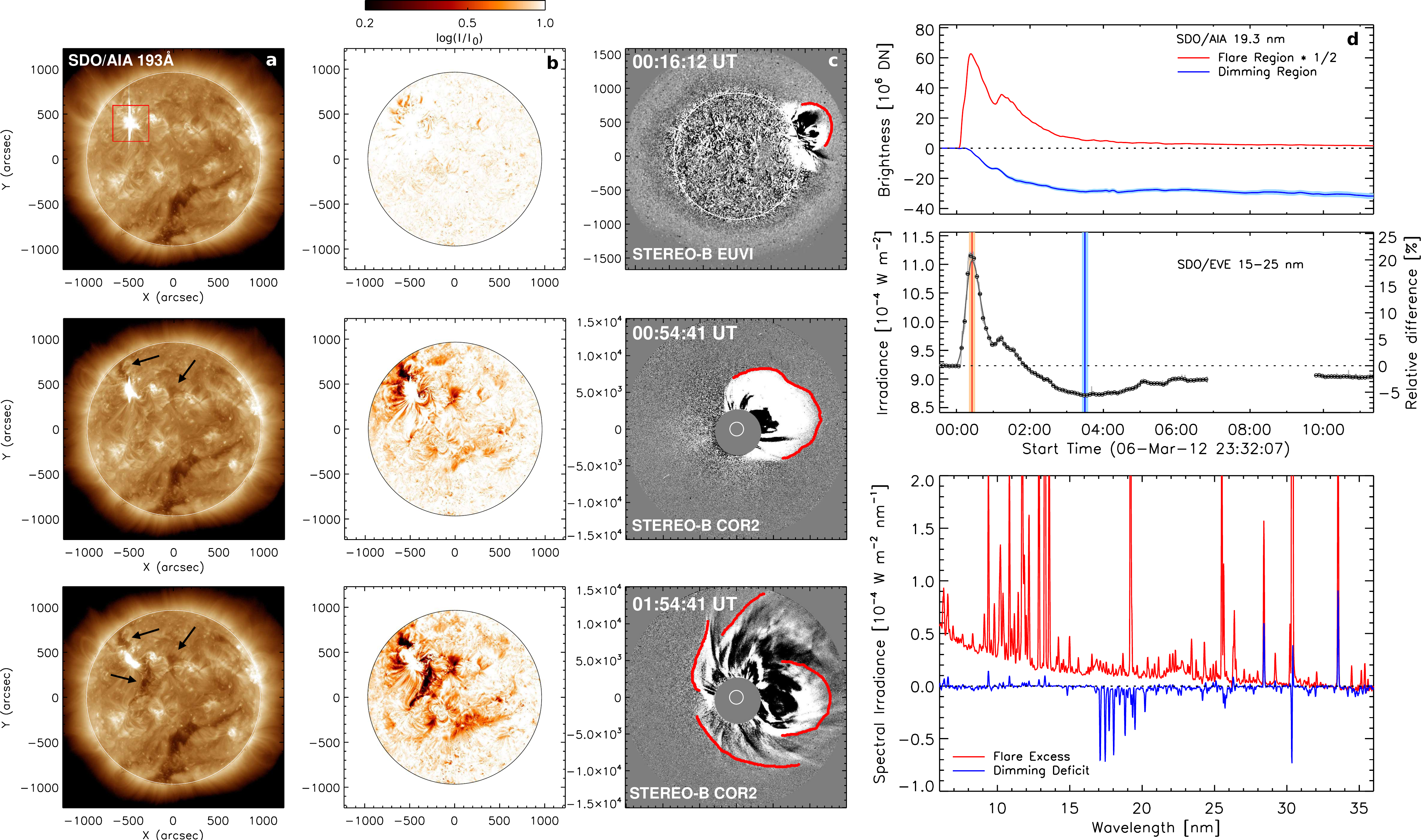}} 
\caption{Coronal dimming event associated with the X5.1 flare/CME SOL2012-03-07. SDO/AIA 193 {\AA} direct (a) and logarithmic base-ratio (b) images showing the flare and the coronal dimming. The red box highlights the region over which the flare light curve shown in d (top panel) was calculated. Black arrows mark the dimming regions. c) Two CMEs associated with this event observed by EUVI and the COR2 coronagraph onboard STEREO-B. Red contours outline
the CME fronts. d) Top panel: pre-event subtracted SDO/AIA 193 {\AA} light curves derived from the flaring region (red curve) and the dimming region (blue curve). Middle panel: SDO/EVE Sun-as-a-Star broad-band 150–-250~{\AA} light curve. 
Bottom: pre-event subtracted SDO/EVE irradiance spectra integrated over 10 min during the flare peak (red) and over the maximum dimming depth (blue), as indicated by orange and blue vertical lines and shaded regions in the middle panel. A movie of this figure is shown in the online supplement. Figure and movie from \cite{Veronig:2021}.} \label{fig:Veronig2021_1}
\end{figure}

\cite{Veronig:2021} did a systematic study of such post-flare coronal dimmings in SDO/EVE Sun-as-a-star observations as a test-bed whether coronal dimmings can serve as a proxy for CME occurrence on solar-like stars. They created broad-band EUV light curves by integrating at each time step the EVE spectra over the wavelengths range from 150--250 {\AA}, binning the original 10-sec light curves to a cadence of 10-min, and comparing the full-Sun EVE observations with the spatially resolved EUV observations from SDO/AIA.
Figure \ref{fig:Veronig2021_1} shows various aspects of the strongest solar dimming event in their sample, which occurred on 7 March 2012.
The SDO/AIA 193~{\AA} direct and difference images reveal the global nature of the dimming (columns a,b). The STEREO-B COR1 and COR2 coronagraph images show the side-on view of two very fast CMEs ($v \approx 3700$~km~s$^{-1}$) that occurred in close succession and caused this dimming (column c). Column d shows spatially integrated quantities.
The top panels shows light curves derived from AIA 193~{\AA} images, integrated over the flare (red) and dimming (blue) regions, whereas the middle panel shows EVE broad-band 150--250~{\AA} full-Sun light curves. The spatially-resolved AIA light curves demonstrate that the emission increase due to the flare and the emission decrease due to the dimming start roughly simultaneously. The full-Sun EVE light curve is first dominated by the flare emission enhancement, but after about 110 min the decrease due to the dimming becomes dominant. This means that the longer time scale of the dimming (which is related to the replenishment of the corona after the mass evacuation due to the CME) than the flare (related to the energy release and plasma cooling time scales) allows us to detect CME-caused dimmings in the Sun-as-a-star EUV light curves. (SDO/EVE light curves of this event in different iron lines are shown in Fig.~\ref{fig:Harra2016_10}.)

The bottom panel of column d in Fig. \ref{fig:Veronig2021_1} shows two 
pre-event subtracted EVE spectra, integrated over the flare peak (red) and over the maximum of the coronal dimming (blue). The flare spectrum shows enhancements both in the continuum and emission lines, whereas the dimming is predominantly due to emission decreases in spectral lines, mostly in the wavelength range from about 170--210 {\AA}. This also demonstrates that the choice of the observing range is crucial for detecting CMEs through post-flare coronal dimmings in the full-Sun lightcurves.

In total, \cite{Veronig:2021} studied a set of 44 large flares of GOES class $\ge$M5 observed by EVE (38 eruptive, 6 confined), using an automatic technique to identify dimmings in the light curves and to test their significance. They found that coronal dimmings are a frequent phenomenon associated with CMEs, with 84\% of the eruptive flares showing a significant dimming in the full-Sun EVE 150--250~{\AA} light curves, and the number of false alerts (i.e., a dimming was detected although there was no CME) being small (17\%). This gives a very high conditional probability for the occurrence of a CME given that a coronal dimming was observed in the Sun-as-a-star EVE data in the aftermath of a large flare,  $P$(CME $\vert$ Dim) = 0.97. These findings were also corroborated by a larger set of flares studied in disk-integrated AIA light curves, and provide strong evidence that dimmings are a robust proxy for CME occurrence.

\begin{figure}
\centering 
{\includegraphics[width=1\textwidth]{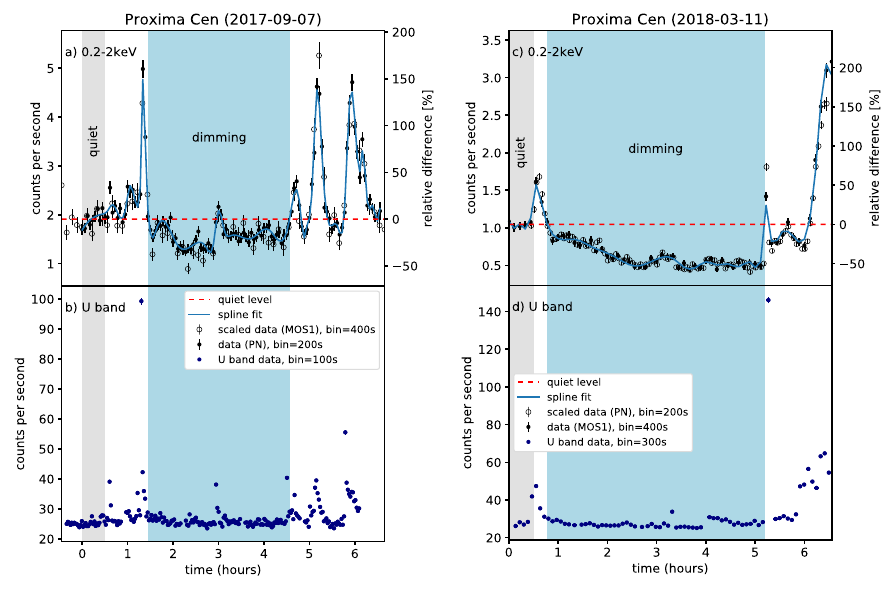}} 
\caption{Two examples of post-flare coronal dimmings detected on Proxima Centauri 
in XMM-Newton data. 
Top panels: Background-subtracted 0.2–2 keV  X-ray light curves together with weighted spline fits (blue). Red horizontal lines show the adopted quiet levels. Dimming and quiet intervals are highlighted by blue and grey shaded areas. Bottom panels:  Simultaneous photometric observations in the U-band shown to estimate the overall variability level. 
Adapted from \cite{Veronig:2021}.} \label{fig:Veronig2021_2}
\end{figure}

\subsection{Post-flare coronal dimmings as indicators of stellar CMEs}
\label{stellardimmings}

Based on the encouraging results obtained for the full-Sun SDO/EVE dimming observations and assuming that the coronae in late-type stars behave similarly to our Sun, \cite{Veronig:2021} applied the same method to search for post-flare coronal dimmings in stellar EUV and X-ray light curves to identify stellar CMEs. 
They focused on Sun-like and late-type main-sequence and pre-main-sequence stars, and compiled suitable data sets (in terms of observation lengths, cadence, availability of a pre-flare baseline) for 201 flaring stars from the XMM-Newton, Chandra and EUVE (Extreme Ultra-Violet Explorer) missions. In total, they report 21 dimmings on 13 different stars (1 from EUVE, 3 from Chandra and 17 from XMM-Newton). Interestingly, 
half of the dimmings identified are from three stars: the rapidly rotating K0V star AB Dor, the young M0Ve star AU Mic and the M5.5Ve star Proxima Centauri.
For one of the five detections for AB Dor, the duration of the dimming was longer than the rotation period. Together with the findings that the rotation axis of AB Dor %
is inclined by $60^{\circ}$ \citep{Kuerster:1994} and the large polar starspot revealed from Doppler imaging \citep{Donati:1999}, these observations suggest that the dimming stems from a CME ejected from a  polar starspot on AB Dor.  

Figure \ref{fig:Veronig2021_2} shows two  examples of stellar dimmings for Proxima Centauri in XMM-Newton data. The 0.2–2 keV X-ray light curves 
show post-flare coronal dimmings, during which the emission significantly decreases below the pre-flare level. The dimming in panel (a) reaches
a maximum decrease of 36\% with respect to the pre-flare level and lasts for approximately three hours
before it is interrupted by several smaller subsequent flares. Panel (c) shows a small flare followed by a very pronounced dimming with an emission decrease of 56\% and a duration of 4.5 hrs, before the dimming is interrupted by another flare.
The bottom panels show the corresponding U-band observations, which clearly reflect the flare enhancements but reveal stable flux levels in the other intervals, i.e.\ making it unlikely that the dimmings are related to rotational or other variabilities.

\begin{figure}[tbp]
\centering {\includegraphics[width=0.9\textwidth]{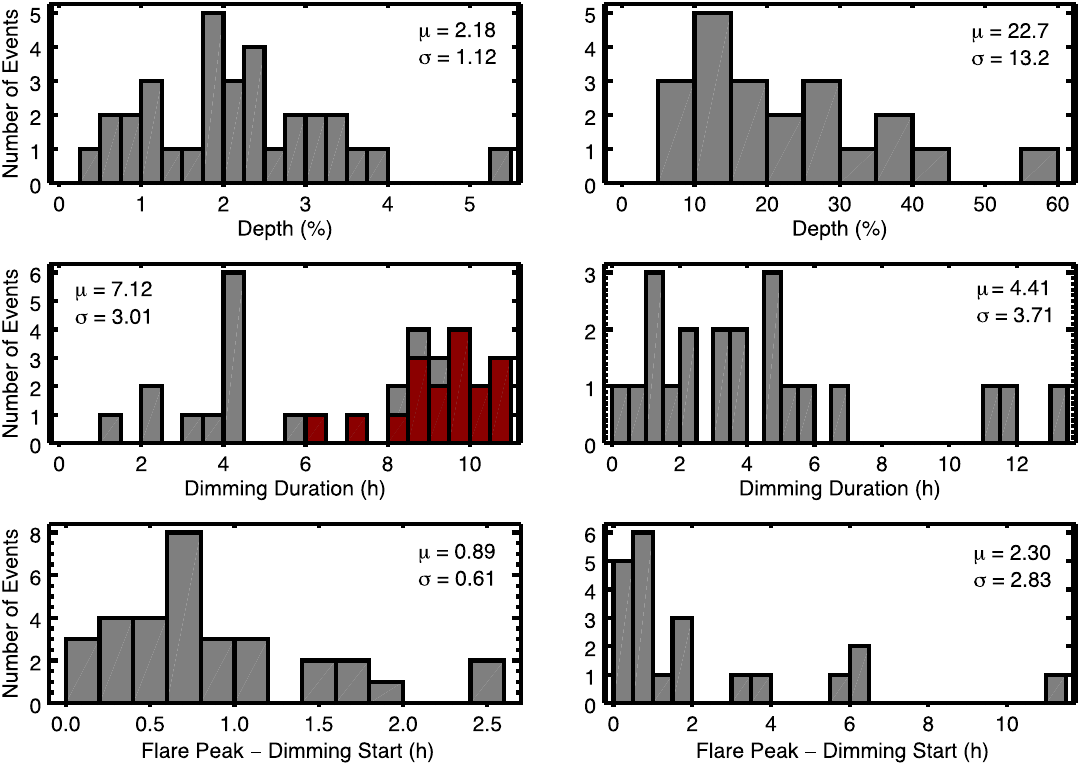}} 
\caption{Distributions of characteristic properties of solar (left) and stellar (right) dimmings. From top to bottom: dimming depth, duration, delay between flare peak and dimming start. The solar dimmings are derived from full-Sun SDO/EVE 150--250 {\AA} light curves, the stellar dimmings from XMM-Newton, Chandra and EUVE. Red-coloured dimming duration and recovery
times indicate solar events where the dimming end was not reached within the 12 h observation interval. In the stellar dimmings, most of the durations are lower estimates. Adapted from \cite{Veronig:2021}. 
}
\label{fig:Veronig2021_4}
\end{figure}

Figure \ref{fig:Veronig2021_4} shows the characteristic dimming properties derived for the stellar (right) and solar (left) dimmings from the \cite{Veronig:2021} study. Notably, the maximum dimming depths in stellar cases are about one order of magnitude larger (about 5--55\%) than the solar dimmings (about 0.5--5\%). The durations are similar in both cases, reaching from about 1 to $\gtrsim$10 hours. However, we note that in many cases the dimming duration may be underestimated due to either end of the observations or interruption by other flares. Important for the observations of coronal dimmings is also the time difference between the flare peak and the start of the dimming. In the solar case, this occurs mostly within 2~hrs after the flare peak. Many of the stellar dimmings also start within two hours but may show a longer delay up to 10 hrs after the flare peak. This time delay is strongly related to the flare duration, which in general shows some correlation with the flare strength. Therefore, it may be easier to identify stellar dimmings that are associated with short-duration, smaller flares than with larger ones. 

\begin{figure}[tbp]
\centering {\includegraphics[width=0.99\textwidth]{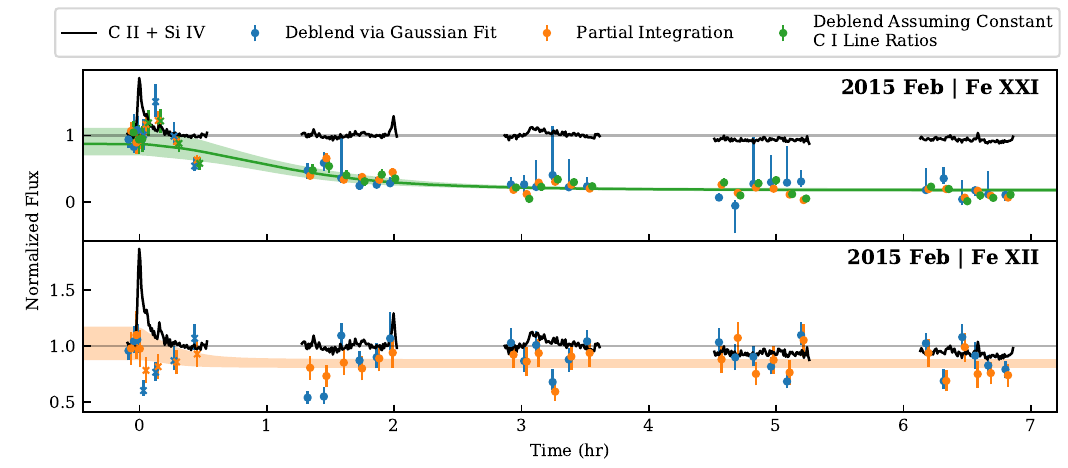}} 
\caption{Fe~{\sc{xxi}} and Fe~{\sc{xii}} lightcurves of $\varepsilon$ Eri  from HST observations (February 2015 epoch). Colored data points indicate different methods used to derive the Fe line fluxes.  Fitted dimming curves are shown by shaded regions; the solid green line the best fit for Fe~{\sc{xxi}}. The black light curves (shown for reference) are derived by summing the flux of two of the brightest FUV lines. From \cite{Loyd:2022}. 
}
\label{fig:Loyd2022_F6}
\end{figure}

In a recent study, \cite{Loyd:2022} applied the method of postflare coronal dimmings for CME detection to Fe~{\sc{xii}} to Fe~{\sc{xxi}}
emission from the young magnetically active K2 dwarf $\varepsilon$ Eri (coronal parameters in Table~\ref{tbl:star_param}), using observations in the far-ultraviolet from the Cosmic Origins Spectrograph (COS) of the Hubble Space Telescope (HST). They find a decrease of $81\pm 5$\% of the Fe~{\sc{xxi}} and $16\pm 8$\% of the Fe~{\sc{xii}} emission in the aftermath of one of the flares under study
(see Figure \ref{fig:Loyd2022_F6}). This dimming is attributed to a flare-associated CME, and a CME mass of $10^{15.2}$ g is estimated from the dimming signature. However, the authors also note that due to the short pre-flare baseline it cannot be excluded that the observed emission decline is due the decay of an earlier, unobserved flare. 

\cite{Loyd:2022} present a method which allows to estimate CME mass $m$ from the observed dimming curves, and to provide also upper mass limits in case of non-detections. 
When a CME is ejected, the fractional decline of the disk-integrated flux from the star (or the maximum dimming depth) is given as
\begin{equation}
\delta_{\rm max} = \frac{ F_{\rm CME}}{ F_{\rm pre}}
\label{eq:loyd1}
\end{equation}
with $F_{\rm CME}$ the flux from the plasma that is ejected with the CME and $F_{\rm pre}$ the flux from the star before the eruption occurred. 
Assuming that the observed coronal emission is optically thin, the flux is given as 
\begin{equation}
F = \int_V G(T,n_e)n_en_H dV
\label{eq:flux1}
\end{equation}
where $V$ is the volume of emitting material, $G(T,n_e)$ is a function describing the emissivity of the plasma depending on temperature $T$ and electron number density $n_e$ of the pre-eruption corona, and $n_H$ is the number density of hydrogen. Making the simplifying assumptions of constant temperature and density throughout the emitting volume~$V$, and setting $n_e \approx n_H = n$, Eq.~\ref{eq:flux1} simplifies to
\begin{equation}
F = G(T,n) n^2 V \, .
\label{eq:flux2}
\end{equation}
The term $n^2 V$ is related to the mass~$m$ of the emitting plasma by $m \approx \mu n V$ with $\mu$ the mean atomic weight. Using this expression in Eq.~\ref{eq:flux2}, Eq.~\ref{eq:loyd1} can be reformulated to  estimate the CME mass from the observations as 
\begin{equation}
m = \frac{ \mu \delta_{\rm max} F_{\rm pre} } {n G(T,n)} \, .
\label{eq:Loyd}
\end{equation}
Or, alternatively, the relation can be rewritten in terms of emission measure ${\rm EM} = n^2 V$, yielding 
\begin{equation}
m = \frac{ \mu \delta_{\rm max} {\rm EM_{pre}} } {n} \, .
\label{eq:Loyd2}
\end{equation}

As is further discussed in \cite{Loyd:2022}, Eq.~\ref{eq:Loyd} describes a linear relationship between the observed dimming flux and the stellar CME mass, $m \propto \delta_{\rm max}$, whereas for the Sun, \cite{Mason:2016} derived an expression $m \propto \delta_{\rm max}^{1/2}$. This difference comes from different assumptions: For the solar case, \cite{Mason:2016} assumed a fixed volume in which the dimming occurs, and variations in dimming depth are mostly related to how effectively plasma is removed from that volume by the CME. The relation derived by \cite{Loyd:2022} in Eqs.~\ref{eq:Loyd} and \ref{eq:Loyd2} intended to be more general to be applicable to stars. Here the assumption is that the volume comprised by the dimming can vary, but that the CME evacuates all
plasma from that volume.

A further recent attempt to search for stellar CMEs using post-flare coronal dimmings was done in \cite{Feinstein:2022}. They studied HST/COS FUV observations of flares observed on the young, active M class dwarf AU Mic. 
Notably, AU Mic is known to host two exoplanets, and the \cite{Veronig:2021} study identified three CME candidates through post-flare dimmings on AU Mic using XMM Newton data. For the one flare in the \cite{Feinstein:2022} study, for which a clear pre-flare baseline was available, no significant post-flare dimming was detected. \cite{Namekata:2024} performed a multi-wavelength campaign to study the young solar-type star EK Draconis. 
They report the first observations of prominence eruptions observed as blueshifted H$\alpha$ line emission that occurred in association with superflares. The faster and more massive of the two prominence eruptions also showed indications of a post-flare X-ray dimming with an amplitude of about $10\%$. These findings provide strong evidence for the occurrence of a CME associated with this superflare.

\subsection{Simulations of stellar CMEs and associated dimmings}

\begin{figure}
\centering {\includegraphics[width=0.6\textwidth]{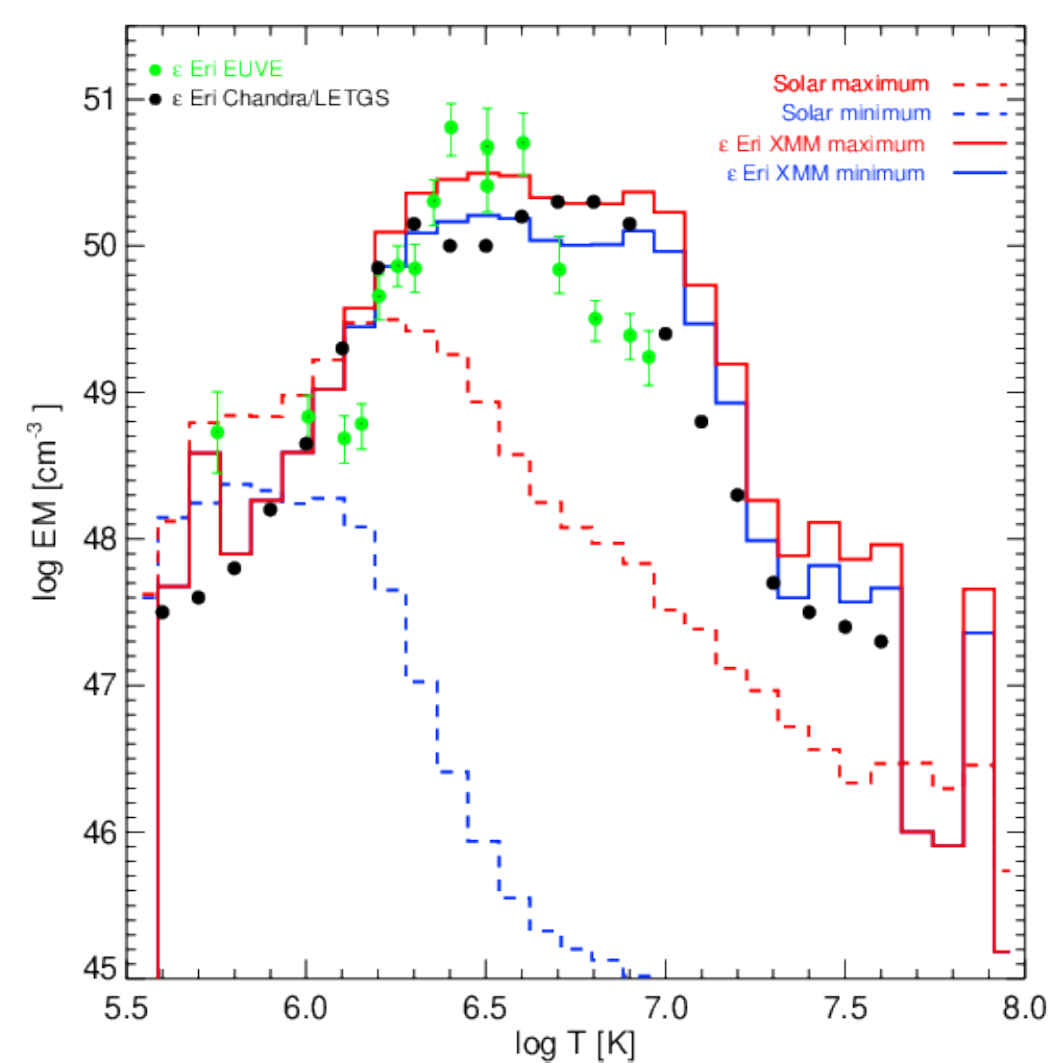}} 
\caption{Emission Measure distributions of the Sun during the minimum (Apr 1996; blue) and maximum (Dec 1991; red) of the solar cycle;
and of $\varepsilon$ Eri during the minimum (Feb 2015; blue) and maximum (July 2018; red) of its X-ray cycle based on the analysis of solar magnetic structures  \citep{Coffaro:2020}. Green and black dots show EM distributions of $\varepsilon$ Eri from EUVE spectra of 1993 \citep{Drake:2000}  and a Chandra/LETGS spectrum of March 2001 \citep{Sanz-Forcada:2004}.
Figure from \cite{Coffaro:2020}. 
}
\label{fig:Coffaro2020_15}
\end{figure}

Moving from solar to stellar coronal dimmings, the higher mean surface magnetic flux density on an M dwarf or young Sun-like stars could lead to different coronal dimming characteristics compared with the solar case. One major influence from the stronger magnetic field, and therefore hotter corona \citep[e.g.,][]{Ribas:2005, Gudel:2007} is that the primary coronal emission lines showing dimming signatures will shift to lines formed at higher temperatures. Because coronal dimmings are mainly due to the mass loss of coronal plasmas, the dimming is supposed to be most evident for spectral lines with formation temperatures near the peak of the plasma emission measure (EM) distribution. For example, \citet{Coffaro:2020} compares the EM distribution of the Sun and the young solar-like star $\varepsilon$ Eri,  
in which the solar corona has a peak temperature around $\sim$1~MK while $\varepsilon$ Eridani’s corona peaks at $>$3~MK (as derived from EUVE data) and a large portion of the plasma has temperature $>$10~MK (see Figure~\ref{fig:Coffaro2020_15}). This result is consistent with the recent detection of stellar coronal dimming signals in soft X-ray observations \citep{Veronig:2021} and the Fe~{\sc{xxi}} line \citep{Loyd:2022} discussed in Sect.~\ref{stellardimmings}. By enhancing the input magnetic flux density, \citet{Jin:2020} reproduces the EM distribution shift with global MHD simulations. 

Confined eruptions in association with large flares do occasionally occur on the Sun \citep[e.g.,][]{Thalmann:2015,Sun:2015}, in which case no CME and coronal dimming are visible although the flare radiative energy could be the same order accompanied by intense magnetic reconnection. About 10\% of the solar X-class and 50\% of the M-class flares are not accompanied by a CME \citep{Yashiro:2005}. In a statistical study of more than 300 solar flares of GOES class $\ge$M1, 
\cite{Li:2020} have shown that as a global parameter the total unsigned magnetic flux of active regions is a key factor whether large flares will be eruptive or confined. ARs with a large magnetic flux have a lower probability that a large flare it produces will be accompanied by a CME, due to the larger confining (strapping) field. 
\citet{Chen:2020} showed that confined flares have a high correlation with the EUV late-phase emissions in the Sun-as-a-star observations of SDO/EVE \citep{Woods:2011}. Therefore, these confined flares (or also failed eruptions) are potentially distinguishable in the stellar observations. The coronal responses under confined stellar eruptions have been modeled recently. For example, by modeling a confined eruption under stellar conditions, \citet{Jin:2020} shows that the synthetic EUV intensity profiles have an evident second peak after the flare that mimics the EUV late-phase emission on the Sun. \citet{AlvaradoGomez:2019} shows that in a fully suppressed stellar CME simulation, the soft X-ray emission has a gradual brightening several hours after the eruption with redshift, which indicates the falling material due to the confinement.\footnote{Note that magnetic reconnection in form of a flare is not included in this simulation; therefore the first flare peak is not shown.}

When a stellar CME has energy that is strong enough to escape the stellar magnetic field confinement, numerical models show that stellar coronal dimming could be generated due to a similar physical process (i.e.\ mass loss) but also with different characteristics \citep[e.g.,][]{AlvaradoGomez:2019, Jin:2020}. \citet{Jin:2020} modelled a stellar CME eruption under a stronger surface magnetic flux density (5~times the solar case). The initial CME flux rope has an energy about 10$^{33}$ ergs, which is an order of magnitude more energetic than the strongest solar eruptions. Due to the strong confinement, the resulting CME speed is $\sim$3000 km s$^{-1}$, which is only slightly higher than a fast solar CME. Also, the primary coronal dimming lines are shifted to higher temperatures (e.g., Fe~{\sc{xv}} 284~\AA, Fe~{\sc{xvi}} 335 \AA).
Furthermore, by applying the instrument performance estimates from the Extreme-ultraviolet Stellar Characterization for Atmospheric Physics and Evolution mission concept (ESCAPE; \citealt{France:2022}) that will provide extreme- and far-UV spectroscopy, the study demonstrates that the coronal dimming generated by the explosive stellar CME can be detected with optimized instrumentation. 

\begin{figure} [tbp]
\centering {\includegraphics[width=1\textwidth]{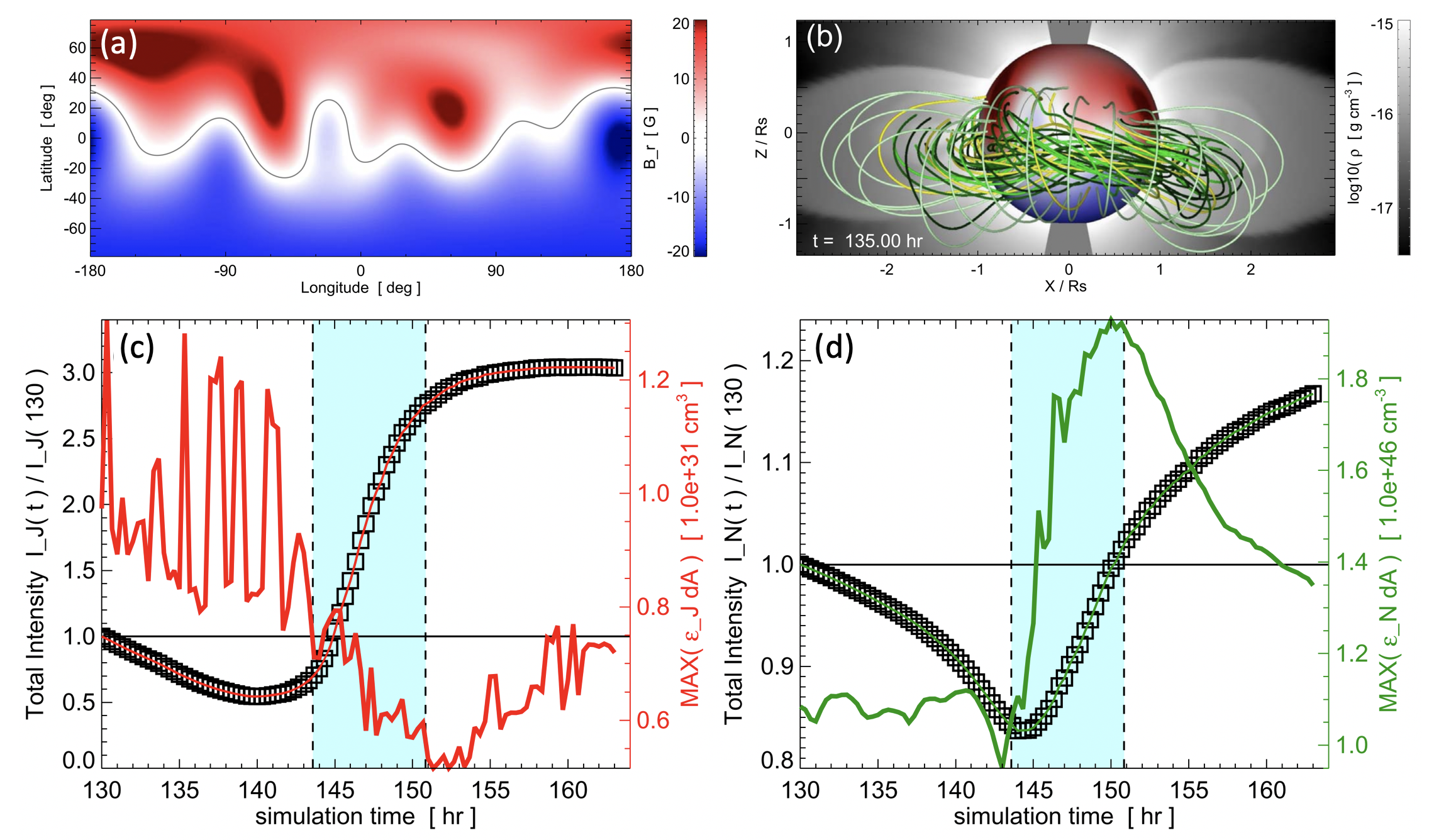}} 
\caption{Simulation of a Carrington-scale superflare/CME and associated coronal dimming. (a)~ZDI magnetic map of Kappa$^1$ Ceti; (b) Energized, pre-eruption magnetic field structure; (c) Area-integrated light curve of mean hot (X-ray) intensity and maximum emissivity; (d) Total light curve of ambient (EUV) intensity and maximum emissivity. Adapted from \cite{Lynch:2019}. A movie of the simulation shown in panel (c) can be found in the online supplement (from \cite{Lynch:2019}).}
\label{fig:Lynch_2019}
\end{figure}

Using the Zeeman Doppler Imaging (ZDI) synoptic magnetic map of Kappa$^1$ Ceti (Fig.~\ref{fig:Lynch_2019}a) calculated by \cite{rosen:2016}, \cite{Lynch:2019} simulated a Carrington-scale X58 superflare and CME by energizing and erupting a 2$\pi$ wide streamer blowout CME (Figure \ref{fig:Lynch_2019}). This model setup is supported by ``slingshot prominences'' in stellar H$\alpha$ observations \citep{jardine:2019,jardine:2020}
which may regularly span the entire stellar disk. To compare with solar flare observations, two synthetic emission proxies (average squared current density over the magnetic field line $<J^2>$ and density square $n^2$) are constructed to represent hot (X-ray) and ambient (EUV) emission. In Fig.~\ref{fig:Lynch_2019}c,d, one can see an evident pre-eruption dimming in both the X-ray and EUV intensities, which is caused by the pre-eruption closed-flux regions expanding and opening up to the stellar wind. This is followed by the rapid CME-related dimming due to the eruption. However, the eruption-related dimming is coincident with the formation of post-eruption flare arcade, which becomes dominant in the intensity light curves.

One relevant issue in terms of dimmings acting as a proxy for stellar CMEs that has been raised from simulations 
is whether confined flares, specifically those associated with failed eruptions, may also produce dimmings. \cite{AlvaradoGomez:2019} showed a simulation where a magnetically confined stellar eruption also produced a dimming in narrowband EUV  emissions. This effect may be due to thermodynamic changes (\emph{thermal dimming}), or due to the rise of the magnetic flux rope and the density decrease by the associated expansion of the coronal plasma. However, one would expect that such cases reveal dimmings that are less strong than those associated with (successful) CMEs, where plasma is expanding over a much larger volume and fully ejected from the corona. To date, even from the solar perspective, a systematic study whether failed eruptions from the Sun produce coronal dimmings 
is still missing but would give important insight on this question.

\section{Conclusions and outlook} 
      \label{sec:conclusions} 
     The aim of this review is to give a comprehensive overview on the properties of coronal dimmings, how they relate to the CME and associated flare, how they contribute toward a better understanding of the physics of solar eruptions and how they inform us on CMEs on other stars. However, we stress that the rich potential of coronal dimmings is only starting to be explored in its full breadth. Here we highlight several topics, where we think that substantial analysis is still missing and new ways forward can be found studying coronal dimmings.

Among the topics that are still underexplored is the use of coronal dimmings for space weather research and forecasting. For Earth-directed CMEs, coronal dimmings give us the first insight into an eruption using observatories in the Sun-Earth line, several tens of minutes before the CME reaches the coronagraphic field-of-view. 
For the majority of events, this allows us to easily distinguish halo CMEs that are propagating towards Earth from those that are backsided. 
Recent studies suggest that the dimming expansion may also contain information on non-radial CME motion \cite[][]{Moestl:2015,Chikunova2023,Jain:2024}. Up to now the relations established between dimming and CME properties \citep[e.g.,][]{Biesecker:2002,Dissauer:2019} have not yet been further developed to obtain additional or proxy information on important CME parameters like speed, mass or propagation direction. 
We also note that these statistical relations were only established during limited time periods of the solar cycle  
and control periods (where no CMEs are occurring) were rarely included \citep[e.g.][]{Veronig:2021}. However, coronal dimmings have proven their unique link to CMEs in basic research as presented in detail in this review and transitioning to applied research and operations (R2O) to further assess its forecasting capabilities is an essential next step. Effort towards this has already started. For instance, Solar Demon \citep[][]{Kraaikamp:2015} provides a real-time coronal dimming detection suite. What is needed in the future is proof-of-concept including large, systematic statistical studies of coronal dimmings and their associated CMEs and flares  
during at least a full solar cycle, including complete contingency tables and updated relations.

We also highlight the relevance of the new categorization scheme of coronal dimmings that we have established here, based on the magnetic flux systems involved in the eruption (Section\ \ref{Terminology}). It provides a framework that connects the observations and the theory, which allows us to obtain a deeper insight into the magnetic connectivities and reconfigurations during an eruption. The process of the ``opening" of the coronal field as traced by dimming regions and the observed plasma flows, may also contain information on how the solar wind is formed.
We also note that up to date only a few studies deal with the recovery phase of coronal dimmings. However, studying the full evolution over the dimming lifetime including the post-event coronal recovery 
provides information on how the large-scale field reforms after a CME. 

As described in Section~\ref{Terminology}, and summarized in Figure~\ref{f:ed_np}, the dimming regions reflect the presence of distinct plasma domains in the corona.
The interactions of these domains constitute an essential aspect of the physics of flares and CMEs, and the dimming regions map out these domains while the coronal reconfiguration develops. 
This involves multiple instances of magnetic reconnection, whereby the plasmas from different domains can mix.
Future observations should be able to identify these plasma domains via 
their elemental abundances \citep[\textit{e.g.}][]{2018ApJ...856...71B}, Doppler signatures, or tools involving X-ray or even $\gamma$-ray imaging spectroscopy \citep[\textit{e.g.}][]{Jin:2022}.

A few solar missions, currently in development, could advance coronal dimming studies.  The Sun Coronal Ejection Tracker \cite[SunCET;][]{Mason:2021} is a NASA CubeSat mission  with an EUV imager for studying CMEs in the middle corona \cite[an underexplored region in the EUV;][]{West:2023}, while imaging the bright solar disk concurrently with short integration times. The Multi-slit Solar Explorer \citep[MUSE;][]{DePontieu:2020,depontieu:2022,Cheung:2022} is a Medium-Class Explorer mission aiming to study the physical processes driving the heating of the solar corona and solar eruptions. MUSE will obtain EUV spectra and images with highest resolution in space (0.33--0.4$''$) and time (1--4 s) ever achieved for solar transition region and corona, along 37 slits with large context FOV (580$''$$\times$290$''$) from context imager. The ESA Vigil mission, scheduled for launch in 2031 to the Sun-Earth Lagrangian L5 point, will operate a full disk EUV imager in addition to a coronagraph and heliospheric imager and thus provide a much-need off Sun-Earth line viewpoint for the study of off-limb dimmings of Earth-directed eruptions.

Recently, first studies on CME-induced coronal dimmings in EUV and X-ray observations on late-type and solar-like stars have been successful \citep{Veronig:2021,Loyd:2022}. This approach opens a new window on the detection and characterization of stellar CMEs, to derive their properties and thus to better constrain their effects on the space weather and habitability of exoplanets orbiting the host star. In this respect, new missions that are being planned to study stars and exoplanets in the EUV, with much higher sensitivity and broad spectral coverage, are of enormous interest.
Extreme-ultraviolet Stellar Characterization for Atmospheric Physics and Evolution \cite[ESCAPE;][]{France:2022} is
a Small Explorer astrophysics mission concept employing spectroscopy in the EUV  (80--825~\AA) and FUV (1280--1650~\AA) with the main aim of exploring the high-energy radiation environment in the habitable zones around nearby stars. One of its aims is to use the same Sun-as-a-star approach as SDO/EVE to use dimmings to detect stellar CMEs and derive the CME occurrence frequencies of nearby stars.
The X-ray probe concept Arcus\footnote{\href{http://www.arcusxray.org/}{http://www.arcusxray.org/}} would provide 10--60~\AA\ coverage at the same time as 970--1350 \AA. While not strictly speaking EUV, the wavelengths spanned would cover stellar chromospheric, transition region, and coronal emission lines, and notably offer the chance to co-observe coronal resonance lines with longer wavelength, coronal forbidden lines, as well as detail the full span of stellar outer atmospheric temperatures. 
This would enable confirmation of some of the trends  seen with recent X-ray and FUV 
observations reported in this review in a comprehensive way. 

The new solar and stellar missions together with a better understanding of the coronal dimming and the connection to the physical processes driving it, are expected to substantially advance the fields of solar and stellar CMEs and their effects on (exo-)planets in the coming decade.

\bmhead{Acknowledgments}
We acknowledge the careful reading of the manuscript by three anonymous reviewers and their constructive comments.
This research was supported by the International Space Science Institute (ISSI) in Bern, through ISSI International Team project \#516 (``Coronal Dimmings and Their Relevance to the Physics of Solar and Stellar Coronal Mass Ejections").
A.M.V. acknowledges support by  the Austrian Space Applications Programme of the Austrian Research Promotion Agency FFG (ASAP-11 4900217) and the Austrian Science Fund FWF 10.55776/I4555. K.D. is supported by NASA/HGI 80NSSC21K0738 and NSF/AGS-ST 2154653.
 B.K. is supported by the DFG and by NASA through grant  
 80NSSC20K1274. 
 A.P. acknowledges the support from the Research Council of Norway through its Centres of Excellence scheme, project number 262622, and Synergy Grant number 810218 459 (ERC-2018-SyG) of the European Research Council.
 J.Q. is supported by NASA/HSR 80NSSC20K1317 and NASA/HGI 80NSSC22K0519. 
 H.T. is supported by NSFC Grant no.\ 12250006. 
 A.V. is supported by NASA Grants 80NSSC19K1261 and 80NSSC19K0069.
 T.P. is supported by the Russian Science Foundation grant 23-22-00242.
 H.H. thanks the School of Physics and Astronomy, University of Glasgow, for support.
M.J. is supported by NASA’s SDO/AIA contract (NNG04EA00C) to LMSAL, HST-AR-15803 Program through Space Telescope Science Institute and grant no.\ 80NSSC24K0136.



%
%


\newpage
\phantomsection
\addcontentsline{toc}{section}{References}
\bibliography{biblio}

\end{document}